\documentclass[twoside,10pt]{article}

\usepackage{amsmath}
\usepackage{amssymb}
\usepackage{subfigure}
\usepackage{graphicx}

\renewcommand{\theequation}{\thesection.\arabic{equation}}

\newtheorem{theorem}{Theorem}[section]
\newtheorem{lemma}[theorem]{Lemma}

\newtheorem{corollary}[theorem]  {Corollary}
\newtheorem{remark}[theorem]  {Remark}

\renewcommand{\section}{\secdef\sct\sect}
\newcommand{\sct}[2][default]{\refstepcounter{section}
\vspace{0.5cm} \setcounter{equation}{0}
\centerline{ 
\scshape \arabic{section}.\ #1} \vspace{0.3cm}}
\newcommand{\sect}[1]{
\vspace{0.5cm} \centerline{\large\scshape #1} \vspace{0.3cm}}

\renewcommand{\subsection}{\secdef \subsct\sbsect}
\newcommand{\subsct}[2][default]{\refstepcounter{subsection}
\nopagebreak \vspace{0.5\baselineskip} {\flushleft\bf
\arabic{section}.\arabic{subsection}~\bf #1  } \nopagebreak}
\newcommand{\sbsect}[1]{\vspace{0.1cm}\noindent
{\bf #1}\vspace{0.1cm}}

\renewcommand{\subsubsection}{%
\secdef \subsubsect\sbsbsect}
\newcommand{\subsubsect}[2][default]{%
\refstepcounter{subsubsection} \nopagebreak
\vspace{0.1\baselineskip} \nopagebreak {\flushleft
\sffamily\slshape
\arabic{section}.\arabic{subsection}.\arabic{subsubsection}
\ %
\sffamily #1\/.}\ }
\newcommand{\sbsbsect}[1]{\vspace{0.1cm}\noindent
{\bf #1}\ }

\input{tcilatex}

\begin{document}

\title{Effect of a Locally Repulsive Interaction on s--wave Superconductors}
\author{J.-B. Bru and W. de Siqueira Pedra}

\maketitle

\begin{abstract}
The thermodynamic impact of the Coulomb repulsion on s--wave superconductors
is analyzed via a rigorous study of equilibrium and ground states of the
strong coupling BCS--Hubbard Hamiltonian. We show that the one--site
electron repulsion can favor superconductivity at fixed chemical potential
by increasing the critical temperature and/or the Cooper pair condensate
density. If the one--site repulsion is not too large, a first or a second
order superconducting phase transition can appear at low temperatures. The
Mei{\ss }ner effect is shown to be rather generic but coexistence of
superconducting and ferromagnetic phases is also shown to be feasible, for
instance near half--filling and at strong repulsion. Our proof of a
superconductor--Mott insulator phase transition implies a rigorous
explanation of the necessity of doping insulators to create superconductors.
These mathematical results are consequences of \textquotedblleft quantum
large deviation\textquotedblright\ arguments combined with an adaptation of
the proof of St{\o }rmer's theorem \cite{Stormer} to even states on the CAR
algebra.\\[1.3ex]
{\small \textit{Keywords:} Superconductivity -- s--wave -- Coulomb interaction -- Hubbard model -- Mei{\ss}ner effect -- Mott insulators -- Equilibrium states -- St{\o }rmer's theorem}%
\end{abstract}

\section{Introduction}

\noindent Since the discovery of mercury superconductivity in 1911 by the
Dutch physicist Onnes, the study of superconductors has continued to
intensify, see, e.g., \cite{Superconductivity1}. Since that discovery, a
significant amount of superconducting materials has been found. This
includes usual metals, like lead, aluminum, zinc or platinum, magnetic
materials, heavy--fermion systems, organic compounds and ceramics. A
complete description of their thermodynamic properties is an entire subject
by itself, see \cite%
{Superconductivity1,Superconductivity2,Superconductivity3} and references
therein. In addition to  zero--resistivity and many other complex phenomena,
superconductors manifest the celebrated Mei{\ss }ner or Mei{\ss }%
ner--Ochsenfeld effect, i.e., they can become perfectly diamagnetic. The
highest\footnote{%
In January 2008, a critical temperature over 180 Kelvin was reported in a
Pb-doped copper oxide.} critical temperature for superconductivity obtained
nowadays is between 100 and 200 Kelvin via doped copper oxides, which are
originally insulators. In contrast to most superconductors, note that
superconduction in
magnetic superconductors only exists
 on a finite range of non--zero temperatures.

Theoretical foundations of superconductivity go back to the celebrated BCS
theory -- appeared in the late fifties (1957) --
 which explains conventional type I superconductors.
This theory is based on the so--called (reduced) BCS\ Hamiltonian
\begin{equation}
\label{BCS Hamilt}
\mathrm{H}_{\Lambda }^{BCS}:=\sum\limits_{k\in \Lambda ^{\ast }}\left(
\varepsilon_{k}-\mu \right) \left(\tilde{ a}_{k,\uparrow }^{\ast }
\tilde{a}_{k,\uparrow}+\tilde{a}_{k,\downarrow }^{\ast }
\tilde{a}_{k,\downarrow }\right) +\frac{1}{|\Lambda |}%
\sum_{k,k^{\prime }\in \Lambda ^{\ast }}\gamma _{k,k^{\prime }}\tilde{a}_{k,\uparrow
}^{\ast }\tilde{a}_{-k,\downarrow }^{\ast }\tilde{a}_{k^{\prime },\downarrow }
\tilde{a}_{-k^{\prime
},\uparrow }
\end{equation}%
defined in a cubic box $\Lambda \subset \mathbb{R}^{3}$ of volume $|\Lambda |
$. Here $\Lambda ^{\ast }$ is the dual group of $\Lambda $ seen as a torus
(periodic boundary condition) and the operator $\tilde{a}_{k,s}^{\ast }$
resp. $\tilde{a}_{k,s}$ creates resp.
annihilates a fermion with spin $s\in \{\uparrow,\downarrow \}$
and momentum $k\in \Lambda ^{\ast }$. The function $\varepsilon _{k}$
represents the kinetic energy, the real number $\mu $ is
the chemical potential and $\gamma _{k,k^{\prime }}$ is the BCS coupling
function. The choice $\gamma _{k,k^{\prime }}=-\gamma <0$ is often used in the Physics literature
and the case $\varepsilon _{k}=0$ is known as the strong coupling limit of the BCS model.

The lattice approximation of the BCS\ Hamiltonian amounts to replace the box
$\Lambda \subset \mathbb{R}^{3}$ by $\Lambda \subset \mathbb{Z}^{3}$ (or
more generally by $\Lambda \subset \mathbb{Z}^{d\geq 1}$) and the strong
coupling limit of the reduced BCS model is in this case known as the strong
coupling (with $\gamma_{k,k'} =-\gamma$) BCS model\footnote{%
See also (\ref{Hamiltonian BCS-Hubbard}) with $\lambda =0$ and $h=0.$}.
The assumptions $\epsilon_k=0$ and $\gamma_{k,k'} = -\gamma$
are of interest, because in this case the BCS Hamiltonian
 can be  explicitly diagonalised.
The
exact solution of the strong coupling BCS model is well--known since the
sixties \cite{ThirWeh67,Thir68}. This model is in a sense unrealistic: among
other things, its representation of the kinetic energy of electrons is
rather poor. Nevertheless it became popular because it displays most of
basic properties of real conventional type I superconductors. See, e.g.,
Chapter VII, Section 4 in \cite{Thou}. Even though the analysis of the
thermodynamics of the BCS Hamiltonian was rigorously performed in the
eighties \cite{BCSrigorous1,BCSrigorous2} (see also the innovating work of
Bernadskii and Minlos in 1972 \cite{Bernadskii-Minlos}), generalizations of
the strong coupling approximation of the BCS model are still subject of
research. For instance, strong coupling--BCS--type models with
superconducting phases at arbitrarily high temperatures are treated in \cite%
{IlieThir07}.

In fact, a general theory of superconductivity is still a subject of debate,
especially for high--$T_{c}$\emph{\ }superconductors. An important
phenomenon ignored in the BCS theory is the Coulomb interaction between
electrons or holes, which can imply strong correlations, for instance in
high--$T_{c}$\emph{\ }superconductors. To study these correlations, most of
theoretical methods, inspired by {Beliaev} \cite{BeliaevI}, use perturbation
theory or renormalization group derived from the diagram approach\ of
Quantum Field Theory. However, even if these approaches have been successful
in explaining many physical properties of superconductors \cite%
{Superconductivity2,Superconductivity3}, only few rigorous results exist on
superconductivity.

For instance, the effect of the Coulomb interaction on superconductivity is
not rigorously known. This problem was of course adressed in theoretical
Physics right after the emergence of the Fr\"{o}hlich model and the BCS
theory, see, e.g., \cite{Bogoliubov-tolman shirkov}. In particular, the
authors explain in \cite[Chapter VI]{Bogoliubov-tolman shirkov}, by means of
diagrammatic pertubation theory, that the effect of the Coulomb interaction
on the Fr\"{o}hlich model should be to lower the critical temperature of the
superconducting phase by lowering the electron density. We rigorously show
that this phenomenology is only true\ -- for our model\ -- in a specific
region of parameters.

Indeed, the aim of the present paper is to understand the possible
thermodynamic impact of the Coulomb repulsion in the strong coupling
approximation. More precisely, we study the thermodynamic properties of the
strong coupling BCS--Hubbard model defined in the box\footnote{%
Without loss of generality we choose $N$ such that $L:=(N^{1/d}-1)/2\in
\mathbb{N}$.} $\Lambda _{N}:=\{\mathbb{Z}\cap \lbrack -L,L]\}^{d\geq 1}$ of
volume $|\Lambda _{N}|=N\geq 2$ by the Hamiltonian%
\begin{eqnarray}
\mathrm{H}_{N}:= &&-\mu \sum\limits_{x\in \Lambda _{N}}\left( n_{x,\uparrow
}+n_{x,\downarrow }\right) -h\sum_{x\in \Lambda _{N}}\left( n_{x,\uparrow
}-n_{x,\downarrow }\right) +2\lambda \sum_{x\in \Lambda _{N}}n_{x,\uparrow
}n_{x,\downarrow }  \notag \\
&&-\frac{\gamma }{N}\sum_{x,y\in \Lambda _{N}}a_{x,\uparrow }^{\ast
}a_{x,\downarrow }^{\ast }a_{y,\downarrow }a_{y,\uparrow }
\label{Hamiltonian BCS-Hubbard}
\end{eqnarray}%
for real parameters $\mu $, $h$, $\lambda $, and $\gamma \geq 0$. The
operator $a_{x,s}^{\ast }$ resp. $a_{x,s}$ creates resp. annihilates a
fermion with spin $s\in \{\uparrow ,\downarrow \}$ at lattice position $x\in
\mathbb{Z}^{d}$ whereas $n_{x,s}:=a_{x,s}^{\ast }a_{x,s}$ is the particle
number operator at position $x$ and spin $s$. The first term of the right
hand side (r.h.s.) of (\ref{Hamiltonian BCS-Hubbard}) represents the strong
coupling limit of the kinetic energy, with $\mu $ being the chemical
potential of the system. Note that this \textquotedblleft strong coupling
limit\textquotedblright\ -- explained above for the BCS\ Hamiltonian
-- is also called \textquotedblleft atomic limit\textquotedblright\ in the
context of the Hubbard model, see, e.g., \cite{atomiclimit1,atomiclimit2}.
The second term in the r.h.s. of (\ref{Hamiltonian BCS-Hubbard}) corresponds
to the interaction between spins and the magnetic field $h$. The one--site
interaction with coupling constant $\lambda $ represents the
(screened) Coulomb repulsion as in the celebrated Hubbard model. So, the
parameter $\lambda $ should be taken as a positive number but our results
are also valid for any real $\lambda $. The last term is the BCS interaction
written in the $x$--space since%
\begin{equation}
\frac{\gamma }{N}\sum_{x,y\in \Lambda _{N}}a_{x,\uparrow }^{\ast
}a_{x,\downarrow }^{\ast }a_{y,\downarrow }a_{y,\uparrow }=\frac{\gamma }{N}%
\sum_{k,q\in \Lambda _{N}^{\ast }}\tilde{a}_{k,\uparrow }^{\ast }\tilde{a}%
_{-k,\downarrow }^{\ast }\tilde{a}_{q,\downarrow }\tilde{a}_{-q,\uparrow },
\label{BCS interaction}
\end{equation}%
with $\Lambda _{N}^{\ast }$ being the reciprocal lattice of quasi--momenta
and where $\tilde{a}_{q,s}$ is the corresponding annihilation operator for $%
s\in \{\uparrow ,\downarrow \}$. Observe that the thermodynamics of the
model for $\gamma =0$ can easily be computed. Therefore we restrict the
analysis to the case $\gamma >0$. Note also that the homogeneous BCS
interaction (\ref{BCS interaction}) can imply a superconducting phase and
the mediator implying this effective interaction does not matter here, i.e.,
it could be due to phonons, as in conventional type I superconductors, or
anything else.

We show that the one--site repulsion suppresses superconductivity for large $%
\lambda \geq 0$. In particular, the repulsive term in (\ref{Hamiltonian
BCS-Hubbard}) cannot imply any superconducting state if $\gamma =0.$
However, the first elementary but nonetheless important property of this
model is that the presence of an electron repulsion is not incompatible with
superconductivity if $|\lambda -\mu |$ and $(\lambda +|h|)$ are not too big
as compared to the coupling constant $\gamma $ of the BCS interaction. In
this case, the superconducting phase appears at low temperatures as either a
first order or a second order phase transition. More surprisingly, the
one--site repulsion can even favor superconductivity at fixed chemical
potential $\mu $ by increasing the critical temperature and/or the Cooper
pair condensate density. This contradicts the naive guess that any one--site
repulsion between electron pairs should at least reduce the formation of
Cooper pairs. It is however important to mention that the physical behavior
 described
by the model depends on which parameter, $\mu $ or $\rho ,$ is fixed. (It
does not mean that the canonical and grand--canonical ensembles are not
equivalent for this model). Indeed, we also analyze the thermodynamic
properties at fixed electron density $\rho $ per site in the
grand--canonical ensemble, as it is done for the perfect Bose gas in the
proof of Bose-Einstein condensation. The analysis of the thermodynamics of
the strong coupling BCS--Hubbard model is performed in details. In
particular, we prove that the Mei{\ss }ner effect is rather generic but also
that the coexistence of superconducting and ferromagnetic phases is possible
(as in the Vonsovkii--Zener model \cite%
{brankov-tonchev1,Bogoliubov-Ermilov-Kurbatov}), for instance at large $%
\lambda >0$ and densities near half--filling. The later situation is related
to a superconductor--Mott insulator phase transition. This transition gives
furthermore a rigorous explanation of the need of doping insulators to
obtain superconductors.\ Indeed, at large enough coupling constant $\lambda $%
, the superconductor--Mott insulator phase transition corresponds to the
breakdown of superconductivity together with the appearance of a gap in the
chemical potential as soon as the electron density per site becomes an
integer, i.e., $0$, $1$ or $2$. If the system has an electron density per
site equal to $1$ without being superconductor, then any non--zero magnetic
field $h\neq 0$ implies a ferromagnetic phase.

Note that the present setting is still too simplified with respect to
(w.r.t.) real superconductors. For instance, the anti--ferromagnetic phase
or the presence of vortices, which can appear in (type II) high--$T_{c}$%
\emph{\ }superconductors \cite{Superconductivity2,Superconductivity3}, are
not modeled. However, the BCS--Hubbard Hamiltonian (\ref{Hamiltonian
BCS-Hubbard}) may be a good model for certain kinds of superconductors or
ultra-cold Fermi gases in optical lattices, where the strong coupling
approximation is experimentally justified. Actually, even if the strong
coupling assumption is a severe simplification, it may be used in order to
analyze the thermodynamic impact of the Coulomb repulsion, as all parameters
of the model have a phenomenological interpretation and can be
directly related to
experiments. See discussions in Section \ref{Section 5}. Moreover, the range
of parameters in which we are interested turns out to be related to a first
order phase transition. This kind of phase transitions are known to be
stable under small perturbations of the Hamiltonian. In particular, by
including a small kinetic part it can be shown by high--low temperature
expansions that the model
\begin{equation*}
\mathrm{H}_{N,\varepsilon }:=\mathrm{H}_{N}+\sum\limits_{x,y\in \Lambda
_{N}}\varepsilon (x-y)\left( a_{y,\downarrow }^{\ast }a_{x,\downarrow
}+a_{y,\uparrow }^{\ast }a_{x,\uparrow }\right)
\end{equation*}%
has essentially the same correlation functions as $\mathrm{H}_{N}$, up to
corrections of order $||\varepsilon ||_{1}$ ($\ell^{1}$--norm of $%
\varepsilon $). This analysis will be the subject of a separated paper. For
any $\varepsilon \neq 0$ notice that the model $\mathrm{H}_{N,\varepsilon }$
is not anymore permutation invariant but only translation invariant. Such
translation invariant models are studied in a systematic way in \cite%
{BruPedra2}. Their detailed analysis is however, generally much more
difficult to perform. Considering first models having more symmetries -- as for
instance, permutation invariance -- is in this case technically easier.

Coming back to the strong coupling BCS--Hubbard model $\mathrm{H}_{N}$, it
turns out that the thermodynamic limit of its (grand--canonical) pressure%
\footnote{%
Our notation for the \textquotedblleft $\mathrm{Trace"}$ does not include
the Hilbert space where it is evaluated but it should be deduced from
operators involved in each statement.}%
\begin{equation}
\mathrm{p}_{N}\left( \beta ,\mu ,\lambda ,\gamma ,h\right) :=\frac{1}{\beta N%
}\ln \mathrm{Trace}\left( e^{-\beta \mathrm{H}_{N}}\right)
\label{BCS pressure}
\end{equation}%
exists at any fixed inverse temperature $\beta >0$. It corresponds to a
variational problem which has minimizers\footnote{%
Because $\omega \mapsto \mathfrak{F}(\omega )$ is lower semicontinuous and $%
E_{\mathcal{U}}^{S,+}$ is compact with respect to the weak$^\ast$--topology.%
} in the set $E_{\mathcal{U}}^{S,+}$ of (even\footnote{%
See Remark \ref{remark-even} in Section \ref{Section main proof}.})
permutation invariant states on the CAR $C^{\ast }$--algebra $\mathcal{U}$
generated by annihilation and creation operators:
\begin{equation}
\mathrm{p}\left( \beta ,\mu ,\lambda ,\gamma ,h\right)
:=\lim\limits_{N\rightarrow \infty }\left\{ \mathrm{p}_{N}\left( \beta ,\mu
,\lambda ,\gamma ,h\right) \right\} =-\inf\limits_{\omega \in E_{\mathcal{U}%
}^{S,+}}\mathfrak{F}\left( \omega \right) .  \label{p.var}
\end{equation}%
Here the map
\begin{equation*}
\omega \mapsto \mathfrak{F}(\omega ):=\mathfrak{e}(\omega )-\beta ^{-1}%
\tilde{S}(\omega )
\end{equation*}%
is the affine (lower weak$^\ast$--semicontinuous) free--energy density
functional defined on $E_{\mathcal{U}}^{S,+}$ from the mean energy per
volume
\begin{equation*}
\mathfrak{e}\left( \omega \right) :=\lim\limits_{N\rightarrow \infty
}\left\{ N^{-1}\omega \left( \mathrm{H}_{N}\right) \right\} <\infty
\end{equation*}%
and the entropy density%
\begin{equation*}
\tilde{S}\left( \omega \right) :=-\lim\limits_{N\rightarrow \infty }\left\{
\frac{1}{N}\mathrm{Trace}\left( D_{\omega |_{{\mathcal{U}}_{N}}}\log
D_{\omega |_{{\mathcal{U}}_{N}}}\right) \right\} <\infty .
\end{equation*}%
Note that $D_{\omega |_{{\mathcal{U}}_{N}}}$ is the density matrix
associated to the state $\omega $ restricted on the local CAR $C^{\ast }$%
--algebra ${\mathcal{U}}_{N}\simeq B\left( \bigwedge \mathbb{C}^{\Lambda
_{N}\times \{\uparrow ,\downarrow \}}\right) $ (isomorphism). Such a
derivation of the pressure as a minimization problem over states on a $%
C^{\ast }$--algebras are also performed for various quantum spin systems,
see, e.g., \cite%
{Petz-Raggio-Verbeure,RaggioWerner1,RaggioWerner2,Petz2008,monsieurremark}.

The minimum of the variational problem (\ref{p.var}) is attained for any
weak$^\ast$--limit point of local Gibbs states%
\begin{equation}
\omega _{N}\left( \cdot \right) :=\frac{\mathrm{Trace}\left( \mathrm{\ }%
\cdot \mathrm{\ }e^{-\beta \mathrm{H}_{N}}\right) }{\mathrm{Trace}\left(
e^{-\beta \mathrm{H}_{N}}\right) }  \label{BCS gibbs state Hn}
\end{equation}%
associated with $\mathrm{H}_{N}$. Similarly to what is done for general
translation invariant models (see \cite{Sewell,BrattelliRobinson}), the set
of \textit{equilibrium states} of the strong coupling BCS--Hubbard model is
naturally defined to be the set $\mathit{\Omega }_{\beta }=\mathit{\Omega }%
_{\beta }(\mu ,\lambda ,\gamma ,h)$ of minimizers of (\ref{p.var}). Note
that $\mathit{\Omega }_{\beta }$ is a non empty convex subset\footnote{%
The map $\omega \mapsto \mathfrak{F}(\omega )$ on the convex set $E_{%
\mathcal{U}}^{S,+}$ is affine and lower semicontinuous, thus $\mathit{\Omega
}_{\beta }$ is a non empty face of $E_{\mathcal{U}}^{S,+}$.} of $E_{\mathcal{%
U}}^{S,+}$ and the extremal decomposition in $\mathit{\Omega }_{\beta }$
coincides with the one in $E_{\mathcal{U}}^{S,+}$, i.e., $\mathit{\Omega }%
_{\beta }$ is a face\footnote{%
A face $\mathrm{F}$ of a compact convex set $\mathrm{K}$ is subset of $%
\mathrm{K}$ with the property that if $\omega =\Sigma _{n=1}^{m}\lambda
_{n}\omega _{n}\in \mathrm{F}$ with $\Sigma _{n=1}^{m}\lambda _{n}=1$ and $%
\{\omega _{n}\}_{n=1}^{m}\subset \mathrm{K}$, then $\{\omega
_{n}\}_{n=1}^{m}\subset \mathrm{F}$.} in $E_{\mathcal{U}}^{S,+}$. So,
\textit{pure} equilibrium states are extremal states of $\mathit{\Omega }%
_{\beta }.$ Meanwhile, any weak$^\ast $ limit point as $n\rightarrow \infty
$ of an equilibrium state sequence $\{\omega ^{(n)}\}_{n\in \mathbb{N}}$
with diverging inverse temperature $\beta _{n}\rightarrow \infty $ is
-- per definition -- a \textit{ground state }$\omega \in E_{\mathcal{U}}^{S,+}$.

Here we have left the Fock space representation of the model to go to a
representation--free formulation of thermodynamic phases. This means that $%
\mathrm{H}_{N}$ is not anymore seen as a Hamiltonian acting on the Fock
space but as a (self--adjoint) element of the CAR $C^{\ast }$--algebra $%
\mathcal{U}$ with thermodynamic phases describes by states on $\mathcal{U}$.
Doing so we take advantage of the non--uniqueness of the representation of
the CAR $C^{\ast }$--algebra $\mathcal{U}$. This property is indeed
necessary to get non--unique equilibrium and ground states which imply phase
transitions. This fact was first observed by R. Haag in 1962 \cite{Haag62},
who established that the non--uniqueness of the ground state of the BCS
model in infinite volume is related to the existence of
 several inequivalent\footnote{%
This means that there is no isomorphism between $\mathfrak{h}_{j_{1}}$ and $%
\mathfrak{h}_{j_{2}}$ whenever $\mathfrak{h}_{j_{1}}$ and $\mathfrak{h}%
_{j_{2}}$ are  the Hilbert spaces corresponding to
two different irreducible representations.} irreducible
representations\footnote{%
This means that the Hamiltonian can be seen as an operator acting on several
Hilbert spaces $\{\mathfrak{h}_{j}\}_{j\in J}$ with no (non-trivial)
 invariant subspace.}
of the Hamiltonian, see also \cite{ThirWeh67,Emch}.

Equilibrium states define tangents to the convex map
\begin{equation*}
\left( \beta ,\mu ,\lambda ,\gamma ,h\right) \mapsto \mathrm{p}\left( \beta
,\mu ,\lambda ,\gamma ,h\right) .
\end{equation*}%
The analysis of the set of tangents of this map gives hence information
about the expectations of many important observables w.r.t. equilibrium
states. The main technical point in the present work is therefore to find an
explicit representation of the pressure by using the permutation invariance
of the model in a crucial way. Indeed, we adapt to our case of fermions on a
lattice the methods of \cite{Petz-Raggio-Verbeure} used to find the pressure
of spin systems of mean--field type. Then, it is proven that it suffices to
minimize the variational problem (\ref{p.var}) w.r.t. the set $\mathcal{E}_{%
\mathcal{U}}^{S,+}$ of extremal states in ${E}_{\mathcal{U}}^{S,+}$. By
adapting the proof of St{\o }rmer's theorem \cite{Stormer} to even states on
the CAR algebra, we show next that extremal, permutation invariant and even
states are product states
\begin{equation*}
\omega _{\zeta }:=\bigotimes\limits_{x\in \mathbb{Z}^{d}}\zeta _{x}
\end{equation*}%
obtained by \textquotedblleft copying\textquotedblright\ some one--site even
state $\zeta $ to all other sites. This result is a non--commutative version
of the celebrated de Finetti Theorem from (classical) probability theory
\cite{De Finetti}. Using this, the variational problem (\ref{p.var}) can be
drastically simplified to a minimization problem on a finite dimensional
manifold. At the end, it yields to another explicit, rather simple,
variational problem on $\mathbb{R}_{0}^{+}$, which can be rigorously
analyzed by analytic or numerical methods to obtain the complete
thermodynamic behavior of the model.

Observe however, that all correlation functions \textit{cannot} be drawn
from an explicit formula for the pressure by taking derivatives combined
with Griffiths arguments \cite{BruZagrebnov8,Griffiths1,HeppLieb} on the
convergence of derivatives of convex functions, unless the (infinite volume)
pressure is shown to be differentiable w.r.t. any perturbation. Showing
differentiability of the pressure as well as the explicit computation of its
corresponding derivative can be a very hard task, for instance for
correlation functions involving many lattice points. By contrast, the method
presented in this paper gives access to all correlation functions at once.
This is one basic (mathematical) message of this method, which is
generalized in \cite{BruPedra2} to all translation invariant Fermi systems
without requiring any quantum spin representation.

In fact, we precisely characterize the sets $\mathit{\Omega }_{\beta }$ for
all $\beta \in (0,\infty ]$, where $\mathit{\Omega }_{\infty }$ is the set
of ground states with parameters $\mu ,$ $\gamma ,$ $\lambda ,$ and $h$.
This detailed study yields our main rigorous results on the strong coupling
BCS--Hubbard model $\mathrm{H}_{N}$, which can be summarized as follows:

\begin{itemize}
\item There is a set of parameters $\mathcal{S}$, defining the
superconducting phase, with equilibrium and ground states breaking the $U(1)$%
--gauge symmetry and showing \emph{off--diagonal long range order} (ODLRO).

\item Depending on the parameters, the superconducting phase transition is
either a first order or a second order phase transition.

\item The superconducting phase $\mathcal{S}$ is characterized by the
formation of Cooper pairs (shown by proving bounds for the density--density
correlations) and a depleted Cooper pair condensate, the density $\mathrm{r}%
_{\beta }\in \lbrack 0,1/4]$ of which is defined by the gap equation.

\item From our proof of St{\o }rmer's theorem \cite{Stormer} for even states
on the CAR algebra, we observe that the superconducting phase $\mathcal{S}$
corresponds to a\ \emph{s--wave} superconductor, i.e., a superconductor with
two--point correlation function, for $x,y\in \mathbb{Z}^{d}$, $%
s_{1},s_{2}\in \{\uparrow ,\downarrow \}$ and within $\mathcal{S}$, equal to
$\omega (a_{x,s_{1}}a_{y,s_{2}})=\mathrm{r}_{\beta }^{1/2}e^{i\phi }\neq 0$
if  $x=y$ and $s_{1}\neq s_{2}$, and $\omega (a_{x,s_{1}}a_{y,s_{2}})=0$
else. (Here $\omega $ is any pure
state of $\mathit{\Omega }_{\beta }$; $\phi \in \lbrack 0,2\pi )$ is
determined by $\omega $.)

\item We observe the Mei{\ss }ner effect\footnote{%
It is mathematically defined here by the absence of magnetization in
presence of superconductivity. Steady surface currents around the bulk of the
superconductor are not analyzed as it is a finite volume effect.} by
analyzing the relation between superconductivity and magnetization.

\item We establish the existence of a superconductor--Mott insulator phase
transition for integer electron density per site.

\item The coexistence of ferromagnetic and superconducting phases is shown
to be feasible at (critical) points of the boundary $\partial \mathcal{S}$
of $\mathcal{S}$, by applying the decomposition theory for states \cite%
{BrattelliRobinsonI} on the weak$^\ast $--compact and convex set $\mathit{%
\Omega }_{\beta }$.

\item The critical temperature $\theta _{c}$ for the superconducting phase
transition w.r.t. $\lambda $, $\gamma $ or $h$ is analyzed in the case of
fixed chemical potential $\mu $ and also in the case of constant electron
density $\rho $. It shows that $\theta _{c}$ can be an \emph{increasing}
function of the \emph{positive} coupling constant $\lambda >0$ at fixed $\mu
\in \mathbb{R}$ but not at fixed $\rho >0$.

\item
 For $\lambda \sim \gamma$
the critical temperature $\theta _{c}$ shows -- as a function of
the electron density $\rho$ -- the typical behavior observed (only) in
high--$T_c$ superconductors: $\theta_c$ is zero or very small for
$\rho \sim 1$ and is much larger for $\rho$ away from $1$. Thus, our model
provides a simple rigorous microscopic explanation for such
 experimentally well--known behavior of high--$T_c$ superconductors.

\item Together with our study of the heat capacity, all these results can be
used to fix experimentally all parameters of $\mathrm{H}_{N}$.
\end{itemize}

Note that our study of equilibrium states is reminiscent of the work of
Fannes, Spohn and Verbeure \cite{S}, performed however within a different
framework. By opposition with our setting, their analysis \cite{S} concerns
symmetric states on an infinite tensor product of one $C^{\ast }$--algebra
and their definition of equilibrium states uses the so--called correlation
inequalities for KMS--states, see \cite[Appendix E]{BruZagrebnov8}.

To conclude, this paper is organized as follows. In Section {\ref{Section 2}
we give the thermodynamic limit of the pressure }$\mathrm{p}_{N}$ (\ref{BCS
pressure}) as well as the gap equation. Then, {our main results concerning
the thermodynamic properties of the model} are formulated in Section {\ref%
{Section phase diagram mu fixed} at fixed chemical potential }$\mu ${\ and
in }Section {\ref{Section phase diagram rho fixed} at fixed electron density
}$\rho ${\ per site. }Section {\ref{Section 5} briefly explains our result
on the level of equilibrium states and gives additional remarks.}
In order to keep the main issues and the physical implications as
transparent as possible, we reduce the technical and formal aspects to a
minimum in Sections {\ref{Section 2}--\ref{Section 5}}. In particular, in
Sections {\ref{Section 2}--\ref{Section phase diagram rho fixed}} we only
stay on the level of pressure and thermodynamic limit of local Gibbs states.
The generalization of the results on the level of equilibrium and ground
states is postponed to Section \ref{equilibirum.paragraph}. Indeed, the
rather long Section {\ref{Section mathematical foundations}} gives the
detailed mathematical foundations of our phase diagrams. In particular, in
Section {\ref{Section main proof}} we introduce the $C^{\ast }$--algebraic
machinery needed in our analysis and prove various technical facts to
conclude in Section \ref{equilibirum.paragraph} with the rigorous study of
equilibrium and ground states. In Section {\ref{section variational problem},%
} we collect some useful properties on the qualitative behavior of the
Cooper pair condensate density, whereas Section {\ref{section proof
griffiths} is an appendix on }Griffiths arguments \cite%
{BruZagrebnov8,Griffiths1,HeppLieb}.

\section{Grand--canonical pressure and gap equation\label{Section 2}}

In order to obtain the thermodynamic behavior of the strong coupling
BCS--Hubbard model $\mathrm{H}_{N}$, it is essential to get first the
thermodynamic limit $N\rightarrow \infty $ of its grand--canonical pressure $%
\mathrm{p}_{N}$ (\ref{BCS pressure}). The rigorous derivation of this limit
is performed in Section \ref{Section main proof}. We explain here the final
result with the heuristic behind it.

The first important remark is that one can guess the correct variational
problem by the so-called approximating Hamiltonian method \cite%
{approx-hamil-method0,approx-hamil-method,approx-hamil-method2} originally
proposed by Bogoliubov Jr. \cite{Bogjunior}. In our case, the correct
approximation of the Hamiltonian $\mathrm{H}_{N}$ is the $c$--dependent
Hamiltonian%
\begin{eqnarray}
H_{N}\left( c\right) := &&-\mu \sum_{x\in \Lambda _{N}}\left( n_{x,\uparrow
}+n_{x,\downarrow }\right) -h\sum_{x\in \Lambda _{N}}\left( n_{x,\uparrow
}-n_{x,\downarrow }\right) +2\lambda \sum_{x\in \Lambda _{N}}n_{x,\uparrow
}n_{x,\downarrow }  \notag \\
&&-\frac{\gamma }{N}\sum_{x\in \Lambda _{N}}\left( \left( Nc\right)
a_{x,\uparrow }^{\ast }a_{x,\downarrow }^{\ast }+\left( N\bar{c}\right)
a_{x,\downarrow }a_{x,\uparrow }\right) ,
\label{Hamiltonian BCS-Hubbard approx}
\end{eqnarray}%
with $c\in \mathbb{C}$, see also \cite{ThirWeh67,Thir68}. The main advantage
of this Hamiltonian in comparison with $\mathrm{H}_{N}$ is the fact that it
is a sum of shifts of the same local operator. For an appropriate order
parameter $c\in \mathbb{C}$, it leads to a good approximation of the
pressure $\mathrm{p}_{N}$ as $N\rightarrow \infty $. This can be partially
seen from the inequality
\begin{equation*}
\gamma N\left\vert c\right\vert ^{2}+H_{N}\left( c\right) -\mathrm{H}_{N}=%
\frac{\gamma }{N}\left( \sum_{x\in \Lambda _{N}}a_{x,\uparrow }^{\ast
}a_{x,\downarrow }^{\ast }-N\bar{c}\right) \left( \sum_{x\in \Lambda
_{N}}a_{x,\uparrow }a_{x,\downarrow }-Nc\right) \geq 0,
\end{equation*}%
which is valid as soon as $\gamma \geq 0$. Observe that the constant term $%
\gamma N|c|^{2}$ is not included in the definition of $H_{N}(c)$. Hence, by
using the Golden-Thompson inequality $\mathrm{Trace}(e^{A+B^{\ast }B})\leq
\mathrm{Trace}(e^{A})$,  the thermodynamic limit $\mathrm{p}(\beta ,\mu
,\lambda ,\gamma ,h)$ of the pressure $\mathrm{p}_{N}$ (\ref{BCS pressure})
is bounded from below by
\begin{equation}
\mathrm{p}\left( \beta ,\mu ,\lambda ,\gamma ,h\right) \geq \underset{c\in
\mathbb{C}}{\sup }\left\{ -\gamma |c|^{2}+p\left( c\right) \right\} .
\label{BCS pressure thermodynamic limit inf}
\end{equation}%
The function $p\left( c\right) =p(\beta ,\mu ,\lambda ,\gamma ,h;c)$ is the
pressure associated with $H_{N}(c)$ for any $N\geq 1$. It can easily\ be
computed since $H_{N}(c)$ is a sum of local operators which commute with
each other. Indeed, for any $N\geq 1,$ this pressure equals\footnote{%
Here $a_{0,\uparrow },$ $a_{0,\downarrow }$ and $n_{0,\uparrow },$ $%
n_{0,\downarrow }$ are replaced respectively by $a_{\uparrow }$, $%
a_{\downarrow }$ and $n_{\uparrow },$ $n_{\downarrow }.$}
\begin{eqnarray}
p\left( c\right) := &&\frac{1}{\beta N}\ln \mathrm{Trace}\left( e^{-\beta
H_{N}\left( c\right) }\right) =\frac{1}{\beta }\ln \mathrm{Trace}\left(
e^{-\beta H_{1}\left( c\right) }\right)  \notag \\
&=&\frac{1}{\beta }\ln \mathrm{Trace}\left( e^{\beta \left\{ (\mu
+h)n_{\uparrow }+(\mu -h)n_{\downarrow }+\gamma (ca_{\downarrow }^{\ast
}a_{\uparrow }^{\ast }+\bar{c}a_{\uparrow }a_{\downarrow })-2\lambda
n_{\uparrow }n_{\downarrow }\right\} }\right) .
\label{BCS pressure approx c}
\end{eqnarray}%
To be useful, the variational problem in (\ref{BCS pressure thermodynamic
limit inf}) should also be an upper bound of $\mathrm{p}(\beta ,\mu ,\lambda
,\gamma ,h).$ By adapting the proof of St{\o }rmer's theorem \cite{Stormer}
to even states on the CAR algebra and by using the Petz--Raggio--Verbeure
proof for spin systems \cite{Petz-Raggio-Verbeure} as a guideline, we prove
this in Section \ref{Section main proof}. Thus, the thermodynamic limit of
the pressure of the model $\mathrm{H}_{N}$ exists and can explicitly be
computed by using the approximating Hamiltonian $H_{N}(c)$:

\begin{theorem}[Grand-canonical pressure]
\label{BCS theorem 1}\mbox{ }\newline
For any $\beta ,\gamma >0$ and $\mu ,\lambda ,h\in \mathbb{R}$, the
thermodynamic limit $\mathrm{p}(\beta ,\mu ,\lambda ,\gamma ,h)$ of the
grand--canonical pressure $\mathrm{p}_{N}$ (\ref{BCS pressure}) equals%
\begin{equation*}
\mathrm{p}\left( \beta ,\mu ,\lambda ,\gamma ,h\right) =\underset{c\in
\mathbb{C}}{\sup }\left\{ -\gamma |c|^{2}+p\left( c\right) \right\} =\beta
^{-1}\ln 2+\mu +\underset{r\geq 0}{\sup }f\left( r\right) <\infty ,
\end{equation*}%
where the real function $f\left( r\right) =f(\beta ,\mu ,\lambda ,\gamma
,h;r)$ is defined by
\begin{equation*}
f\left( r\right) :=-\gamma r+\frac{1}{\beta }\ln \left\{ \cosh \left( \beta
h\right) +e^{-\lambda \beta }\cosh \left( \beta g_{r}\right) \right\} ,
\end{equation*}%
with $g_{r}:=\{(\mu -\lambda )^{2}+\gamma ^{2}r\}^{1/2}.$
\end{theorem}

%
%
\begin{figure*}[hbtp]
\begin{center}
\mbox{
\leavevmode
\subfigure
{ \includegraphics[angle=0,scale=1,clip=true,width=3.8cm]{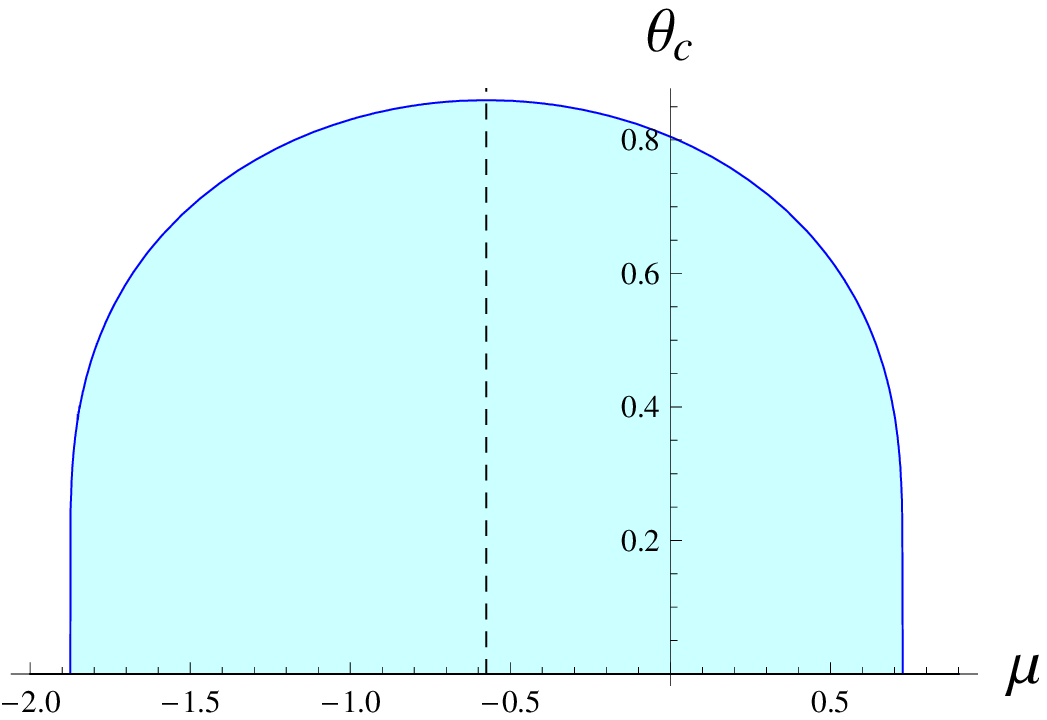} }

\leavevmode
\subfigure
{ \includegraphics[angle=0,scale=1,clip=true,width=3.8cm]{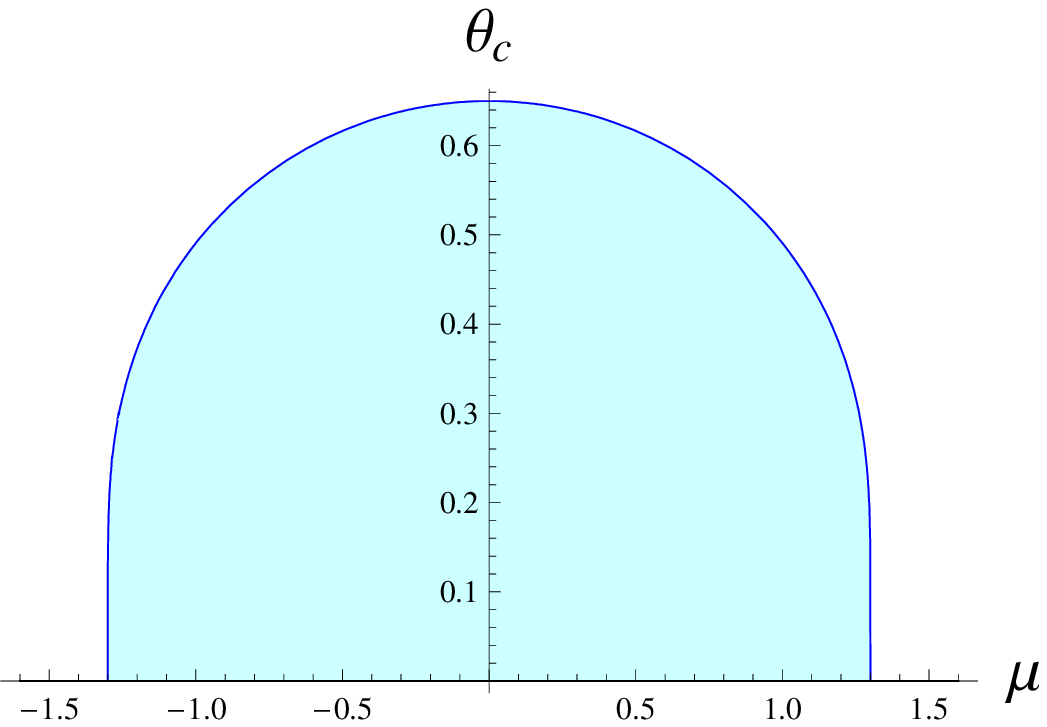} }

\leavevmode
\subfigure
{ \includegraphics[angle=0,scale=1,clip=true,width=3.8cm]{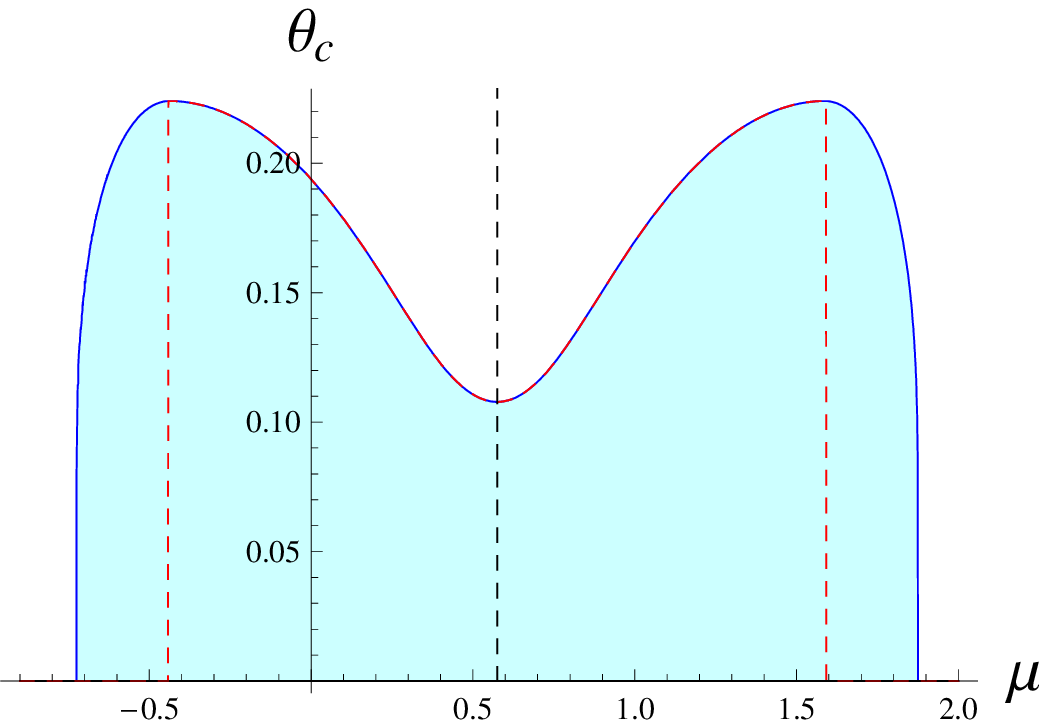} }
}
\end{center}
\caption{\emph{Illustration, as a function of $\mu $, of the critical temperature $\theta
_{c}=\theta _{c}(\mu ,\lambda ,\gamma ,h)$ such that $\mathrm{r}_{\beta }>0$
if and only if $\beta >\theta _{c}^{-1}$ (blue area) for $\gamma
=2.6,$ $h=0$ and with $\lambda =-0.575$ (left figure), $0$ (figure on the center) and $0.575$
(right figure). The blue line corresponds to a second order phase transition,
whereas the red dashed line represents the domain of $\mu$ with a first order phase transition.
The black dashed line is the chemical potential $\mu=\lambda$ corresponding to an electron density
per site equal to $1$, see Section \ref{Section phase diagram mu fixed}.}}
\label{domain-temp-critique-mu.eps}
\end{figure*}%

\begin{remark}
The fact that the pressure $\mathrm{p}_{N}$ coincides as $N\rightarrow
\infty $ with the variational problem given by the so-called approximating
Hamiltonian (here $H_{N}(c)$) was previously proven via completely different
methods in \cite{approx-hamil-method0} for a large class of Hamiltonian
(including $\mathrm{H}_{N}$) with BCS--type interaction. However, as
explained in the introduction, our proof gives deeper results, not expressed
in Theorem \ref{BCS theorem 1}, on the level of states, cf. (\ref{p.var})
and (\ref{p.varbis}). In contrast to the approximating Hamiltonian method
\cite%
{approx-hamil-method0,approx-hamil-method,approx-hamil-method2,Bogjunior},
it leads to a natural notion of equilibrium and ground states and allows the
direct analysis of correlation functions. For more details, we recommend
Section \ref{Section mathematical foundations}, particularly Section \ref%
{equilibirum.paragraph}.
\end{remark}

From the gauge invariance of the map $c\mapsto p(c)$ observe that any
maximizer $\mathrm{c}_{\beta }\in \mathbb{C}$ of the first variational
problem given in Theorem \ref{BCS theorem 1} has the form $\mathrm{r}_{\beta
}^{1/2}e^{i\phi }$ with $\mathrm{r}_{\beta }\geq 0$ being solution of
\begin{equation}
\underset{r\geq 0}{\sup }\,f(r)=f(\mathrm{r}_{\beta })
\label{BCS pressure 2}
\end{equation}%
and $\phi \in \lbrack 0,2\pi )$. For any $\beta ,\gamma >0$ and real numbers
$\mu ,\lambda ,h,$ it is also clear that the order parameter $\mathrm{r}%
_{\beta }$ is always bounded since $f(r)$ diverges to $-\infty $ when $%
r\rightarrow \infty .$ Up to (special) points $(\beta ,\mu ,\lambda ,\gamma
,h)$ corresponding to a phase transition of first order, it is always unique
and continuous w.r.t. each parameter (see Section \ref{section variational
problem}).

\begin{figure*}[hbtp]
\begin{center}
\mbox{
\leavevmode
\subfigure
{ \includegraphics[angle=0,scale=1,clip=true,width=3.8cm]{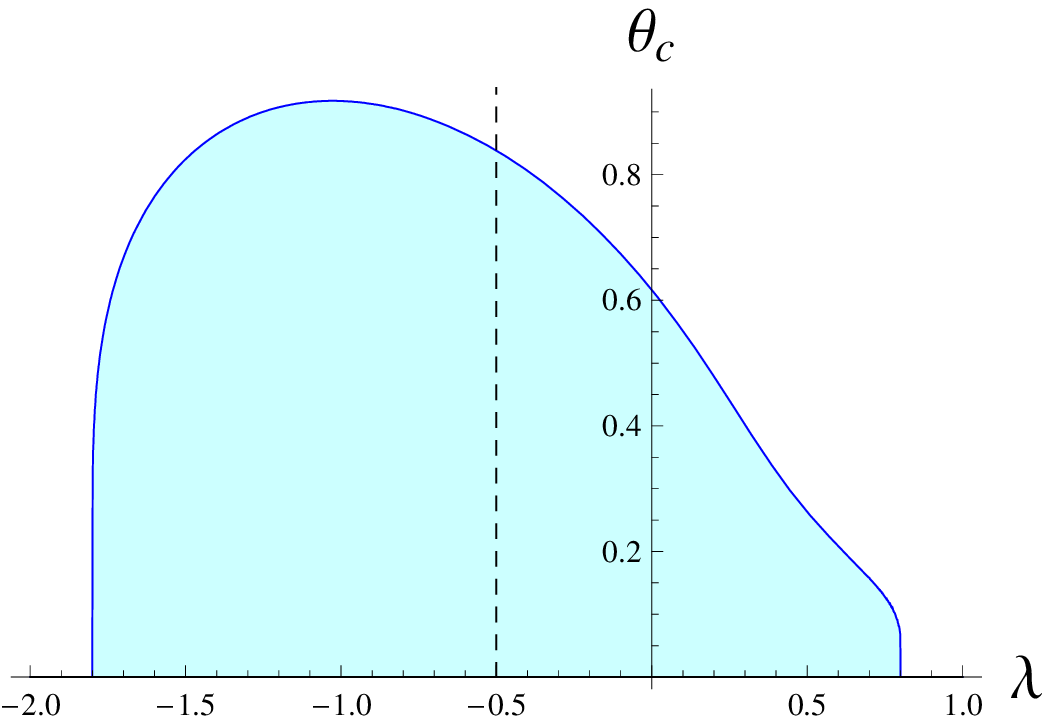} }

\leavevmode
\subfigure
{ \includegraphics[angle=0,scale=1,clip=true,width=3.8cm]{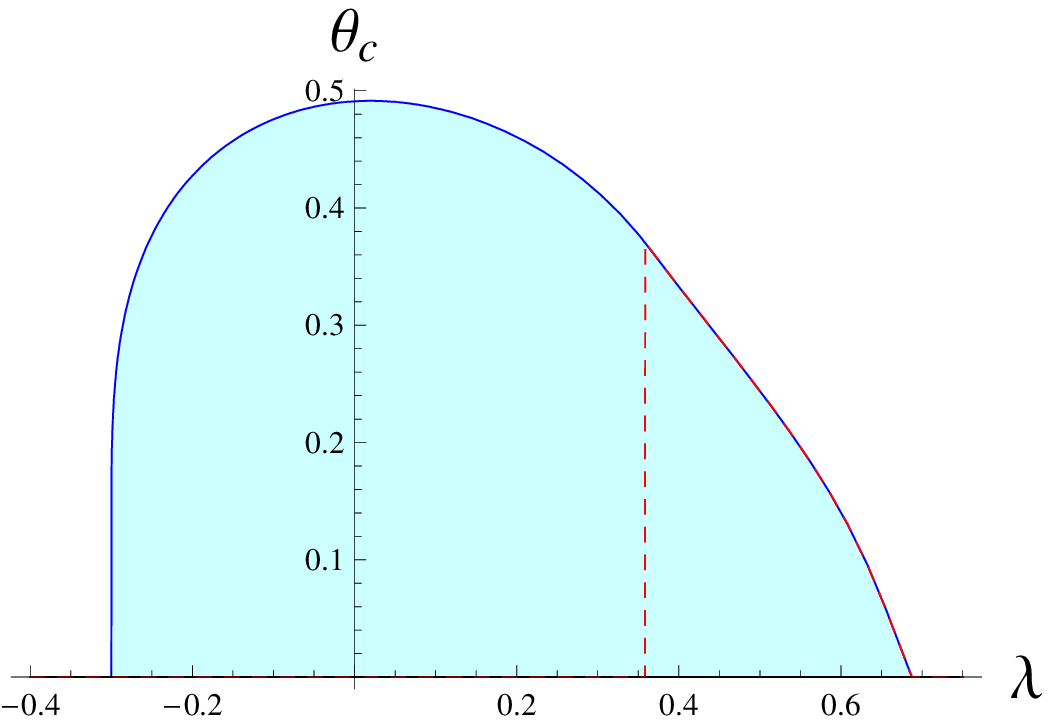} }

\leavevmode
\subfigure
{ \includegraphics[angle=0,scale=1,clip=true,width=3.8cm]{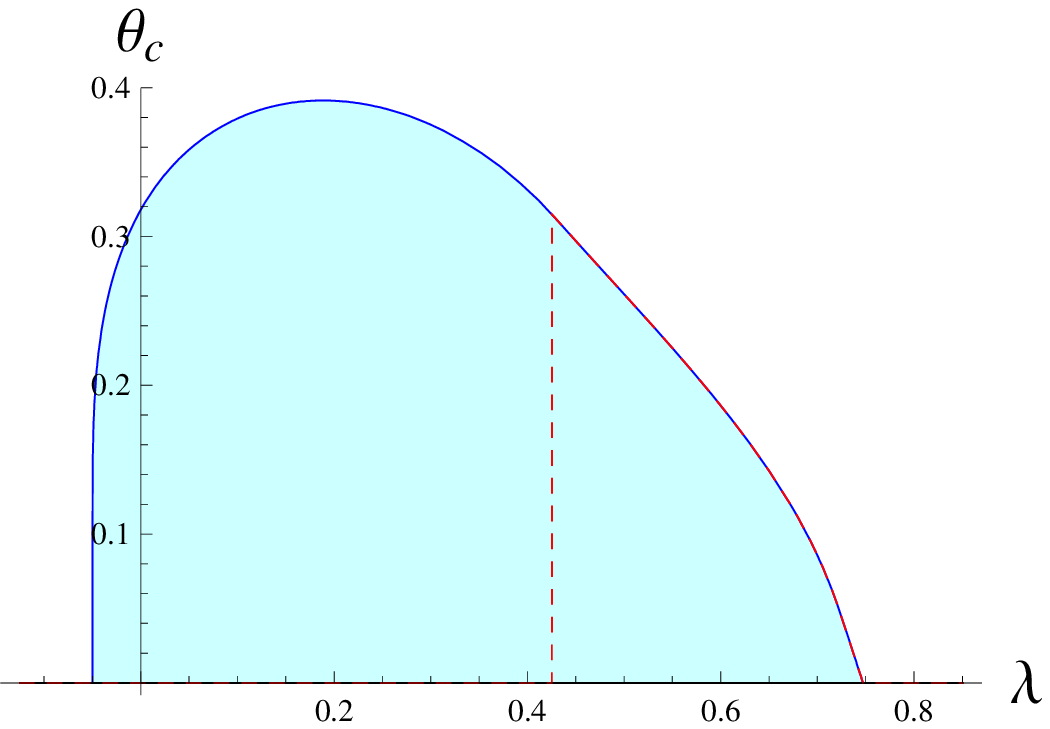} }
}
\end{center}
\caption{\emph{Illustration, as a function of $\lambda ,$ of the critical temperature $\theta _{c}=\theta _{c}(\mu ,\lambda ,\gamma ,h)$
for $\gamma =2.6,$ $h=0$ and with $\mu =-0.5$ (left figure), $\mu =1$
(figure at the center) and $\mu =1.25$ (right figure).
The blue line corresponds to a second order phase transition,
whereas the red dashed line represents the domain of $\lambda$ with first order phase transition.
The black dashed line is the coupling constant $\lambda=\mu$ corresponding to an electron density per site equal to $1$, see Section \ref{Section phase diagram mu fixed}.}}
\label{domain-temp-critique-lamb.eps}
\end{figure*}

For low inverse temperatures $\beta $ (high temperature regime) $\mathrm{r}%
_{\beta }=0$.\ Indeed, straightforward computations at low enough $\beta $
show that the function $f(r)$ is concave as a function of $r\geq 0$ whereas $%
\partial _{r}f(0)<0$, see Section \ref{section variational problem}. On the
other hand, any non--zero solution $\mathrm{r}_{\beta }$ of the variational
problem (\ref{BCS pressure 2}) has to be solution of the gap equation (or
Euler--Lagrange equation)
\begin{equation}
\tanh \left( \beta g_{\mathrm{r}_{\beta }}\right) =\frac{2g_{\mathrm{r}%
_{\beta }}}{\gamma }\left( 1+\frac{e^{\lambda \beta }\cosh \left( \beta
h\right) }{\cosh \left( \beta g_{\mathrm{r}_{\beta }}\right) }\right) .
\label{BCS gap equation}
\end{equation}%
If $g_{r}=0,$ observe that one uses in (\ref{BCS gap equation}) the
asymptotics $x^{-1}\tanh x\sim 1$ as $x\rightarrow 0,$ see also (\ref{BCS
gap equationbis}). Because $\tanh (x)\leq 1$ for $x\geq 0,$ we then conclude
that
\begin{equation}
0\leq \mathrm{r}_{\beta }\leq \max \left\{ 0,\mathrm{r}_{\max }\right\} ,%
\mathrm{\ with\;r}_{\max }:=\frac{1}{4}-\gamma ^{-2}\left( \mu -\lambda
\right) ^{2}.  \label{definition r max}
\end{equation}%
In particular, if $\gamma \leq 2|\mu -\lambda |,$ then $\mathrm{r}_{\beta
}=0 $ for any $\beta >0$. However, at large enough $\beta >0$ (low
temperature regime) and at fixed $\lambda ,h,\mu \in \mathbb{R}$, there is a
unique $\gamma _{c}>2|\lambda -\mu |$ such that $\mathrm{r}_{\beta }>0$ for
any $\gamma \geq \gamma _{c}$. In other words, the domain of parameters $%
(\beta ,\mu ,\lambda ,\gamma ,h)$ where $\mathrm{r}_{\beta }$ is strictly
positive is non--empty, see figures \ref{domain-temp-critique-mu.eps}--\ref%
{domain-temp-critique-lamb.eps} and Section \ref{section variational problem}%
. Observe in figure \ref{domain-temp-critique-lamb.eps} that a positive $%
\lambda $, i.e., a one--site repulsion, can significantly increase (right
figure) the critical temperature $\theta _{c}=\theta _{c}(\mu ,\lambda
,\gamma ,h),$ which is defined such that $\mathrm{r}_{\beta }>0$ if and only
if $\beta >\theta _{c}^{-1}$.%
%
%

From Lemma \ref{lemma cardinality}, the set of maximizers of the variational
problem (\ref{BCS pressure 2}) has at most two elements in $[0,1/4]$. It
follows by continuity of $(\beta ,\mu ,\lambda ,\gamma ,h,r)\mapsto f(\beta
,\mu ,\lambda ,\gamma ,h;r)$, and from the fact that the interval $[0,1/4]$
is compact, that the set
\begin{equation}
\mathcal{S}:=\Big\{(\beta ,\mu ,\lambda ,\gamma ,h)\;:\;\beta ,\gamma >0\;%
\mbox{and $\mathrm{r}_{\beta }>0$ is the unique maximizer
of (\ref{BCS pressure 2})}\Big\}  \label{critical point open set}
\end{equation}%
is open. In Section \ref{Section BCS phase transition}, we prove that the
set $\mathcal{S}$ corresponds to the superconducting phase since the order
parameter solution of (\ref{BCS pressure 2}) can be interpreted as the
Cooper pair condensate density. The boundary $\partial \mathcal{S}$ of the
set $\mathcal{S}$ is called the set of \textit{critical points} of our
model. By definition, if (\ref{BCS pressure 2}) has more than one maximizer,
then $(\beta ,\mu ,\lambda ,\gamma ,h)\in \partial \mathcal{S}$, whereas if $%
(\beta ,\mu ,\lambda ,\gamma ,h)\not\in \overline{\mathcal{S}}$, then $r=0$
is the unique maximizer of (\ref{BCS pressure 2}).

For more details on the study of the variational problem (\ref{BCS pressure
2}), we recommend Section \ref{section variational problem}.

\section{Phase diagram at fixed chemical potential\label{Section phase
diagram mu fixed}}

By using our main theorem, i.e., Theorem \ref{BCS theorem 1}, we can now
explain the thermodynamic behavior of the strong coupling BCS--Hubbard model
$\mathrm{H}_{N}$. The rigorous proofs are however given in Section \ref%
{equilibirum.paragraph}. Actually, we concentrate here on the physics of the
model extracted from the (finite volume) grand--canonical Gibbs state $%
\omega _{N}$ (\ref{BCS gibbs state Hn}) associated with $\mathrm{H}_{N}$. We
start by showing the existence of a superconducting phase transition in the
thermodynamic limit.

\subsection{Existence of a s--wave superconducting phase transition\label%
{Section BCS phase transition}}

The solution $\mathrm{r}_{\beta }$ of (\ref{BCS pressure 2}) can be
interpreted as an order parameter related to the Cooper pair condensate
density $\omega _{N}(\mathfrak{c}_{0}^{\ast }\mathfrak{c}_{0})/N$, where
\begin{equation*}
\mathfrak{c}_{0}:=\frac{1}{\sqrt{N}}\sum_{x\in \Lambda _{N}}a_{x,\downarrow
}a_{x,\uparrow }=\frac{1}{\sqrt{N}}\sum\limits_{k\in \Lambda _{N}^{\ast }}%
\tilde{a}_{k,\downarrow }\tilde{a}_{-k,\uparrow }
\end{equation*}%
resp. $\mathfrak{c}_{0}^{\ast }$ annihilates resp. creates one Cooper pair
within the condensate, i.e., in the zero-mode for electron pairs. Indeed, in
Section \ref{equilibirum.paragraph} (see Theorem \ref{Theorem equilibrium
state 3}) we prove, by using a notion of equilibrium states, the following.

\begin{theorem}[Cooper pair condensate density]
\label{BCS theorem 2-0}\mbox{}\newline
For any $\beta ,\gamma >0$ and real numbers $\mu ,\lambda ,h$ away from any
critical point, the (infinite volume) Cooper pair condensate density equals%
\begin{eqnarray*}
\underset{N\rightarrow \infty }{\lim }\left\{ \frac{1}{N}\omega _{N}\left(
\mathfrak{c}_{0}^{\ast }\mathfrak{c}_{0}\right) \right\} &=&\underset{%
N\rightarrow \infty }{\lim }\left\{ \dfrac{1}{N^{2}}\sum_{x,y\in \Lambda
_{N}}\omega _{N}\left( a_{x,\uparrow }^{\ast }a_{x,\downarrow }^{\ast
}a_{y,\downarrow }a_{y,\uparrow }\right) \right\} \\
&=&\mathrm{r}_{\beta }\leq \max \left\{ 0,\mathrm{r}_{\max }\right\} ,
\end{eqnarray*}
with $\mathrm{r}_{\max }\leq 1/4$ defined in (\ref{definition r max}). The
(uniquely defined) order parameter $\mathrm{r}_{\beta }=\mathrm{r}_{\beta
}(\mu ,\lambda ,\gamma ,h)$ is an increasing function of $\gamma >0$.
\end{theorem}

\begin{remark}
In fact, Theorem \ref{BCS theorem 2-0} is not anymore satisfied only if the
order parameter $\mathrm{r}_{\beta }$ is discontinuous w.r.t. $\gamma >0$ at
fixed $(\beta ,\mu ,\lambda ,h)$. In this case, the thermodynamic limit of
the Cooper pair condensate density is bounded by the left and right limits
of the corresponding (infinite volume) density, see Section \ref{section
proof griffiths}, in particular (\ref{critical point}). Similar remarks can
be done for Theorems \ref{BCS theorem 2-1}, \ref{BCS theorem 2-2}, \ref{BCS
theorem 2-3} and \ref{BCS theorem 2-4}.
\end{remark}

\begin{figure}[h]
\centerline{\includegraphics[angle=0,scale=1.,clip=true,width=12.5cm]{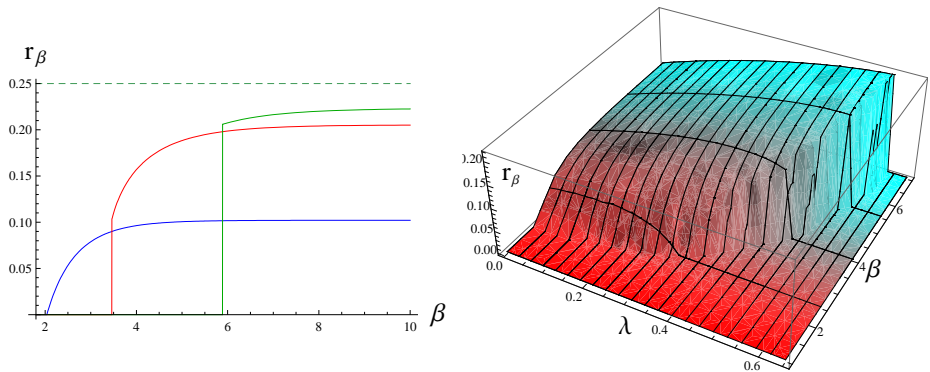}}
\caption{\emph{In the figure on the left, we have three illustrations of the  Cooper pair condensate density $\mathrm{r}_{\beta } $ as a function of the inverse temperature $\beta$ for $\lambda =0$ (blue line), $\lambda =0.45$ (red line) and $\lambda =0.575\,$ (green line). The figure on the right represents a 3D
illustration of $\mathrm{r}_{\beta }$ as a function of $\lambda$ and $\beta$. The color from red to blue reflects the decrease of the
temperature. In all figures, $\mu=1$, $\gamma=2.6$ and $h=0$.}}
\label{order-parameter-temp-lambda.eps}
\end{figure}

At least for large enough $\beta $ and $\gamma $, we have explained that $%
\mathrm{r}_{\beta }>0,$ see figures \ref{domain-temp-critique-mu.eps}--\ref%
{domain-temp-critique-lamb.eps}. Illustrations of the Cooper pair condensate
density $\mathrm{r}_{\beta }$ as a function of $\beta $ and $\lambda $ are
given in figure \ref{order-parameter-temp-lambda.eps}. In other words, a
superconducting phase transition can appear in our model. Its order depends
on parameters: it can be a first order or a second order superconducting
phase transition, cf. figure \ref{order-parameter-temp-lambda.eps} and
Section \ref{section variational problem} for more details. From numerical
investigations, note that $\mathrm{r}_{\beta }$ was always found to be an
increasing function of $\beta >0$. Unfortunately we are able to prove only a
part of this fact in Section \ref{section variational problem}. Therefore, a
superconducting phase appearing only in a range of non--zero temperatures as
for magnetic superconductors cannot not rigorously been excluded. But we
conjecture that our model can \textit{never} show this phenomenon, i.e., $%
\mathrm{r}_{\beta }$ should always be an increasing function of $\beta >0.$

Observe that a non--trivial solution $\mathrm{r}_{\beta }\neq 0$ is a
manifestation of the breakdown of the $U(1)$--gauge symmetry. To see this
phenomenon, we need to perturb the Hamiltonian $\mathrm{H}_{N}$ with the
external field%
\begin{equation*}
\alpha \sqrt{N}\left( e^{-i\phi }\mathfrak{c}_{0}+e^{i\phi }\mathfrak{c}%
_{0}^{\ast }\right) \mathrm{\ for\ any\ }\alpha \geq 0\mathrm{\ and\ }\phi
\in \left[ 0,2\pi \right) .
\end{equation*}%
This leads to the perturbed Gibbs state $\omega _{N,\alpha ,\phi }\left(
\cdot \right) $ defined by (\ref{BCS gibbs state Hn}) with $\mathrm{H}_{N}$
replaced by
\begin{equation}
\mathrm{H}_{N,\alpha ,\phi }:=\mathrm{H}_{N}-\alpha \sum_{x\in \Lambda
_{N}}\left( e^{-i\phi }a_{x,\downarrow }a_{x,\uparrow }+e^{i\phi
}a_{x,\uparrow }^{\ast }a_{x,\downarrow }^{\ast }\right) ,
\label{perturbed hamiltonian}
\end{equation}%
see (\ref{local Gibbs states avec alpha}). We then obtain the following
result for the so--called Bogoliubov quasi--averages (cf. Theorem \ref%
{Theorem equilibrium state 4 U(1) broken}).

\begin{theorem}[Breakdown of the $U(1)$-gauge symmetry]
\label{BCS theorem 3}\mbox{ }\newline
For any $\beta ,\gamma >0$ and real numbers $\mu ,\lambda ,h$ away from any
critical point, and for any $\phi \in \left[ 0,2\pi \right) $, one gets for
the Bogoliubov quasi--average below:
\begin{equation*}
\underset{\alpha \downarrow 0}{\lim }\mathrm{\ }\underset{N\rightarrow
\infty }{\lim }\omega _{N,\alpha ,\phi }(\mathfrak{c}_{0}/\sqrt{N})=\underset%
{\alpha \downarrow 0}{\lim }\mathrm{\ }\underset{N\rightarrow \infty }{\lim }%
\left\{ \dfrac{1}{N}\sum_{x\in \Lambda _{N}}\omega _{N,\alpha ,\phi }\left(
a_{x,\uparrow }a_{x,\downarrow }\right) \right\} =\mathrm{r}_{\beta
}^{1/2}e^{i\phi },
\end{equation*}%
with $\mathrm{r}_{\beta }\geq 0$ being the unique solution of (\ref{BCS
pressure 2}), see Theorem \ref{BCS theorem 1}.
\end{theorem}

Note that the breakdown of the $U(1)$--gauge symmetry should be
\textquotedblleft seen\textquotedblright\ in experiments via the so--called
\textit{off diagonal long range order} (ODLRO) property of the correlation
functions \cite{ODLRO}, see Section \ref{equilibirum.paragraph}. In fact,
because of the permutation invariance, Theorem \ref{BCS theorem 2-0} still
holds if we remove the space average, i.e., for any lattice sites $x$ and $%
y\neq x$,
\begin{equation*}
\underset{N\rightarrow \infty }{\lim }\omega _{N}(a_{y,\downarrow }^{\ast
}a_{y,\uparrow }^{\ast }a_{x,\uparrow }a_{x,\downarrow })=\mathrm{r}_{\beta
},
\end{equation*}%
see Theorem \ref{Theorem equilibrium state 3}. Similar remarks can be done
for Theorems \ref{BCS theorem 2-1}, \ref{BCS theorem 2-2}, \ref{BCS theorem
2-3} and \ref{BCS theorem 2-4}.

Observe also that the type of superconductivity described here is the \emph{%
s--wave} superconductivity, which is defined via the two--point correlation
function.

\begin{theorem}[s--wave superconductivity]
\label{BCS s-wave-thm}\mbox{ }\newline
For any $\beta ,\gamma >0$ and real numbers $\mu ,\lambda ,h$ away from any
critical point, and for any $\phi \in \left[ 0,2\pi \right) $, $x,y\in
\mathbb{Z}^{d}$ and $s_{1},s_{2}\in \{\uparrow ,\downarrow \}$, the
two--point correlation function defined from the Bogoliubov quasi--averages
equals%
\begin{equation*}
\underset{\alpha \downarrow 0}{\lim }\mathrm{\ }\underset{N\rightarrow
\infty }{\lim }\omega _{N,\alpha ,\phi }(a_{x,s_{1}}a_{y,s_{2}})=\mathrm{r}%
_{\beta }^{1/2}e^{i\phi }\delta _{x,y}\left( 1-\delta _{s_{1},s_{2}}\right) ,
\end{equation*}%
with $\mathrm{r}_{\beta }\geq 0$ being the unique solution of (\ref{BCS
pressure 2}), see Theorem \ref{BCS theorem 1}. Here $\delta _{x,y}=1$ if and
only if $x=y$.
\end{theorem}

In other words, for $x,y\in \mathbb{Z}^{d}$ and $s_{1},s_{2}\in \{\uparrow
,\downarrow \}$ the two--point correlation function inside the
superconducting phase is non--zero if and only if $x=y$ and $s_{1}\neq s_{2}$%
. More generally, for any infinite volume equilibrium state $\omega$, we
have $\omega(a_{x,s_{1}}a_{y,s_{2}})=\omega(a_{0,s_{1}}a_{0,s_{2}})\delta
_{x,y}$, see Section \ref{Section mathematical foundations}.

We conclude now this analysis by giving the zero--temperature limit $\beta
\rightarrow \infty $ of the Cooper pair condensate density $\mathrm{r}%
_{\beta }$ proven in Section \ref{section variational problem}.

\begin{figure}[h]
\centerline{\includegraphics[angle=0,scale=1.,clip=true,width=12.5cm]{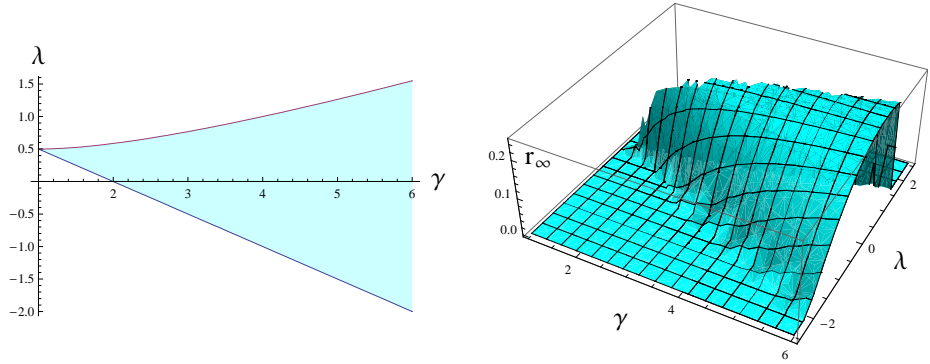}}
\caption{\emph{In the figure on the left, the blue area represents the domain of $(\lambda ,\gamma )$ with $1\leq \gamma \leq 6$, where the (zero--temperature) Cooper pair condensate density
$\mathrm{r}_{\infty }$ is non--zero at $\mu= 1$ and $h=0$. The
figure on the right represents a 3D illustration of $\mathrm{r}_{\infty }$ when  $1\leq \gamma
\leq 6$ and $-2.5\leq \lambda \leq 2.5$ with again $\mu =1,$ $h=0$.}}
\label{order-parameter-zerp-temp.eps}
\end{figure}

\begin{corollary}[Cooper pair condensate density at zero--temperature]
\label{BCS theorem 2-0bis}\mbox{}\newline
The Cooper pair condensate density $\mathrm{r}_{\infty }=\mathrm{r}_{\infty
}(\mu ,\lambda ,\gamma ,h)$ is equal at zero--temperature to
\begin{equation*}
\mathrm{r}_{\infty }:=\underset{\beta \rightarrow \infty }{\lim }\mathrm{r}%
_{\beta }=\left\{
\begin{array}{l}
\mathrm{r}_{\max }\mathrm{\ \ \ \ for\ any}\text{ }\gamma >\Gamma _{|\mu
-\lambda |,\lambda +|h|} \\
0\mathrm{\ \ \ \ \ \ \ \ for\ any}\text{ }\gamma <\Gamma _{|\mu -\lambda
|,\lambda +|h|}%
\end{array}%
\right.
\end{equation*}%
with $\mathrm{r}_{\max }\leq 1/4$ (cf. (\ref{definition r max}) and figure %
\ref{order-parameter-zerp-temp.eps}) and
\begin{equation*}
\Gamma _{x,y}:=2\left( y+\left\{ y^{2}-x^{2}\right\} ^{1/2}\right) \chi
_{\lbrack 0,y)}\left( x\right) \chi _{(0,\infty )}\left( y\right) +2x\chi
_{\lbrack y,\infty )}\left( x\right) \geq 0
\end{equation*}%
be defined for any $x\in \mathbb{R}_{+}$ and $y\in \mathbb{R}$. Here $\chi _{%
\mathcal{K}}$ is the characteristic function of the set $\mathcal{K}\subset
\mathbb{R}$.
\end{corollary}

\begin{remark}
If $\gamma =\Gamma _{|\mu -\lambda |,\lambda +|h|}$, straightforward
estimations show that the order parameter $\mathrm{r}_{\beta }$ converges to
$\mathrm{r}_{\infty }=0,$ see Section \ref{section variational problem}.
This special case is a critical point at sufficiently large $\beta $. We
exclude it in our discussion since all thermodynamic limits of densities in
Section \ref{Section phase diagram mu fixed} are performed away from any
critical point, see for instance Theorem \ref{BCS theorem 2-0}.
\end{remark}

The result of Corollary \ref{BCS theorem 2-0bis} is in accordance with
Theorem \ref{BCS theorem 2-0} in the sense that the order parameter $\mathrm{%
r}_{\infty }$ is an increasing function of $\gamma \geq 0$. Observe also
that
\begin{equation*}
\underset{\lambda \in \mathbb{R}}{\sup }\left\{ \text{\textrm{r}}_{\infty
}\left( \mu ,\lambda ,\gamma ,h\right) \right\} =\mathrm{r}_{\infty }\left(
\mu ,\mu ,\gamma ,h\right) =\frac{1}{4}
\end{equation*}%
for any fixed $\gamma >\Gamma _{0,\mu +|h|},$ whereas for any real numbers $%
\mu ,\lambda ,h,$
\begin{equation*}
\underset{\gamma \rightarrow \infty }{\lim }\mathrm{r}_{\infty }\left( \mu
,\lambda ,\gamma ,h\right) =\frac{1}{4}.
\end{equation*}%
In other words, the superconducting phase for $\mu =\lambda $ is as perfect
as for $\gamma =\infty $. In particular, in order to optimize the Cooper
pair condensate density, if $\mu >0$, then it is necessary to increase the
one--site repulsion by tuning in $\lambda $ to $\mu .$ Consequently, the
direct repulsion between electrons can favor the superconductivity at fixed $%
\mu $. This phenomenon is confirmed by the following analysis.

First observe that the equation (\ref{BCS gap equation}) has no solution if $%
\gamma \leq 2|\mu |$ and $\lambda =0.$ In other words, the strong coupling
BCS theory has no phase transition as soon as $\gamma \leq 2|\mu |$ and $\mu
\neq 0.$ However, even if $\gamma \leq 2|\mu |$, there is a range of $%
\lambda $ where a superconducting phase takes place. For instance, take $\mu
>0$ and note that $\gamma >\Gamma _{|\mu -\lambda |,\lambda +|h|}$ when%
\begin{equation}
0\leq \mu -\frac{\gamma }{2}<\lambda <\mu +\frac{\gamma }{2}-\sqrt{\gamma
\left( \mu +|h|\right) }.  \label{petit inequality new}
\end{equation}%
This last inequality can always be satisfied for some $\lambda >0$, if $\mu
+|h|<\gamma \leq 2\mu $. Therefore, although there is no superconductivity
for $\gamma \leq 2|\mu |$ and $\lambda =0,$ there is a range of positive $%
\lambda \geq 0$ defined by (\ref{petit inequality new}) for $\mu +|h|<\gamma
\leq 2\mu ,$ where the superconductivity appears at low enough temperature,
see Corollary \ref{BCS theorem 2-0bis} and figure \ref%
{order-parameter-zerp-temp.eps}. In the region $\gamma \geq 2\mu >0$ where
the superconducting phase can occur for $\lambda =0,$ observe also that the
critical temperature $\theta _{c}$ for $\lambda >0$ can sometimes be larger
as compared with the one for $\lambda =0$, cf. figure \ref%
{domain-temp-critique-lamb.eps}.

\begin{remark} The effect of a one--site repulsion on the superconducting phase transition
may be surprising since one would naively guess that any repulsion between
pairs of electrons should destroy the formation of Cooper pairs. In fact,
the one--site and BCS interactions in (\ref{Hamiltonian BCS-Hubbard}) are
not diagonal in the same basis, i.e., they do not commute. In particular,
the Hubbard interaction cannot be directly interpreted as a repulsion
between Cooper pairs. This interpretation is only valid for large $\lambda
\geq 0.$ Indeed, at fixed $\mu $ and $\gamma >0$, if $\lambda $ is large
enough, there is no superconducting phase.
\end{remark}

\subsection{Electron density per site and electron--hole symmetry\label%
{Section electron hole sym}}

We give next the grand--canonical density of electrons per site in the
system (cf. Theorem \ref{Theorem equilibrium state 4}).

\begin{figure*}[hbtp]
\begin{center}
\mbox{
\leavevmode
\subfigure
{ \includegraphics[angle=0,scale=1,clip=true,width=3.8cm]{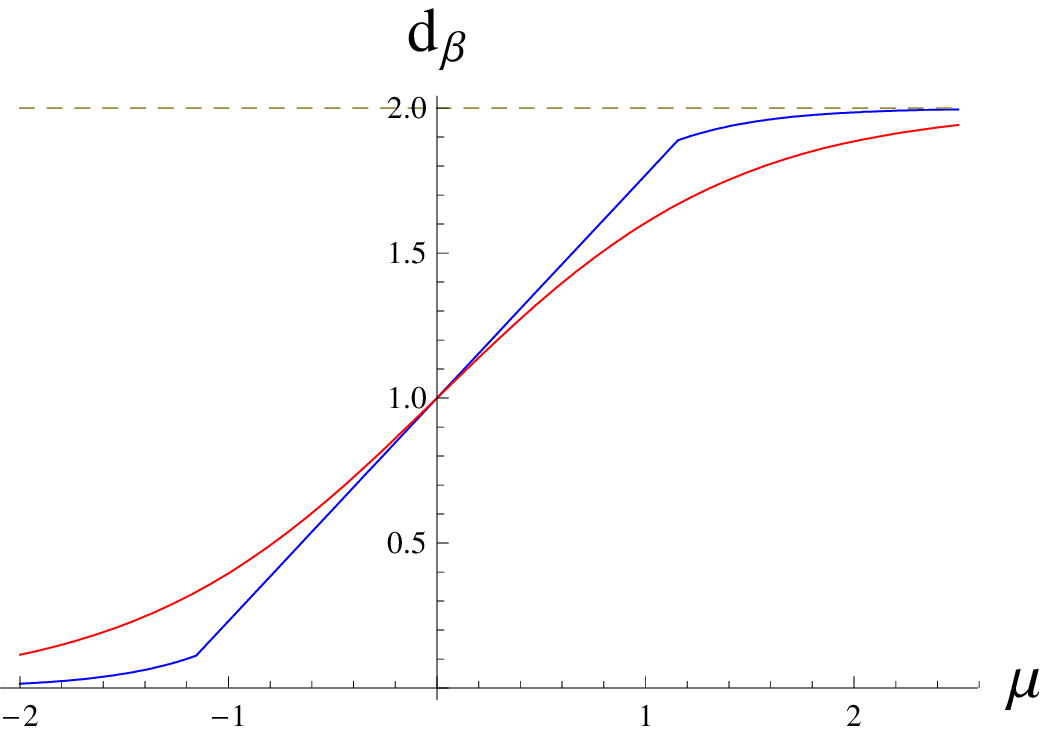} }

\leavevmode
\subfigure
{ \includegraphics[angle=0,scale=1,clip=true,width=3.8cm]{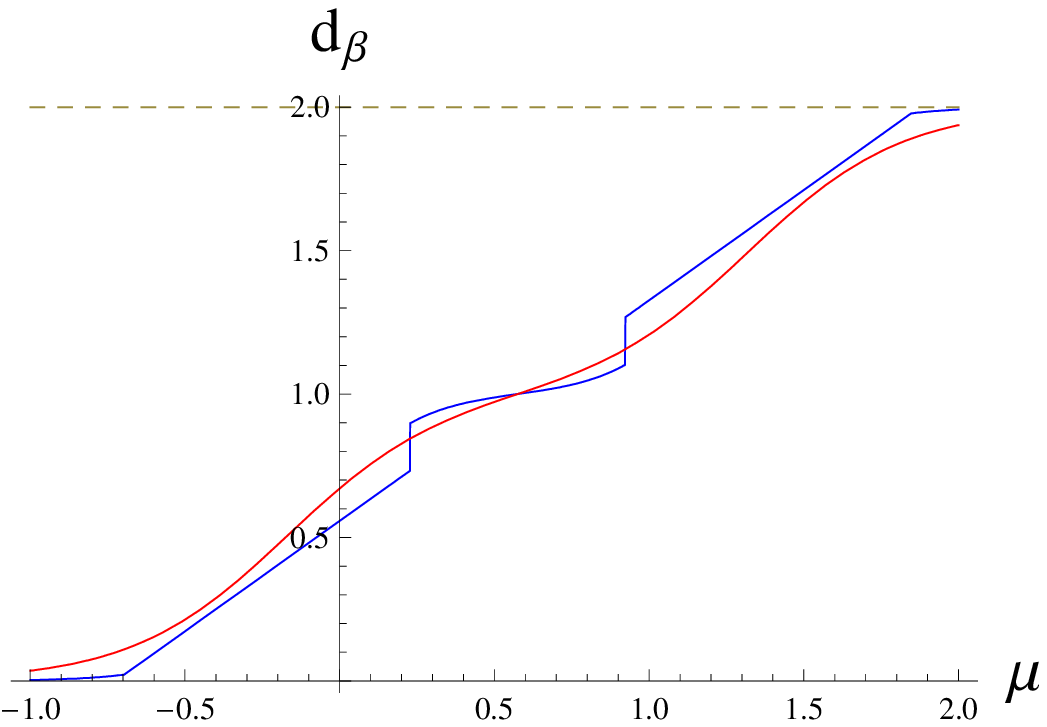} }

\leavevmode
\subfigure
{ \includegraphics[angle=0,scale=1,clip=true,width=3.8cm]{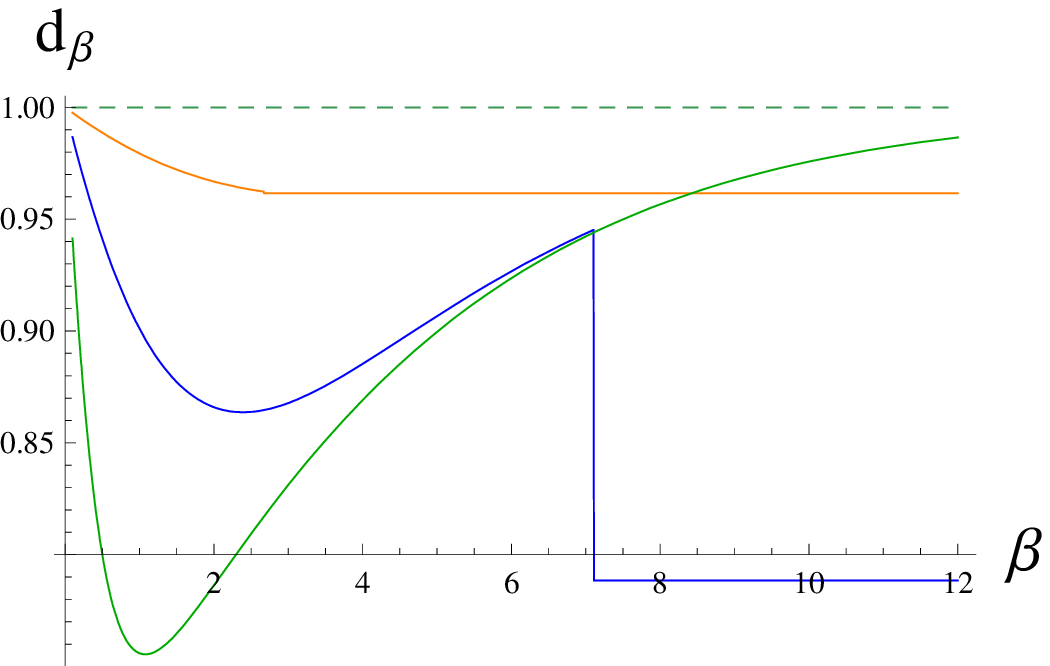} }
}
\end{center}
\caption{\emph{In the figures on the left, we give illustrations of the electron density $\mathrm{d}_{\beta }$ as a function of the chemical potential $\mu $ for $\beta <\beta _{c}$ (red line) and $\beta >\beta _{c}$ (blue line) at
coupling constant $\lambda =0$ (figure on the left, $\beta =1.4,$ $2.45$)
and $\lambda =0.575$ (figure on the center, $\beta =4,$ $6.45$). In the
figure on the right, $\mathrm{d}_{\beta }$ is given as a function of $\beta $
at $\mu =0.3$ with $\lambda > \mu $ equal to $0.35$ (orange line, second order phase transition), $0.575$ (blue
line, first order phase transition) and $1.575$ (green line, no phase transition). In all figures, $\gamma=2.6$, $h=0$ and $\beta
_{c}=\theta _{c}^{-1}$ is the critical inverse temperature.
}}
\label{density-mu.eps}
\end{figure*}

\begin{theorem}[Electron density per site]
\label{BCS theorem 2-1}\mbox{}\newline
For any $\beta ,\gamma >0$ and real numbers $\mu ,\lambda ,h$ away from any
critical point, the (infinite volume) electron density equals
\begin{equation*}
\underset{N\rightarrow \infty }{\lim }\left\{ \dfrac{1}{N}\sum_{x\in \Lambda
_{N}}\omega _{N}\left( n_{x,\uparrow }+n_{x,\downarrow }\right) \right\} =%
\mathrm{d}_{\beta }:=1+\frac{\left( \mu -\lambda \right) \sinh \left( \beta
g_{\mathrm{r}_{\beta }}\right) }{g_{\mathrm{r}_{\beta }}\left( e^{\beta
\lambda }\cosh \left( \beta h\right) +\cosh \left( \beta g_{\mathrm{r}%
_{\beta }}\right) \right) },
\end{equation*}%
with $\mathrm{d}_{\beta }=\mathrm{d}_{\beta }(\mu ,\lambda ,\gamma ,h)\in
\lbrack 0,2],$ $\mathrm{r}_{\beta }\geq 0$ being the unique solution of (\ref%
{BCS pressure 2}) and $g_{r}:=\{(\mu -\lambda )^{2}+\gamma ^{2}r\}^{1/2},$
see Theorem \ref{BCS theorem 1} and figure \ref{density-mu.eps}.
\end{theorem}

%
%
At low enough temperature and for $\gamma >\Gamma _{|\mu -\lambda |,\lambda
+|h|}$, Corollary \ref{BCS theorem 2-0bis} tells us that a superconducting
phase appears, i.e., $\mathrm{r}_{\beta }>0.$ In this case, it is important
to note that the electron density becomes \textit{independent} of the
temperature. Indeed, by combining Theorem \ref{BCS theorem 2-1} with (\ref%
{BCS gap equation}) one gets that
\begin{equation}
\mathrm{d}_{\beta }=1+2\gamma ^{-1}\left( \mu -\lambda \right)
\label{suprenante inequality}
\end{equation}%
is linear as a function of $\mu $ in the domain of $(\beta ,\mu ,\lambda
,\gamma ,h)$ where $\mathrm{r}_{\beta }>0,$ i.e., in the presence of
superconductivity, see figure \ref{density-mu.eps}.

We give next the electron density per site in the zero--temperature limit $%
\beta \rightarrow \infty ,$ which straightforwardly follows from Theorem \ref%
{BCS theorem 2-1} combined with Corollary \ref{BCS theorem 2-0bis}.

\begin{corollary}[Electron density per site at zero--temperature]
\label{BCS theorem 2-1bis}\mbox{}\newline
The (infinite volume) electron density $\mathrm{d}_{\infty }=\mathrm{d}%
_{\infty }(\mu ,\lambda ,\gamma ,h)\in \lbrack 0,2]$ at zero--temperature is
equal to
\begin{equation*}
\mathrm{d}_{\infty }:=\underset{\beta \rightarrow \infty }{\lim }\mathrm{d}%
_{\beta }=1+\dfrac{\mathrm{sgn}\left( \mu -\lambda \right) }{1+\delta _{|\mu
-\lambda |,\lambda +|h|}\left( 1+\delta _{h,0}\right) }\chi _{\left[ \lambda
+|h|,\infty \right) }\left( \left\vert \mu -\lambda \right\vert \right)
\end{equation*}%
for $\gamma <\Gamma _{|\mu -\lambda |,\lambda +|h|},$ whereas within the
superconducting phase, i.e., for $\gamma >\Gamma _{|\mu -\lambda |,\lambda
+|h|}$ (Corollary \ref{BCS theorem 2-0bis}), $\mathrm{d}_{\infty }=1+2\gamma
^{-1}(\mu -\lambda )$. Recall that $\mathrm{sgn}(0):=0$.
\end{corollary}

To conclude, observe that $(2-\mathrm{d}_{\beta })$ is the density of holes
in the system. So, if $\mu >\lambda ,$ then $\mathrm{d}_{\beta }\in (1,2],$
i.e., there are more electrons than holes in the system, whereas $\mathrm{d}%
_{\beta }\in \lbrack 0,1)$ for $\mu <\lambda $, i.e., there are more holes
than electrons. This phenomenon can directly be seen in the Hamiltonian $%
\mathrm{H}_{N}$, where there is a symmetry between electrons and holes as in
the Hubbard model. Indeed, by replacing the creation operators $%
a_{x,\downarrow }^{\ast }$ and $a_{x,\uparrow }^{\ast }$ of electrons by the
annihilation operators $-b_{x,\downarrow }$ and $-b_{x,\uparrow }$ of holes,
we can map the Hamiltonian $\mathrm{H}_{N}$ (\ref{Hamiltonian BCS-Hubbard})
for electrons to another strong coupling BCS--Hubbard model for holes
defined via the Hamiltonian
\begin{eqnarray*}
\widehat{H}_{N} &:= &-\mu _{\mathrm{hole}}\sum_{x\in \Lambda _{N}}\left(
\hat{n}_{x,\uparrow }+\hat{n}_{x,\downarrow }\right) -h_{\mathrm{hole}%
}\sum_{x\in \Lambda _{N}}\left( \hat{n}_{x,\uparrow }-\hat{n}_{x,\downarrow
}\right) +2\lambda \sum_{x\in \Lambda _{N}}\hat{n}_{x,\uparrow }\hat{n}%
_{x,\downarrow } \\
&&-\frac{\gamma }{N}\sum_{x,y\in \Lambda _{N}}b_{y,\uparrow }^{\ast
}b_{y,\downarrow }^{\ast }b_{x,\downarrow }b_{x,\uparrow }+2\left( \lambda
-\mu \right) N-\gamma ,
\end{eqnarray*}%
with
\begin{equation*}
\hat{n}_{x,\downarrow }:=b_{x,\downarrow }^{\ast }b_{x,\downarrow },\mathrm{%
\ }\hat{n}_{x,\uparrow }:=b_{x,\uparrow }^{\ast }b_{x,\uparrow },\mathrm{\ }%
h_{\mathrm{hole}}:=-h\mathrm{\ and\ }\mu _{\mathrm{hole}}:=2\lambda -\mu
-\gamma N^{-1}.
\end{equation*}%
Therefore, if one knows the thermodynamic behavior of $\mathrm{H}_{N}$ for
any $h\in \mathbb{R}$ and $\mu \geq \lambda $ (regime with more electrons
than holes), we directly get the thermodynamic properties for $\mu <\lambda $
(regime with more holes than electrons), which correspond to the one given
by $\widehat{H}_{N}$ with $h_{\mathrm{hole}}=-h$ and a chemical potential
for holes $\mu _{\mathrm{hole}}>\lambda $ at large enough $N$. Note that the
last constant term in $\widehat{H}_{N}$ shifts the grand--canonical pressure
by a constant,\textit{\ }but also the (infinite volume) mean--energy per
site $\mathrm{\epsilon }_{\beta }$ (Section \ref{Section specific heat}).

\subsection{Superconductivity versus magnetization: Mei{\ss }ner effect\label%
{section Meissner effect}}

It is well--known that for magnetic fields $h$ with $|h|$ below some
critical value $\mathrm{h}_{\beta }^{(c)}$, type I superconductors become
perfectly diamagnetic in the sense that the magnetic induction in the bulk
is zero. Magnetic fields with strength above $\mathrm{h}_{\beta }^{(c)}$
destroy the superconducting phase completely. This property is the
celebrated Mei{\ss }ner or Mei{\ss }ner--Ochsenfeld effect.
For small fields $h$ (i.e., $|h|<\mathrm{h}_{\beta }^{(c)}$) the magnetic
field in the bulk of the superconductor is (almost) cancelled by the
presence of steady surface currents. As we do not analyze transport here, we
only give the magnetization density explicitly as a function of the external
magnetic field $h$ for the strong coupling BCS--Hubbard model. Note that
type II superconductors cannot be covered in the strong coupling regime
since the vortices appearing in presence of magnetic fields come from the
magnetic kinetic energy.

\begin{figure*}[hbtp]
\begin{center}
\mbox{
\leavevmode
\subfigure
{ \includegraphics[angle=0,scale=1,clip=true,width=6cm]{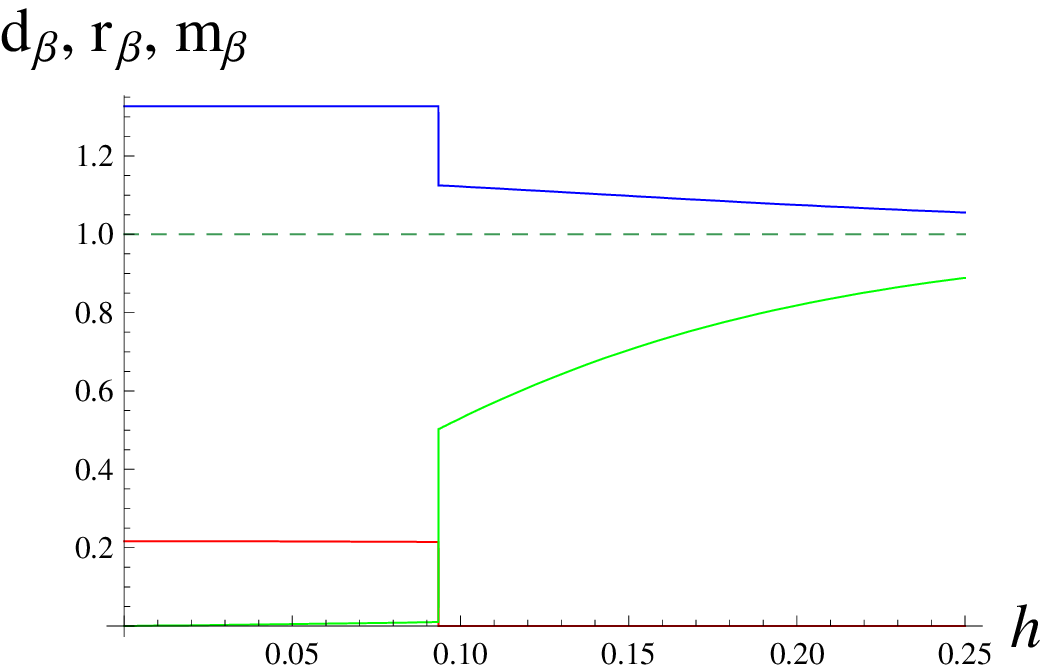} }

\leavevmode
\subfigure
{ \includegraphics[angle=0,scale=1,clip=true,width=6cm]{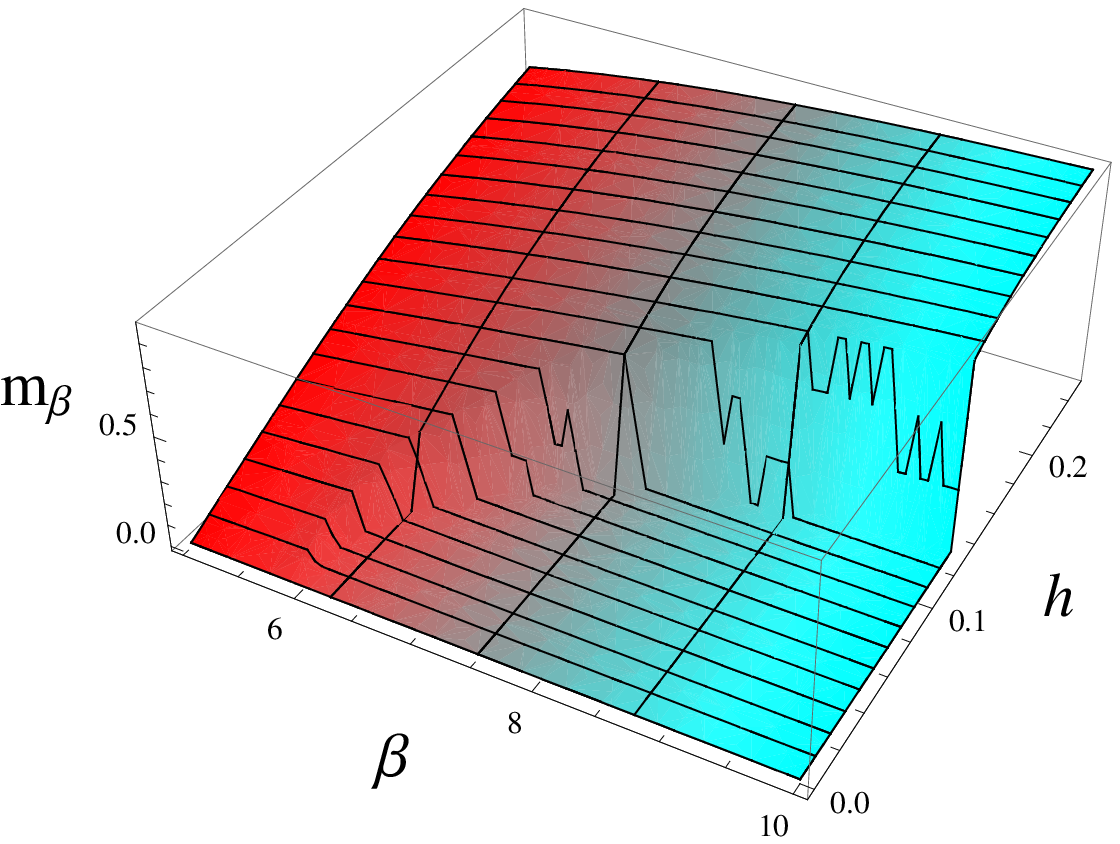} }
}
\end{center}
\caption{\emph{In the figure on the left, we have an illustration of the electron density $\mathrm{d}_{\beta }$ (blue line), the  Cooper pair condensate density $\mathrm{r}_{\beta }$
(red line) and the magnetization density $\mathrm{m}_{\beta }$ (green line) as functions of the magnetic
field $h$ at $\beta=7$, $\mu=1$, $\lambda=0.575$ and $\gamma=2.6$. The figure on the right represents a 3D
illustration of $\mathrm{m}_{\beta }=\mathrm{m}_{\beta }\left(
1,0.575,2.6,h\right) $ as a function of $h$ and $\beta$. The color from red to blue reflects the decrease of the
temperature. In both figures, we can see the Mei{\ss}ner effect (In the 3D
illustration, the area with no magnetization corresponds to $\mathrm{r}_{\beta }>0$).}}
\label{meismer.eps}
\end{figure*}

\begin{theorem}[Magnetization density]
\label{BCS theorem 2-2}\mbox{}\newline
For any $\beta ,\gamma >0$ and real numbers $\mu ,\lambda ,h$ away from any
critical point, the (infinite volume) magnetization density equals
\begin{equation*}
\underset{N\rightarrow \infty }{\lim }\left\{ \dfrac{1}{N}\sum_{x\in \Lambda
_{N}}\omega _{N}\left( n_{x,\uparrow }-n_{x,\downarrow }\right) \right\} =%
\mathrm{m}_{\beta }:=\frac{\sinh \left( \beta h\right) e^{\lambda \beta }}{%
e^{\lambda \beta }\cosh \left( \beta h\right) +\cosh \left( \beta g_{\mathrm{%
r}_{\beta }}\right) },
\end{equation*}%
with $\mathrm{m}_{\beta }=\mathrm{m}_{\beta }(\mu ,\lambda ,\gamma ,h)\in
\lbrack -1,1]$, $\mathrm{r}_{\beta }\geq 0$ being the unique solution of (%
\ref{BCS pressure 2}) and $g_{r}:=\{(\mu -\lambda )^{2}+\gamma ^{2}r\}^{1/2}$%
, see Theorem \ref{BCS theorem 1} and figure \ref{meismer.eps}.
\end{theorem}

%
This theorem deduced from Theorem \ref{Theorem equilibrium state 4} does not
seem to show any Mei{\ss }ner effect since $\mathrm{m}_{\beta }>0$ as soon
as $h\neq 0.$ However, when the Cooper pair condensate density $\mathrm{r}%
_{\beta }$ is strictly positive, from Theorem \ref{BCS theorem 2-2} combined
with (\ref{BCS gap equation}) note that
\begin{equation}
\mathrm{m}_{\beta }=\frac{2g_{\mathrm{r}_{\beta }}e^{\lambda \beta }\sinh
\left( \beta h\right) }{\gamma \sinh \left( \beta g_{\mathrm{r}_{\beta
}}\right) }.  \label{BCS equation sup 1}
\end{equation}%
In particular, it decays exponentially as $\beta \rightarrow \infty $ when $%
\mathrm{r}_{\beta }\rightarrow \mathrm{r}_{\infty }>0,$ see figure \ref%
{meismer.eps}. We give therefore the zero--temperature limit $\beta
\rightarrow \infty $ of $\mathrm{m}_{\beta }$ in the next corollary.

\begin{corollary}[Magnetization density at zero--temperature]
\label{BCS theorem 2-2bis}\mbox{}\newline
The (infinite volume) magnetization density $\mathrm{m}_{\infty }=\mathrm{m}%
_{\infty }(\mu ,\lambda ,\gamma ,h)\in \lbrack -1,1]$ at zero--temperature
is equal to
\begin{equation*}
\mathrm{m}_{\infty }:=\underset{\beta \rightarrow \infty }{\lim }\mathrm{m}%
_{\beta }=\frac{\mathrm{sgn}(h)}{1+\delta _{|\mu -\lambda |,\lambda +|h|}}%
\chi _{\lbrack 0,\lambda +|h|]}\left( \left\vert \mu -\lambda \right\vert
\right) ,
\end{equation*}%
for $\gamma <\Gamma _{|\mu -\lambda |,\lambda +|h|}$ (see Corollary \ref{BCS
theorem 2-0bis}), whereas for $\gamma >\Gamma _{|\mu -\lambda |,\lambda
+|h|} $ there is no magnetization at zero--temperature since $\mathrm{m}%
_{\beta }$ decays exponentially\footnote{%
Actually, $\mathrm{m}_{\beta }=\mathcal{O}(e^{-(\gamma -2(\lambda
+|h|))\beta /2})$ for $\gamma >\Gamma _{|\mu -\lambda |,\lambda +|h|}\geq
2(\lambda +|h|).$} as $\beta \rightarrow \infty $ to $\mathrm{m}_{\infty }=0$%
.
\end{corollary}

Consequently, there is
no superconductivity, i.e. $\mathrm{r}_{\infty }=0$,  when
 $\gamma <\Gamma _{|\mu -\lambda |,\lambda +|h|}$ and, as soon as $h\neq
0$ with $|\mu -\lambda |<\lambda +|h|$, there is a perfect magnetization at
zero--temperature, i.e., $\mathrm{m}_{\infty }=\mathrm{sgn}(h).$ Observe
that the condition $|\mu -\lambda |>\lambda +|h|$ implies from Corollary \ref%
{BCS theorem 2-1bis} that either $\mathrm{d}_{\infty }=0$ or $\mathrm{d}%
_{\infty }=2,$ which implies that $\mathrm{m}_{\infty }$ must be zero.

On the other hand, if $\gamma >\Gamma _{|\mu -\lambda |,\lambda },$ we can
define the critical magnetic field at zero--temperature by the unique
positive solution
\begin{equation}
\mathrm{h}_{\infty }^{(c)}:=\gamma \left( \frac{1}{4}+\gamma ^{-2}\left( \mu
-\lambda \right) ^{2}\right) -\lambda >0  \label{critical magnetic field}
\end{equation}%
of the equation $\Gamma _{|\mu -\lambda |,\lambda +y}=\gamma $ for $y\geq 0$%
. Then, by increasing $|h|$ up to $\mathrm{h}_{\infty }^{(c)},$ the
(zero--temperature) Cooper pair condensate density $\mathrm{r}_{\infty }$
stays constant, whereas the (zero--temperature) magnetization density $%
\mathrm{m}_{\infty }$ is zero, i.e., $\mathrm{r}_{\infty }=\mathrm{r}_{\max
} $ and $\mathrm{m}_{\infty }=0$ for $|h|<\mathrm{h}_{\infty }^{(c)}$, see
Corollary \ref{BCS theorem 2-0bis}. However, as soon as $|h|>\mathrm{h}%
_{\infty }^{(c)},$ $\mathrm{r}_{\infty }=0$ and $\mathrm{m}_{\infty }=%
\mathrm{sgn}(h)$, i.e., there is no Cooper pair and a pure magnetization
takes place. In
other words, the model manifests a pure Mei{\ss }ner effect at
zero--temperature corresponding to a superconductor of type I, cf. figure %
\ref{meismer.eps}.

Finally, note that we give an energetic interpretation of the critical
magnetic field $\mathrm{h}_{\infty }^{(c)}$ after Corollary \ref{BCS theorem
2-4bis}. Observe also that a measurement of $\mathrm{h}_{\infty }^{(c)}$ (%
\ref{critical magnetic field}) implies, for instance, a measurement of the
chemical potential $\mu $ if one would know $\gamma $ and $\lambda $, which
could be found via the asymptotic (\ref{asymptotics of specific heat}) of
the specific heat, see discussions in Section \ref{Section 5}.

\subsection{Coulomb correlation density\label{section Coulom correlation
density}}

The space distribution of electrons is still unknown and for such a
consideration, we need the (infinite volume) Coulomb correlation density%
\begin{equation}
\underset{N\rightarrow \infty }{\lim }\left\{ \frac{1}{N}\sum_{x\in \Lambda
_{N}}\omega _{N}\left( n_{x,\uparrow }n_{x,\downarrow }\right) \right\} .
\label{Coulomb correlation density}
\end{equation}%
Together with the electron and magnetization densities $\mathrm{d}_{\beta }$
and $\mathrm{m}_{\beta }$, the knowledge of (\ref{Coulomb correlation
density}) allows us in particular to explain in detail the difference
between superconducting and non--superconducting phases in terms of space
distributions of electrons.

\begin{figure*}[hbtp]
\begin{center}
\mbox{
\leavevmode
\subfigure
{ \includegraphics[angle=0,scale=1,clip=true,width=3.8cm]{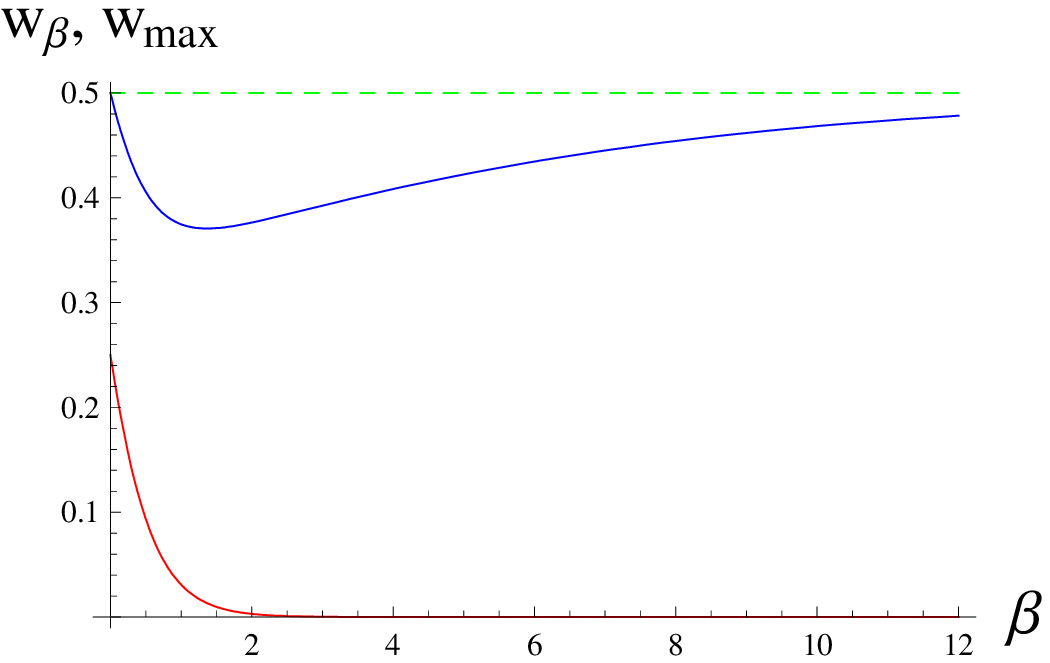} }

\leavevmode
\subfigure
{ \includegraphics[angle=0,scale=1,clip=true,width=3.8cm]{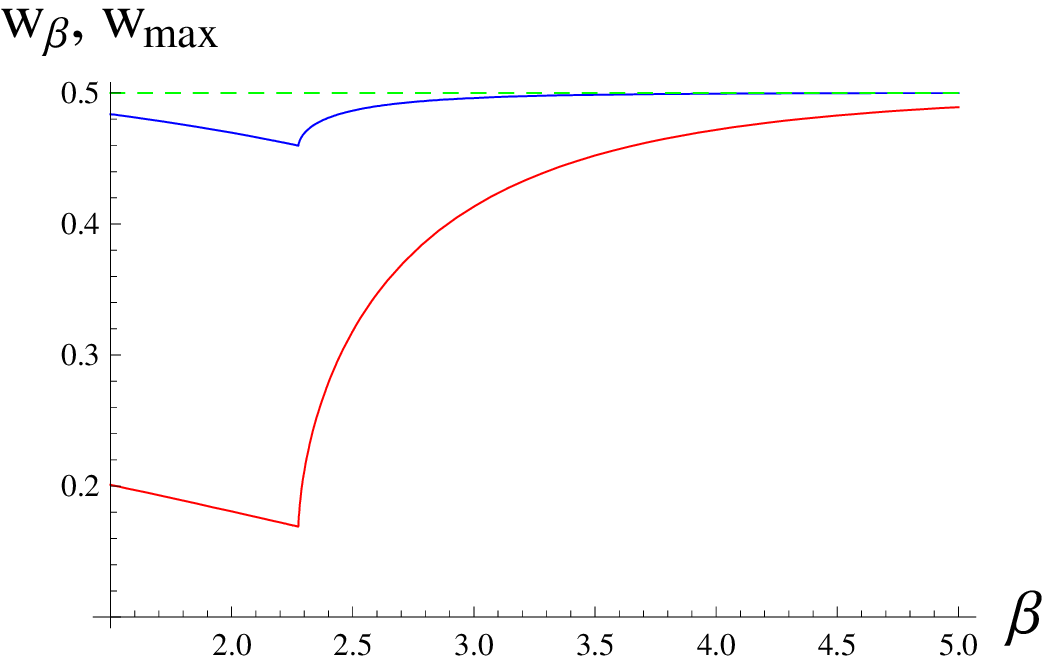} }

\leavevmode
\subfigure
{ \includegraphics[angle=0,scale=1,clip=true,width=3.8cm]{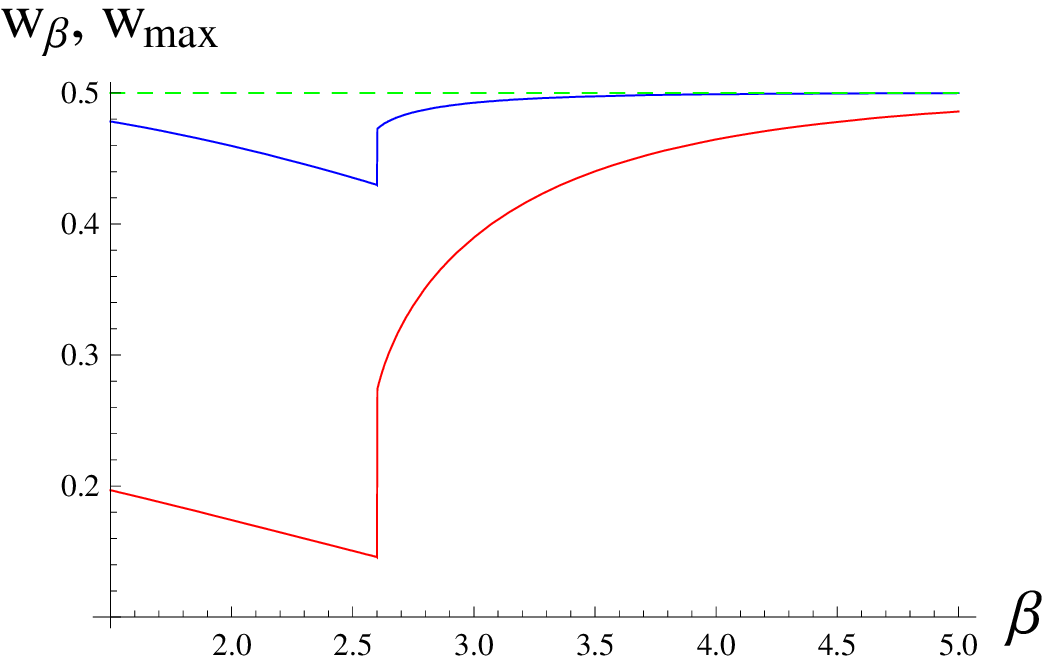} }
}
\end{center}
\caption{\emph{Illustration of the Coulomb correlation density $\mathrm{w}_{\beta }$ (red lines) and its corresponding upper
bound $\mathrm{w}_{\max }$ (blue lines) as a function of $\beta >0$ at $\mu=0.2$, $\gamma=2.6$, for $\lambda =1.305<\mu$ (left figure, $\mathrm{d}_{\beta }<1$), $\lambda  =0.2=\mu$
(two right figures, $\mathrm{d}_{\beta }=1$), and from the left to the right,
with $h=0$ ($\mathrm{m}_{\beta }=0$), and $h=0.3$, $0.35$ (where $\mathrm{m}_{\beta
}>0$). The dashed green lines indicate that $\mathrm{d}_{\infty }/2=0.5$ in
the three cases. In the figure on the left there is no superconducting phase
in opposition to the right figures where we see
a phase transition for $\beta >2.3$ (second order) or $2.6$
(first order).}}
\label{Coulomb-densite1.eps}
\end{figure*}

Actually, by the Cauchy--Schwarz inequality for the states one gets that%
\begin{equation}
\frac{1}{N}\sum_{x\in \Lambda _{N}}\omega _{N}\left( n_{x,\uparrow
}n_{x,\downarrow }\right) \leq \sqrt{\frac{1}{N}\sum_{x\in \Lambda
_{N}}\omega _{N}\left( n_{x,\uparrow }\right) }\sqrt{\frac{1}{N}\sum_{x\in
\Lambda _{N}}\omega _{N}\left( n_{x,\downarrow }\right) }.
\label{schwartz 1}
\end{equation}%
From Theorems \ref{BCS theorem 2-1} and \ref{BCS theorem 2-2}, the densities
of electrons with spin up $\uparrow $ and down $\downarrow $ equal
respectively%
\begin{equation*}
\underset{N\rightarrow \infty }{\lim }\left\{ \dfrac{1}{N}\sum_{x\in \Lambda
_{N}}\omega _{N}\left( n_{x,\uparrow }\right) \right\} =\dfrac{\mathrm{d}%
_{\beta }+\mathrm{m}_{\beta }}{2}\in \left[ 0,1\right]
\end{equation*}%
and%
\begin{equation*}
\underset{N\rightarrow \infty }{\lim }\left\{ \dfrac{1}{N}\sum_{x\in \Lambda
_{N}}\omega _{N}\left( n_{x,\downarrow }\right) \right\} =\dfrac{\mathrm{d}%
_{\beta }-\mathrm{m}_{\beta }}{2}\in \left[ 0,1\right]
\end{equation*}%
for any $\beta ,\gamma >0\ $and $\mu ,\lambda $, $h$ away from any critical
point. Consequently, by using (\ref{schwartz 1}) in the thermodynamic limit,
the (infinite volume) Coulomb correlation density is always bounded by%
\begin{equation}
0\leq \underset{N\rightarrow \infty }{\lim }\left\{ \frac{1}{N}\sum_{x\in
\Lambda _{N}}\omega _{N}\left( n_{x,\uparrow }n_{x,\downarrow }\right)
\right\} \leq \mathrm{w}_{\max }:=\frac{1}{2}\sqrt{\mathrm{d}_{\beta }^{2}-%
\mathrm{m}_{\beta }^{2}}.  \label{schwartz 2}
\end{equation}%
If for instance (\ref{Coulomb correlation density}) equals zero, then as
soon as an electron is on a definite site, the probability to have a second
electron with opposite spin at the same place goes to zero as $N\rightarrow
\infty $. In this case, there would be no formation of pairs of
electrons on a single site. This phenomenon  does not appear exactly in finite
temperature due to thermal fluctuations. Indeed, we can explicitly compute
the Coulomb correlation in the thermodynamic limit (cf. Theorem \ref{Theorem
equilibrium state 4}):

\begin{theorem}[Coulomb correlation density]
\label{BCS theorem 2-3}\mbox{}\newline
For any $\beta ,\gamma >0$ and real numbers $\mu ,\lambda ,h$ away from any
critical point, the (infinite volume) Coulomb correlation density equals%
\footnote{%
If $h=0$, then $\mathrm{w}_{\beta }(\mu ,\lambda ,\gamma ,0):=\underset{%
h\rightarrow 0}{\lim }\mathrm{w}_{\beta }(\mu ,\lambda ,\gamma ,h).$}%
\begin{equation*}
\underset{N\rightarrow \infty }{\lim }\left\{ \frac{1}{N}\sum_{x\in \Lambda
_{N}}\omega _{N}\left( n_{x,\uparrow }n_{x,\downarrow }\right) \right\} =%
\mathrm{w}_{\beta }:=\frac{1}{2}\left( \mathrm{d}_{\beta }-\mathrm{m}_{\beta
}\coth \left( \beta h\right) \right) ,
\end{equation*}%
with $\mathrm{w}_{\beta }=\mathrm{w}_{\beta }(\mu ,\lambda ,\gamma ,h)\in (0,%
\mathrm{w}_{\max })$, see figure \ref{Coulomb-densite1.eps}. Here $\mathrm{d}%
_{\beta }$ and $\mathrm{m}_{\beta }$ are respectively defined in Theorems %
\ref{BCS theorem 2-1} and \ref{BCS theorem 2-2}.
\end{theorem}

%
%
Consequently, because $g_{\mathrm{r}_{\beta }}\geq |\lambda -\mu |,$ for any
inverse temperature $\beta >0\ $the Coulomb correlation density is never
zero, i.e., $\mathrm{w}_{\beta }>0,$ even if the electron density $\mathrm{d}%
_{\beta }$ is exactly $1$, i.e., if $\lambda =\mu $. Moreover, the upper
bound in (\ref{schwartz 2}) is also never attained. However, for low
temperatures, $\mathrm{w}_{\beta }$ goes exponentially fast w.r.t. $\beta $
to one of the bounds in (\ref{schwartz 2}), cf. figure \ref%
{Coulomb-densite1.eps}. Indeed, one has the following zero--temperature
limit:

\begin{corollary}[Coulomb correlation density at zero--temperature]
\label{BCS theorem 2-3bis}\mbox{}\newline
The (infinite volume) Coulomb correlation density $\mathrm{w}_{\infty }=%
\mathrm{w}_{\infty }(\mu ,\lambda ,\gamma ,h)\in \lbrack 0,1]$ at
zero--temperature is equal to
\begin{equation*}
\mathrm{w}_{\infty }:=\underset{\beta \rightarrow \infty }{\lim }\mathrm{w}%
_{\beta }=\dfrac{1+\mathrm{sgn}\left( \mu -\lambda \right) }{2\left(
1+\delta _{|\mu -\lambda |,\lambda +|h|}\left( 1+\delta _{h,0}\right)
\right) }\chi _{\left[ \lambda +|h|,\infty \right) }\left( \left\vert \mu
-\lambda \right\vert \right)
\end{equation*}%
for $\gamma <\Gamma _{|\mu -\lambda |,\lambda +|h|}$ whereas $\mathrm{w}%
_{\infty }=\mathrm{d}_{\infty }/2$ for $\gamma>\Gamma _{|\mu -\lambda
|,\lambda +|h|}$, see Corollaries \ref{BCS theorem 2-0bis}-\ref{BCS theorem
2-1bis}.
\end{corollary}

If $\left\vert \mu -\lambda \right\vert >\lambda +|h|,$ the interpretation
of this asymptotics is clear since either $\mathrm{d}_{\infty }=0$ for $\mu
<\lambda $ or $\mathrm{d}_{\infty }=2$ for $\mu >\lambda $. The interesting
phenomena are when $\left\vert \mu -\lambda \right\vert <\lambda +|h|$. In
this case, if there is no superconducting phase, i.e., $\gamma <\Gamma
_{|\mu -\lambda |,\lambda +|h|}$, then $\mathrm{w}_{\beta }$ converges
towards $\mathrm{w}_{\infty }=0$ as $\beta \rightarrow \infty $. In
particular, as explained above, if an electron is on a definite site, the
probability to have a second electron with opposite spin at the same place
goes to zero as $N\rightarrow \infty $ and $\beta \rightarrow \infty $.

\begin{figure*}[hbtp]
\begin{center}
\mbox{
\leavevmode
\subfigure
{ \includegraphics[angle=0,scale=1,clip=true,width=3.8cm]{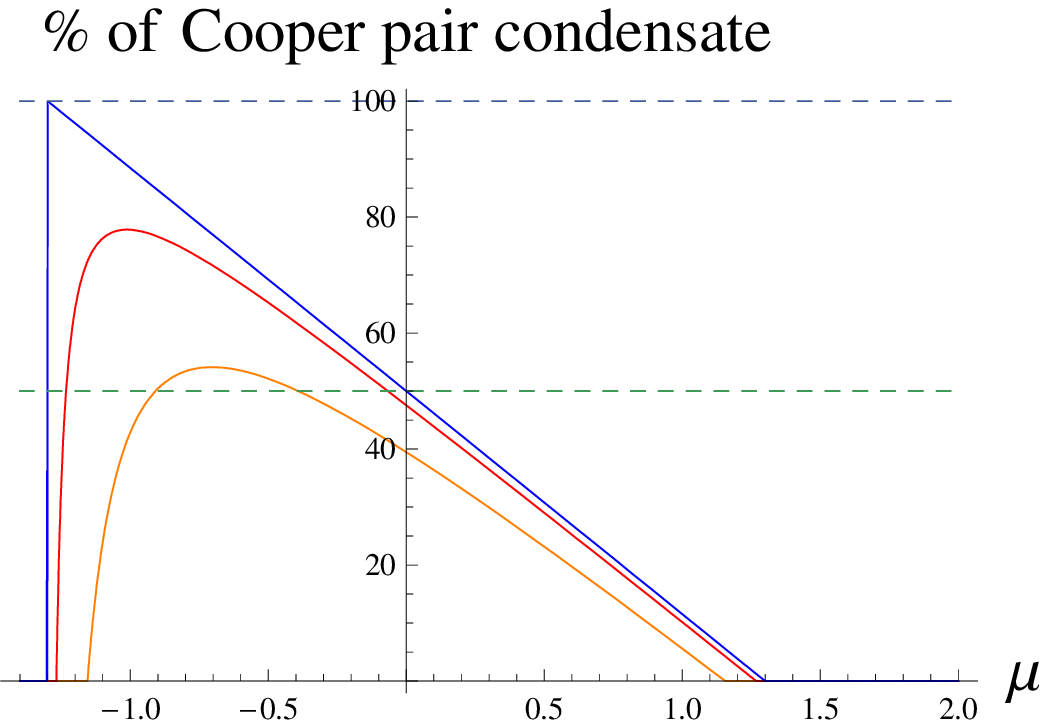} }

\leavevmode
\subfigure
{ \includegraphics[angle=0,scale=1,clip=true,width=3.8cm]{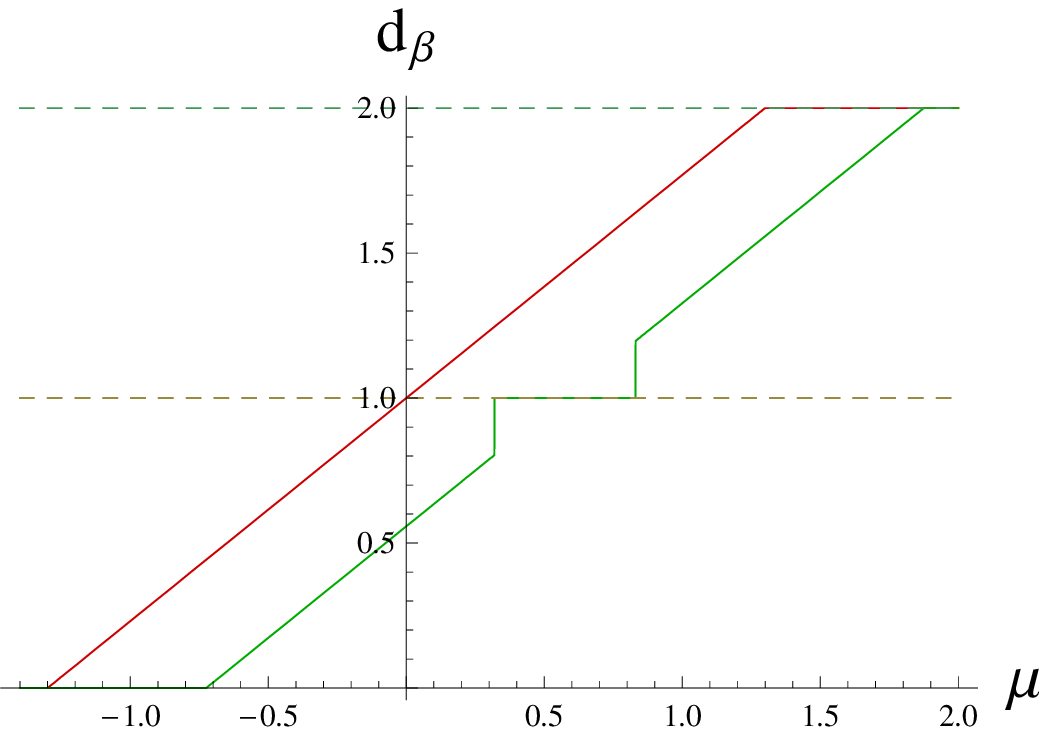} }

\leavevmode
\subfigure
{ \includegraphics[angle=0,scale=1,clip=true,width=3.8cm]{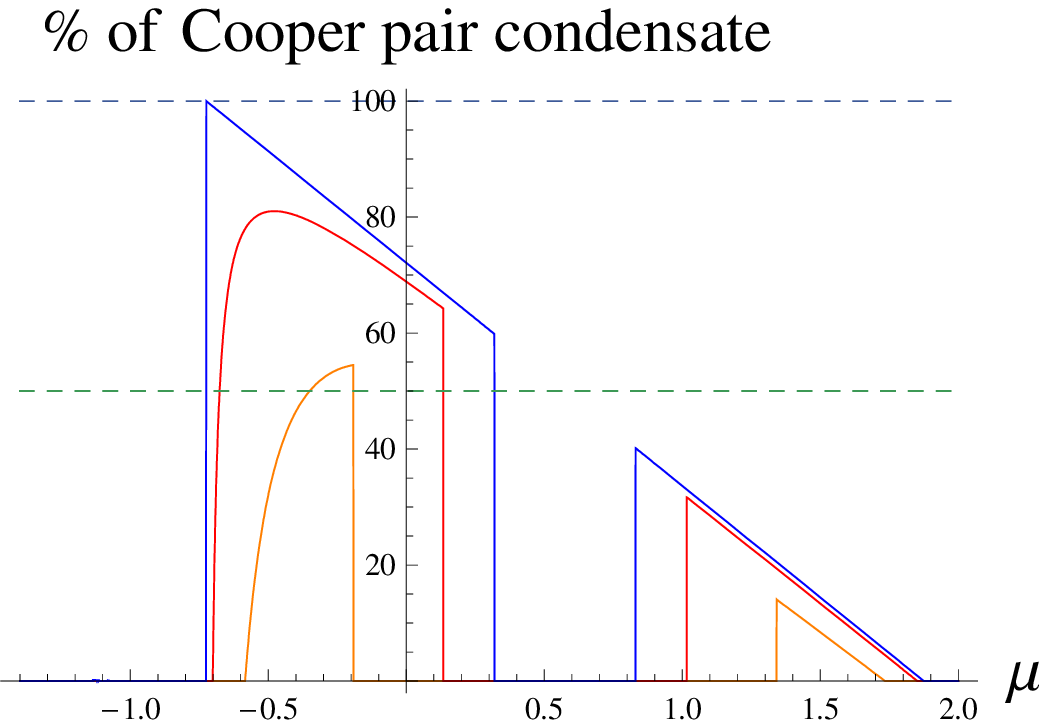} }
}
\end{center}
\caption{\emph{The fraction of electron pairs in the condensate is given in right and left figures
as a function of $\mu$. In the figure on the left, $\lambda=h=0$, with inverse temperatures $\beta =2.45$
(orange line), $3.45$ (red line) and $30$ (blue line). In the figure on the
right, $\lambda=0.575$ and $h=0.1$ with $\beta =5$ (orange line),
$7$ (red line) and $30$ (blue line). The figure on the center illustrates the
electron density $\mathrm{d}_{\beta }$ also as a function of $\mu$ at $\beta=30$ (low temperature regime) for $\lambda=h=0$ (red line)
and for $\lambda=0.575$ and $h=0.1$ (green line). In all figures, $\gamma=2.6$.}}
\label{fraction1.eps}
\end{figure*}

However, in the superconducting phase, i.e., for $\gamma >\Gamma _{|\mu
-\lambda |,\lambda +|h|}$, the upper bound $\mathrm{w}_{\max }$ (\ref%
{schwartz 2}) is asymptotically attained. Since $\mathrm{w}_{\max }=\mathrm{d%
}_{\infty }/2$ as $\beta \rightarrow \infty $, it means that 100\% of
electrons form Cooper pairs in the limit of zero--temperature, which is in
accordance with the fact that the magnetization density must disappear,
i.e., $\mathrm{m}_{\infty }=0$, cf. Corollary \ref{BCS theorem 2-2bis}. As
explained in Section \ref{Section BCS phase transition}, the highest Cooper
pair condensate density is $1/4$, which corresponds to an electron density $%
\mathrm{d}_{\infty }=1$. Actually, although all electrons form Cooper pairs
at small temperatures, there are never $100\%$ of electron pairs in the
condensate, see figure \ref{fraction1.eps}. In the special case where $%
\mathrm{d}_{\infty }=1$, only $50\%$ of Cooper pairs are in the condensate.

The same analysis can be done for hole pairs by changing $a_{x}$ by $%
-b_{x}^{\ast }$ in the definition of extensive quantities. Define the
electron and hole pair condensate fractions respectively by $\mathrm{v}%
_{\beta }:=2\mathrm{r}_{\beta }/\mathrm{d}_{\beta }$ and $\mathrm{\hat{v}}%
_{\beta }:=2\mathrm{\hat{r}}_{\beta }/\mathrm{\hat{d}}_{\beta }$, where $%
\mathrm{\hat{r}}_{\beta }$ and $\mathrm{\hat{d}}_{\beta }$ are the hole
condensate density and the hole density respectively. Because of the
electron--hole symmetry, $\mathrm{\hat{r}}_{\beta }=\mathrm{r}_{\beta }$ and
$\mathrm{\hat{d}}_{\beta }=2-\mathrm{d}_{\beta }$. In particular, when $%
\mathrm{r}_{\beta }>0$, we asymptotically get that $\mathrm{\hat{v}}_{\beta
}+\mathrm{v}_{\beta }\rightarrow 1$ as $\beta \rightarrow \infty $. Hence,
in the superconducting phase, an electron pair condensate fraction below $%
50\%$ means in fact that there are more than $50\%$ of hole pair condensate
and conversely at low temperatures. For more details concerning ground
states in relation with this phenomenon, see discussions around (\ref{ineq
paring}) in Section \ref{equilibirum.paragraph}.%
%
%

\subsection{Superconductor--Mott insulator phase transition\label{Section
Mott insulator}}

By Corollary \ref{BCS theorem 2-1bis}, if $\lambda >0$ and the system is not
in the superconducting phase (i.e., if $r_{\beta }=0$), then the electron
density converges to either $0$, $1$ or $2$ as $\beta \rightarrow \infty $
since
\begin{equation}
\mathrm{d}_{\infty }=1+\mathrm{sgn}\left( \mu -\lambda \right) .
\label{Mott equation 0}
\end{equation}%
We define the phase where the system does not form a pair condensate and the
electron density is around $1$, as a \emph{Mott insulator} phase. More
precisely, we say that the system forms a Mott insulator, if for some $%
\epsilon <1$, some $0<\beta _{0}<\infty $, some $\mu _{0}\in \mathbb{R}$ and
some $\delta \mu >0$, the electron density%
\begin{equation*}
\mathrm{d}_{\beta }\in (1-\epsilon ,1+\epsilon )\mathrm{\ and\ r}_{\beta }=0%
\mathrm{\ for\ all\ }(\beta ,\mu )\in (\beta _{0},\infty )\times (\mu
_{0}-\delta \mu ,\mu _{0}+\delta \mu ).
\end{equation*}%
As discussed in Section \ref{section Coulom correlation density}, observe
that we have, in this phase, exactly one electron (or hole) localized in
each site at the low temperature limit since $\mathrm{d}_{\beta }\rightarrow
1$ and $\mathrm{w}_{\beta }\rightarrow 0$ as $\beta \rightarrow \infty $.

To extract the whole region of parameters where such a thermodynamic phase
takes place, a preliminary analysis of the function $\Gamma _{x,y}$ defined
in Corollary \ref{BCS theorem 2-0bis} is first required. Observe that $%
\Gamma _{0,y}>0$ if and only if $y>0.$ Consequently, for any real numbers $%
\lambda $ and $h$ such that $\lambda +|h|\leq 0$ we have $\Gamma _{0,\lambda
+|h|}=0.$ However, if $\lambda +|h|>0$ then $\Gamma _{0,\lambda +|h|}>0.$
Meanwhile, at fixed $y>0$, the continuous function $\Gamma _{x,y}$ of $x\geq
0$ is convex with minimum for $x=y,$ i.e.,
\begin{equation}
\underset{x\geq 0}{\inf }\left\{ \Gamma _{x,y}\right\} =\Gamma _{y,y}=2y>0.
\label{minimum of gamma}
\end{equation}%
In particular, $\Gamma _{x,y}$ is strictly decreasing as a function of $x\in %
\left[ 0,y\right] $ and strictly increasing for $x\geq y.$
\begin{figure}[h]
\centerline{\includegraphics[angle=0,scale=1.,clip=true,width=12.5cm]{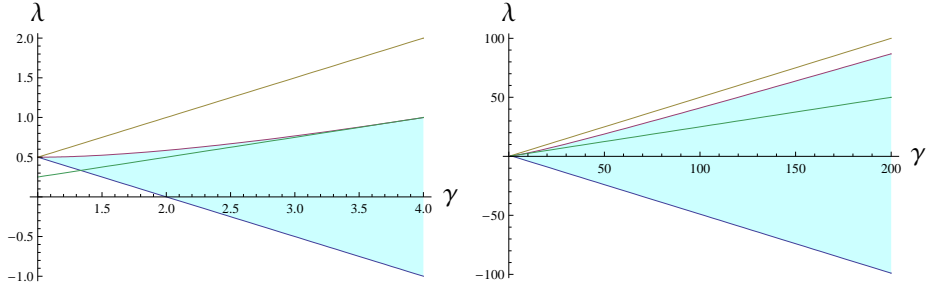}}
\caption{\emph{In both figures, the blue area represents
the domain of $(\lambda ,\gamma )$, where there is a superconducting phase at zero
temperature for $\mu =1$ and $h=0$. The two increasing straight lines (green and brown) are $\gamma
=4\lambda $ and $\gamma =2\lambda$ for $\gamma \geq 1$. In particular, between these two lines
($2\lambda <\gamma <4\lambda $), there is a superconducting-Mott-Insulator phase transition
by tuning $\mu $.}}
\label{Mott-Insulator-2.eps}
\end{figure}%

Now, by combining Corollaries \ref{BCS theorem 2-0bis}, \ref{BCS theorem
2-1bis}, \ref{BCS theorem 2-2bis} and \ref{BCS theorem 2-3bis}, we are in
position to extract the set of parameters corresponding to insulating or
superconducting phases:\newline

\noindent \textbf{1.} For any $\gamma >0$ and $\mu ,\lambda \in \mathbb{R}$
such that
\begin{equation*}
|\mu -\lambda |>\max \{\gamma /2,\lambda +|h|\},
\end{equation*}%
observe first that there are no superconductivity ($\mathrm{r}_{\infty }=0$%
), either no electrons or no holes (see (\ref{Mott equation 0})) and, in any
case, no magnetization since $\mathrm{m}_{\infty }=0$. It is a standard (non
ferromagnetic) insulator.

The next step is now to analyze the thermodynamic behavior for
\begin{equation}
|\mu -\lambda |<\max \{\gamma /2,\lambda +|h|\},  \label{Mott equation 1}
\end{equation}%
which depends on the strength of $\gamma >0.$ From \textbf{2.} to \textbf{4.}%
, we assume that (\ref{Mott equation 1}) is satisfied. \newline

\noindent \textbf{2.} If the BCS coupling constant $\gamma $ satisfies%
\begin{equation*}
0<\gamma \leq \Gamma _{\lambda +|h|,\lambda +|h|}=2(\lambda +|h|),
\end{equation*}%
then from (\ref{minimum of gamma}) combined with Corollary \ref{BCS theorem
2-0bis} there is no Cooper pair for any $\mu $ and any $\lambda .$ In
particular, under the condition (\ref{Mott equation 1}) there are a perfect
magnetization, i.e., $\mathrm{m}_{\infty }=\mathrm{sgn}(h)$, and exactly one
electron or one hole per site since $\mathrm{d}_{\infty }=1$ and $\mathrm{w}%
_{\infty }=0$. In other words, we obtain a ferromagnetic Mott insulator
phase.\newline

\noindent \textbf{3.} Now, if $\gamma >0$ becomes too strong, i.e.,
\begin{equation*}
\gamma >\Gamma _{0,\lambda +|h|}=4(\lambda +|h|),
\end{equation*}
then for any $\mu \in \mathbb{R}$ such that $|\mu -\lambda |<\gamma /2$
there are Cooper pairs because $\mathrm{r}_{\infty }=\mathrm{r}_{\max }>0,$
an electron density $\mathrm{d}_{\infty }$ equal to (\ref{suprenante
inequality}) and no magnetization ($\mathrm{m}_{\infty }=0$). In this case,
observe that all quantities are continuous at $|\mu -\lambda |=\gamma /2$.
This is a superconducting phase. \newline

\begin{figure*}[hbtp]
\begin{center}
\mbox{
\leavevmode
\subfigure
{ \includegraphics[angle=0,scale=1,clip=true,width=3.8cm]{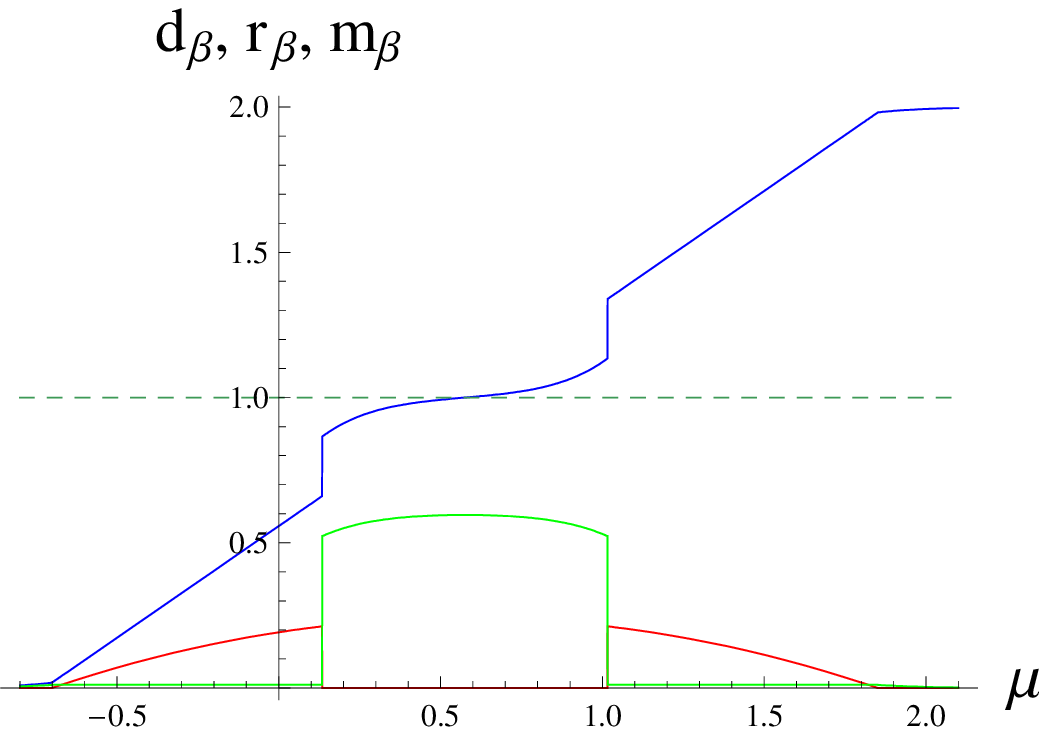} }

\leavevmode
\subfigure
{ \includegraphics[angle=0,scale=1,clip=true,width=3.8cm]{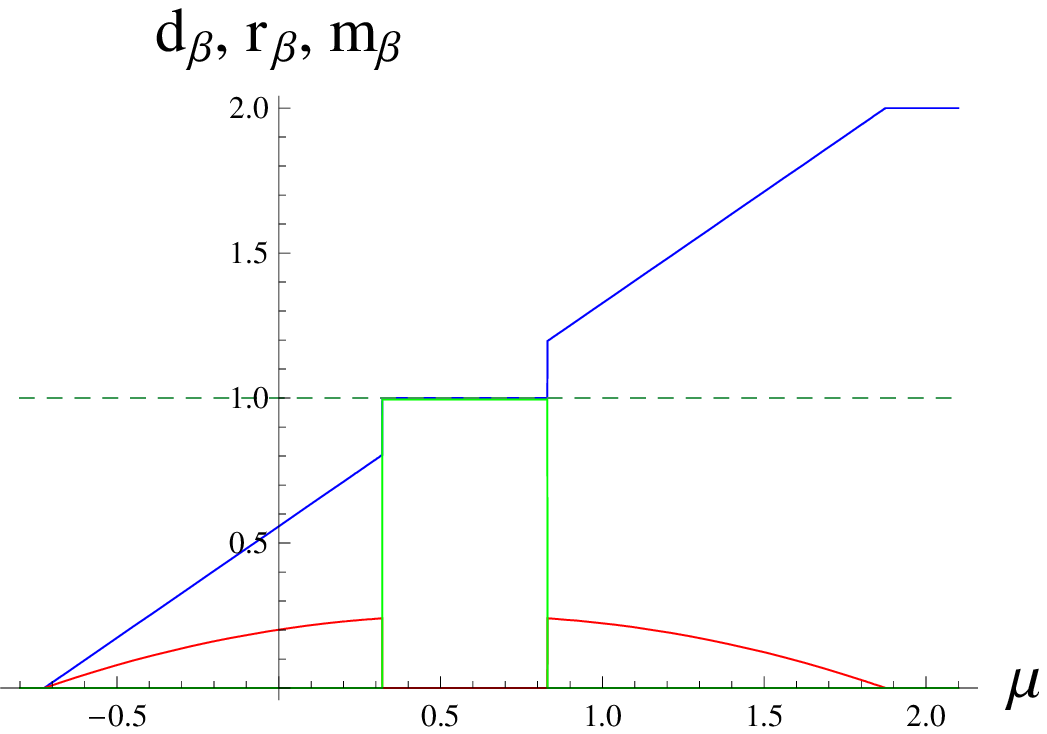} }

\leavevmode
\subfigure
{ \includegraphics[angle=0,scale=1,clip=true,width=3.8cm]{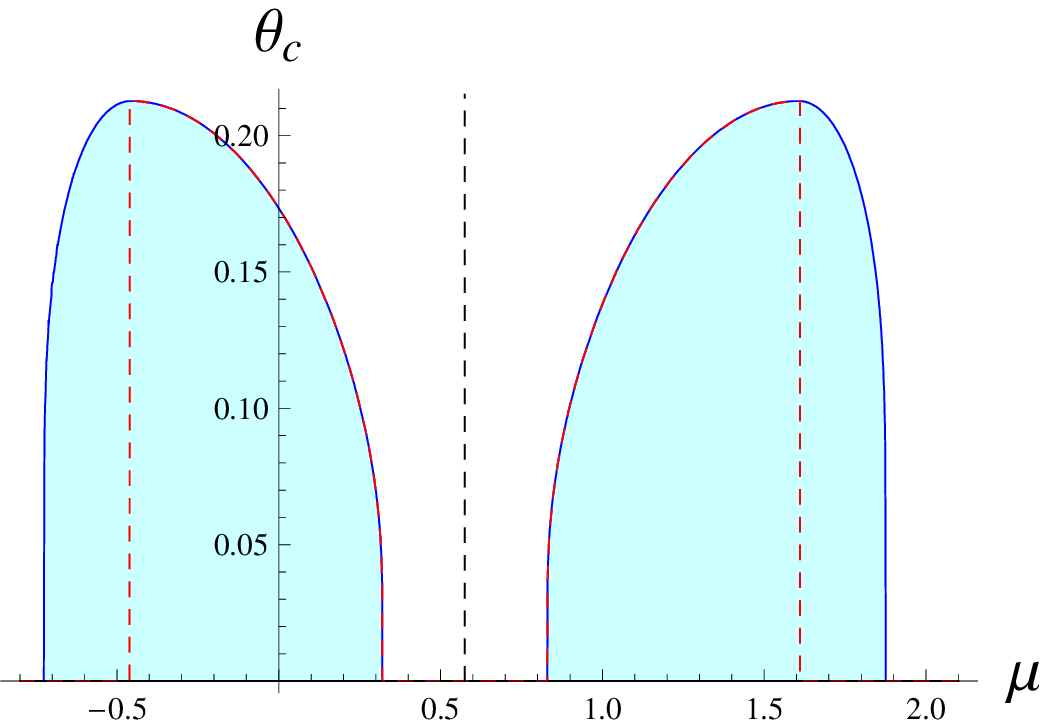} }
}
\end{center}
\caption{\emph{Here $\lambda =0.575,$ $\gamma =2.6$, and $h=0.1$. In the two figures on the left,
we plot the electron density $\mathrm{d}_{\beta }$ (blue line), the
Cooper pair condensate density $\mathrm{r}_{\beta }$ (red line) and the
magnetization density $\mathrm{m}_{\beta }$ (green line) as functions
of $\mu$ for $\beta =7$ (left figure) or $30$ (low temperature regime, figure on the center).
Observe the superconducting-Mott Insulator phase transition which appears in
both cases. In the right figure, we illustrate as a function of $\mu$
the corresponding critical temperature $\theta _{c}$.
The blue line corresponds to a second order phase transition,
whereas the red dashed line represents the domain of $\mu$ with first order phase transition.
The black dashed line is the chemical potential $\mu=\lambda$
corresponding to an electron density per site equal to $1$.}}
\label{Mott-Insulator-1.eps}
\end{figure*}

\noindent \textbf{4.} The superconducting--Mott insulator phase transition
only appears in the intermediary regime where
\begin{equation}
\Gamma _{\lambda +|h|,\lambda +|h|}=2\left( \lambda +\left\vert h\right\vert
\right) <\gamma <\Gamma _{0,\lambda +|h|}=4\left( \lambda +\left\vert
h\right\vert \right) ,  \label{BCS mott insulator 1}
\end{equation}%
cf. figure \ref{Mott-Insulator-2.eps}. Indeed, the function $\Gamma
_{x,\lambda +|h|}=\gamma $ has two solutions%
\begin{equation*}
\mathrm{x}_{1}:=\frac{\gamma ^{1/2}}{2}\{4\left( \lambda +|h|\right) -\gamma
\}^{1/2}\mathrm{\ and\;x}_{2}:=\frac{\gamma }{2}>\mathrm{x}_{1}.
\end{equation*}%
In particular, for any $\mu \in \mathbb{R}$ such that $|\mu -\lambda |\in (%
\mathrm{x}_{1},\gamma /2)$, the BCS coupling constant $\gamma $ is strong
enough to imply the superconductivity ($\mathrm{r}_{\infty }=\mathrm{r}%
_{\max }>0$), with an electron density $\mathrm{d}_{\infty }$ equal to (\ref%
{suprenante inequality}) and no magnetization ($\mathrm{m}_{\infty }=0$). We
are in the superconducting phase. However, for any $\mu \in \mathbb{R}$ such
that $|\mu -\lambda |<\mathrm{x}_{1},$ the BCS coupling constant $\gamma $
becomes too weak and there is no superconductivity ($\mathrm{r}_{\infty }=0$%
), exactly one electron per site, i.e., $\mathrm{d}_{\infty }=1$ and $%
\mathrm{w}_{\infty }=0$, and a pure magnetization if $h\neq 0,$ i.e., $%
\mathrm{m}_{\infty }=\mathrm{sgn}(h)$. In this regime, one gets a
ferromagnetic Mott insulator phase. All quantities are continuous at $|\mu
-\lambda |=\gamma /2$ but not for $|\mu -\lambda |=\mathrm{x}_{1}.$ In other
words, we get a superconductor--Mott insulator phase transition by tuning in
the chemical potential $\mu $. An illustration of this phase transition is
given in figure \ref{Mott-Insulator-1.eps}, see also figure \ref%
{fraction1.eps}.%
%
%

\subsection{Mean--energy per site and the specific heat\label{Section
specific heat}}

To conclude, low--\textrm{T}$_{c}$ superconductors and high--\textrm{T}$_{c}$
superconductors differ by the behavior of their specific heat. The first one
shows a discontinuity of the specific heat at the critical point whereas the
specific heat for high--\textrm{T}$_{c}$ superconductors is continuous. It
is therefore interesting to give now the mean--energy per site in the
thermodynamic limit in order to compute next the specific heat.

\begin{figure*}[hbtp]
\begin{center}
\mbox{
\leavevmode
\subfigure
{ \includegraphics[angle=0,scale=1,clip=true,width=3.8cm]{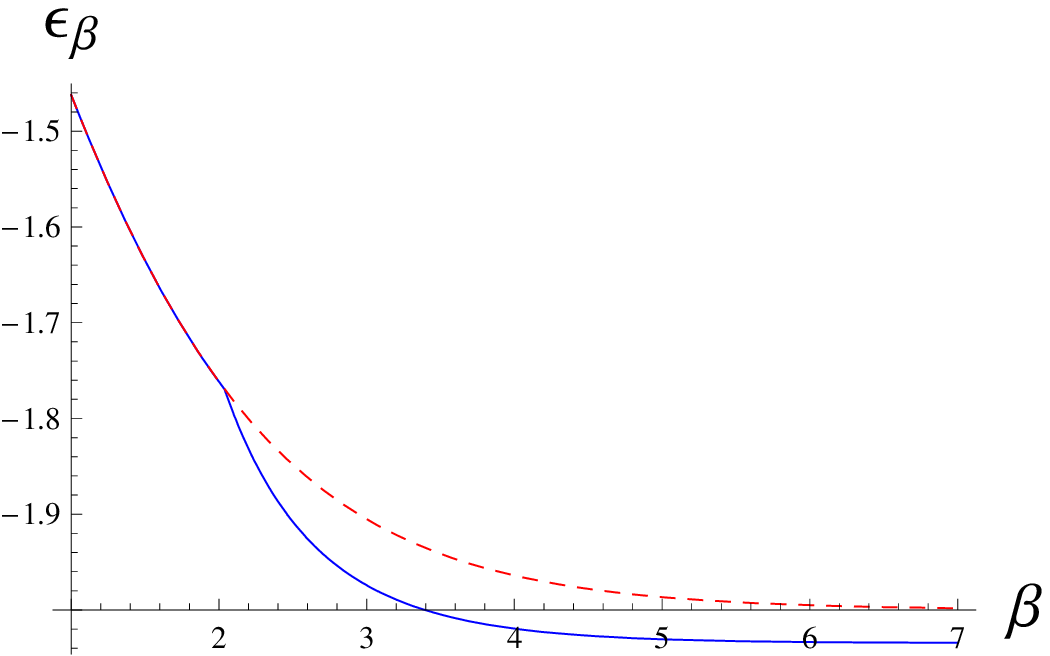} }

\leavevmode
\subfigure
{ \includegraphics[angle=0,scale=1,clip=true,width=3.8cm]{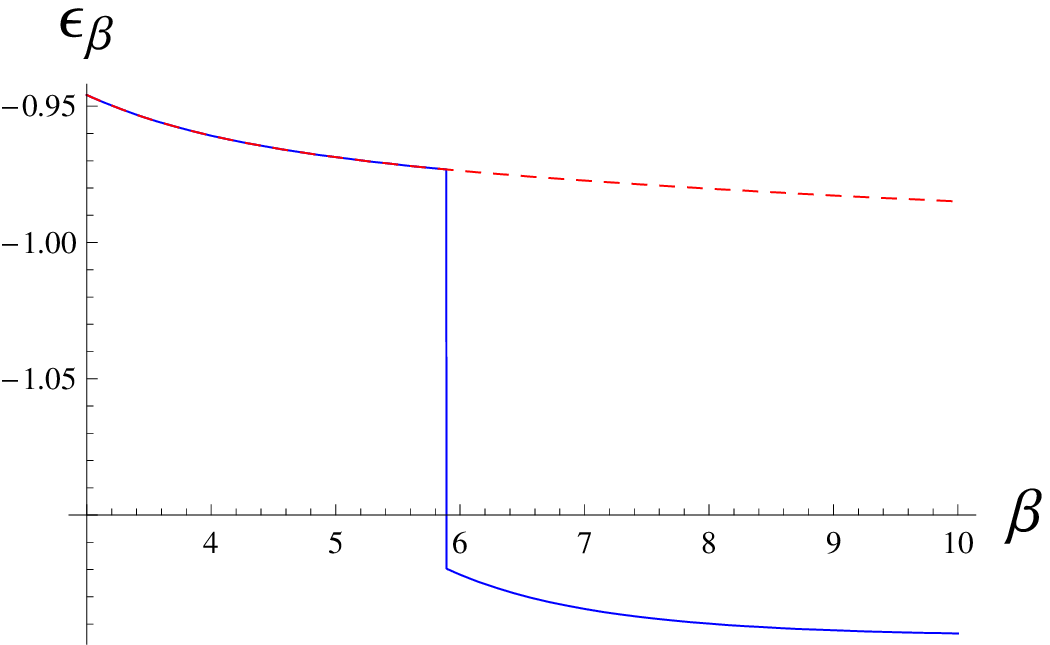} }

\leavevmode
\subfigure
{ \includegraphics[angle=0,scale=1,clip=true,width=3.8cm]{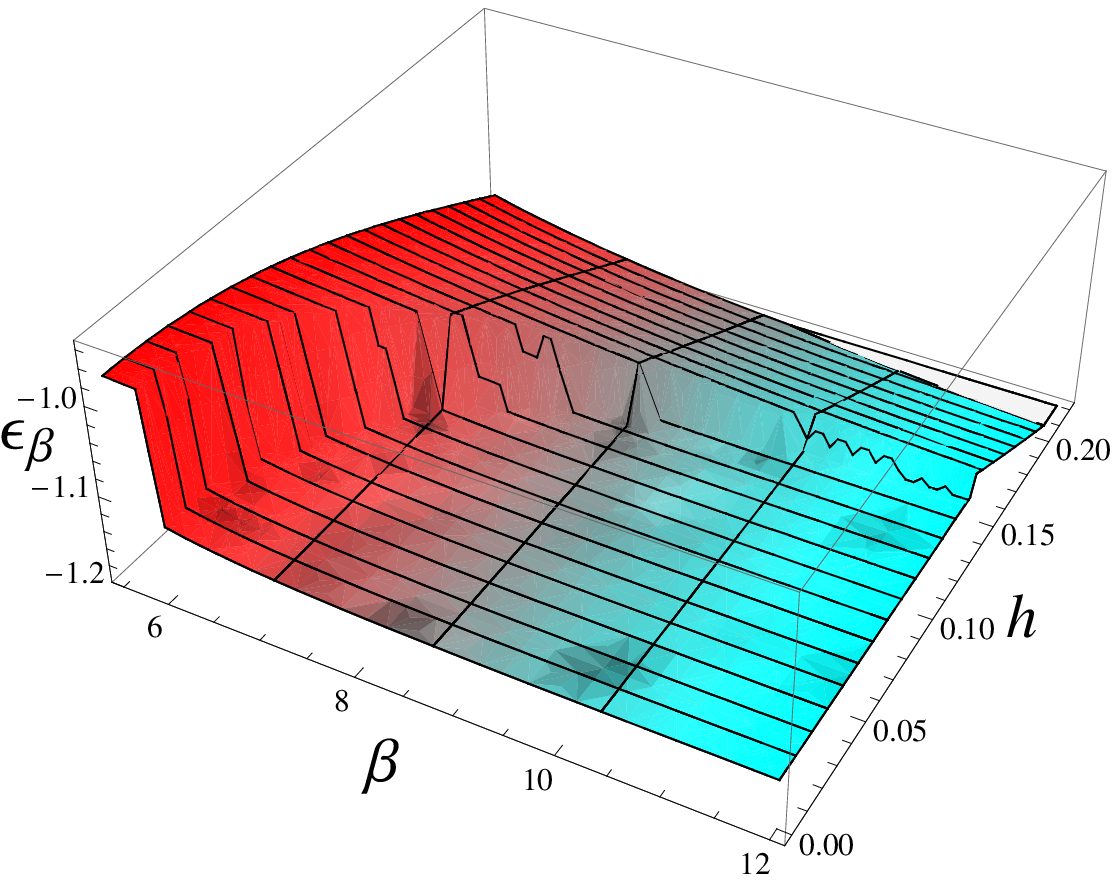} }
}
\end{center}
\caption{\emph{In the two figures on the left, we give the mean energy per site $\mathrm{\epsilon }_{\beta }$ as a
function of $\beta $ at $h=0$ for $\lambda =0$ (figure on the
left, second order BCS phase transition) or $\lambda =0.575$ (figure on the
center, first order phase transition). The dashed line in both figures is the
mean energy per site with zero Cooper pair condensate density. On the right figure, $\mathrm{\epsilon }_{\beta }$ is given as a function of $\beta $ and $h$ at $\lambda=0.575$. The color
from red to blue reflects the decrease of the temperature and the plateau
corresponds to the superconducting phase. In all figures, $\mu=1$ and $\gamma=2.6$.}}
\label{Energiemean.eps}
\end{figure*}

\begin{theorem}[Mean-energy per site]
\label{BCS theorem 2-4}\mbox{}\newline
For any $\beta ,\gamma >0$ and real numbers $\mu ,\lambda ,h$ away from any
critical point, the (infinite volume) mean energy per site is equal to
\begin{equation*}
\underset{N\rightarrow \infty }{\lim }\left\{ N^{-1}\omega _{N}\left(
\mathrm{H}_{N}\right) \right\} =\mathrm{\epsilon }_{\beta }:=-\mu \mathrm{d}%
_{\beta }-h\mathrm{m}_{\beta }+2\lambda \mathrm{w}_{\beta }-\gamma \mathrm{r}%
_{\beta },
\end{equation*}%
see Theorems \ref{BCS theorem 2-0}, \ref{BCS theorem 2-1}, \ref{BCS theorem
2-2}, \ref{BCS theorem 2-3} and figure \ref{Energiemean.eps}.
\end{theorem}

%
%
At zero--temperature, Corollaries \ref{BCS theorem 2-0bis}, \ref{BCS theorem
2-1bis}, \ref{BCS theorem 2-2bis} and \ref{BCS theorem 2-3bis} imply an
explicit computation of the mean energy per site:

\begin{corollary}[Mean-energy per site at zero--temperature]
\label{BCS theorem 2-4bis}\mbox{}\newline
The (infinite volume) mean energy per site $\mathrm{\epsilon }_{\infty }=%
\mathrm{\epsilon }_{\infty }(\mu ,\lambda ,\gamma ,h)$ at zero--temperature
is equal to
\begin{eqnarray*}
\mathrm{\epsilon }_{\infty }&:=&\underset{\beta \rightarrow \infty }{\lim }%
\mathrm{\epsilon }_{\beta }=-\mu +\dfrac{\lambda +\left\vert \lambda -\mu
\right\vert }{1+\delta _{|\mu -\lambda |,\lambda +|h|}\left( 1+\delta
_{h,0}\right) }\chi _{\left[ \lambda +|h|,\infty \right) }\left( \left\vert
\mu -\lambda \right\vert \right) \\
&&-\frac{\left\vert h\right\vert }{1+\delta _{|\mu -\lambda |,\lambda +|h|}}%
\chi _{\lbrack 0,\lambda +|h|]}\left( \left\vert \mu -\lambda \right\vert
\right) ,
\end{eqnarray*}%
for $\gamma <\Gamma _{|\mu -\lambda |,\lambda +|h|}$ whereas for $\gamma
>\Gamma _{|\mu -\lambda |,\lambda +|h|}$
\begin{equation*}
\mathrm{\epsilon }_{\infty }:=\underset{\beta \rightarrow \infty }{\lim }%
\mathrm{\epsilon }_{\beta }=-\frac{\gamma }{4}+\left( \lambda -\mu \right)
\left( 1+\gamma ^{-1}\left( \mu -\lambda \right) \right) ,
\end{equation*}%
cf. Corollary \ref{BCS theorem 2-0bis}.
\end{corollary}

Note that the critical magnetic field $\mathrm{h}_{\infty }^{(c)}$ (\ref%
{critical magnetic field}) has a direct interpretation in terms of the
zero--temperature mean energy per site $\mathrm{\epsilon }_{\infty }.$
Indeed, if $\left\vert \mu -\lambda \right\vert <\lambda +|h|,$ i.e., $%
\mathrm{d}_{\infty }\notin \{0,2\},$ by equating $\mathrm{\epsilon }_{\infty
}$ in the superconducting phase with the mean energy $\mathrm{\epsilon }%
_{\infty }=-\mu -|h|$ in the non--superconducting (ferromagnetic) state, we
directly get that the magnetic field should be equal to $|h|=\mathrm{h}%
_{\infty }^{(c)}$ (\ref{critical magnetic field}). In other words, the
critical magnetic field $\mathrm{h}_{\infty }^{(c)}$ corresponds to the
point where the mean energies at zero-temperature in both cases are equal to
each other, as it should be. Note that this phenomenon is not true at
non--zero temperature since the mean energy per site can be discontinuous as
a function of $h$ (even if $\lambda =0$), see figure \ref{Energiemean.eps}.

Now, the specific heat at finite volume equals
\begin{equation}
\mathrm{c}_{N,\beta }:=-\beta ^{2}\partial _{\beta }\left\{ N^{-1}\omega
_{N}\left( \mathrm{H}_{N}\right) \right\} =N^{-1}\beta ^{2}\omega _{N}\left( %
\left[ \mathrm{H}_{N}-\omega _{N}\left( \mathrm{H}_{N}\right) \right]
^{2}\right) .  \label{specific heat at finite volume}
\end{equation}%
However, its thermodynamic limit
\begin{equation}
\mathrm{c}_{\beta }:=\underset{N\rightarrow \infty }{\lim }\mathrm{c}%
_{N,\beta }=-\beta ^{2}\partial _{\beta }\mathrm{\epsilon }_{\beta }+%
\mathfrak{C}_{\beta }  \label{specific heat at infinite volume}
\end{equation}%
cannot be easily computed because one cannot exchange the limit $%
N\rightarrow \infty $ and the derivative $\partial _{\beta },$ i.e., $%
\mathfrak{C}_{\beta }=\mathfrak{C}_{\beta }(\mu ,\lambda ,\gamma ,h)$ may be
non--zero. For instance, Griffiths arguments \cite%
{BruZagrebnov8,Griffiths1,HeppLieb} (Section \ref{section proof griffiths})
would allow to exchange any derivative of the pressure $\mathrm{p}_{N}$ and
the limit $N\rightarrow \infty $ by using the convexity of $\mathrm{p}_{N}$.
To compute (\ref{specific heat at infinite volume}) in this way, we would
need to prove the (piece--wise) convexity of $\mathrm{\epsilon }_{N,\beta
}:=N^{-1}\omega _{N}\left( \mathrm{H}_{N}\right) $ as a function $\beta >0.$
As suggested by figure \ref{Energiemean.eps}, this property of convexity
might be right but it is not proven here.
%
%
\begin{figure*}[hbtp]
\begin{center}
\mbox{
\leavevmode
\subfigure
{ \includegraphics[angle=0,scale=1,clip=true,width=3.8cm]{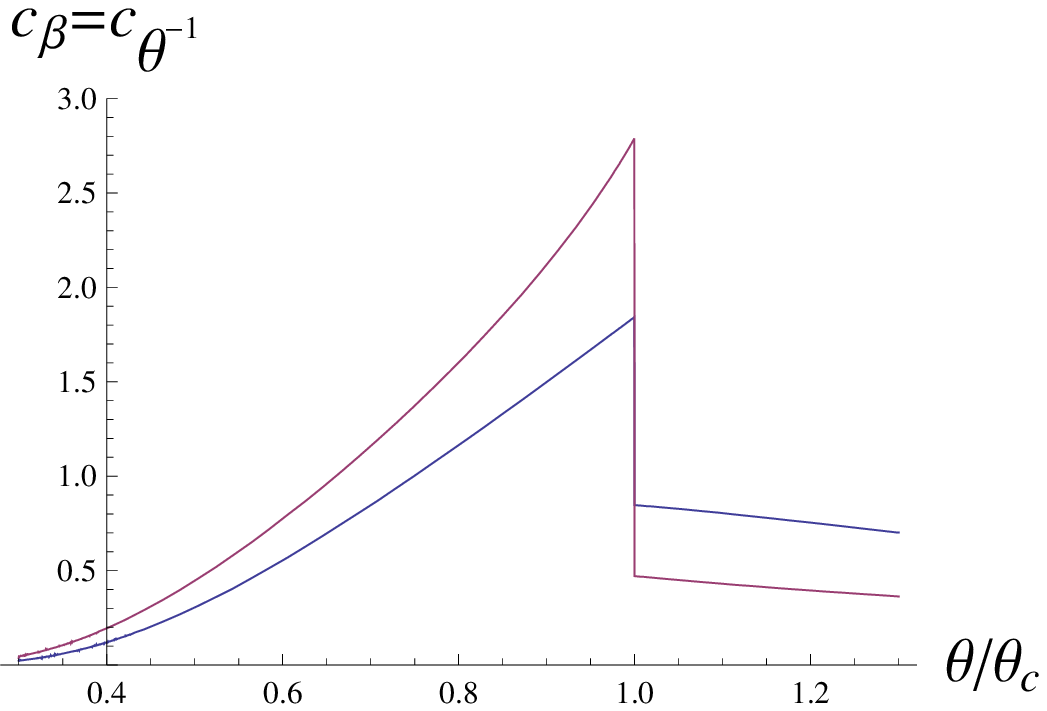} }

\leavevmode
\subfigure
{ \includegraphics[angle=0,scale=1,clip=true,width=3.8cm]{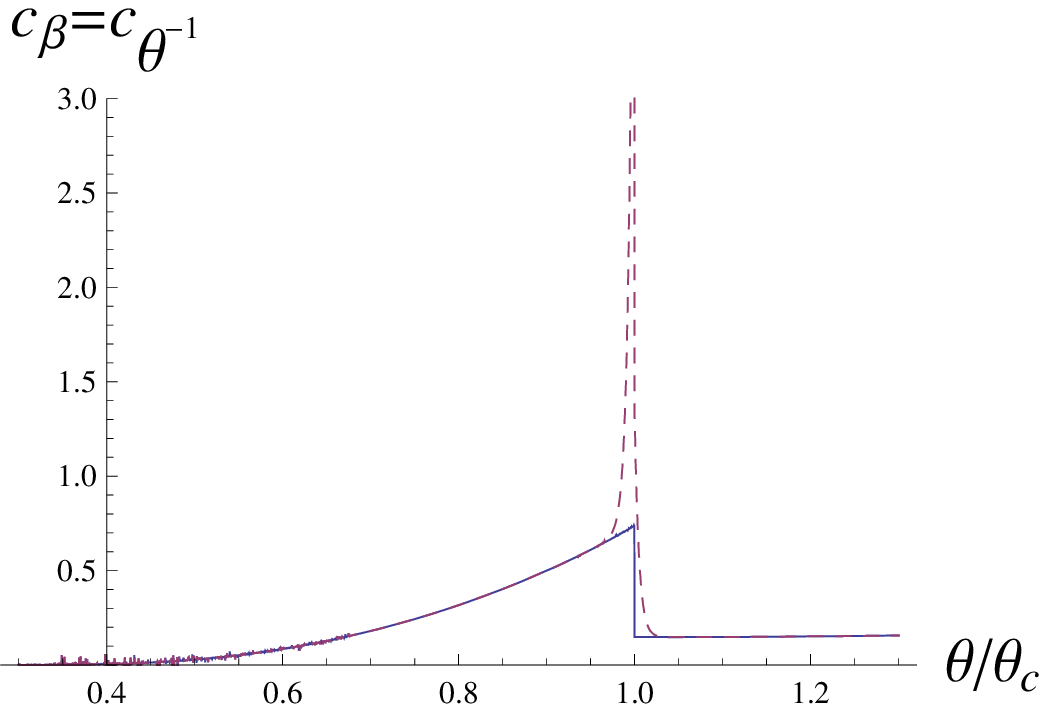} }

\leavevmode
\subfigure
{ \includegraphics[angle=0,scale=1,clip=true,width=3.8cm]{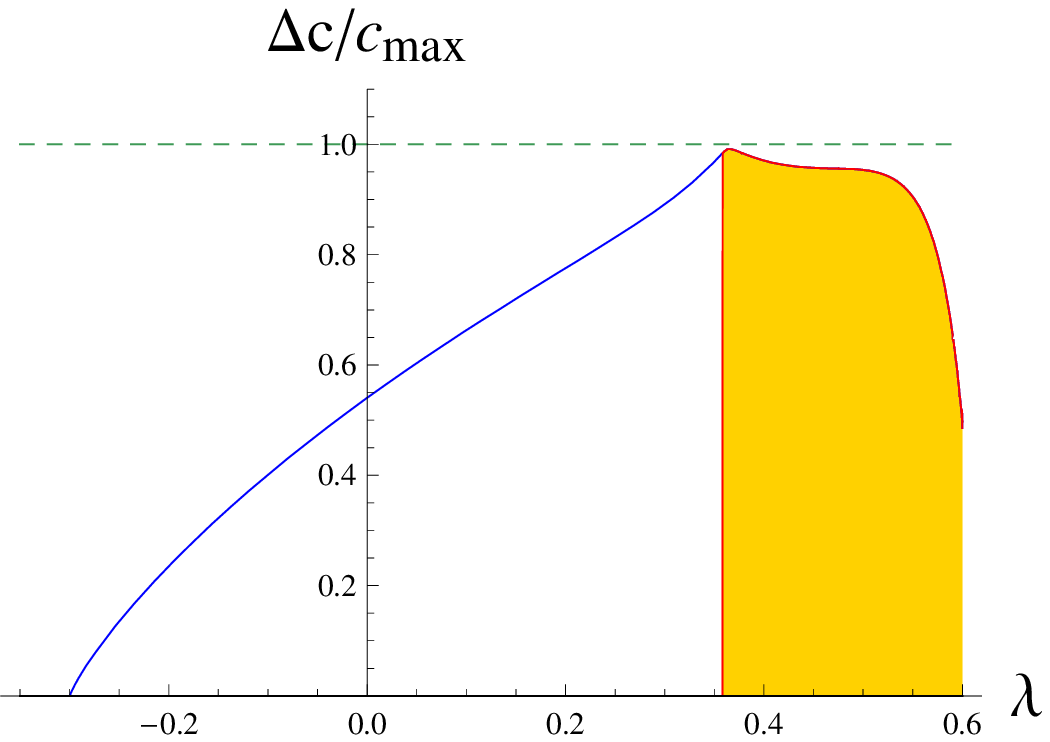} }
}
\end{center}
\caption{\emph{Here  $\mu=1$, $\gamma=2.6$ and $h=0$. Assuming $\mathfrak{C}_{\beta }=0,$ we give 3 plots of the specific heat $\mathrm{c}_{\beta }$ as a function of the ratio $\theta
/\theta _{c}$ between $\theta :=\beta ^{-1}$ and the critical temperature $\theta _{c}$ for $\lambda =0$, $0.5$ (both left figure,
respectively blue and red lines, second order phase transition), and $\lambda =0.575$ (figure on the center, blue line, first order phase
transition). The dashed red line in the figure on the center indicates what the specific heat at finite
volume might be since $\mathrm{c}_{\theta _{c}^{-1}}=+\infty$.
The right figure is a plot as a function of $\lambda $ of the relative specific heat jump, i.e., the ratio $\Delta c/c_{\max }$ between the jump $\Delta c$ at $\theta =\theta _{c}$ and
the maximum value $c_{\max }$ of $\mathrm{c}_{\theta _{c}^{-1}}$ at the same point. The yellow colored area indicates that this ratio
numerically computed is formally infinite due to a first order phase
transition.}}
\label{specific-heat.eps}
\end{figure*}%

Notice however that if experimental measurements of the specific heat comes
from a \textit{discrete} derivative of the mean energy per site $\mathrm{%
\epsilon }_{\beta },$ it is then clear that it corresponds to forget about
the term $\mathfrak{C}_{\beta }$. In this case, i.e., assuming $\mathfrak{C}%
_{\beta }=0$, we find again the well--known BCS--type behavior of the
specific heat in presence of a second order phase transition, see figure \ref%
{specific-heat.eps}. In addition, if $\mathfrak{C}_{\beta }=0$, then for any
$\mu ,$ $\lambda $, $h$ and $\gamma >\Gamma _{|\mu -\lambda |,\lambda +|h|}$
(Corollary \ref{BCS theorem 2-0bis}), we explicitly obtain via direct
computations the well--known exponential decay of the specific heat at
zero-temperature for s--wave superconductors:
\begin{equation}
\mathrm{c}_{\beta }=\frac{1}{4}\left( 2\lambda \gamma +\gamma ^{2}-4\lambda
^{2}\right) \beta ^{2}e^{-\beta \gamma }+o\left( \beta ^{2}e^{-\beta \gamma
}\right) \mathrm{\ as\ }\beta \rightarrow \infty .
\label{asymptotics of specific heat}
\end{equation}%
(Note that this asymptotic could give access to $\gamma $ and also $\lambda $%
, see discussions in Section \ref{Section 5}.) However, if a first order
phase transition appears, then the (infinite volume) mean energy per site $%
\mathrm{\epsilon }_{\beta }$ is discontinuous at the critical temperature $%
\theta _{c}$ (cf. figure \ref{Energiemean.eps}) and the specific heat $%
\mathrm{c}_{\theta _{c}^{-1}}$ is infinite. In figure \ref{specific-heat.eps}
we give an illustration of the ratio $\Delta c/c_{\max }$ between the jump $%
\Delta c$ at $\theta =\theta _{c}$ and the maximum value $c_{\max }$ of $%
\mathrm{c}_{\theta _{c}^{-1}}.$ For most of standard superconductors%
\footnote{%
\bigskip at least for the following elements: Hg, In, Nb, Pb, Sn, Ta, Tl, V.}
note that the measured values are between 0.6 and 0.7. Numerical
computations suggest that this ratio $\Delta c/c_{\max }$ may always be
bounded in our model by one as soon as a second order phase transition
appears.

\section{Phase diagram at fixed electron density per site\label{Section
phase diagram rho fixed}}

In any finite volume, the electron density per site is strictly increasing
as a function of the chemical potential $\mu $ by strict convexity of the
pressure. Therefore, for any fixed electron density $\rho \in (0,2)$ there
exists a unique $\mu _{N,\beta }=\mu _{N,\beta }(\rho ,\lambda ,\gamma ,h)$
such that
\begin{equation}
\rho =\dfrac{1}{N}\sum_{x\in \Lambda _{N}}\omega _{N}\left( n_{x,\uparrow
}+n_{x,\downarrow }\right) ,  \label{mu fixed particle density}
\end{equation}%
where $\omega _{N}$ represents the (finite volume) grand--canonical Gibbs
state (\ref{BCS gibbs state Hn}) associated with $\mathrm{H}_{N}$ and taken
at inverse temperature $\beta $ and chemical potential $\mu =\mu _{N,\beta }$%
. The aim of this section is now to analyze the thermodynamic properties of
the model for a fixed $\rho $ instead of a fixed chemical potential $\mu .$
We start by investigating it away from any critical point.

\subsection{Thermodynamics away from any critical point\label{Section phase
diagram rho fix sans critical}}

In the thermodynamic limit and away from any critical point, the chemical
potential $\mu _{N,\beta }$ converges to a solution $\mathrm{\mu }_{\beta }=%
\mathrm{\mu }_{\beta }(\rho ,\lambda ,\gamma ,h)$ of the equation%
\begin{equation}
\rho =\mathrm{d}_{\beta }\left( \mu ,\lambda ,\gamma ,h\right) ,
\label{mu fixed particle density 2}
\end{equation}%
see Theorem \ref{BCS theorem 2-1}. For instance, if $\rho =1,\ $the chemical
potential $\mathrm{\mu }_{\beta }$ is simply given by $\lambda $, i.e., $%
\mathrm{\mu }_{\beta }(1,\lambda ,\gamma ,h)=\lambda $. At least away from any
critical point, this chemical potential $\mathrm{\mu }_{\beta }$ is always
uniquely defined.

Indeed, outside the superconducting phase (see Section \ref{Section BCS
phase transition}), the electron density $\mathrm{d}_{\beta }$ given by
Theorem \ref{BCS theorem 2-1} is a strictly increasing continuous function
of the chemical potential $\mu $ at fixed $\beta >0$. In other words, for
any fixed electron density $\rho \in \left( 0,2\right) $, the equation (\ref%
{mu fixed particle density 2}) has a unique solution $\mathrm{\mu }_{\beta }$%
, i.e., the chemical potential $\mathrm{\mu }_{\beta }$ is the inverse of
the electron density $\mathrm{d}_{\beta }$ taken as a function of $\mu \in
\mathbb{R}.$

On the other hand, inside the superconducting phase, from (\ref{suprenante
inequality}) the chemical potential $\mathrm{\mu }_{\beta }$ is also unique
and equals%
\begin{equation}
\mathrm{\mu }_{\beta }=\frac{\gamma }{2}\left( \rho -1\right) +\lambda ,
\label{renormalized chemical potential}
\end{equation}%
see figures \ref{density-mu.eps} and \ref{Mott-Insulator-1.eps}. In
particular, $\mathrm{\mu }_{\beta }$ does not depend on $h$ or $\beta $ as
soon as $\mathrm{r}_{\beta }>0$. The gap equation (\ref{BCS gap equation})
then equals
\begin{equation*}
\tanh \left( \beta \gamma \mathfrak{g}_{r}\right) =2\mathfrak{g}_{r}\left( 1+%
\frac{e^{\lambda \beta }\cosh \left( \beta h\right) }{\cosh \left( \beta
\gamma \mathfrak{g}_{r}\right) }\right) ,\mathrm{\ with\;}\mathfrak{g}_{r}:=%
\frac{1}{2}\{(\rho -1)^{2}+4r\}^{1/2},
\end{equation*}%
and
\begin{equation*}
0\leq \mathrm{r}_{\beta }\leq \max \left\{ 0,\rho \left( 2-\rho \right)
/4\right\} ,
\end{equation*}%
for any fixed electron density $\rho \in (0,2)$.

Hence, the thermodynamic behavior of the strong coupling BCS--Hubbard model $%
\mathrm{H}_{N}$ is simply given for any $\rho \in (0,2)$, away from any
critical point, by setting $\mu =\mathrm{\mu }_{\beta }$ in Section \ref%
{Section phase diagram mu fixed}. In particular, the superconducting phase
can appear by tuning in each parameter: the BCS coupling constant $\gamma $
(see (\ref{definition r max})), the inverse temperature $\beta >0$ (see
Corollary \ref{BCS theorem 2-0bis}), the coupling constant $\lambda $, the
magnetic field $h$ (see Section \ref{section Meissner effect}), the chemical
potential $\mu $ or the electron density $\rho $ (see Section \ref{Section
Mott insulator}). Therefore, to explain the phase diagram at fixed electron
density, it is sufficient to give the behavior of the Cooper pair condensate
density $\mathrm{r}_{\beta }$ as a function of $\rho \in (0,2)$. Everything
can be easily performed via numerical methods, see figure \ref%
{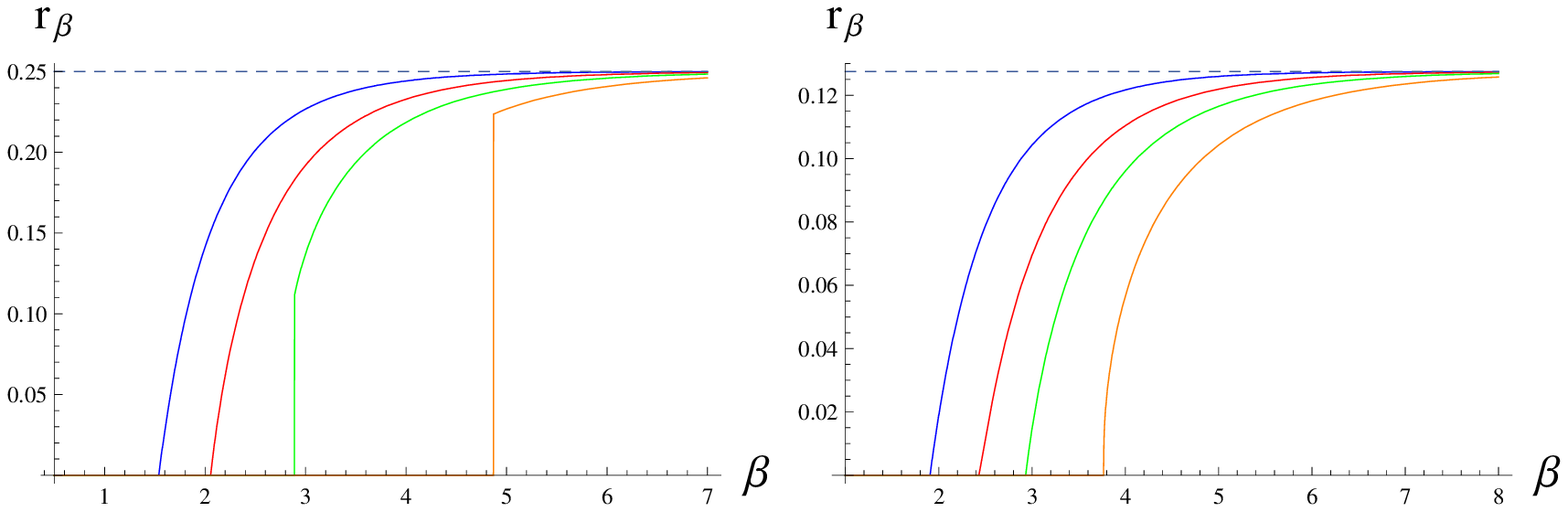}. We restrict our rigorous analysis to the
zero--temperature limit of $\mathrm{r}_{\beta }$, which is a straightforward
consequence of Corollary \ref{BCS theorem 2-0bis} and (\ref{renormalized
chemical potential}).

\begin{figure}[h]
\centerline{\includegraphics[angle=0,scale=1.,clip=true,width=12.5cm]{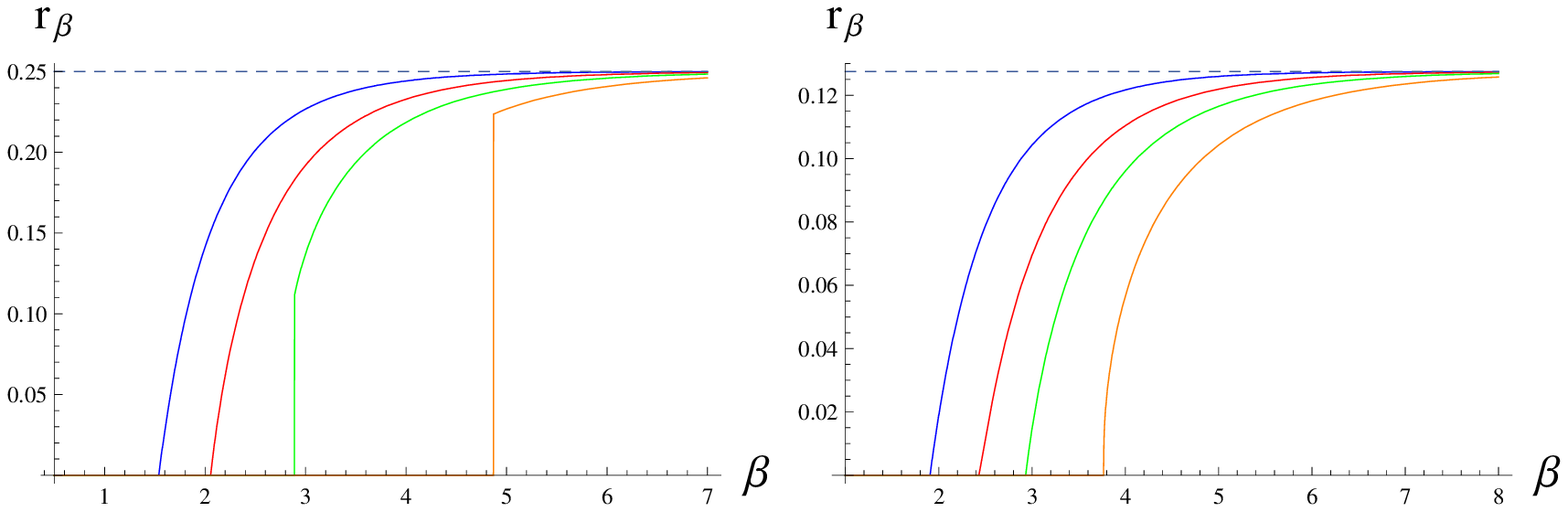}}
\caption{\emph{Illustrations of the Cooper pair condensate density $\mathrm{r}_{\beta }$ as a
function of the inverse temperature $\beta $ for $\gamma =2.6$, $h=0,$ and densities $\rho =1$, $1.7$  (respectively left and right figures), with $\lambda =0$
(blue line), $0.5$ (red line), $0.75$ (green line), and $1$ (orange line). The dashed line indicates the value of $\mathrm{r}_{\infty }.$}}
\label{order-parameter-densite-2.eps}
\end{figure}%

\begin{corollary}[Zero--temperature Cooper pair condensate density]
\label{BCS theorem 2-6}\mbox{}\newline
At zero--temperature, fixed electron density $\rho \in (0,2)$ and $\lambda
,h\in \mathbb{R}$, the Cooper pair condensate density $\mathrm{r}_{\beta }$
converges as $\beta \rightarrow \infty $ towards $\mathrm{r}_{\infty }=\rho
(2-\rho )/4$ when $\gamma >\max \{\tilde{\Gamma}_{\rho ,\lambda +|h|},0\}$.
Here%
\begin{equation*}
\tilde{\Gamma}_{x,y}:=\frac{4y}{x\left( x-2\right) +2}\chi _{\lbrack
0,\infty )}\left( y\right)
\end{equation*}%
is a function defined for any $x,y\in \mathbb{R}$.
\end{corollary}

\begin{remark}
\label{remark subtil}The case $0<\gamma <\tilde{\Gamma}_{\rho ,\lambda +|h|}$
is more subtle than its analogous with a fixed chemical potential $\mu $,
because phase mixtures can take place. See Section \ref{coexistence}.
\end{remark}

As explained above, as soon as $\gamma >\tilde{\Gamma}_{\rho ,\lambda +|h|}$
we can extract from this corollary all the zero--temperature thermodynamics
of the strong coupling BCS--Hubbard model by using Corollaries \ref{BCS
theorem 2-0bis}, \ref{BCS theorem 2-1bis}, \ref{BCS theorem 2-2bis}, and \ref%
{BCS theorem 2-3bis}.

If $\lambda +|h|>0$ and $\gamma $ satisfy the inequalities%
\begin{equation*}
\gamma >\underset{\rho \in \left( 0,2\right) }{\min }\left\{ \tilde{\Gamma}%
_{\rho ,\lambda +|h|}\right\} =\tilde{\Gamma}_{0,\lambda +|h|}=\tilde{\Gamma}%
_{2,\lambda +|h|}=2\left( \lambda +|h|\right)
\end{equation*}%
and%
\begin{equation*}
\gamma <\underset{\rho \in \left( 0,2\right) }{\max }\left\{ \tilde{\Gamma}%
_{\rho ,\lambda +|h|}\right\} =\tilde{\Gamma}_{1,\lambda +|h|}=4\left(
\lambda +|h|\right) ,
\end{equation*}%
it is also clear that the superconductor--Mott insulator phase transition
appears by tuning the electron density $\rho $ in the same way as described
in Section \ref{Section Mott insulator} for $\mu$. See figures \ref%
{Mott-Insulator-1.eps}. In this case however, we recommend Section \ref%
{coexistence} for more details because of the subtlety mentioned in Remark %
\ref{remark subtil}. See figures \ref{order-parameter-densite-1-1bis.eps}-%
\ref{specific-heat-densite.eps} below.

From (\ref{renormalized chemical potential}) combined with Corollary \ref%
{BCS theorem 2-6}, note that the asymptotics (\ref{asymptotics of specific
heat}) of the specific heat at zero-temperature is still valid at fixed
electron density $\rho $ as soon as $\gamma >\max \{\tilde{\Gamma}_{\rho
,\lambda +|h|},0\}$. Meanwhile, from Corollary \ref{BCS theorem 2-6} the
zero--temperature Cooper pair condensate density $\mathrm{r}_{\infty }$ does
not depend on $\lambda $, $\gamma $, or $h$, as soon as $\gamma >\tilde{%
\Gamma}_{\rho ,\lambda +|h|}$ is satisfied. Indeed, the chemical potential $%
\mathrm{\mu }_{\beta }$ in the case where $\mathrm{r}_{\beta }>0$ is
renormalized, cf. (\ref{renormalized chemical potential}). In other words,
at zero--temperature, the thermodynamic behavior of the strong coupling
BCS--Hubbard model for $\gamma >\tilde{\Gamma}_{\rho ,\lambda +|h|}$ is
equal to the well--known behavior of the BCS theory in the strong coupling
approximation ($\lambda =h=0$). This phenomenon is also seen by using
renormalization methods where it is believed that the Coulomb interaction
simply modifies the mass of electrons by creating quasi--particles (which
however do not exist in our model).

\subsection{Coexistence of ferromagnetic and superconducting phases\label%
{coexistence}}

Observe that the electron density $\mathrm{d}_{\beta }$ given by Theorem \ref%
{BCS theorem 2-1} can have discontinuities as a function of the chemical
potential $\mu $. This phenomenon appears at the superconductor--Mott
insulator phase transition, see Section \ref{Section Mott insulator} and
figure \ref{Mott-Insulator-1.eps}. Because of electron--hole symmetry
(Section \ref{Section electron hole sym}), without loss of generality we can
restrict our study to the case where $\mathrm{d}_{\beta }\in \left[ 0,1%
\right] $, i.e., $\rho \in \left[ 0,1\right] $ and $\mathrm{\mu }_{\beta
}\leq \lambda $.

In this regime, the electron density $\mathrm{d}_{\beta }$\ has, at most,
one discontinuity point at the so-called \textit{critical chemical potential}
$\mathrm{\mu }_{\beta }^{(c)}\leq \lambda $. In particular, there are two
critical electron densities
\begin{equation*}
\mathrm{d}_{\beta }^{\pm }:=\mathrm{d}_{\beta }(\mathrm{\mu }_{\beta
}^{(c)}\pm 0,\lambda ,\gamma ,h)\quad \mathrm{\ with\ d}_{\beta }^{+}>%
\mathrm{d}_{\beta }^{-}.
\end{equation*}%
Similarly, we can also define two critical Cooper pair condensate densities $%
\mathrm{r}_{\beta }^{\pm }$, two critical magnetization densities\footnote{%
If $h=0$, then $\mathrm{m}_{\beta }^{\pm }=0$.} $\mathrm{m}_{\beta }^{\pm }$
and two critical Coulomb correlation density $\mathrm{w}_{\beta }^{\pm }$.
Of course, since $\mathrm{r}_{\beta }^{+}>\mathrm{r}_{\beta }^{-}=0,$ we are
here on a critical point, i.e.,
\begin{equation*}
(\beta ,\mathrm{\mu }_{\beta }^{(c)},\lambda ,\gamma ,h)\in \partial
\mathcal{S}
\end{equation*}%
(see (\ref{critical point open set})), with $\beta ,\gamma >0$ and $\lambda
,h\in \mathbb{R}$ such that this critical chemical potential $\mathrm{\mu }%
_{\beta }^{(c)}=\mathrm{\mu }_{\beta }^{(c)}(\lambda ,\gamma ,h)$ exists.

The thermodynamics of the model for $\rho \not\in \lbrack \mathrm{d}_{\beta
}^{-},\mathrm{d}_{\beta }^{+}]$ is already explained in Section \ref{Section
phase diagram rho fix sans critical} because the solution $\mathrm{r}_{\beta
}$ of (\ref{BCS pressure 2}) is unique at $\mu =\mathrm{\mu }_{\beta }$.
The chemical potential $\mu _{N,\beta }$ converges to $\mathrm{\mu }_{\beta
}=\mathrm{\mu }_{\beta }^{(c)}$, if  $\rho \in \lbrack \mathrm{d}_{\beta }^{-},\mathrm{d}_{\beta }^{+}]$. In this case the variational problem (\ref{BCS
pressure 2}) has exactly two maximizers $\mathrm{r}_{\beta }^{\pm }$. The
thermodynamic behavior of the system in this regime is not, a priori, clear
except from the obvious fact that
\begin{equation*}
\lim\limits_{N\rightarrow \infty }\frac{1}{N}\sum\limits_{x\in \Lambda
_{N}}\omega _{N}(n_{x,\uparrow }+n_{x,\downarrow })=\rho
\end{equation*}%
per definition. In particular, it cannot be deduced from the above results.
We handle this situation within a much more general framework in Theorem \ref%
{Theorem equilibrium state 4bis}. As a consequence of this study (see
discussions after Theorem \ref{Theorem equilibrium state 4bis}), all the
extensive quantities can be obtained in the thermodynamic limit:

\begin{theorem}[Densities in coexistent phases]
\label{phase.mix.Th}\mbox{}\newline
Take $\beta ,\gamma >0$ and real numbers $\lambda ,h$ in the domain of
definition of the critical chemical potential $\mathrm{\mu }_{\beta }^{(c)}$%
. For any $\rho \in \lbrack \mathrm{d}_{\beta }^{-},\mathrm{d}_{\beta }^{+}]$%
, all densities are uniquely defined:\newline
(i) The Cooper pair condensate density equals%
\begin{equation*}
\underset{N\rightarrow \infty }{\lim }\left\{ \dfrac{1}{N^{2}}\sum_{x,y\in
\Lambda _{N}}\omega _{N}\left( a_{x,\uparrow }^{\ast }a_{x,\downarrow
}^{\ast }a_{y,\downarrow }a_{y,\uparrow }\right) \right\} =\tau _{\rho }%
\mathrm{r}_{\beta }^{+},\mathrm{\ with\ }\tau _{\rho }:=\frac{\rho -\mathrm{d%
}_{\beta }^{-}}{\mathrm{d}_{\beta }^{+}-\mathrm{d}_{\beta }^{-}}\in \lbrack
0,1].
\end{equation*}%
(ii) The magnetization density equals
\begin{equation*}
\underset{N\rightarrow \infty }{\lim }\left\{ \frac{1}{N}\sum\limits_{x\in
\Lambda _{N}}\omega _{N}\left( n_{x,\uparrow }-n_{x,\downarrow }\right)
\right\} =\left( 1-\tau _{\rho }\right) \mathrm{m}_{\beta }^{-}+\tau _{\rho }%
\mathrm{m}_{\beta }^{+}.
\end{equation*}%
(iii) The Coulomb correlation density equals%
\begin{equation*}
\underset{N\rightarrow \infty }{\lim }\left\{ \frac{1}{N}\sum\limits_{x\in
\Lambda _{N}}\omega _{N}\left( n_{x,\uparrow }n_{x,\downarrow }\right)
\right\} =\left( 1-\tau _{\rho }\right) \mathrm{w}_{\beta }^{-}+\tau _{\rho }%
\mathrm{w}_{\beta }^{+}.
\end{equation*}%
(iv) The mean energy per site equals
\begin{equation*}
\underset{N\rightarrow \infty }{\lim }\left\{ N^{-1}\omega _{N}\left(
\mathrm{H}_{N}\right) \right\} =\left( 1-\tau _{\rho }\right) \mathrm{%
\epsilon }_{\beta }^{-}+\tau _{\rho }\mathrm{\epsilon }_{\beta }^{+},
\end{equation*}%
with $\mathrm{\epsilon }_{\beta }^{\pm }:=-\mathrm{\mu }_{\beta }^{(c)}\rho
-h\mathrm{m}_{\beta }^{\pm }+2\lambda \mathrm{w}_{\beta }^{\pm }-\gamma
\mathrm{r}_{\beta }^{\pm }.$
\end{theorem}

As a consequence of this theorem, as soon as the magnetic field $h\neq 0$,
there is a coexistence of ferromagnetic and superconducting phases at low
temperatures for $\rho \in (\mathrm{d}_{\beta }^{-},\mathrm{d}_{\beta }^{+})$%
. In other words, the Mei{\ss }ner effect is not valid in this interval of
electron densities. An illustration of this is given in figure \ref%
{coexistence.eps}. Such phenomenon was also observed in experiments and from
our results, it should occur rather near half--filling (but \textit{not
exactly} at half--filling) and at strong repulsion $\lambda >0$.
Additionally, observe
that this coexistence of thermodynamic phases can also appear at the
critical magnetic field $\mathrm{h}_{\beta }^{(c)}$ (see Section \ref%
{section Meissner effect}).
%
%
%
\begin{figure*}[hbtp]
\begin{center}
\mbox{
\leavevmode
\subfigure
{ \includegraphics[angle=0,scale=1,clip=true,width=3.8cm]{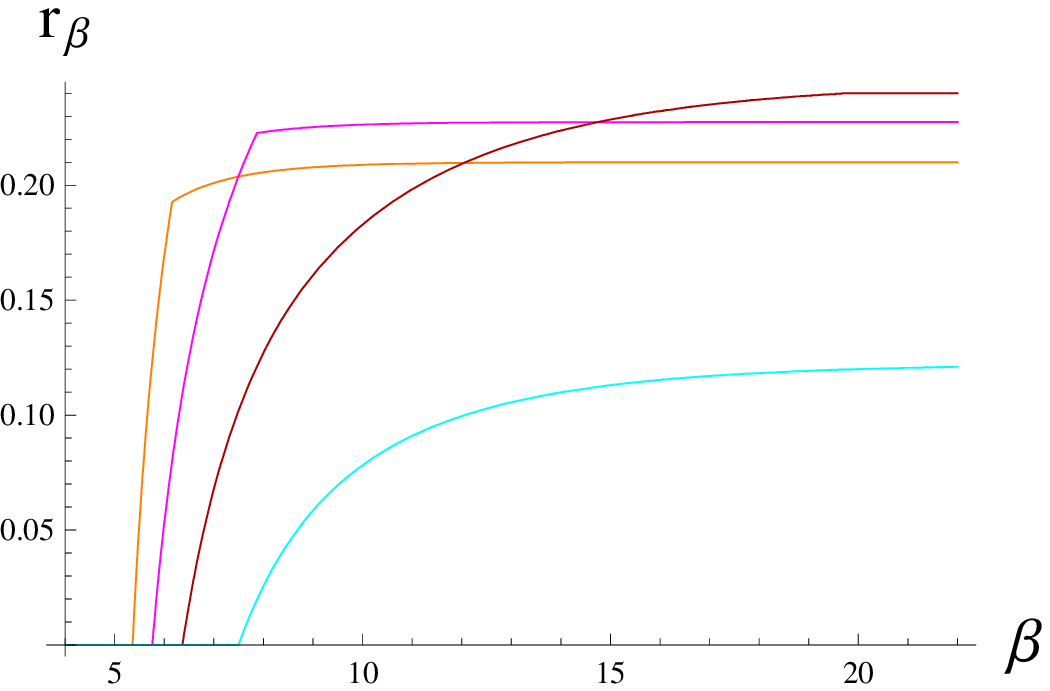} }

\leavevmode
\subfigure
{ \includegraphics[angle=0,scale=1,clip=true,width=3.8cm]{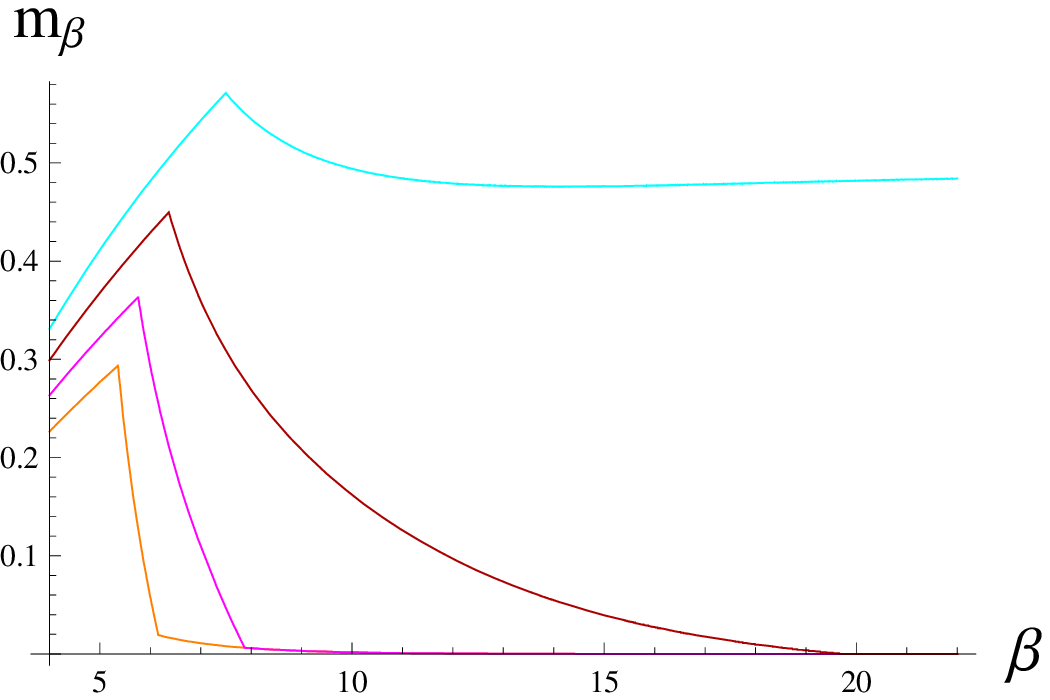} }

\leavevmode
\subfigure
{ \includegraphics[angle=0,scale=1,clip=true,width=3.8cm]{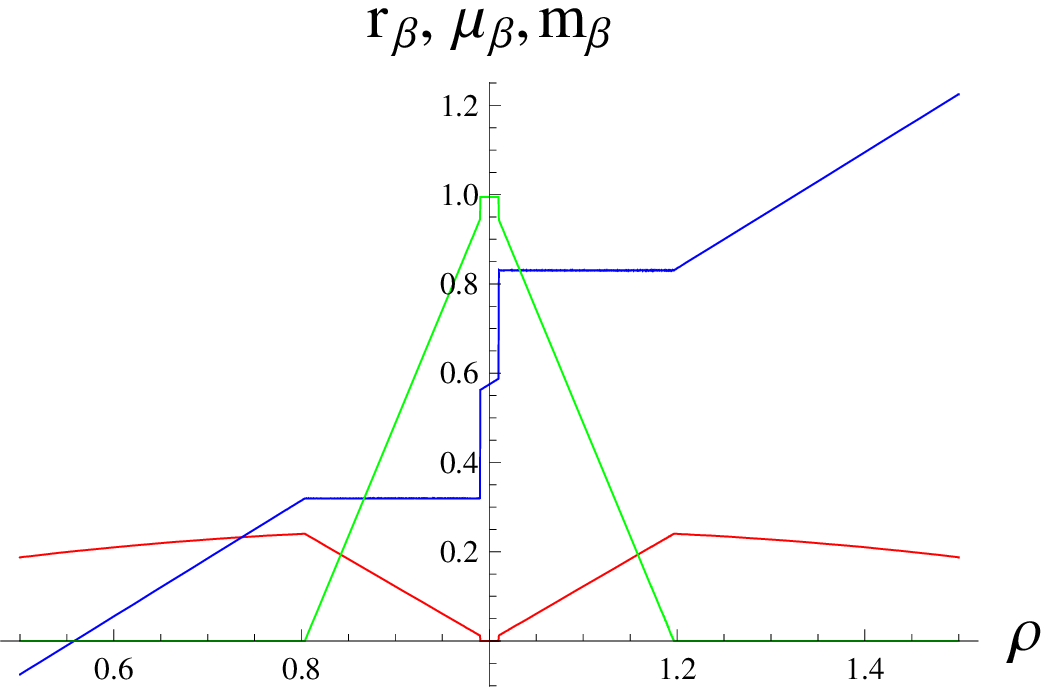} }

}
\end{center}
\caption{\emph{In the two figures on the left, we give illustrations of the Cooper pair
condensate density $\mathrm{r}_{\beta }$ and the magnetization density $\mathrm{m}_{\beta }$
as functions of the inverse temperature $\beta $ for densities $\rho =0.6$ (orange
line), $0.7$  (magenta line), $0.8$ (red line), $0.9$ (cyan line). In the
figure on the right, we illustrate the coexistence of ferromagnetic and
superconducting phases via graphs of $\mathrm{r}_{\beta }$, $\mathrm{m}_{\beta }$ and the chemical potential $\mathrm{\mu }_{\beta }$ as functions
of $\rho $ for $\beta=30$ (low temperature regime). In all figures, $\lambda =0.575$, $\gamma =2.6$, and $h=0.1$. (The small
discontinuities around $\rho =1$ in the right figure are numerical anomalies)}}
\label{coexistence.eps}
\end{figure*}%

\begin{remark}
Coexistence of ferromagnetic and superconducting phases has already been
rigorously investigated, see, e.g., \cite%
{brankov-tonchev1,Bogoliubov-Ermilov-Kurbatov}. For instance, in \cite%
{brankov-tonchev1} such phenomenon is shown to be impossible in the ground
state of the Vonsovkii--Zener model applied to s--wave superconductors%
\footnote{%
It is a combination of the BCS interaction (\ref{BCS interaction}) with the
Zener s--d exchange interaction.}, whereas at finite temperature, numerical
computations \cite{Bogoliubov-Ermilov-Kurbatov} suggests the contrary. This
last analysis \cite{Bogoliubov-Ermilov-Kurbatov} is however not performed in
details.
\end{remark}

The second interesting physical aspect related to densities $\rho $ between
the critical densities $\mathrm{d}_{\beta }^{-}$ and $\mathrm{d}_{\beta
}^{+} $ is a smoothing effect of the extensive quantities (magnetization
density, Cooper pair condensate density, etc.) as functions of the inverse
temperature $\beta $. Indeed, since the critical chemical potential $\mathrm{%
\mu }_{\beta }^{(c)}$ only exists when a first order phase transition
occurs, one could expect that the extensive quantities are not continuous as
functions of $\beta >0$. In fact, for $\rho \in (\mathrm{d}_{\beta }^{-},%
\mathrm{d}_{\beta }^{+})$, there is a convex interpolation between
quantities related to the solutions $\mathrm{r}_{\beta }^{-}=0$ and $\mathrm{%
r}_{\beta }^{+}>0$ of (\ref{BCS pressure 2}), see Theorem \ref{phase.mix.Th}%
. The continuity of the extensive quantities then follows, see figure \ref%
{coexistence.eps}. It does not imply however, that all densities become
always continuous at fixed $\rho $ as a function of the inverse temperature $%
\beta $. For instance, in figure \ref{order-parameter-densite-2.eps}, the
green and orange graphs give two illustrations of a discontinuity of the
order parameter $\mathrm{r}_{\beta }$ at fixed electron density $\rho =1$
where $\mathrm{\mu }_{\beta }=\lambda $. To understand this first order
phase transition, other extensive quantity should be additionally fixed, see
discussions in Section \ref{Section 5} and figure \ref{pointspecial.eps}.

\begin{figure*}[hbtp]
\begin{center}
\mbox{
\leavevmode
\subfigure
{ \includegraphics[angle=0,scale=1,clip=true,width=3.8cm]{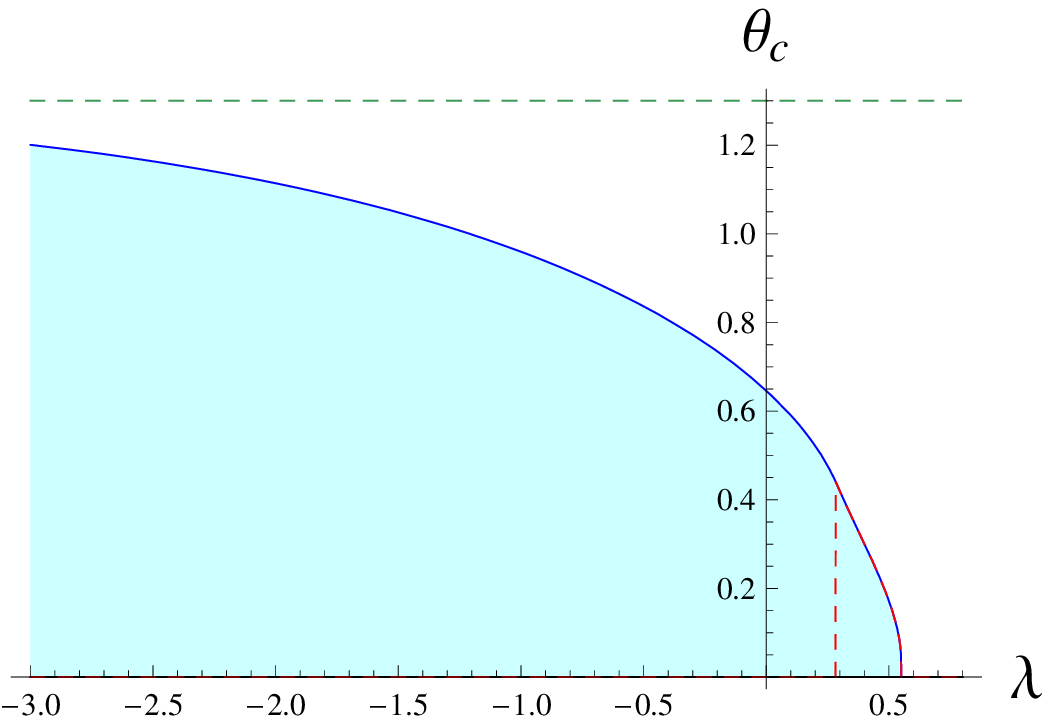} }

\leavevmode
\subfigure
{ \includegraphics[angle=0,scale=1,clip=true,width=3.8cm]{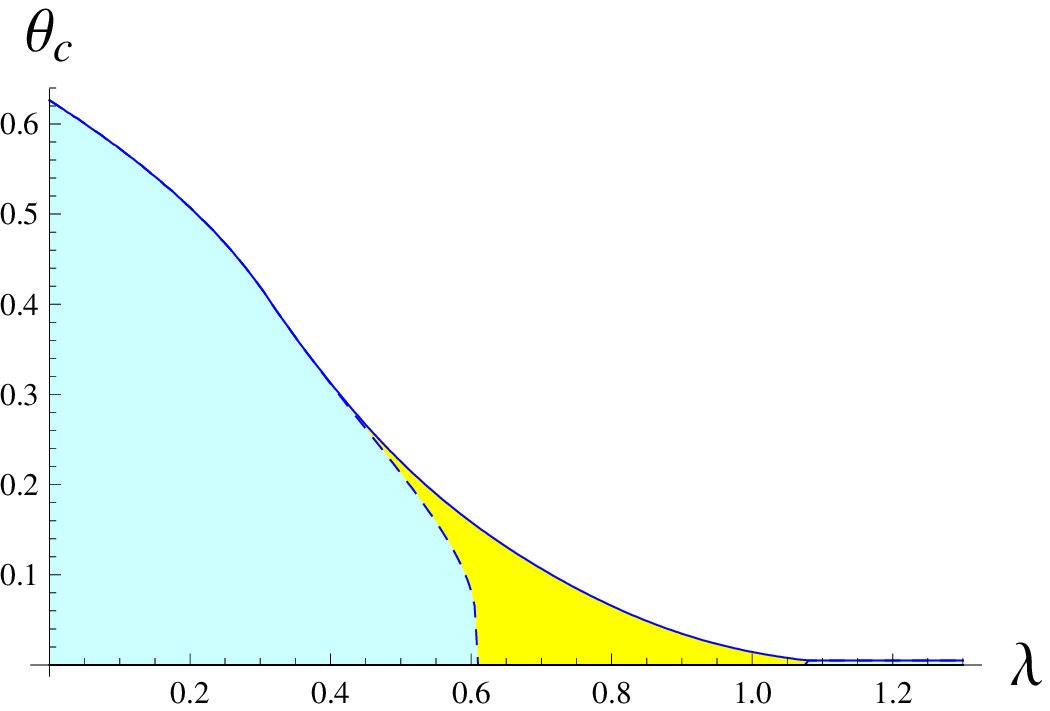} }

\leavevmode
\subfigure
{ \includegraphics[angle=0,scale=1,clip=true,width=3.8cm]{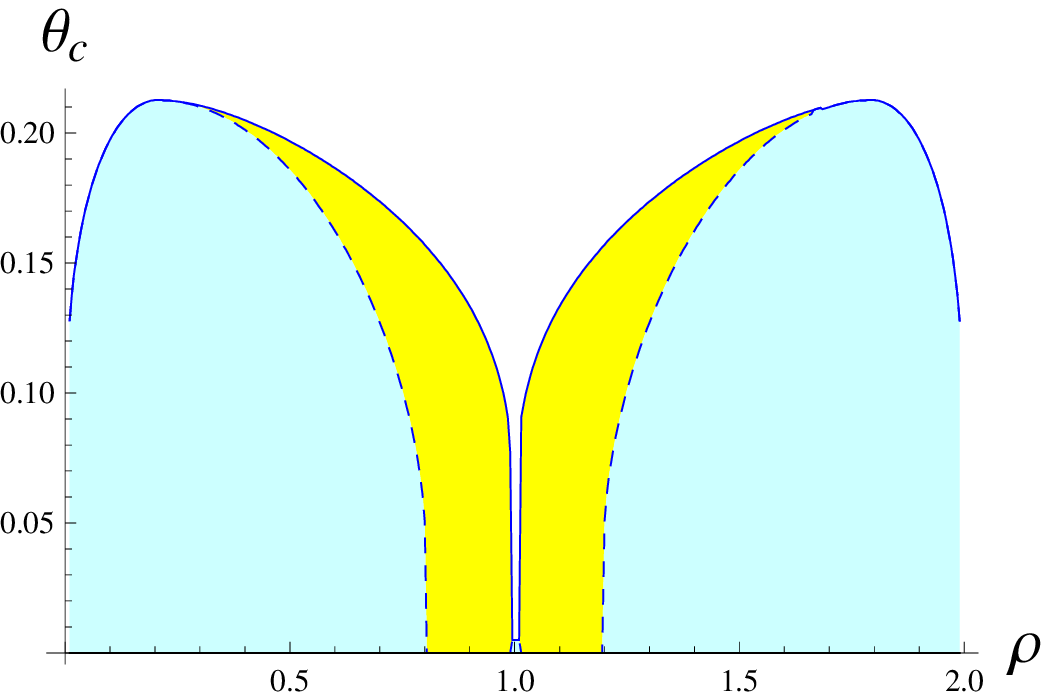} }
}
\end{center}
\caption{\emph{Illustration, as a function of $\lambda$ (the two figures on the left) or $\rho$ (figure on the right), of the critical temperature $\theta _{c}=\theta _{c}(\rho ,\lambda ,\gamma ,h)$
for $\gamma =2.6,$ $h=0.1$ and with $\rho =1$ (left figure), $\rho =0.7$
(figure on the center) and $\lambda =0.575$ (right figure).
The blue and yellow area correspond respectively to the superconducting and ferromagnetic--superconducting phases,
whereas the red dashed line indicates the domain of $\lambda$ with a first order phase transition as a function of
$\beta$ or the temperature $\theta:=\beta^{-1}$ (It only exists in the left figure). The dashed green line (left figure) is the asymptote when $\lambda \rightarrow -\infty$. In the right figure, observe that there is no phase transition for $\rho=1$.}}
\label{order-parameter-densite-1-1bis.eps}
\end{figure*}

Following these last results, we give now in figure \ref%
{order-parameter-densite-1-1bis.eps} other plots of the critical temperature
$\theta _{c}=\theta _{c}(\rho ,\lambda ,\gamma ,h),$ which is defined as
usual such that $\mathrm{r}_{\beta }>0$ if and only if $\beta >\theta
_{c}^{-1}$. In this figure, observe that a positive $\lambda $, i.e., a
one--site repulsion, can never increase the critical temperature if the
electron density $\rho $ is fixed instead of the chemical potential $\mu $,
compare with figure \ref{domain-temp-critique-lamb.eps}. We also show in
figure \ref{order-parameter-densite-1-1bis.eps} (right figure) that if the
density of holes equals the density of electrons, i.e., $\rho =1$, then we
have a Mott insulator, whereas a small doping of electrons or holes implies
either a superconducting phase (blue area) or a superconductor--Mott
insulator (ferromagnetic) phase (yellow area) related to the
superconductor--Mott insulator phase transition described in Section \ref%
{Section Mott insulator} and figure \ref{Mott-Insulator-1.eps}.
%
%

\begin{figure*}[hbtp]
\begin{center}
\mbox{
\leavevmode
\subfigure
{ \includegraphics[angle=0,scale=1,clip=true,width=3.8cm]{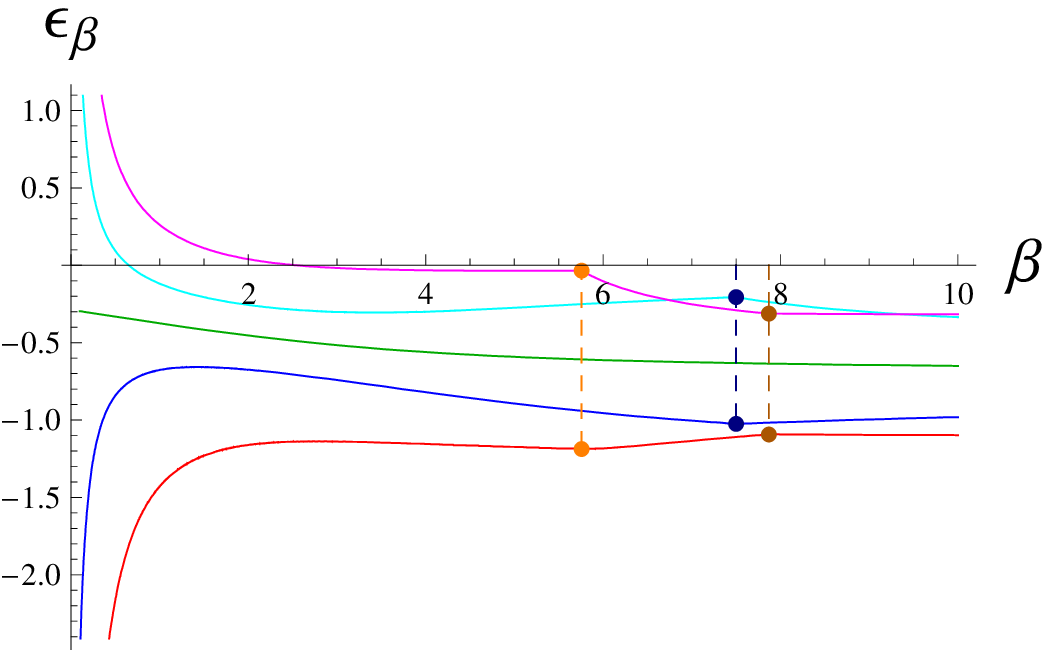} }

\leavevmode
\subfigure
{ \includegraphics[angle=0,scale=1,clip=true,width=3.8cm]{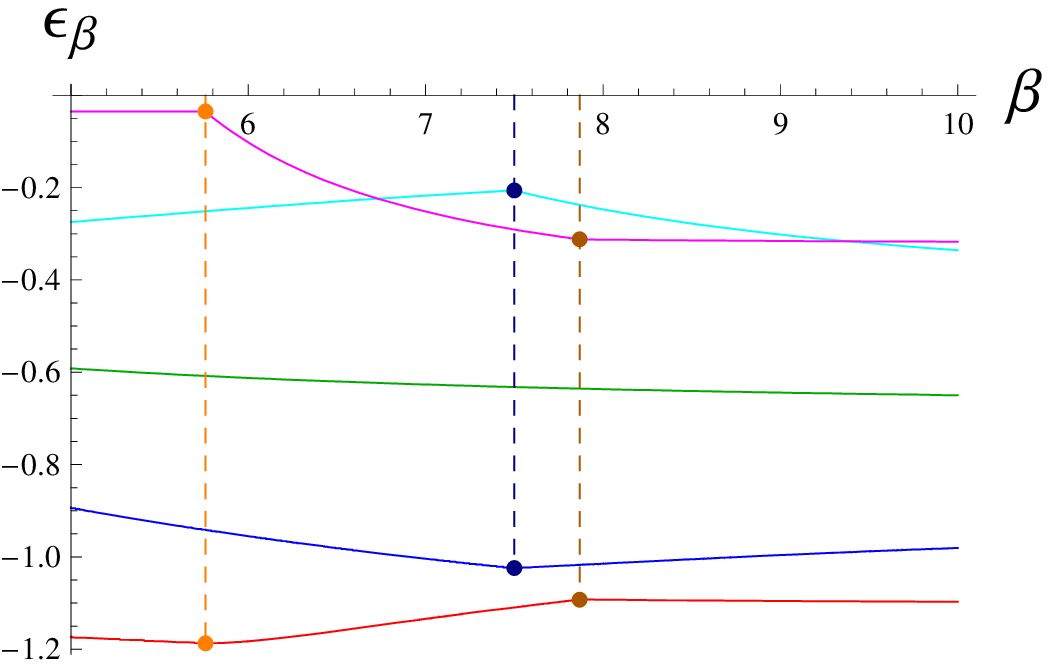} }

\leavevmode
\subfigure
{ \includegraphics[angle=0,scale=1,clip=true,width=3.8cm]{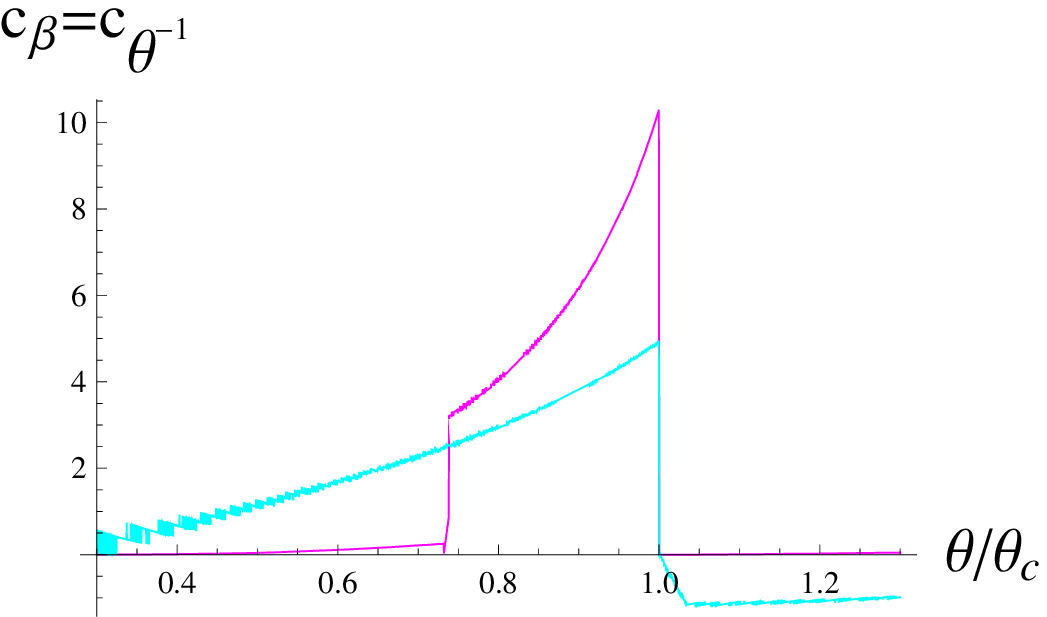} }

}
\end{center}
\caption{\emph{In the two figures on the left, we give illustrations of the mean energy per site
$\mathrm{\epsilon }_{\beta }$ as a function of the inverse temperature $\beta$
for densities $\rho =0.7$ (magenta line), $0.9$ (cyan line),
$1$ (green line), $1.1$ (blue line) and $1.3$ (red line). For $\rho=1$, there
is no phase transition and for $\rho=0.9$
or $1.1$ only a ferromagnetic-superconducting phase appears,  whereas for $\rho=0.7$ or $1.3$ this last phase
is followed for larger $\beta$ by
a superconducting phase. In the figure on the right, assuming $\mathfrak{C}_{\beta }=0$, we give two plots of the specific heat $\mathrm{c}_{\beta }$ as a function of the ratio $\theta
/\theta _{c}$ between $\theta :=\beta ^{-1}$ and the critical temperature $\theta _{c}$ for densities $\rho =0.7$ (magenta line) and $0.9$ (cyan line).
In all figures, $\lambda =0.575$, $\gamma =2.6$, and $h=0.1$.}}
\label{specific-heat-densite.eps}
\end{figure*}

To conclude, the figure \ref{specific-heat-densite.eps} illustrates various
thermodynamic features of the system at fixed $\rho $. First, as a function
of $\beta >0$, $\mathrm{\epsilon }_{\beta }$\ is continuously differentiable
only for $\rho =1$. In other words, there is no phase transition by
opposition to the cases with $\rho =0.7$, $0.9$ or $\rho =1.1$, $1.3$. This
is the Mott insulator phase transition illustrated in figure \ref%
{Mott-Insulator-1.eps}. As in figure \ref{Mott-Insulator-1.eps}, we also
observe the electron--hole symmetry implying that $\rho =0.7$ and $\rho =1.3$%
, or $\rho =0.9$ and $\rho =1.1$, has same phase transitions at exactly the
same critical points. As explained in Section \ref{Section BCS phase
transition}, the mean energy per site $\mathrm{\epsilon }_{\beta }$ for $%
\rho =0.7$, $1.3$, or $\rho =0.9$, $1.1$, differs by a constant, i.e., in
absolute value by $|2\lambda -\mathrm{\mu }_{\beta }|$. At high
temperatures, i.e., when $\beta \rightarrow 0$, the function $\mathrm{%
\epsilon }_{\beta }$ diverges to $\pm \infty $ if $\rho =1\mp \varepsilon $
with $\varepsilon \in (0,1)$ whereas it stays finite at $\rho =1$. Indeed,
when $\beta \rightarrow 0$ the electron density $\mathrm{d}_{\beta }$
converges to $1$ at fixed $\mu $, $\lambda $, $\gamma $, $h$, see Theorem %
\ref{BCS theorem 2-1} and figure \ref{density-mu.eps}. If $\rho =1\mp
\varepsilon $, it follows that the chemical potential $\mathrm{\mu }_{\beta
} $ diverges to $\mp \infty $ as $\beta \rightarrow 0$, implying that $%
\mathrm{\epsilon }_{\beta }\rightarrow \pm \infty $. In other words, it is
energetically unfavorable to fix an election density $\rho \neq 1$ at high
temperatures. Finally, the specific heat $\mathrm{c}_{\beta }$ has only one
jump in the case of one phase transition and two jumps when there are two
phase transitions, namely when the superconductor--Mott insulator
(ferromagnetic) phase and the purely superconducting phase appear.%
%
%
%

\section{Concluding remarks\label{Section 5}}

\noindent \textbf{1.} First, it is important to note that two different
physical behaviors
 can be extracted from the strong coupling BCS--Hubbard model
$\mathrm{H}_{N}$: a first one at fixed chemical potential $\mu $ and a
second one at fixed electron density $\rho \in (0,2)$. This does \textit{not}
mean that the canonical and grand--canonical ensembles are not equivalent
for this model. But, the influence of the direct interaction with coupling
constant $\lambda $ drastically changes from the case at fixed $\mu $ to the
other one at fixed $\rho $. For instance, via Corollary \ref{BCS theorem 2-6}
(see also figure \ref{order-parameter-densite-1-1bis.eps}), any one--site
repulsion between pairs of electrons is in any case unfavorable to the
formation of Cooper pairs, as soon as the electron density $\rho $ is fixed.
This property is however wrong at fixed chemical potential $\mu $, see
figure \ref{domain-temp-critique-lamb.eps}. In other words, fixing the
electron density $\rho $ is \textit{not} equivalent\footnote{%
"Equivalent" is not taken here in the sense of the equivalence of ensembles.}
to fixing the chemical potential $\mu $ in the model. Physically, a fixed
electron density can be modified by doping the superconductor. Changing the
chemical potential may be more difficult.\ One naive proposition would be to
impose an electric potential on a superconductor which is coupled to an
additional conductor serving as a reservoir of electrons or holes at fixed
chemical potential.\newline

\noindent \textbf{2.} A measurement of the asymptotics as $\beta \rightarrow
\infty $ of the specific heat $\mathrm{c}_{\beta }$ (see (\ref{specific heat
at infinite volume}) with $\mathfrak{C}_{\beta }=0$)\ in a superconducting
phase would determine, by using (\ref{asymptotics of specific heat}), first
the parameter $\gamma >0$ via the exponential decay and then the coupling
constant $\lambda $. Next, the measurement of the critical magnetic field at
very low temperature would allow to obtain by (\ref{critical magnetic field}%
) the chemical potential $\mu $ and hence the electron density at
zero--temperature. Since the inverse temperature $\beta $ as well as the
magnetic field $h$ can directly be measured, all parameters of the strong
coupling BCS--Hubbard model $\mathrm{H}_{N}$ (\ref{Hamiltonian BCS-Hubbard})
would be experimentally found. In particular, its thermodynamic behavior,
explained in Sections \ref{Section 2}--\ref{Section phase diagram rho fixed}%
, could finally be confronted to the real system. One could for instance
check if the critical temperature $\theta _{c}$ given by $\mathrm{H}_{N}$ in
appropriate dimension corresponds to the one measured in the real
superconductor. Such studies would highlight the thermodynamic impact of the
kinetic energy.\newline

\noindent \textbf{3.} In Section \ref{Section phase diagram rho fixed}, the
electron density is fixed but one could have fixed each extensive quantity:
the Cooper pair condensate density, the magnetization density, the Coulomb
correlation density or the mean--energy per site. For instance, if the
magnetization density $\mathrm{m}\in \mathbb{R}$ is fixed, by strict
convexity of the pressure there is a unique magnetic field $h_{N,\beta
}=h_{N,\beta }(\mu ,\lambda ,\gamma ,m)$ such that
\begin{equation*}
\mathrm{m}=\dfrac{1}{N}\sum_{x\in \Lambda _{N}}\omega _{N}\left(
n_{x,\uparrow }-n_{x,\downarrow }\right) .
\end{equation*}%
In the thermodynamic limit, we then have $h_{N,\beta }$ converging to $%
\mathrm{h}_{\beta }$ solution of the equation $\mathrm{m}_{\beta }=\mathrm{m}
$ at fixed $\beta ,\gamma >0$ and $\mu ,\lambda \in \mathbb{R}$. By using
Theorem \ref{Theorem equilibrium state 4bis}, we would obtain the
thermodynamics of the system for any $\beta ,\gamma >0$ and $\mu ,\lambda ,%
\mathrm{m}\in \mathbb{R}$. More generally, when one of the extensive
quantities $\mathrm{r}_{\beta }$, $\mathrm{d}_{\beta }$, $\mathrm{m}_{\beta
} $, $\mathrm{w}_{\beta }$, or $\mathrm{\epsilon }_{\beta }$ is
discontinuous at a critical point, then the thermodynamic limit of the local
Gibbs states $\omega _{N}$ can be uniquely determined by fixing one of the
corresponding extensive quantity between its critical values. The other
extensive quantities are determined in this case by an obvious transcription
of Theorem \ref{phase.mix.Th} for the considered discontinuous quantity at
the critical point. Observe, however, that $\mathrm{r}_{\beta }$, $\mathrm{d}%
_{\beta }$, $\mathrm{m}_{\beta }$, $\mathrm{w}_{\beta }$, and $\mathrm{%
\epsilon }_{\beta }$ should be related respectively to the parameters $%
\gamma $, $\mu $, $h$, $\lambda $ and $\beta $. For instance, the existence
of a magnetic field $h_{N,\beta }$ solution of (\ref{mu fixed particle
density}) at fixed $\rho \in (0,2)$ is not clear at finite volume.

\begin{figure*}[hbtp]
\begin{center}
\mbox{
\leavevmode
\subfigure
{ \includegraphics[angle=0,scale=1,clip=true,width=3.8cm]{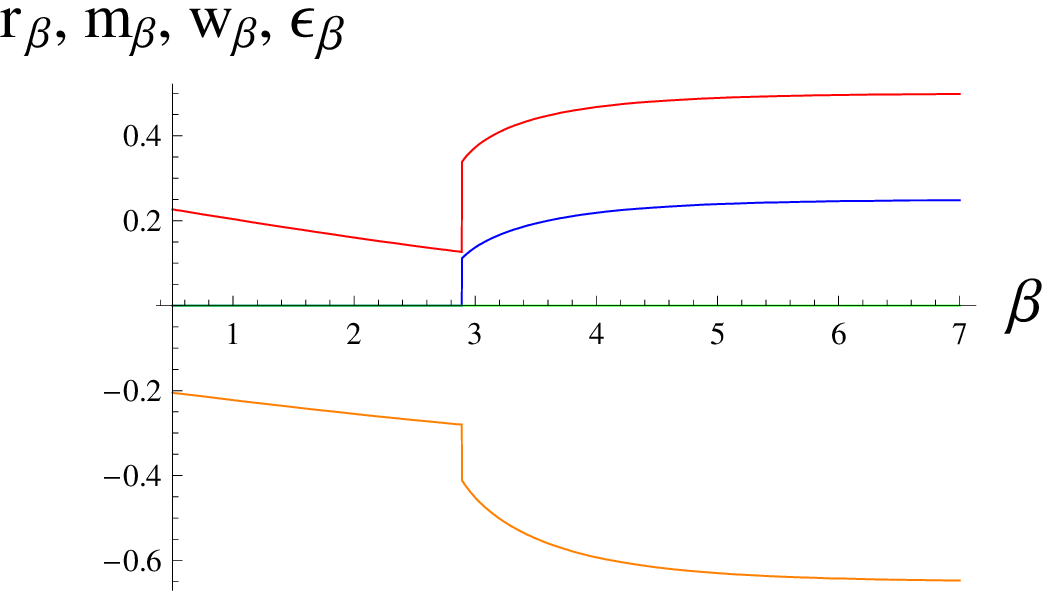} }

\leavevmode
\subfigure
{ \includegraphics[angle=0,scale=1,clip=true,width=3.8cm]{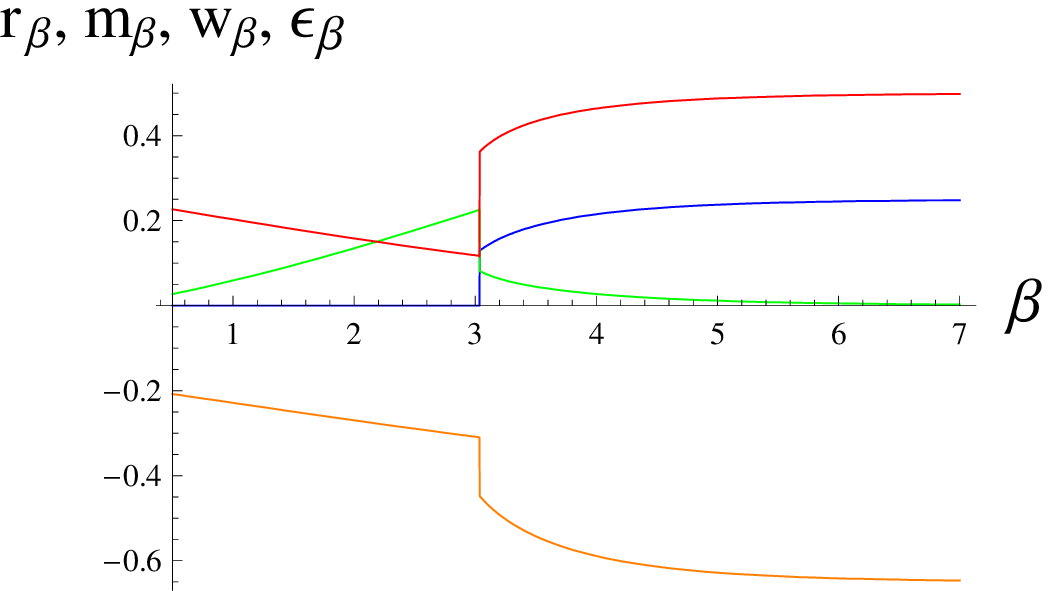} }

\leavevmode
\subfigure
{ \includegraphics[angle=0,scale=1,clip=true,width=3.8cm]{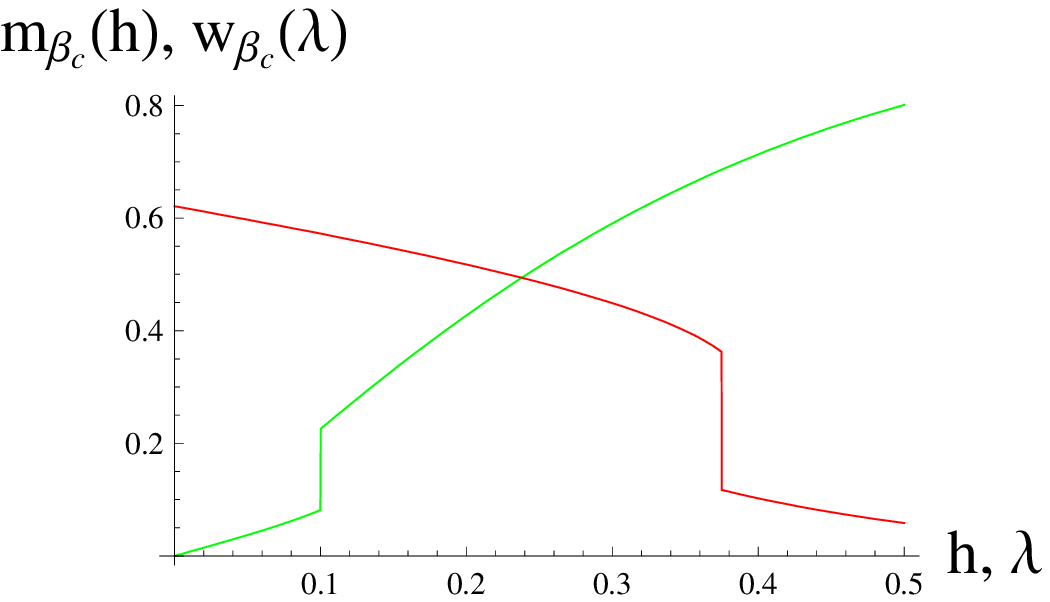} }

}
\end{center}
\caption{\emph{In the two figures on the left, we give illustrations of the
Cooper pair condensate density $\mathrm{r}_{\beta }$ (blue line), the
magnetization density $\mathrm{m}_{\beta }$ (green line), the Coulomb
correlation density $\mathrm{w}_{\beta }$ (red line), and the mean--energy
per site $\mathrm{\epsilon }_{\beta }$ (orange line) as functions of the
inverse temperature $\beta $ for $h=0$ (figure on the left) and $h=0.1$
(figure on the center) whereas $\mathrm{\mu }_{\beta }=\lambda =0.375$, i.e., $\rho =1$. In the figure on the right, we illustrate $\mathrm{m}_{\beta _{c}}$ (green line) and $\mathrm{w}_{\beta _{c}}$ (red line) respectively as functions of $h$ with $\mu=\lambda=0.375$ and $\lambda $ with $(\mu,h)=(0.375,0.1)$ at
the critical inverse temperature $\beta _{c}:=\theta _{c}^{-1}\simeq 3.04$.}}
\label{pointspecial.eps}
\end{figure*}

Figure \ref{pointspecial.eps} gives an example of an electron density always
equal to $1$ for $\mu =\lambda $ together with discontinuity of all other
extensive quantities. In order to get well--defined quantities at the
thermodynamic limit in this example for parameters allowing a first order
phase transition, it is not sufficient to have the electron density fixed.
At the critical point we could for instance fix the magnetization density $%
\mathrm{m}\in \mathbb{R}$ in the ferromagnetic case ($h=0.1$) or in any
case, the Coulomb correlation density $\mathrm{w}\geq 0$ which determines a
coupling constant $\lambda _{N,\beta }$ converging to $\mathrm{\lambda }%
_{\beta }$, see the right illustrations of figure \ref{pointspecial.eps}
with the existence of a critical magnetic field and a critical coupling
constant.
%
%
%

\noindent \textbf{4.} To conclude, as explained in the introduction, for a
suitable space of states it is possible to define a free energy density
functional $\mathfrak{F}$ (\ref{p.var})\ associated with the Hamiltonians $%
\mathrm{H}_{N}$. The states minimizing this functional are equilibrium
states and implies all the thermodynamics of the strong coupling
BCS--Hubbard model discussed in Sections \ref{Section phase diagram mu fixed}%
--\ref{Section phase diagram rho fixed}. Indeed, the weak$^\ast $--limit $%
\omega _{\infty }$ of the local Gibbs state $\omega _{N}$ as $N\rightarrow
\infty $ exists and belongs to our set of equilibrium states for any $\beta
,\gamma >0$ and $\mu ,\lambda ,h\in \mathbb{R}$, cf. Theorem \ref{Theorem
equilibrium state 4bis}. In Section \ref{equilibirum.paragraph}, we prove in
particular the following properties of equilibrium states:

\begin{enumerate}
\item[(i)] Any pure equilibrium state $\omega $ satisfies $\omega
(a_{x,\downarrow }a_{x,\uparrow })=\mathrm{r}_{\beta }^{1/2}e^{i\phi }$ for
some $\phi \in \lbrack 0,2\pi )$. In particular, if $\mathrm{r}_{\beta }\neq
0$ they are not $U(1)$--gauge invariant and show \emph{off diagonal long
range order} \cite{ODLRO} (ODLRO), cf. Theorems \ref{Theorem equilibrium
state 1}, \ref{Theorem equilibrium state 3} and Corollary \ref{Theorem
equilibrium state 2}.

\item[(ii)] All densities are uniquely defined: the electron density of any
equilibrium states $\omega $ is given by $\omega (n_{x,\uparrow
}+n_{x,\downarrow })=\mathrm{d}_{\beta }$\textrm{,} its magnetization
density by $\omega (n_{x,\uparrow }-n_{x,\downarrow })=\mathrm{m}_{\beta }$,
and its Coulomb correlation density equals $\omega (n_{x,\uparrow
}n_{x,\downarrow })=\mathrm{w}_{\beta }$, cf. Theorem \ref{Theorem
equilibrium state 4}.

\item[(iii)] The Cooper fields $\Phi _{x}:=a_{x,\downarrow }^{\ast
}a_{x,\uparrow }^{\ast }+a_{x,\uparrow }a_{x,\downarrow }$ and $\Psi
_{x}:=i(a_{x,\downarrow }^{\ast }a_{x,\uparrow }^{\ast }-a_{x,\uparrow
}a_{x,\downarrow })$ for pure states become classical in the limit $\gamma
\beta \rightarrow \infty $, i.e., their fluctuations go to zero in this
limit, cf. Theorem \ref{Theorem equilibrium state 5}.
\end{enumerate}

Any weak$^\ast$ limit point of equilibrium states with diverging inverse
temperature is (by definition) a ground state. For $\gamma >0$ and $\mu
,\lambda ,h\in \mathbb{R}$, most of ground states inherit the properties
(i)-(iii) of equilibrium states. In particular, within the
GNS--representation \cite{BrattelliRobinsonI} of pure ground states, Cooper
fields are exactly $c$--numbers, see Corollary \ref{coro c-number}. In this
case, correlation functions can explicitly be computed at any order in
Cooper fields. Furthermore, notice that even in the case $h=0$ where the
Hamiltonian $\mathrm{H}_{N}$ is spin invariant, there exist ground states
breaking the spin $SU(2)$--symmetry. For more details including a precise
formulation of these results, we recommend Section \ref{Section mathematical
foundations}, in particular Section \ref{equilibirum.paragraph}.

\section{Mathematical foundations of the thermodynamic results\label{Section
mathematical foundations}}

The aim of this section is to give all the detailed proofs of the
thermodynamics of the strong coupling BCS--Hubbard model $\mathrm{H}_{N}$ (%
\ref{Hamiltonian BCS-Hubbard}). The central result of this section is the
thermodynamic limit of the pressure, i.e., the proof of Theorem \ref{BCS
theorem 1}. The main ingredient in this analysis is the celebrated St{\o }%
rmer Theorem \cite{Stormer}, which we adapt here for the CAR algebra (see
Lemma \ref{BCS Lemma 5}). We orient our approach on the
Petz--Raggio--Verbeure results in \cite{Petz-Raggio-Verbeure}, but we would
like to mention that the analysis of permutation invariant quantum systems
in the thermodynamic limit (with St{\o }rmer's theorem as the background) is
carried out for different classes of systems also by other authors. See,
e.g., \cite{S,AdamsDorlas}. Finally, we introduce in Section \ref%
{equilibirum.paragraph} a notion of equilibrium and ground states by a usual
variational principle for the free energy density. The thermodynamics of the
strong coupling BCS-Hubbard model described in Sections \ref{Section phase
diagram mu fixed}--\ref{Section phase diagram rho fixed} is encoded in this
notion and the thermodynamic limits of local Gibbs states used above for
simplicity are special cases of equilibrium and ground states defined in
Section \ref{equilibirum.paragraph}. Before we proceed, we first define some
basic mathematical objects needed in our analysis.

Let $I$ be the set of finite subsets of $\mathbb{Z}^{d\geq 1}.$ For any $%
\Lambda \in I$ we then define $\mathcal{U}_{\Lambda }$ as the $C^{\ast }$%
--algebra generated by $\{a_{x,\uparrow },a_{x,\downarrow }\}_{x\in \Lambda
} $ and the identity. Choosing some fixed bijective map $\kappa :\mathbb{N}%
\rightarrow \mathbb{\ Z}^{d}$, $\mathbb{N}:=\{1,2,\ldots \}$, $\mathcal{U}%
_{N}$ denotes the local $C^{\ast }$--algebra $\mathcal{U}_{\{\kappa
(1),...,\kappa (N)\}}$ at fixed $N\in \mathbb{N}$, whereas $\mathcal{U}$ is
the full $C^{\ast }$--algebra, i.e., the closure of the union of all $%
\mathcal{U}_{N}$ for any integer $N\geq 1$. Note that
\begin{equation*}
n_{\kappa (l),\uparrow }:=a_{\kappa (l),\uparrow }^{\ast }a_{\kappa
(l),\uparrow }\quad \mathrm{\ and\ }\quad n_{\kappa (l),\downarrow
}:=a_{\kappa (l),\downarrow }^{\ast }a_{\kappa (l),\downarrow }
\end{equation*}%
are the electron number operators on the site $\kappa (l)$, respectively
with spin up $\uparrow $ and down $\downarrow $. To simplify the notation,
as soon as a statement clearly concerns the one--site algebra $\mathcal{U}%
_{1}=\mathcal{U}_{\{\kappa (1)\}}$, we replace $a_{\kappa (1),\uparrow },$ $%
a_{\kappa (1),\downarrow }$ and $n_{\kappa (1),\uparrow },$ $n_{\kappa
(1),\downarrow }$ respectively by $a_{\uparrow }$, $a_{\downarrow }$ and $%
n_{\uparrow },$ $n_{\downarrow }$, whereas any state on $\mathcal{U}_{1}$ is
denoted by $\zeta $ and not by $\omega ,$ which is by definition a state on
more than one site (on $\mathcal{U}_{\Lambda },$ $\mathcal{U}_{N}$ or $%
\mathcal{U}$). Important one--site Gibbs states in our analysis are the
states $\mathrm{\zeta }_{c}$ associated for any $c\in \mathbb{C}$ with the
Hamiltonian $H_{1}(c)$ (\ref{Hamiltonian BCS-Hubbard approx}) and defined by
\begin{equation}
\mathrm{\zeta }_{c}(A):=\frac{\mathrm{Trace}\left( Ae^{\beta \left\{ \left(
\mu -h\right) n_{\uparrow }+\left( \mu +h\right) n_{\downarrow }+\gamma
\left( ca_{\downarrow }^{\ast }a_{\uparrow }^{\ast }+\bar{c}a_{\uparrow
}a_{\downarrow }\right) -2\lambda n_{\uparrow }n_{\downarrow }\right\}
}\right) }{\mathrm{Trace}\left( e^{\beta \left\{ \left( \mu -h\right)
n_{\uparrow }+\left( \mu +h\right) n_{\downarrow }+\gamma \left(
ca_{\downarrow }^{\ast }a_{\uparrow }^{\ast }+\bar{c}a_{\uparrow
}a_{\downarrow }\right) -2\lambda n_{\uparrow }n_{\downarrow }\right\}
}\right) },  \label{Gibbs.state nu}
\end{equation}%
for any $A\in \mathcal{U}_{1}.$ Finally, note that our notation for the
\textquotedblleft $\mathrm{Trace"}$ does not include the Hilbert space where
it is evaluated. Using the isomorphisms $\mathcal{U}_{\Lambda }\simeq
B\left( \bigwedge \mathbb{C}^{\Lambda \times \{\uparrow ,\downarrow
\}}\right) $ of $C^{\ast }$--algebras, the corresponding Hilbert space is
deduced from the local algebra where the operators involved in each
statement are living.

Now, we are in position to start the proof of Theorem \ref{BCS theorem 1}.
It is followed by a rigorous analysis of the corresponding equilibrium and
ground states.

\subsection{Thermodynamic limit of the pressure: proof of Theorem \protect
\ref{BCS theorem 1}\label{Section main proof}}

Since we have already shown the lower bound (\ref{BCS pressure thermodynamic
limit inf}) in section 2, to finish the proof of Theorem \ref{BCS theorem 1}
it remains to obtain
\begin{equation}
\underset{N\rightarrow \infty }{\lim \sup }\left\{ \mathrm{p}_{N}\left(
\beta ,\mu ,\lambda ,\gamma ,h\right) \right\} \leq \underset{c\in \mathbb{C}%
}{\sup }\left\{ -\gamma |c|^{2}+p\left( c\right) \right\} .
\label{BCS pressure upper bond}
\end{equation}%
We split this proof into several lemmata. But first, we need some additional
definitions.

We define the set of all $S$--invariant even states. Let $S$ be the set of
bijective maps from $\mathbb{N}$ to $\mathbb{N}$ which leaves invariant all
but finitely many elements. It is a group w.r.t. the composition. The condition
\begin{equation}
\eta _{s}:a_{\kappa (l),\#}\mapsto a_{\kappa (s(l)),\#},\quad s\in S, \;
l\in \mathbb{N},  \label{definition eta}
\end{equation}%
defines a group homomorphism $\eta :S\rightarrow \mathrm{Aut}(\mathcal{U})$,
$s\mapsto \eta _{s}$ uniquely. Here, $\#$ stands for a spin up $\uparrow $
or down $\downarrow $. Then, let%
\begin{eqnarray*}
E_{\mathcal{U}}^{S,+}&:=&\big\{\omega \in E_{\mathcal{U}}:\mathrm{\ }\omega
\circ \eta _{s}=\omega \; \mathrm{for\ any\ }s\in S, \mathrm{\ and\ } \\
&&\omega (a_{\kappa (l_{1}),\#}^{\ast }...a_{\kappa (l_{t}),\#}^{\ast
}a_{\kappa (m_{1}),\#}...a_{\kappa (m_{\tau }),\#})=0\mathrm{\ if\ }t+\tau
\mathrm{\;is\;odd}\big\}
\end{eqnarray*}%
be the set of all $S$--invariant even states, where $E_{\mathcal{U}}$ is the
set of all states on $\mathcal{U}$. The set $E_{\mathcal{U}}^{S,+}$ is
weak$^\ast $--compact and convex. In particular, the set of extremal points of $E_{%
\mathcal{U}}^{S,+}$, denoted by $\mathcal{E}_{\mathcal{U}}^{S,+}$, is not
empty.

\begin{remark}
\label{remark-even}Any permutation invariant (p.i.) state on $\mathcal{U}$ is in
fact automatically\textit{\ }even, see, e.g., Example 5.2.21 of \cite%
{BrattelliRobinson}. We explicitly write the evenness of states in the
definition of $E_{\mathcal{U}}^{S,+}$ because this property is essential in
our arguments below.
\end{remark}

Now, to fix the notation and for the reader convenience, we collect
well--known results about the so--called relative entropy, cf. \cite%
{BrattelliRobinson,Araki-Moriya}. Let $\omega ^{(1)}$ and $\omega ^{(2)}$ be
two states on the local algebra $\mathcal{U}_{\Lambda }$, with $\omega
^{(1)} $ being faithful. Define the relative entropy\footnote{%
As in \cite{Araki-Moriya} we use the Araki--Kosaki definition, which has
opposite sign than the one given in \cite{BrattelliRobinson}.}%
\begin{equation*}
S(\omega ^{(1)}|\omega ^{(2)}):=\mathrm{Trace}\,\left( D_{\omega ^{(2)}}\ln
D_{\omega ^{(2)}}\right) -\mathrm{Trace}\,\left( D_{\omega ^{(2)}}\ln
D_{\omega ^{(1)}}\right) ,
\end{equation*}%
where $D_{\omega ^{(j)}}$ is the density matrix associated to the state $%
\omega ^{(j)}$ with $j=1,2$. The relative entropy is super--additive: for
any $\Lambda _{1},\Lambda _{2}\in I$, $\Lambda _{1}\cap \Lambda
_{2}=\emptyset $, and for any even states $\omega ^{(1)},\omega
^{(2)},\omega ^{(1,2)}$ respectively on $\mathcal{U}_{\Lambda _{1}}$, $%
\mathcal{U}_{\Lambda _{2}}$ and $\mathcal{U}_{\Lambda _{1}\cup \Lambda _{2}}$%
, $\omega ^{(1)}$ and $\omega ^{(2)}$ faithful, we have
\begin{equation}
S(\omega ^{(1)}\otimes \omega ^{(2)}\,|\,\omega ^{(1,2)})\geq S(\omega
^{(1)}|\,\omega ^{(1,2)}|_{\mathcal{U}_{\Lambda _{1}}})+S(\omega
^{(2)}\,|\,\omega ^{(1,2)}|_{\mathcal{U}_{\Lambda _{2}}}).
\label{rel.ent.supadd}
\end{equation}%
For even states $\omega ^{(1)}$ and $\omega ^{(2)}$, respectively on $%
\mathcal{U}_{\Lambda _{1}}$ and $\mathcal{U}_{\Lambda _{2}}$ with $\Lambda
_{1}\cap \Lambda _{2}=\emptyset $, the even state $\omega ^{(1)}\otimes
\omega ^{(2)}$ is the unique extension of $\omega ^{(1)}$ and $\omega ^{(2)}$
on $\mathcal{U}_{\Lambda _{1}\cup \Lambda _{2}}$ satisfying for all $A\in
\mathcal{U}_{\Lambda _{1}}$ and all $B\in \mathcal{U}_{\Lambda _{2}}$,
\begin{equation*}
\omega ^{(1)}\otimes \omega ^{(2)}(AB)=\omega ^{(1)}(A)\omega ^{(2)}(B).
\end{equation*}%
The state $\omega ^{(1)}\otimes \omega ^{(2)}$ is called the\textit{\
product }of $\omega ^{(1)}$ and $\omega ^{(2)}$. The product of even states
is an associative operation. In particular, products of even states can be
defined w.r.t. any countable set $\{\mathcal{U}_{\Lambda _{n}}\}_{n\in
\mathbb{N}}$ of subalgebras of $\mathcal{U}$ with $\Lambda _{m}\cap \Lambda
_{n}=\emptyset $ for $m\not=m$.

Observe that the relative entropy becomes additive w.r.t. product states: if
$\omega ^{(1,2)}=\hat{\omega}^{(1)}\otimes \hat{\omega}^{(2)}$, where $\hat{%
\omega}^{(1)}$ and $\hat{\omega}^{(2)}$ are two even states respectively on $%
\mathcal{U}_{\Lambda _{1}}$ and $\mathcal{U}_{\Lambda _{2}}$, then (\ref%
{rel.ent.supadd}) is satisfied with equality. The relative entropy is also
convex: for any states $\omega ^{(1)},$ $\omega ^{(2)},$ and $\omega ^{(3)}$
on $\mathcal{U}_{\Lambda }$, $\omega ^{(1)}$ faithful, and for any $\tau \in
(0,1)$
\begin{equation}
S(\omega ^{(1)}\,|\,\tau \omega ^{(2)}+(1-\tau )\omega ^{(3)})\leq \tau
S(\omega ^{(1)}\,|\,\omega ^{(2)})+(1-\tau )S(\omega ^{(1)}\,|\,\omega
^{(3)}).  \label{rel.ent.convex}
\end{equation}%
Meanwhile
\begin{eqnarray}
S(\omega ^{(1)}\,|\,\tau \omega ^{(2)}+(1-\tau )\omega ^{(3)}) &\geq &\tau
\log \tau +(1-\tau )\log (1-\tau )+\tau S(\omega ^{(1)}\,|\,\omega ^{(2)})
\notag \\
&&+(1-\tau )S(\omega ^{(1)}\,|\,\omega ^{(3)}),  \label{rel.ent.convex+}
\end{eqnarray}%
for any $\tau \in (0,1)$. Note that the relative entropy makes sense in a
class of states on $\mathcal{U}$ much larger than that of even states on $%
\mathcal{U}_\Lambda$ (cf. \cite{Araki-Moriya}), but this is not needed here.

The condition
\begin{equation*}
\sigma :a_{\kappa (l),\#}\mapsto a_{\kappa (l+1),\#}
\end{equation*}%
uniquely defines a homomorphism $\sigma $ on $\mathcal{U}$ called \emph{%
right--shift} homomorphism. Any state $\omega $ on $\mathcal{U}$ such that $%
\omega =\omega \circ \sigma $ is called shift--invariant and we denote by $%
E_{\mathcal{U}}^{\sigma }$ the set of shift--invariant states on $\mathcal{U}
$. An important class of shift--invariant states are product states $\omega
_{\zeta }$ obtained by \textquotedblleft copying\textquotedblright\ some
even state $\zeta $ of the one--site algebra $\mathcal{U}_{1}$ on all other
sites, i.e.,
\begin{equation}
\omega _{\zeta }:=\bigotimes\limits_{k=0}^{\infty }\zeta \circ \sigma ^{k}.
\label{shifft-invariant gibbs states}
\end{equation}%
Such product states are important and used below as reference states. More
generally, a state $\omega $ is $L$--periodic with $L\in \mathbb{N}$ if $%
\omega =\omega \circ \sigma ^{L}$. For each $L\in \mathbb{N}$, the set of
all $L$--periodic states from $E_{\mathcal{U}}$ is denoted by $E_{\mathcal{U}%
}^{\sigma ^{L}}.$

Let $\zeta $ be any faithful even state on $\mathcal{U}_{1}$ and let $\omega
$ be any $L$--periodic state on $\mathcal{U}$. It immediately follows from
super--additivity (\ref{rel.ent.supadd}) that for any $N,M\in \mathbb{N}$%
\begin{equation*}
S(\omega _{\zeta }|_{\mathcal{U}_{(M+N)L}}\,|\,\omega |_{\mathcal{U}%
_{(M+N)L}})\geq S(\omega _{\zeta }|_{\mathcal{U}_{ML}}\,|\,\omega |_{%
\mathcal{U}_{ML}})+S(\omega _{\zeta }|_{\mathcal{U}_{NL}}\,|\,\omega |_{%
\mathcal{U}_{NL}}).
\end{equation*}%
In particular, the following limit exists
\begin{equation}
\tilde{S}(\zeta ,\omega ):=\lim\limits_{N\rightarrow \infty }\frac{S(\omega
_{\zeta }|_{\mathcal{U}_{NL}}\,|\,\omega |_{\mathcal{U}_{NL}})}{NL}%
=\sup\limits_{N\in \mathbb{N}}\frac{S(\omega _{\zeta }|_{\mathcal{U}%
_{NL}}\,|\,\omega |_{\mathcal{U}_{NL}})}{NL}  \label{ent.density}
\end{equation}%
and is the relative entropy density of $\omega $ w.r.t. the reference state $%
\zeta $. This functional has the following important properties:

\begin{lemma}[Properties of the relative entropy density]
\label{lemma Section main proof 1}\mbox{ }\newline
At any fixed $L\in \mathbb{N}$, the relative entropy density functional $%
\omega \mapsto \tilde{S}(\zeta ,\omega )$ is lower
 weak$^\ast$--semicontinuous, i.e., for any faithful even state
$\zeta \in E_{\mathcal{U}_{1}}$ and any $r\in \mathbb{R}$, the set
\begin{equation*}
M_{r}:=\left\{ \omega \in E_{\mathcal{U}}^{\sigma ^{L}}\,:\,\tilde{S}(\zeta
,\omega )>r\right\}
\end{equation*}%
is open w.r.t. the weak$^\ast $--topology. It is also affine, i.e., for any
faithful state $\zeta \in E_{\mathcal{U}_{1}}$ and states $\omega ,\omega
^{\prime }\in E_{\mathcal{U}}^{\sigma ^{L}}$
\begin{equation*}
\tilde{S}(\zeta ,\tau \omega +(1-\tau )\omega ^{\prime })=\tau \tilde{S}%
(\zeta ,\omega )+(1-\tau )\tilde{S}(\zeta ,\omega ^{\prime }),
\end{equation*}%
with $\tau \in (0,1).$
\end{lemma}

\noindent \textit{Proof: }Without loss of generality, let $L=1$. From the
second equality of (\ref{ent.density}),
\begin{equation*}
M_{r}=\bigcup\limits_{N\in \mathbb{N}}\left\{ \omega \in E_{\mathcal{U}%
}^{\sigma }\,:\,S(\omega _{\zeta }|_{\mathcal{U}_{N}}\,|\,\omega |_{\mathcal{%
U}_{N}})>rN\right\} .
\end{equation*}%
As the maps $\omega \mapsto S(\omega _{\zeta }|_{\mathcal{U}_{N}}\,|\,\omega
|_{\mathcal{U}_{N}})$ are weak$^\ast $--continuous for each $N$, it follows
that $M_{r}$ is the union of open sets, which implies the
 lower weak$^\ast$--semicontinuity of the
relative entropy density functional. Moreover
from (\ref{rel.ent.convex}) and (\ref{rel.ent.convex+}) we directly obtain
that $\tilde{S}(\zeta ,\omega )$ is affine.\hfill $\Box $

Notice that any p.i. state is automatically shift--invariant. Thus, the mean
relative entropy density is a well--defined functional on $E_{\mathcal{U}%
}^{S,+}$. Now, we need to define on $E_{\mathcal{U}}^{S,+}$ the functional $%
\Delta \left( \omega \right) $ relating to the mean BCS interaction energy
per site:

\begin{lemma}[BCS energy per site for p.i. states]
\label{BCS Lemma 0}\mbox{ }\newline
For any $\omega \in E_{\mathcal{U}}^{S,+}$, the mean BCS interaction energy
per site in the thermodynamic limit
\begin{eqnarray*}
\Delta \left( \omega \right) & := &\underset{N\rightarrow \infty }{\lim }%
\frac{\gamma }{N^{2}}\overset{N}{\sum_{l,m=1}}\omega \left( a_{\kappa
(l),\uparrow }^{\ast }a_{\kappa (l),\downarrow }^{\ast }a_{\kappa
(m),\downarrow }a_{\kappa (m),\uparrow }\right) \\
&=&\gamma \omega \left( a_{\kappa (1),\uparrow }^{\ast }a_{\kappa
(1),\downarrow }^{\ast }a_{\kappa (2),\downarrow }a_{\kappa (2),\uparrow
}\right)
\end{eqnarray*}%
is well--defined and the affine map $\Delta :E_{\mathcal{U}%
}^{S,+}\rightarrow \mathbb{C}$, $\omega \mapsto \Delta \left( \omega \right)
$ is weak$^\ast $--continuous.
\end{lemma}

\noindent \textit{Proof:} First,
\begin{eqnarray}
\overset{N}{\sum_{l,m=1}}\omega \left( a_{\kappa (l),\uparrow }^{\ast
}a_{\kappa (l),\downarrow }^{\ast }a_{\kappa (m),\downarrow }a_{\kappa
(m),\uparrow }\right) &=&\overset{N}{\sum_{l=1}}\omega \left( a_{\kappa
(l),\uparrow }^{\ast }a_{\kappa (l),\downarrow }^{\ast }a_{\kappa
(l),\downarrow }a_{\kappa (l),\uparrow }\right)  \notag \\
&&+\overset{N}{\sum_{\substack{ l,m=1  \\ l\neq m}}}\omega \left( a_{\kappa
(l),\uparrow }^{\ast }a_{\kappa (l),\downarrow }^{\ast }a_{\kappa
(m),\downarrow }a_{\kappa (m),\uparrow }\right) .  \notag \\
&&  \label{BCS equation 1}
\end{eqnarray}%
Since $\omega \in E_{\mathcal{U}}^{S,+}$, for any $l\neq m$ observe that
\begin{equation}
\omega \left( a_{\kappa (l),\uparrow }^{\ast }a_{\kappa (l),\downarrow
}^{\ast }a_{\kappa (m),\downarrow }a_{\kappa (m),\uparrow }\right) =\omega
\left( a_{\kappa (1),\uparrow }^{\ast }a_{\kappa (1),\downarrow }^{\ast
}a_{\kappa (2),\downarrow }a_{\kappa (2),\uparrow }\right) ,
\label{BCS equation 2}
\end{equation}%
whereas
\begin{equation}
\omega \left( a_{\kappa (l),\uparrow }^{\ast }a_{\kappa (l),\downarrow
}^{\ast }a_{\kappa (l),\downarrow }a_{\kappa (l),\uparrow }\right) =\omega
\left( a_{\kappa (1),\uparrow }^{\ast }a_{\kappa (1),\downarrow }^{\ast
}a_{\kappa (1),\downarrow }a_{\kappa (1),\uparrow }\right) .
\label{BCS equation 3}
\end{equation}%
Therefore, by combining (\ref{BCS equation 1}) with (\ref{BCS equation 2})
and (\ref{BCS equation 3}), the lemma follows.\hfill $\Box $

Now, we define by
\begin{equation}
\omega ^{H}\left( A\right) :=\frac{\mathrm{Trace}\left( A\,e^{-\beta
H}\right) }{\mathrm{Trace}\left( e^{-\beta H}\right) },\quad A\in \mathcal{U}%
_{\Lambda },  \label{Gibbs.state}
\end{equation}%
the Gibbs state associated with any self--adjoint element $H$ of $\mathcal{U}%
_{\Lambda }$ at inverse temperature $\beta >0$. This definition is of course
in accordance with the Gibbs state $\omega _{N}$ (\ref{BCS gibbs state Hn})
associated with the Hamiltonian\footnote{%
with the appropriate numbering of sites defined by the bijective map $\kappa
$.} $\mathrm{H}_{N}$ (\ref{Hamiltonian BCS-Hubbard}) since $\omega
_{N}=\omega ^{\mathrm{H}_{N}}$ for any $N\in \mathbb{N}$. Note however, that
the state $\omega _{N}$ is seen either as defined on the local algebra $%
\mathcal{U}_{N}$ or as defined on the whole algebra $\mathcal{U}$ by
periodically extending it (with period $N$).

Next we give an important property of Gibbs states (\ref{Gibbs.state}):

\begin{lemma}[Passivity of Gibbs states]
\label{passivity.Gibbs}\mbox{ }\newline
Let $H_{0}$, $H_{1}$ be self--adjoint elements from $\mathcal{U}_{\Lambda }$
and define for any state $\omega $ on $\mathcal{U}_{\Lambda }$
\begin{equation*}
F_{\Lambda }(\omega ):=-\omega (H_{1})-\beta ^{-1}S(\omega ^{H_{0}}|\omega
)+P^{H_{0}},
\end{equation*}%
where $P^{H}:=\beta ^{-1}\ln \mathrm{Trace}\left( e^{-\beta H}\right) $ for
any self--adjoint $H\in \mathcal{U}_{\Lambda }$. Then $P^{H_{1}+H_{0}}\geq
F_{\Lambda }(\omega )$ for any state $\omega $ on $\mathcal{U}_{\Lambda }$
with equality if $\omega =\omega ^{H_{0}+H_{1}}$. Note that $-F_{\Lambda
}(\omega )$ is the free energy associated with the state $\omega $.
\end{lemma}

\noindent \textit{Proof:} For any self--adjoint $H\in \mathcal{U}_{\Lambda }$
and any state $\omega $ on $\mathcal{U}_{\Lambda }$ observe that
\begin{equation}
\mathrm{Trace}\left( D_{\omega }\ln D_{\omega ^{H}}\right) =\mathrm{Trace}%
\left( D_{\omega }\ln \left( \exp \left( -\beta P^{H}-\beta H\right) \right)
\right) =-\beta \omega (H)-\beta P^{H},  \label{petit equality}
\end{equation}%
which implies that
\begin{eqnarray}
P^{H_{1}+H_{0}} &=&-\beta ^{-1}\left( \mathrm{Trace}\left( D_{\omega
^{H_{0}+H_{1}}}\ln D_{\omega ^{H_{0}+H_{1}}}\right) -\mathrm{Trace}\left(
D_{\omega ^{H_{0}+H_{1}}}\ln D_{\omega ^{H_{0}}}\right) \right)  \notag \\
&&-\omega ^{H_{0}+H_{1}}(H_{1})+P^{H_{0}},  \label{apres petit equality}
\end{eqnarray}%
i.e., $P^{H_{1}+H_{0}}=F_{\Lambda }(\omega ^{H_{0}+H_{1}})$. Without loss of
generality take any faithful state $\omega $ on $\mathcal{U}_{\Lambda }$. In
this case, there are positive numbers $\lambda _{j}$ with $\sum_{j}\lambda
_{j}=1$ and vectors $\left\langle j\right\vert $ from the Hilbert space $%
\bigwedge {\mathcal{H}}_{\Lambda }$ such that $\omega (\cdot
)=\sum_{j}\lambda _{j}\left\langle j\right\vert \cdot \left\vert
j\right\rangle $. In particular, from (\ref{petit equality}) we have
\begin{equation*}
-\beta \omega (H_{1})-S(\omega ^{H_{0}}|\omega )+\beta
P^{H_{0}}=\sum_{j}\lambda _{j}\left( -\ln \lambda _{j}-\beta \left\langle
j\right\vert H_{0}+H_{1}\left\vert j\right\rangle \right) .
\end{equation*}%
Consequently, by convexity of the exponential function combined with Jensen
inequality we obtain that
\begin{eqnarray*}
&&\exp \Big(-\beta \omega (H_{1})-S(\omega ^{H_{0}}|\omega )+\beta P^{H_{0}}%
\Big) \\
&\leq &\sum_{j}\lambda _{j}\exp \left( -\ln \lambda _{j}-\beta \left\langle
j\right\vert H_{0}+H_{1}\left\vert j\right\rangle \right) \\
&\leq &\mathrm{Trace}\,\left( \exp \left( -\beta (H_{0}+H_{1})\right)
\right) =\exp \left( \beta P^{H_{1}+H_{0}}\right) .
\end{eqnarray*}%
Note that the last inequality uses the so--called Peierls--Bogoliubov
inequality which is again a consequence of Jensen inequality.\hfill $\Box $%
\newline
This proof is standard (see, e.g., \cite{BrattelliRobinson}). It is only
given in detail here, because we also need later equations (\ref{petit
equality}) and (\ref{apres petit equality}).

Observe that Lemma \ref{passivity.Gibbs} applied to $\omega =\omega ^{H_{0}}$
gives the Bogoliubov (convexity) inequality \cite{BruZagrebnov8}. We can
also deduce from this lemma that the pressure $\mathrm{p}_{N}\left( \beta
,\mu ,\lambda ,\gamma ,h\right) $ (\ref{BCS pressure}) associated with $%
\mathrm{H}_{N}$ equals%
\begin{eqnarray}
\mathrm{p}_{N}\left( \beta ,\mu ,\lambda ,\gamma ,h\right) &=&\frac{\gamma }{%
N^{2}}\overset{N}{\sum_{l,m=1}}\omega _{N}\left( a_{\kappa (l),\uparrow
}^{\ast }a_{\kappa (l),\downarrow }^{\ast }a_{\kappa (m),\downarrow
}a_{\kappa (m),\uparrow }\right)  \notag \\
&&-\frac{1}{\beta N}S\left( \omega _{\mathrm{\zeta }_{0}}|_{\mathcal{U}%
_{N}}|\omega _{N}\right) +\mathrm{p}_{N}\left( \beta ,\mu ,\lambda
,0,h\right) ,  \label{BCS equation 4}
\end{eqnarray}%
for any $\beta ,\gamma >0$ and real numbers $\mu ,\lambda ,h.$ Recall that $%
\omega _{\mathrm{\zeta }_{0}}$ is the shift--invariant state obtained by
\textquotedblleft copying\textquotedblright\ the state $\mathrm{\zeta }_{0}$
(\ref{Gibbs.state nu}) of the one--site algebra $\mathcal{U}_{1}$, see (\ref%
{shifft-invariant gibbs states}).

\begin{lemma}[From $S$ to the relative entropy density $\tilde{S}$ at finite
$N$]
\label{BCS Lemma 1}\mbox{ }\newline
Let $\tilde{\omega}_{N}$ be the shift--invariant state defined by
\begin{equation*}
\tilde{\omega}_{N}:=\frac{1}{N}\left( \omega _{N}+\omega _{N}\circ \sigma
+\cdots +\omega _{N}\circ \sigma ^{N-1}\right) ,
\end{equation*}%
where $\sigma $ is the right--shift homomorphism. Then $S\left( \omega _{%
\mathrm{\zeta }_{0}}|_{\mathcal{U}_{N}}|\omega _{N}\right) =N\tilde{S}\left(
\mathrm{\zeta }_{0},\tilde{\omega}_{N}\right) $, cf. (\ref{ent.density}).
\end{lemma}

\noindent \textit{Proof:} By Lemma \ref{lemma Section main proof 1} combined
with (\ref{ent.density}), the relative entropy density $\tilde{S}\left(
\mathrm{\zeta }_{0},\tilde{\omega}_{N}\right) $ equals
\begin{equation}
\tilde{S}\left( \mathrm{\zeta }_{0},\tilde{\omega}_{N}\right)
=\lim\limits_{M\rightarrow \infty }\left\{ \frac{1}{MN}\sum%
\limits_{k=0}^{N-1}\frac{1}{N}S\left( \omega _{\mathrm{\zeta }_{0}}|_{%
\mathcal{U}_{MN}}\,|\,\omega _{N}\circ \sigma ^{k}|_{\mathcal{U}%
_{MN}}\right) \right\} ,  \label{petit equality 2}
\end{equation}%
for any fixed $N\in \mathbb{N}$. By using now the additivity of the relative
entropy for product states observe that
\begin{eqnarray}
S\left( \omega _{\mathrm{\zeta }_{0}}|_{\mathcal{U}_{MN}}\,|\,\omega
_{N}\circ \sigma ^{k}|_{\mathcal{U}_{MN}}\right) &=&(M-1)S\left( \omega _{%
\mathrm{\zeta }_{0}}|_{\mathcal{U}_{N}}\,|\,\omega _{N}|_{\mathcal{U}%
_{N}}\right) +S\left( \omega _{\mathrm{\zeta }_{0}}|_{\mathcal{U}%
_{k}}\,|\,\omega _{N}|_{\mathcal{U}_{k}}\right)  \notag \\
&&+S\left( \omega _{\mathrm{\zeta }_{0}}|_{\mathcal{U}_{N-k}}\,|\,\omega
_{N}|_{\mathcal{U}_{N-k}}\right) ,  \label{petit equality 3}
\end{eqnarray}%
for any $k\in \left\{ 0,\cdots ,N-1\right\} ,$ with $S\left( \omega _{%
\mathrm{\zeta }_{0}}|_{\mathcal{U}_{0}}\,|\,\omega _{N}|_{\mathcal{U}%
_{0}}\right) :=0$ by definition. Therefore the equality $S\left( \omega _{%
\mathrm{\zeta }_{0}}|_{\mathcal{U}_{N}}|\omega _{N}\right) =N\tilde{S}\left(
\mathrm{\zeta }_{0},\tilde{\omega}_{N}\right) $ directly follows from (\ref%
{petit equality 2}) combined with (\ref{petit equality 3}).\hfill $\Box $

We are now in position to give a first general upper bound for the pressure $%
\mathrm{p}_{N}\left( \beta ,\mu ,\lambda ,\gamma ,h\right) $ by using the
equality (\ref{BCS equation 4}) together with Lemmata \ref{BCS Lemma 0} and %
\ref{BCS Lemma 1}.

\begin{lemma}[General upper bound of the pressure $p_{N}$]
\label{BCS Lemma 3}\mbox{ }\newline
For any $\beta ,\gamma >0$ and $\mu ,\lambda ,h\in \mathbb{R}$, one gets
that
\begin{equation*}
\underset{N\rightarrow \infty }{\lim \sup }\left\{ \mathrm{p}_{N}\left(
\beta ,\mu ,\lambda ,\gamma ,h\right) \right\} \leq \mathrm{p}\left( \beta
,\mu ,\lambda ,0,h\right) +\underset{\omega \in \mathcal{E}_{\mathcal{U}%
}^{S,+}}{\sup }\left\{ \Delta \left( \omega \right) -\beta ^{-1}\tilde{S}%
\left( \mathrm{\zeta }_{0},\omega \right) \right\} ,
\end{equation*}%
where we recall that $\mathcal{E}_{\mathcal{U}}^{S,+}$ is the non empty set
of extremal points of $E_{\mathcal{U}}^{S,+}$.
\end{lemma}

\noindent \textit{Proof: }By (\ref{BCS equation 4}) combined with Lemma \ref%
{BCS Lemma 1} one gets
\begin{eqnarray}
\mathrm{p}_{N}\left( \beta ,\mu ,\lambda ,\gamma ,h\right) &=&\frac{\gamma }{%
N^{2}}\overset{N}{\sum_{l,m=1}}\omega _{N}\left( a_{\kappa (l),\uparrow
}^{\ast }a_{\kappa (l),\downarrow }^{\ast }a_{\kappa (m),\downarrow
}a_{\kappa (m),\uparrow }\right)  \notag \\
&&-\beta ^{-1}\tilde{S}\left( \mathrm{\zeta }_{0},\tilde{\omega}_{N}\right) +%
\mathrm{p}_{N}\left( \beta ,\mu ,\lambda ,0,h\right) .
\label{BCS equation 5}
\end{eqnarray}%
The last term of this equality is independent of $N\in \mathbb{N}$ since
\begin{equation}
\mathrm{p}_{N}\left( \beta ,\mu ,\lambda ,0,h\right) =\frac{1}{\beta }\ln
\mathrm{Trace}\left( e^{\beta \left[ \left( \mu -h\right) n_{\uparrow
}+\left( \mu +h\right) n_{\downarrow }-2\lambda n_{\uparrow }n_{\downarrow }%
\right] }\right) =:\mathrm{p}\left( \beta ,\mu ,\lambda ,0,h\right) ,
\label{BCS equation 5bis}
\end{equation}%
cf. (\ref{BCS pressure approx c}).

However, the other terms require the knowledge of the states $\omega _{N}$
and $\tilde{\omega}_{N}$ in the limit $N\rightarrow \infty $. Actually,
because the unit ball in $\mathcal{U}$ is a metric space w.r.t. the
weak$^\ast $--topology, the sequence $\{\tilde{\omega}_{N}\}$ converges in the
weak$^\ast $--topology along a subsequence towards $\omega _{\infty }$.
Meanwhile, it is easy to see that for all $A\in \mathcal{U}_{\Lambda }$, $%
\Lambda \in I$,
\begin{equation*}
\underset{N\rightarrow \infty }{\lim }\left\{ \omega _{N}\left( A\right) -%
\tilde{\omega}_{N}\left( A\right) \right\} =0.
\end{equation*}%
Thus, the sequences of states $\omega _{N}$ and $\tilde{\omega}_{N}$ have
the same limit points. Since $\omega _{N}$ is even and permutation invariant
w.r.t. the $N$ first sites, the state $\omega _{\infty }$ belongs to $E_{%
\mathcal{U}}^{S,+}$. We now estimate the first term (\ref{BCS equation 5})
as in Lemma \ref{BCS Lemma 0} to get
\begin{eqnarray}
\underset{N\rightarrow \infty }{\lim \sup }\left\{ \mathrm{p}_{N}\left(
\beta ,\mu ,\lambda ,\gamma ,h\right) \right\} &\leq &\mathrm{p}\left( \beta
,\mu ,\lambda ,0,h\right) +\gamma \omega _{\infty }\left( a_{\kappa
(1),\uparrow }^{\ast }a_{\kappa (1),\downarrow }^{\ast }a_{\kappa
(2),\uparrow }a_{\kappa (2),\downarrow }\right)  \notag \\
&&+\beta ^{-1}\underset{N\rightarrow \infty }{\lim \sup }\left\{ -\tilde{S}%
\left( \mathrm{\zeta }_{0},\tilde{\omega}_{N}\right) \right\} .
\label{BCS equation 6}
\end{eqnarray}%
From Lemma \ref{lemma Section main proof 1} the relative entropy density is
lower semicontinuous in the weak$^\ast $--topology, which implies that
\begin{equation*}
\underset{N\rightarrow \infty }{\lim \sup }\left\{ -\tilde{S}\left( \mathrm{%
\zeta }_{0},\tilde{\omega}_{N}\right) \right\} \leq -\tilde{S}\left( \mathrm{%
\zeta }_{0},\omega _{\infty }\right) .
\end{equation*}%
By combining this last inequality with (\ref{BCS equation 6}) we then find
that
\begin{equation}
\underset{N\rightarrow \infty }{\lim \sup }\left\{ \mathrm{p}_{N}\left(
\beta ,\mu ,\lambda ,\gamma ,h\right) \right\} \leq \mathrm{p}\left( \beta
,\mu ,\lambda ,0,h\right) +\Delta \left( \omega _{\infty }\right) -\beta
^{-1}\tilde{S}\left( \mathrm{\zeta }_{0},\omega _{\infty }\right) ,
\label{petit equality 4}
\end{equation}%
with $\omega _{\infty }\in E_{\mathcal{U}}^{S,+}.$

Now, from Lemma \ref{BCS Lemma 0} the functional $\omega \mapsto \Delta
(\omega )$ is affine and weak$^\ast $--continuous, whereas by Lemma \ref%
{lemma Section main proof 1} the map $\omega \mapsto \tilde{S}(\mathrm{\zeta
}_{0},\omega )$ is affine and  lower weak$^\ast $--semicontinuous. The free
energy functional $\omega \mapsto \Delta (\omega )-\beta ^{-1}\tilde{S}(%
\mathrm{\zeta }_{0},\omega )$ is, in particular, convex and
upper weak$^\ast$--semicontinuous. Meanwhile recall that $E_{\mathcal{U}}^{S,+}$ is a
weak$^\ast $--compact and convex set. Therefore, from the Bauer maximum
principle \cite[Lemma 4.1.12]{BrattelliRobinsonI} it follows that
\begin{equation}
\underset{\omega \in E_{\mathcal{U}}^{S,+}}{\sup }\left\{ \Delta \left(
\omega \right) -\beta ^{-1}\tilde{S}\left( \mathrm{\zeta }_{0},\omega
\right) \right\} =\underset{\omega \in \mathcal{E}_{\mathcal{U}}^{S,+}}{\sup
}\left\{ \Delta \left( \omega \right) -\beta ^{-1}\tilde{S}\left( \mathrm{%
\zeta }_{0},\omega \right) \right\} .  \label{BCS Lemma equation 3bis}
\end{equation}%
Together with (\ref{petit equality 4}), this last inequality implies the
upper bound stated in the lemma.\hfill $\Box $

Since even states on $\mathcal{U}$ are entirely determined by their action
on even elements from $\mathcal{U}$, observe that we can identify the set of
even p.i. states of $\mathcal{U}$ with the set of p.i. states on the even
sub--algebra $\mathcal{U}^{+}$. We want to show next that the set of
extremal points $\mathcal{E}_{\mathcal{U}}^{S,+}$ belongs to the set of
strongly clustering states on the even sub--algebra $\mathcal{U}^{+}$ of $%
\mathcal{U}$. By strongly clustering states $\omega $ w.r.t. $\mathcal{U}%
^{+} $, we mean that for any $B$ in $\mathcal{U}^{+}$, there exists a net $%
\{B_{j}\}\subseteq \mathrm{Co}\{\eta _{s}(B):s\in S\}$ such that for any $%
A\in \mathcal{U}^{+}$,
\begin{equation*}
\underset{j}{\lim }\left\vert \omega \left( A\,\eta _{s}\left( B_{j}\right)
\right) -\omega \left( A\right) \omega \left( B\right) \right\vert =0
\end{equation*}%
uniformly in $s\in S$. Here, $\mathrm{Co}\, M$ denotes the convex hull of
the set $M$.

\begin{lemma}[Characterization of the set of extremal states of $E_{\mathcal{%
U}}^{S,+}$]
\label{BCS Lemma 4}\mbox{ }\newline
Any extremal state $\omega \in \mathcal{E}_{\mathcal{U}}^{S,+}$ is strongly
clustering w.r.t. the even sub--algebra $\mathcal{U}^{+}$ and conversely.
\end{lemma}

\noindent \textit{Proof:} We use some standard facts about extremal
decompositions of states which can be found in \cite[Theorems 4.3.17 and
4.3.22]{BrattelliRobinsonI}. To satisfy the requirements of these theorems,
we need to prove that the $C^{\ast }$--algebra $\mathcal{U}^{+}$ of even
elements of $\mathcal{U}$ is asymptotically abelian w.r.t. the action of the
group $S$. This is proven as follows. For each $l\in \mathbb{N}$ define the
map $\pi ^{(l)}:\mathbb{N}\rightarrow \mathbb{N}$ by
\begin{equation}
\pi ^{(l)}(k):=\left\{
\begin{array}{lcl}
k+2^{l-1} & , & \mathrm{\ if\ }1\leq k\leq 2^{l-1}. \\
k-2^{l-1} & , & \mathrm{\ if\ }2^{l-1}+1\leq k\leq 2^{l}. \\
k & , & \mathrm{\ if\ }k>2^{l}.%
\end{array}%
\right.  \label{definition de pi}
\end{equation}%
In other words, the map $\pi ^{(l)}$ exchanges the block $\{1,\cdots
,2^{l-1}\}$ with $\{2^{l-1}+1,\cdots ,2^{l}\},$ and leaves the rest
invariant. For any $A,B\in \mathcal{U}_{\Lambda }\cap \mathcal{U}^{+}$ with $%
\Lambda \in I$, it is then not difficult to see that
\begin{equation*}
\lim\limits_{l\rightarrow \infty }\left[ A,\eta _{\pi ^{(l)}}\left( B\right) %
\right] =0
\end{equation*}%
in the norm sense. Recall that the map $\eta _{\pi ^{(l)}}$ is defined via (%
\ref{definition eta}). By density of local elements of $\mathcal{U}^{+}$ the
limit above equals zero for all $A,B\in \mathcal{U}^{+}$. Therefore, by
using now \cite[Theorems 4.3.17 and 4.3.22]{BrattelliRobinsonI} all states $%
\omega \in \mathcal{E}_{\mathcal{U}}^{S,+}$ are then strongly clustering
w.r.t. $\mathcal{U}^{+}$ and conversely.\hfill $\Box $

We show next that p.i. states, which are strongly clustering w.r.t. the even
sub--algebra $\mathcal{U}^{+},$ have clustering properties w.r.t. the whole
algebra $\mathcal{U}$.

\begin{lemma}[Extension of the strongly clustering property]
\label{clust}\mbox{ }\newline
Let $\omega \in E_{\mathcal{U}}^{S,+}$ be any strongly clustering state
w.r.t. $\mathcal{U}^{+}$. Then, for any $A,B\in \mathcal{U}$ and $%
\varepsilon>0$, there are $B_{\varepsilon }\in \mathrm{Co}\{\eta
_{s}(B)\;:\;s\in S\}$ and $l_{\varepsilon }$ such that for any $l\geq
l_{\varepsilon }$,
\begin{equation*}
|\omega (A\eta _{\pi ^{(l)}}(B_{\varepsilon }))-\omega (A)\omega
(B)|<\varepsilon.
\end{equation*}
\end{lemma}

\noindent \textit{Proof:} By density of local elements it suffices to prove
the lemma for any $A,B\in \mathcal{U}_{N}$ and $N\in \mathbb{N}$. The
operators $A$ and $B$ can always be written as sums $A=A^{+}+A^{-}$ and $%
B=B^{+}+B^{-}$, where $A^{+}$ and $B^{+}$ are in the even sub--algebra $%
\mathcal{U}^{+}$ whereas $A^{-}$ and $B^{-}$ are odd elements, i.e., they
are sums of monomials of odd degree in annihilation and creation operators.
Since $\omega $ is assumed to be strongly clustering w.r.t. $\mathcal{U}^{+}$%
, for any $\varepsilon >0$ there are positive numbers $\lambda _{1},\ldots
,\lambda _{k}$ with $\lambda _{1}+\cdots +\lambda _{k}=1,$ and maps $%
s_{1},\ldots ,s_{k}\in S$ such that for any $l\in \mathbb{N}$,
\begin{equation}
\left\vert \omega \Big(A^{+}\,\eta _{\pi ^{(l)}}\Big(\sum_{j=1}^{k}\lambda
_{k}\eta _{s_{j}}(B^{+})\Big)\Big)-\omega (A^{+})\omega (B^{+})\right\vert
<\varepsilon .  \label{petit equality 7}
\end{equation}%
By parity and linearity of $\omega $ observe that $\omega (A^{+})\omega
(B^{+})=\omega (A)\omega (B)$, whereas
\begin{equation}
\omega (A\eta _{\pi ^{(l)}}(B_{\varepsilon }))=\omega \Big(A^{+}\,\eta _{\pi
^{(l)}}\Big(\sum_{j=1}^{k}\lambda _{k}\eta _{s_{j}}(B^{+})\Big)\Big)
\label{petit equality 8}
\end{equation}%
for large enough $l$ with the operator $B_{\varepsilon }\in \mathrm{Co}%
\{\eta _{s}(B)\;:\;s\in S\}$ defined by
\begin{equation}
B_{\varepsilon }:=\sum_{j=1}^{k}\lambda _{k}\eta _{s_{j}}(B).
\label{petit equality 9}
\end{equation}

The equality (\ref{petit equality 8}) follows from parity and the statement
\begin{equation*}
\omega (A\eta _{\pi ^{(l)}}(\tilde{B}^{-}))=0
\end{equation*}%
for any $\omega \in E_{\mathcal{U}}^{S,+},$ $A,\tilde{B}^{-}\in \mathcal{U}%
_{N}$, $\tilde{B}^{-}$ odd, and sufficiently large $l$. This can be seen as
follows. Since any element of $\mathcal{U}_{N}$ with defined parity can be
written as a linear combination of two self--adjoint elements with same
parity, we assume without loss of generality that $(\tilde{B}^{-})^{\ast }=%
\tilde{B}^{-}$. Choose $l^{\prime }\in \mathbb{N}$ large enough such that
the support of $\tilde{B}_{l}^{-}:=\pi ^{(l)}(\tilde{B}^{-})$ does not
intersect $\{\kappa (1),...,\kappa (N)\}$ for all $l\geq l^{\prime }$. The
map $\pi ^{(l)}:\mathbb{N}\rightarrow \mathbb{N} $ is defined by (\ref%
{definition de pi}). Define $\tilde{B}_{l,m}^{-}:=\sigma ^{m2^{l+1}}(\tilde{B%
}_{l}^{-})$, $m\in \mathbb{N}_{0} := \{0,1,2,\ldots\}$, where $\sigma $ is
the right--shift homomorphism. For any $J\in \mathbb{N}$
\begin{equation*}
\omega \Big(\sum\limits_{m=0}^{J}A\tilde{B}_{l,m}^{-}\Big)=(J+1)\omega (A%
\tilde{B}_{l,0}^{-})
\end{equation*}%
by symmetry of $\omega $. Use now the Cauchy--Schwarz inequality for states
to get
\begin{equation*}
(J+1)|\omega (A\tilde{B}_{l,0}^{-})|\leq \sqrt{\omega (A^{\ast }A)}\sqrt{%
\sum\limits_{m,m^{\prime }=0}^{J}\omega (\tilde{B}_{l,m}^{-}\tilde{B}%
_{l,m^{\prime }}^{-}}).
\end{equation*}%
Since per construction, $\tilde{B}_{l,m}^{-}$ and $\tilde{B}_{l,m^{\prime
}}^{-}$ anti--commute if $m\not=m^{\prime }$,
\begin{equation*}
\sum\limits_{m,m^{\prime }=0}^{J}\omega (B_{l,m}B_{l,m^{\prime
}})=\sum\limits_{m=0}^{J}\omega (B_{l,m}B_{l,m}).
\end{equation*}%
By symmetry of $\omega $, the right--hand side of the equation above equals $%
(J+1)\omega ((\tilde{B}_{l,0}^{-})^{2})$. Hence, we conclude that
\begin{equation*}
|\omega (A\tilde{B}_{l,0}^{-})|\leq (J+1)^{-1/2}\sqrt{\omega (|A|^{2})\omega
((\tilde{B}_{l,0}^{-})^{2})},
\end{equation*}%
for any $J\in \mathbb{N}$, i.e., $\omega (A\tilde{B}_{l,0}^{-})=0$ for all $%
l\geq l^{\prime }$.

Therefore, the lemma follows from (\ref{petit equality 7})--(\ref{petit
equality 8}) with $B_{\varepsilon }\in \mathrm{Co}\{\eta _{s}(B)\;:\;s\in
S\} $ defined by (\ref{petit equality 9}) for any $\varepsilon >0$.\hfill $%
\Box $

We now identify the set of clustering states on $\mathcal{U}$ with the set
of product states by the following lemma, which is a non--commutative
version of de Finetti Theorem of probability theory \cite{De Finetti}. St{\o
}rmer \cite{Stormer} was the first to show the corresponding result for
infinite tensor products of $C^{\ast }$--algebras.

\begin{lemma}[Strongly clustering p.i. states are product states]
\label{BCS Lemma 5}\mbox{ }\newline
Any p.i. and strongly clustering (in the sense of Lemma \ref{clust})
state $\omega $ is a product state (\ref{shifft-invariant gibbs states})
with the one--site state $\zeta =\zeta _{\omega }:=\omega |_{\mathcal{U}%
_{1}} $ being the restriction of $\omega $ on the local (one--site) algebra $%
\mathcal{U}_{1}$.
\end{lemma}

\noindent \textit{Proof:} Let $l_{1},\ldots ,l_{k}\in \mathbb{N}$ with $%
l_{i}\not=l_{j}$ whenever $i\not=j$, and for any $j\in \{1,\ldots ,k\}$ take
$A_{j}\in \mathcal{U}_{1}$. To prove the lemma we need to show that
\begin{equation}
\omega (\sigma ^{l_{1}}(A_{1})\ldots \sigma ^{l_{k}}(A_{k}))=\zeta _{\omega
}(A_{1})\ldots \zeta _{\omega }(A_{k}).  \label{petit equality 5}
\end{equation}%
The proof of this last equality for any $k\geq 1$ is performed by induction.
First, for $k=1$ the equality (\ref{petit equality 5}) immediately follows
by symmetry of the state $\omega $. Now, assume the equality (\ref{petit
equality 5}) verified at fixed $k\geq 1.$ The state $\omega $ is strongly
clustering in the sense of Lemma \ref{clust}. Therefore for each $%
\varepsilon >0$ there are $q\in \mathbb{N}$, positive numbers $\lambda
_{1},\ldots ,\lambda _{q}$ with $\lambda _{1}+\cdots +\lambda _{q}=1,$ and
maps $s_{1},\ldots ,s_{q}\in S$ such that
\begin{equation}
\begin{array}{l}
\Big |\mathop{\displaystyle \sum }_{j=1}^{q}\lambda _{j}\omega \left( \sigma
^{l_{1}}\left( A_{1}\right) \ldots \sigma ^{l_{k}}\left( A_{k}\right) \eta
_{\pi ^{(l)}\circ s_{j}}\left( \sigma ^{l_{k+1}}\left( A_{k+1}\right)
\right) \right) \\
-\omega \left( \sigma ^{l_{1}}\left( A_{1}\right) \ldots \sigma
^{l_{k}}\left( A_{k}\right) \right) \omega \left( \sigma ^{l_{k+1}}\left(
A_{k+1}\right) \right) \Big |<\varepsilon ,%
\end{array}
\label{petit equality 6bis}
\end{equation}%
for any $l\in \mathbb{N}$. Fix $N$ sufficiently large such that the
operators $\sigma ^{l_{m}}(A_{m})$ and $\eta _{s_{j}}(\sigma
^{l_{k+1}}\left( A_{k+1}\right) )$ belong to $\mathcal{U}_{N}$ for any $m\in
\left\{ 1,\cdots ,k+1\right\} $ and $j\in \left\{ 1,\cdots ,q\right\} $. We
can choose $l$ sufficiently large such that $\eta _{\pi ^{(l)}\circ
s_{j}}(\sigma ^{l_{k+1}}\left( A_{k+1}\right) )\notin \mathcal{U}_{N}$ for
any $j\in \left\{ 1,\cdots ,q\right\} $, which by symmetry of $\omega $
implies that%
\begin{eqnarray*}
&&\omega \left( \sigma ^{l_{1}}\left( A_{1}\right) \ldots \sigma
^{l_{k}}\left( A_{k}\right) \eta _{\pi ^{(l)}\circ s_{j}}\left( \sigma
^{l_{k+1}}\left( A_{k+1}\right) \right) \right) \\
&=&\omega \left( \sigma ^{l_{1}}\left( A_{1}\right) \ldots \sigma
^{l_{k}}\left( A_{k}\right) \sigma ^{l_{k+1}}\left( A_{k+1}\right) \right) .
\end{eqnarray*}%
Combined with (\ref{petit equality 6bis}) and $\lambda _{1}+\cdots +\lambda
_{q}=1,$ it yields
\begin{equation*}
\left\vert \omega \left( \sigma ^{l_{1}}\left( A_{1}\right) \ldots \sigma
^{l_{k}}\left( A_{k}\right) \sigma ^{l_{k+1}}\left( A_{k+1}\right) \right)
-\omega \left( \sigma ^{l_{1}}\left( A_{1}\right) \ldots \sigma
^{l_{k}}\left( A_{k}\right) \right) \zeta _{\omega }\left( A_{k+1}\right)
\right\vert <\varepsilon .
\end{equation*}%
Since the equality (\ref{petit equality 5}) is assumed to be verified at
fixed $k\geq 1,$ it follows that
\begin{equation*}
\left\vert \omega \left( \sigma ^{l_{1}}(A_{1})\ldots \sigma
^{l_{k+1}}(A_{k+1})\right) -\zeta _{\omega }(A_{1})\ldots \zeta _{\omega
}(A_{k+1})\right\vert <\varepsilon ,
\end{equation*}%
for any $\varepsilon >0$. In other words, by induction the equality (\ref%
{petit equality 5}) is proven for any $k\geq 1.$\hfill $\Box $

As soon as the upper bound is concerned, we combine Lemma \ref{BCS Lemma 3}
with Lemmata \ref{BCS Lemma 4}--\ref{BCS Lemma 5} to obtain that
\begin{equation}
\underset{N\rightarrow \infty }{\lim \sup }\left\{ \mathrm{p}_{N}\left(
\beta ,\mu ,\lambda ,\gamma \right) \right\} \leq \mathrm{p}\left( \beta
,\mu ,\lambda ,0,h\right) +\underset{\zeta \in E_{\mathcal{U}_{1}}^{+}}{\sup
}\left\{ \gamma |\zeta (a_{\uparrow }^{\ast }a_{\downarrow }^{\ast
})|^{2}-\beta ^{-1}S(\mathrm{\zeta }_{0}|\zeta )\right\} .
\label{BCS equation 7}
\end{equation}%
Here $E_{\mathcal{U}_{1}}^{+}$ denotes the set of even states on the
(one--site) algebra $\mathcal{U}_{1}$. Now the proof of the upper bound (\ref%
{BCS pressure upper bond}) easily follows from the passivity of Gibbs states
on $\mathcal{U}_{1}$. Indeed, we apply Lemma \ref{passivity.Gibbs} to the
one--site Hamiltonians 
$H_{0}=H_{1}(0)$ (see (\ref{Hamiltonian BCS-Hubbard approx})) and
\begin{equation*}
H_{1}=-\frac{c}{2}a_{\uparrow }^{\ast }a_{\downarrow }^{\ast }-\frac{\bar{c}%
}{2}a_{\uparrow }a_{\downarrow }
\end{equation*}%
in order to bound the relative entropy $S(\mathrm{\zeta }_{0}\,|\,\zeta )$.
More precisely, it follows that
\begin{eqnarray}  \label{bound.xi(aa)2}
\mathrm{p}\left( \beta ,\mu ,\lambda ,0,h\right) -\beta ^{-1}S(\mathrm{\zeta
}_{0}\,|\,\zeta ) &\leq &p\left( c/(2\gamma )\right) -x\mathrm{\mathop{\rm
Re}}\left\{ \zeta \left( a_{\uparrow }a_{\downarrow }\right) \right\}  \notag
\\
&&-y\mathop{\rm Im}\left\{ \zeta \left( a_{\uparrow }a_{\downarrow }\right)
\right\} ,
\end{eqnarray}%
for any state $\zeta \in E_{\mathcal{U}_{1}}^{+}$ and any $c\in \mathbb{C}$
with $x:=\mathop{\rm Re}\{c\}$ and $y:=\mathop{\rm Im}\{c\}.$ Consequently,
from (\ref{BCS equation 7}) we deduce that
\begin{eqnarray*}
\underset{N\rightarrow \infty }{\lim \sup }\left\{ \mathrm{p}_{N}\left(
\beta ,\mu ,\lambda ,\gamma ,h\right) \right\} &\leq &\underset{\zeta \in E_{%
\mathcal{U}_{1}}^{+}}{\sup }\Big\{\underset{x,y\in \mathbb{R}}{\inf }\Big\{%
\gamma (\mathop{\rm Re}\{\zeta (a_{\uparrow }a_{\downarrow })\}^{2}+%
\mathop{\rm Im}\{\zeta (a_{\uparrow }a_{\downarrow })\}^{2}) \\
&&-x\mathop{\rm Re}\{\zeta (a_{\uparrow }a_{\downarrow })\}-y\mathop{\rm Im}%
\{\zeta (a_{\uparrow }a_{\downarrow })\} \\
&&+p\left( (x+iy)/(2\gamma )\right) \Big\}\Big\} \\
&\leq &\underset{t,s\in \mathbb{R}}{\sup }\Big\{\underset{x,y\in \mathbb{R}}{%
\inf }\Big\{\gamma \left( t^{2}+s^{2}\right) -tx-sy \\
&&+p\left( (x+iy)/(2\gamma )\right) \Big\}\Big\}.
\end{eqnarray*}%
In particular, by fixing $x=2t\gamma $ and $y=2s\gamma $ in the infimum we
finally obtain
\begin{equation*}
\underset{N\rightarrow \infty }{\lim \sup }\left\{ \mathrm{p}_{N}\left(
\beta ,\mu ,\lambda ,\gamma ,h\right) \right\} \leq \underset{t,s\in \mathbb{%
R}}{\sup }\left\{ -\gamma \left( t^{2}+s^{2}\right) +p\left( t+is\right)
\right\} ,
\end{equation*}%
i.e., the upper bound (\ref{BCS pressure upper bond})\ for any $\beta
,\gamma >0$ and $\mu ,\lambda ,h\in \mathbb{R}.$

\subsection{Equilibrium and ground states of the strong coupling BCS-Hubbard
model\label{equilibirum.paragraph}}

It follows immediately from the passivity of Gibbs states that
\begin{equation}
\mathrm{p}\left( \beta ,\mu ,\lambda ,\gamma ,h\right) \geq \Delta (\omega
)-\beta ^{-1}\tilde{S}\left( \mathrm{\zeta }_{0},\omega \right) +\mathrm{p}%
\left( \beta ,\mu ,\lambda ,0,h\right) ,  \label{ineq.eq.states}
\end{equation}%
for any $\omega \in E_{\mathcal{U}}^{S,+}$, cf. (\ref{Gibbs.state nu}) and
Lemmata \ref{BCS Lemma 0}--\ref{passivity.Gibbs}. Therefore, by using Lemma %
\ref{BCS Lemma 3} with (\ref{BCS Lemma equation 3bis}) the (infinite volume)
pressure can be written as
\begin{equation*}
\mathrm{p}\left( \beta ,\mu ,\lambda ,\gamma ,h\right) =\underset{\omega \in
E_{\mathcal{U}}^{S,+}}{\sup }\left\{ \Delta \left( \omega \right) -\beta
^{-1}\tilde{S}\left( \mathrm{\zeta }_{0},\omega \right) \right\} +\mathrm{p}%
\left( \beta ,\mu ,\lambda ,0,h\right) .
\end{equation*}%
Moreover, as shown above (see the upper bound in the proof of Lemma \ref{BCS
Lemma 3}), any weak$^\ast$ limit point $\omega _{\infty }$ of local Gibbs
states $\omega _{N}$ (\ref{BCS gibbs state Hn}) when $N\rightarrow \infty $
satisfies (\ref{ineq.eq.states}) with equality.

Indeed, by using (\ref{petit equality}) one obtains for any state $\omega $
that
\begin{eqnarray}
\frac{1}{N}\left( -\omega \left( \mathrm{H}_{N}\right) -\beta ^{-1}S\left(
\mathrm{tr}_{N}\,|\omega |_{{\mathcal{U}}_{N}}\right) \right) &=&\frac{%
\gamma }{N^{2}}\overset{N}{\sum_{l,m=1}}\omega \left( a_{\kappa (l),\uparrow
}^{\ast }a_{\kappa (l),\downarrow }^{\ast }a_{\kappa (m),\downarrow
}a_{\kappa (m),\uparrow }\right)  \notag \\
&&-\frac{1}{\beta N}S\left( \omega _{\mathrm{\zeta }_{0}}|_{\mathcal{U}%
_{N}}|\omega |_{\mathcal{U}_{N}}\right)  \notag \\
&&+\mathrm{p}_{N}\left( \beta ,\mu ,\lambda ,0,h\right) ,
\label{petit equalitybis}
\end{eqnarray}%
with $\mathrm{p}_{N}$ being the (finite volume) pressure (\ref{BCS pressure}%
) associated with the Hamiltonian $\mathrm{H}_{N}$ (\ref{Hamiltonian
BCS-Hubbard}), $\omega _{\mathrm{\zeta }_{0}}$ being the product state
obtained by \textquotedblleft copying\textquotedblright\ the state $\mathrm{%
\zeta }_{0}$ (\ref{Gibbs.state nu}) on the one--site algebra $\mathcal{U}%
_{1} $ (see (\ref{shifft-invariant gibbs states})), and with the trace state
$\mathrm{tr}_{N}$ defined on the local algebra $\mathcal{U}_{N}$ for $N\in
\mathbb{N}$ by
\begin{equation*}
\mathrm{tr}_{N}(\,\cdot \,):=\frac{\mathrm{Trace}(\,\cdot \,)}{\mathrm{Trace}%
(\mathbb{I}_{\mathcal{U}_{N}})}.
\end{equation*}%
For any permutation invariant state $\omega $ it is straightforward to check
that the limits
\begin{equation*}
\lim\limits_{N\rightarrow \infty }\left\{ N^{-1}S\left( \omega _{\mathrm{%
\zeta }_{0}}|_{{\mathcal{U}}_{N}}\,|\omega |_{{\mathcal{U}}_{N}}\right)
\right\}
\end{equation*}%
and
\begin{equation*}
\mathfrak{e}\left( \omega \right) :=\lim\limits_{N\rightarrow \infty
}\left\{ N^{-1}\omega \left( \mathrm{H}_{N}\right) \right\} =\omega \left(
H_{1}(0)\right) -\Delta \left( \omega \right)
\end{equation*}%
exist for any fixed parameters $\beta ,\gamma >0$ and $\mu ,\lambda ,h\in
\mathbb{R}$, see respectively (\ref{Hamiltonian BCS-Hubbard approx}) and
Lemma \ref{BCS Lemma 0} for the definitions of $H_{1}(0)$ and $\Delta
(\omega )$. Combined with (\ref{BCS equation 5bis}) and (\ref{petit
equalitybis}) it then follows that the usual entropy density
\begin{eqnarray*}
\tilde{S}\left( \omega \right) := &&-\lim\limits_{N\rightarrow \infty
}\left\{ N^{-1}S\left( \mathrm{tr}_{N}\,|\omega |_{{\mathcal{U}}_{N}}\right)
\right\} \\
&=&-\lim\limits_{N\rightarrow \infty }\left\{ \frac{1}{N}\mathrm{Trace}%
\left( D_{\omega |_{{\mathcal{U}}_{N}}}\log D_{\omega |_{{\mathcal{U}}%
_{N}}}\right) \right\} <\infty
\end{eqnarray*}%
of the permutation invariant state $\omega $ also exists and
\begin{equation*}
\lim\limits_{N\rightarrow \infty }\frac{1}{\beta N}S\left( \omega _{\mathrm{%
\zeta }_{0}}|_{{\mathcal{U}}_{N}}\,|\omega |_{{\mathcal{U}}_{N}}\right) =%
\mathfrak{e}(\omega )+\Delta (\omega )-\beta ^{-1}\tilde{S}\left( \omega
\right) +p(\beta ,\mu ,\lambda ,0,h).
\end{equation*}%
The set $\mathit{\Omega }_{\beta }=\mathit{\Omega }_{\beta }(\mu ,\lambda
,\gamma ,h)$ of \textit{equilibrium states} of the strong coupling
BCS--Hubbard model is defined by
\begin{eqnarray*}
\mathit{\Omega }_{\beta }:= &&\Big\{\omega \in E_{\mathcal{U}}^{S,+}:-%
\mathfrak{e}\left( \omega \right) +\beta ^{-1}\tilde{S}\left( \omega \right)
=\mathrm{p}\left( \beta ,\mu ,\lambda ,\gamma ,h\right) \\
&&\phantom{\Big\{}=\Delta \left( \omega \right) -\beta ^{-1}\tilde{S}\left(
\mathrm{\zeta }_{0},\omega \right) +\mathrm{p}\left( \beta ,\mu ,\lambda
,0,h\right) \Big\}.
\end{eqnarray*}%
Note that $\mathit{\Omega }_{\beta }$ contains per construction all
weak$^\ast $ limit points of local Gibbs states $\omega _{N}$ as $N\rightarrow
\infty $.

Consequently, the equilibrium states are, as usual, the minimizers of the
free energy functional
\begin{equation}
\omega \mapsto \mathfrak{F}(\omega ):=\mathfrak{e}(\omega )-\beta ^{-1}%
\tilde{S}(\omega )  \label{p.varbis}
\end{equation}%
on the convex and weak$^\ast $--compact set $E_{{\mathcal{U}}}^{S,+}$, cf. (%
\ref{p.var}). They also maximize the upper semicontinuous affine functional $%
\omega \mapsto \Delta (\omega )-\beta ^{-1}\tilde{S}(\mathrm{\zeta }%
_{0},\omega ).$ It follows that $\mathit{\Omega }_{\beta }$ is a closed face
of $E_{\mathcal{U}}^{S,+}$ and we have in this set a notion of pure and
mixed thermodynamic phases (equilibrium states) by identifying purity with
extremality. In particular, it is convex and weak$^\ast $--compact. Each
weak$^\ast $--limit $\omega $ of equilibrium states $\omega ^{(n)}\in
\mathit{\Omega }_{\beta _{n}}(\mu _{n},\lambda _{n},\gamma _{n},h_{n})$ such
that $(\mu _{n},\lambda _{n},\gamma _{n},h_{n})\rightarrow (\mu ,\lambda
,\gamma ,h)$ and $\beta _{n}\rightarrow \infty $ is called a \textit{ground
state} of the strong coupling BCS--Hubbard model. The set of all ground
states with parameters $\gamma >0$ and $\mu ,\lambda ,h\in \mathbb{R}$ is
denoted by $\mathit{\Omega }_{\infty }=\mathit{\Omega }_{\infty }(\mu
,\lambda ,\gamma ,h)$. Extremal states of the weak$^\ast $--compact convex
set $\mathit{\Omega }_{\infty }$ are called \textit{pure} ground states.

We analyze now the set of pure equilibrium states, i.e., the equilibrium
states $\omega \in \mathit{\Omega }_{\beta }$ belonging to the set $\mathcal{%
E}_{\mathcal{U}}^{S,+}$ of extremal points of $E_{\mathcal{U}}^{S,+}$, cf. (%
\ref{BCS Lemma equation 3bis}). First, from Lemmata \ref{BCS Lemma 4}--\ref%
{BCS Lemma 5} recall that any extremal state is a product state $\omega
_{\zeta }$ (\ref{shifft-invariant gibbs states}), i.e., it is obtained by
\textquotedblleft copying\textquotedblright\ a state $\zeta $ on the
one--site algebra $\mathcal{U}_{1}$ to the other sites. In particular, by
combining (\ref{BCS Lemma equation 3bis}) with (\ref{ineq.eq.states})
observe that
\begin{equation}
\mathrm{p}\left( \beta ,\mu ,\lambda ,\gamma ,h\right) =\underset{\zeta \in
E_{\mathcal{U}_{1}}^{+}}{\sup }\left\{ \gamma |\zeta (a_{\uparrow }^{\ast
}a_{\downarrow }^{\ast })|^{2}-\beta ^{-1}S(\mathrm{\zeta }_{0}|\zeta
)\right\} +\mathrm{p}\left( \beta ,\mu ,\lambda ,0,h\right) .
\label{ineq.eq.statesbisbis}
\end{equation}%
Therefore, a product state $\omega _{\zeta }$ is a pure equilibrium state if
and only if $\zeta $ belongs to the set $\mathcal{G}_{\beta }=\mathcal{G}%
_{\beta }(\mu ,\lambda ,\gamma ,h)$ of one--site equilibrium states defined
by
\begin{equation}
\mathcal{G}_{\beta }:=\big\{\zeta \in E_{\mathcal{U}_{1}}^{+}:\gamma |\zeta
(a_{\uparrow }^{\ast }a_{\downarrow }^{\ast })|^{2}-\beta ^{-1}S(\mathrm{%
\zeta }_{0}|\zeta )=\mathrm{p}\left( \beta ,\mu ,\lambda ,\gamma ,h\right) -%
\mathrm{p}\left( \beta ,\mu ,\lambda ,0,h\right) \big\}.
\label{definition of one-site equilibrium state}
\end{equation}%
In other words, the study of pure states of $\mathit{\Omega }_{\beta }$ can
be reduced, without loss of generality, to the analysis of $\mathcal{G}%
_{\beta }.$ The first important statement concerns the characterization of
the set $\mathcal{G}_{\beta }$ in relation with the variational problems (%
\ref{BCS pressure 2}) and (\ref{ineq.eq.statesbisbis}).

\begin{theorem}[Explicit description of one-site equilibrium states]
\label{Theorem equilibrium state 1}\mbox{ }\newline
For any $\beta ,\gamma >0$ and $\mu ,\lambda ,h\in \mathbb{R}$, the set $%
\mathcal{G}_{\beta }$ of one--site equilibrium states are given by the
states $\mathrm{\zeta }_{\mathrm{c}_{\beta }}$ (\ref{Gibbs.state nu}) with $%
\mathrm{c}_{\beta }:=\mathrm{r}_{\beta }^{1/2}e^{i\phi }$ for any order
parameter $\mathrm{r}_{\beta }$ solution of (\ref{BCS pressure 2}) and any
phase $\phi \in \lbrack 0,2\pi )$.
\end{theorem}

\noindent \textit{Proof:} Take any solution $\mathrm{r}_{\beta }$ of (\ref%
{BCS pressure 2}) and any $\phi \in \lbrack 0,2\pi )$. Then, from (\ref%
{apres petit equality}) observe that
\begin{equation}
-\beta ^{-1}S(\mathrm{\zeta }_{0}\,|\,\mathrm{\zeta }_{\mathrm{c}_{\beta }})+%
\mathrm{p}\left( \beta ,\mu ,\lambda ,0,h\right) =-\gamma \mathrm{\zeta }_{%
\mathrm{c}_{\beta }}(\mathrm{c}_{\beta }a_{\uparrow }^{\ast }a_{\downarrow
}^{\ast }+\mathrm{\bar{c}}_{\beta }a_{\downarrow }a_{\uparrow })+p(\mathrm{c}%
_{\beta }).  \label{eq equilibrium state 00}
\end{equation}%
Since $\mathrm{\zeta }_{\mathrm{c}_{\beta }}(a_{\downarrow }a_{\uparrow })=%
\mathrm{c}_{\beta }$ and $\mathrm{\zeta }_{\mathrm{c}_{\beta }}(a_{\uparrow
}^{\ast }a_{\downarrow }^{\ast })=\mathrm{\bar{c}}_{\beta },$ the last
equality combined with Theorem \ref{BCS theorem 1} implies that
\begin{equation}
\gamma |\mathrm{\zeta }_{\mathrm{c}_{\beta }}(a_{\downarrow }a_{\uparrow
})|^{2}-\beta ^{-1}S(\mathrm{\zeta }_{0}\,|\,\mathrm{\zeta }_{\mathrm{c}%
_{\beta }})=\mathrm{p}\left( \beta ,\mu ,\lambda ,\gamma ,h\right) -\mathrm{p%
}\left( \beta ,\mu ,\lambda ,0,h\right) .  \label{eq equilibrium state 0}
\end{equation}%
In other words, $\mathrm{\zeta }_{\mathrm{c}_{\beta }}$ is a maximizer of
the variational problem defined in (\ref{ineq.eq.statesbisbis}) and hence, $%
\mathrm{\zeta }_{\mathrm{c}_{\beta }}\in \mathcal{G}_{\beta }$.

On the other hand, any state $\zeta \in \mathcal{G}_{\beta }$ satisfies (\ref%
{eq equilibrium state 0}) and by combining Theorem \ref{BCS theorem 1} with
the inequality (\ref{bound.xi(aa)2}) for $c= 2\gamma \zeta(a_{\downarrow
}a_{\uparrow})$ it follows that
\begin{equation*}
-\gamma|\zeta(a_{\downarrow }a_{\uparrow})|^2 + p(\zeta(a_{\downarrow
}a_{\uparrow})) \geq \sup\limits_{c \in \mathbb{C}}\{ - \gamma|c|^2 + p(c)\}.
\end{equation*}
Hence, $\zeta (a_{\downarrow }a_{\uparrow })=\mathrm{r}_{\beta
}^{1/2}e^{i\phi }=\mathrm{c}_{\beta }$ for some $\phi \in \lbrack 0,2\pi )$.
It remains to prove that the equality $\zeta (a_{\downarrow }a_{\uparrow })=%
\mathrm{c}_{\beta }$ uniquely defines the one--site equilibrium state $\zeta
\in \mathcal{G}_{\beta }$. It follows from $\zeta (a_{\downarrow
}a_{\uparrow })=\mathrm{\zeta }_{\mathrm{c}_{\beta }}(a_{\downarrow
}a_{\uparrow })=\mathrm{c}_{\beta }$ with $\zeta ,\mathrm{\zeta }_{\mathrm{c}%
_{\beta }}\in \mathcal{G}_{\beta }$ that $S(\zeta _{0}|\mathrm{\zeta }_{%
\mathrm{c}_{\beta }})=S(\mathrm{\zeta }_{0}|\zeta )$ and
\begin{equation}
\gamma \zeta (\mathrm{c}_{\beta }a_{\uparrow }^{\ast }a_{\downarrow }^{\ast
}+\mathrm{\bar{c}}_{\beta }a_{\downarrow }a_{\uparrow })-\beta ^{-1}S(%
\mathrm{\zeta }_{0}|\zeta )=P^{H_{1}(\mathrm{c}_{\beta })}-P^{H_{1}(0)}
\label{e.p.1}
\end{equation}%
because of (\ref{eq equilibrium state 00}), see (\ref{Hamiltonian
BCS-Hubbard approx}) for the definition of $H_{1}(c)$. By Lemma \ref%
{passivity.Gibbs}, one obtains for any self--adjoint $A\in {\mathcal{U}}_{1}$
that
\begin{equation}
-\zeta (A)+\gamma \zeta (\mathrm{c}_{\beta }a_{\uparrow }^{\ast
}a_{\downarrow }^{\ast }+\mathrm{\bar{c}}_{\beta }a_{\downarrow }a_{\uparrow
})-\beta ^{-1}S(\mathrm{\zeta }_{0}|\zeta )\leq P^{H_{1}(\mathrm{c}_{\beta
})+A}-P^{H_{1}(0)}.  \label{e.p.2}
\end{equation}%
Consequently, we obtain by combining (\ref{e.p.1}) and (\ref{e.p.2}) that
\begin{equation*}
P^{H_{1}(\mathrm{c}_{\beta })+A}-P^{H_{1}(\mathrm{c}_{\beta })}\geq -\zeta
(A),
\end{equation*}%
for any self--adjoint $A\in {\mathcal{U}}_{1}$ and $\zeta \in \mathcal{G}%
_{\beta }$ such that $\zeta (a_{\downarrow }a_{\uparrow })=\mathrm{c}_{\beta
}$. In other words, the functional $\{-\zeta \}$ is tangent to the pressure
at $H_{1}(\mathrm{c}_{\beta })$. Since the convex map $A\mapsto P^{H_{1}(%
\mathrm{c}_{\beta })+A}$ is continuously differentiable and self--adjoint
elements separate states, the tangent functional is unique and $\zeta =%
\mathrm{\zeta }_{\mathrm{c}_{\beta }}$.\hfill $\Box $

It follows immediately from the theorem above that pure states of $\mathit{%
\Omega }_{\beta }$ solve the gap equation:

\begin{corollary}[Gap equation for pure equilibrium states]
\label{Theorem equilibrium state 2}\mbox{ }\newline
For any $\beta ,\gamma >0$ and $\mu ,\lambda ,h\in \mathbb{R}$, pure states
from $\mathit{\Omega }_{\beta }$ are precisely the product states $\omega _{%
\mathrm{\zeta }_{\mathrm{c}_{\beta }}}$ satisfying the gap equation $\omega
_{\mathrm{\zeta }_{\mathrm{c}_{\beta }}}(a_{\kappa (l),\uparrow },a_{\kappa
(l),\downarrow })=\mathrm{c}_{\beta }$ for any $l\in \mathbb{N}$ and with $%
\mathrm{c}_{\beta }:=\mathrm{r}_{\beta }^{1/2}e^{i\phi }$ being any
maximizer of the first variational problem given in Theorem \ref{BCS theorem
1}.
\end{corollary}

If $\mathrm{c}_{\beta }\neq 0,$ observe that the gap equation $\omega _{%
\mathrm{\zeta }_{\mathrm{c}_{\beta }}}(a_{\kappa (l),\uparrow },a_{\kappa
(l),\downarrow })=\mathrm{c}_{\beta }$ with $\mathrm{\zeta }_{c}$ defined in
(\ref{Gibbs.state nu}) corresponds to the Euler--Lagrange equation satisfied
by the solutions $\mathrm{c}_{\beta }:=\mathrm{r}_{\beta }^{1/2}e^{i\phi }$
of the first variational problem given in Theorem \ref{BCS theorem 1}. The
phase $\phi \in \lbrack 0,2\pi )$ is arbitrarily taken because of the gauge
invariance of the map $c\mapsto p(c)$, and the gap equation $\omega _{%
\mathrm{\zeta }_{\mathrm{c}_{\beta }}}(a_{\kappa (l),\uparrow },a_{\kappa
(l),\downarrow })=\mathrm{c}_{\beta }$ can be reduced to (\ref{BCS gap
equation}). In other words, if $\mathrm{c}_{\beta }\neq 0$, the gap equation
can be written in two different ways: either $\omega _{\mathrm{\zeta }_{%
\mathrm{c}_{\beta }}}(a_{\kappa (l),\uparrow },a_{\kappa (l),\downarrow })=%
\mathrm{c}_{\beta }$ in the view point of extremal equilibrium states or (%
\ref{BCS gap equation}) in the view point of the order parameter $\mathrm{r}%
_{\beta }$.

From this last corollary observe also that the existence of non--zero
maximizers $\mathrm{c}_{\beta } \neq 0$ implies the existence of equilibrium
states breaking the $U(1)$--gauge symmetry satisfied by $\mathrm{H}_{N}$ (%
\ref{Hamiltonian BCS-Hubbard}). This breakdown of the $U(1)$--gauge symmetry
for $\mathrm{c}_{\beta }$ $\neq 0$ is already explained by Theorem \ref{BCS
theorem 3}, which can be proven by our notion of equilibrium states as
follows.

Consider the upper semicontinuous convex map on $E_{\mathcal{U}}^{S,+}$
defined for any $\alpha \geq 0$ and $\phi \in \lbrack 0,2\pi )$ by%
\begin{equation}
\omega \mapsto -\mathfrak{e}\left( \omega \right) +\beta ^{-1}\tilde{S}%
\left( \omega \right) +2\alpha \mathop{\rm Re}\left\{ e^{i\phi }\omega
\left( a_{\downarrow }^{\ast }a_{\uparrow }^{\ast }\right) \right\} .
\label{U(1).broken.var.probl}
\end{equation}%
From Section \ref{Section main proof} it is straightforward to check that
\begin{eqnarray}
\mathrm{p}_{\alpha ,\phi }\left( \beta ,\mu ,\lambda ,\gamma ,h\right) &:=&
\lim\limits_{N\rightarrow \infty }\left\{ \frac{1}{\beta N}\ln \mathrm{Trace}%
\left( e^{-\beta \mathrm{H}_{N,\alpha ,\phi }}\right) \right\}  \notag \\
&=&\underset{\omega \in E_{\mathcal{U}}^{S,+}}{\sup }\left\{ -\mathfrak{e}%
\left( \omega \right) +\beta ^{-1}\tilde{S}\left( \omega \right) +2\alpha %
\mathop{\rm Re}\left\{ e^{i\phi }\omega \left( a_{\downarrow }^{\ast
}a_{\uparrow }^{\ast }\right) \right\} \right\} ,  \notag \\
&&  \label{pressure avec alpha}
\end{eqnarray}%
with the Hamiltonian $\mathrm{H}_{N,\alpha ,\phi }$ defined in (\ref%
{perturbed hamiltonian}). Moreover, any weak$^\ast $--limits $\omega
_{\infty ,\alpha ,\phi }$ of local Gibbs states%
\begin{equation}
\omega _{N,\alpha ,\phi }\left( \cdot \right) :=\frac{\mathrm{Trace}\left(
\mathrm{\ }\cdot \mathrm{\ }e^{-\beta \mathrm{H}_{N,\alpha ,\phi }}\right) }{%
\mathrm{Trace}\left( e^{-\beta \mathrm{H}_{N,\alpha ,\phi }}\right) }
\label{local Gibbs states avec alpha}
\end{equation}%
are equilibrium states (see the proof of Lemma \ref{BCS Lemma 3} applied to $%
\mathrm{H}_{N,\alpha ,\phi }$), i.e., the state $\omega _{\infty ,\alpha
,\phi }$ belongs to the (non-empty) convex set $\mathit{\Omega }_{\beta
,\alpha ,\phi }=\mathit{\Omega }_{\beta ,\alpha ,\phi }(\mu ,\lambda ,\gamma
,h)$ of maximizers of (\ref{U(1).broken.var.probl}) at fixed $\alpha \geq 0$
and $\phi \in \lbrack 0,2\pi )$. In fact, one gets the following statement,
which implies Theorem \ref{BCS theorem 3}.

\begin{theorem}[Breakdown of the $U(1)$-gauge symmetry]
\label{Theorem equilibrium state 4 U(1) broken}\mbox{ }\newline
Take $\beta ,\gamma >0$ and real numbers $\mu ,\lambda ,h$ away from any
critical point. Then at fixed phase $\phi \in \lbrack 0,2\pi )$,
\begin{equation*}
\lim\limits_{\alpha \downarrow 0}\lim\limits_{N\rightarrow \infty }\frac{1}{N%
}\sum\limits_{l=1}^{N}\omega _{N,\alpha ,\phi }\left( a_{\kappa
(l),\downarrow }a_{\kappa (l),\uparrow }\right) =\lim\limits_{\alpha
\downarrow 0}\omega _{\infty ,\alpha ,\phi }(a_{\kappa (1),\downarrow
}a_{\kappa (1),\uparrow })=\mathrm{r}_{\beta }^{1/2}e^{i\phi },
\end{equation*}%
with $\omega _{\infty ,\alpha ,\phi }\in \mathit{\Omega }_{\beta ,\alpha
,\phi }$ being the unique maximizer of (\ref{U(1).broken.var.probl}) for
sufficiently small $\alpha \geq 0$.
\end{theorem}

\noindent \textit{Proof}: First we need to characterize pure states of $%
\mathit{\Omega }_{\beta ,\alpha ,\phi }$ as it is done in Corollary \ref%
{Theorem equilibrium state 2} for $\alpha =0.$ By convexity and upper
semicontinuity, note that maximizers of (\ref{U(1).broken.var.probl}) are
taken on the set of extremal states whereas the set of extremal maximizers
is a face. Since extremal states are product states (cf. Lemma \ref{BCS
Lemma 4}-\ref{BCS Lemma 5}), we get that
\begin{eqnarray}
&&\underset{\omega \in E_{\mathcal{U}}^{S,+}}{\sup }\left\{ -\mathfrak{e}%
\left( \omega \right) +\beta ^{-1}\tilde{S}\left( \omega \right) +\alpha %
\mathop{\rm Re}\left\{ e^{i\phi }\omega \left( a_{\downarrow }^{\ast
}a_{\uparrow }^{\ast }\right) \right\} \right\}  \notag \\
&=&\sup\limits_{c\in \mathbb{C}}\left\{ -\gamma |c|^{2}+p\left( c+\alpha
\gamma ^{-1}e^{i\phi }\right) \right\} ,  \label{pressure avec alphabis}
\end{eqnarray}%
as in the case $\alpha =0$ (see (\ref{BCS pressure approx c}) for the
definition of $p(c)$). If $\mathrm{c}_{\beta ,\alpha ,\phi }=\mathrm{c}%
_{\beta ,\alpha ,\phi }(\mu ,\lambda ,\gamma ,h)\in \mathbb{C}$ is a
maximizer of
\begin{equation}
-\gamma |c|^{2}+p(c+\alpha \gamma ^{-1}e^{i\phi }),
\label{var.c.broken.U(1)}
\end{equation}%
then observe that \textrm{$z$}$_{\beta ,\alpha ,\phi }:=\mathrm{c}_{\beta
,\alpha ,\phi }+\alpha \gamma ^{-1}e^{i\phi }$ maximizes the function
\begin{equation*}
-\gamma |z-\alpha \gamma ^{-1}e^{i\phi }|^{2}+p(z)
\end{equation*}%
of the complex variable $z\in \mathbb{C}$. By gauge invariance of the map $%
z\mapsto p(\beta ,\mu ,\lambda ,h;z)$, it follows that \textrm{$z$}$_{\beta
,\alpha ,\phi }\in e^{i\phi }\mathbb{R}$ and thus $\mathrm{c}_{\beta ,\alpha
,\phi }\in e^{i\phi }\mathbb{R}$. Using this, we extend Corollary \ref%
{Theorem equilibrium state 2} to $\alpha \geq 0$ and $\phi \in \lbrack
0,2\pi )$.$\ $In other words, for any $\beta ,\gamma >0$, $\alpha \geq 0$, $%
\phi \in \lbrack 0,2\pi )$ and $\mu ,\lambda ,h\in \mathbb{R}$, pure states
of $\mathit{\Omega }_{\beta ,\alpha ,\phi }$ are product states $\omega _{%
\mathrm{\zeta }_{\mathrm{c}_{\beta ,\alpha ,\phi }}}$ satisfying the gap
equation
\begin{equation}
\omega _{\mathrm{\zeta }_{\mathrm{c}_{\beta ,\alpha ,\phi }}}(a_{\kappa
(l),\uparrow },a_{\kappa (l),\downarrow })=\mathrm{c}_{\beta ,\alpha ,\phi },
\label{Theorem equilibrium state 2 U(1) broken}
\end{equation}%
for any $l\in \mathbb{N}$ and with $\mathrm{c}_{\beta ,\alpha ,\phi }\in
e^{i\phi }\mathbb{R}$ being any maximizer of (\ref{var.c.broken.U(1)}).

As $|c|\rightarrow \infty $, notice that $p\left( c\right) =\mathcal{O}%
\left( |c|\right) $. So, by gauge invariance we obtain%
\begin{eqnarray*}
\sup\limits_{c\in \mathbb{C}}\{-\gamma |c|^{2}+p(c+\alpha \gamma
^{-1}e^{i\phi })\} &=&\max\limits_{s\in \lbrack -M,M]}\left\{ -\gamma
|s\,e^{i\phi }|^{2}+p\left( [s+\alpha \gamma ^{-1}]e^{i\phi }\right) \right\}
\\
&=&\max\limits_{s\in \lbrack -M,M]}\{-\gamma s^{2}+p(s+\alpha \gamma
^{-1})\},
\end{eqnarray*}%
for any $\alpha \in (0,1)$ and $M<\infty $ sufficiently large. Consequently,
if the parameters $\beta ,$ $\mu ,$ $\lambda ,$ $\gamma ,$ and $h$ are such
that the maximizer $\mathrm{r}_{\beta }$ (\ref{BCS pressure 2}) is unique,
then the maximizer $\mathrm{c}_{\beta ,\alpha ,\phi }\in e^{i\phi }\mathbb{R}
$ of (\ref{var.c.broken.U(1)}) is also unique as soon as $\alpha >0$ is
sufficiently small. Indeed the map $s\mapsto p\left( s\right) $ is
continuous on the compact interval $[-M,M].$ In particular, from (\ref%
{Theorem equilibrium state 2 U(1) broken}) there is a unique maximizer of (%
\ref{U(1).broken.var.probl}), i.e.,
\begin{equation}
\mathit{\Omega }_{\beta ,\alpha ,\phi }=\{\omega _{\mathrm{\zeta }_{\mathrm{c%
}_{\beta ,\alpha ,\phi }}}\}.  \label{gauge invariance equation 0}
\end{equation}%
Moreover, $\mathrm{c}_{\beta ,\alpha ,\phi }$ converges to $\mathrm{r}%
_{\beta }^{1/2}e^{i\phi }$ as $\alpha \rightarrow 0$. Therefore, it follows
from (\ref{Theorem equilibrium state 2 U(1) broken}) that%
\begin{equation}
\lim\limits_{\alpha \downarrow 0}\omega _{\mathrm{\zeta }_{\mathrm{c}_{\beta
,\alpha ,\phi }}}\left( a_{\kappa (l),\downarrow }a_{\kappa (l),\uparrow
}\right) =\mathrm{r}_{\beta }^{1/2}e^{i\phi }
\label{gauge invariance equation 1}
\end{equation}%
for any $l\in \mathbb{N}$.

By permutation invariance
\begin{equation*}
\frac{1}{N}\sum\limits_{l=1}^{N}\omega _{N,\alpha ,\phi }\left( a_{\kappa
(l),\uparrow }^{\ast }a_{\kappa (l),\downarrow }^{\ast }\right) =\omega
_{N,\alpha ,\phi }\left( a_{\kappa (1),\uparrow }^{\ast }a_{\kappa
(1),\downarrow }^{\ast }\right) .
\end{equation*}%
Now, let $\{N_{j}^{(1)}\}$ and $\{N_{j}^{(2)}\}$ be two subsequences in $%
\mathbb{N}$ such that
\begin{eqnarray*}
\lim\limits_{j\rightarrow \infty }\omega _{N_{j}^{(1)},\alpha ,\phi }\left(
a_{\kappa (1),\uparrow }^{\ast }a_{\kappa (1),\downarrow }^{\ast }\right) &=&%
\underset{N\rightarrow \infty }{\lim \sup }\;\omega _{N,\alpha ,\phi }\left(
a_{\kappa (1),\uparrow }^{\ast }a_{\kappa (1),\downarrow }^{\ast }\right) ,
\\
\lim\limits_{j\rightarrow \infty }\omega _{N_{j}^{(2)},\alpha ,\phi }\left(
a_{\kappa (1),\uparrow }^{\ast }a_{\kappa (1),\downarrow }^{\ast }\right) &=&%
\underset{N\rightarrow \infty }{\lim \inf }\;\omega _{N,\alpha ,\phi }\left(
a_{\kappa (1),\uparrow }^{\ast }a_{\kappa (1),\downarrow }^{\ast }\right) .
\end{eqnarray*}%
We can assume without loss of generality that $\omega _{N_{j}^{(2)}}$ and $%
\omega _{N_{j}^{(1)}}$ both converge w.r.t. the weak$^\ast $--topology as $%
j\rightarrow \infty $. Since any weak$^\ast $--limits $\omega _{\infty
,\alpha ,\phi }$ of local Gibbs states $\omega _{N,\alpha ,\phi }$ (\ref%
{local Gibbs states avec alpha}) are equilibrium states (see again the proof
of Lemma \ref{BCS Lemma 3}), i.e., $\omega _{\infty ,\alpha ,\phi }\in
\mathit{\Omega }_{\beta ,\alpha ,\phi }$, the theorem then follows from (\ref%
{gauge invariance equation 0}) and (\ref{gauge invariance equation 1}).
Indeed, for any $\beta ,\gamma >0$ and $\mu ,\lambda ,h\in \mathbb{R}$
away from any critical point, the sequence $\omega _{N,\alpha ,\phi }$ of
local Gibbs state converges towards $\omega _{\infty ,\alpha ,\phi }=\omega
_{\mathrm{\zeta }_{\mathrm{c}_{\beta ,\alpha ,\phi }}}$ in the
weak$^\ast $--topology as soon as $\alpha \geq 0$
is sufficiently small.\hfill $\Box $

From Corollary \ref{Theorem equilibrium state 2} note that the expectation
values of Cooper fields
\begin{equation}
\begin{array}{l}
\Phi _{\kappa (l)}:=a_{\kappa (l),\downarrow }^{\ast }a_{\kappa (l),\uparrow
}^{\ast }+a_{\kappa (l),\uparrow }a_{\kappa (l),\downarrow } \\
\Psi _{\kappa (l)}:=i(a_{\kappa (l),\downarrow }^{\ast }a_{\kappa
(l),\uparrow }^{\ast }-a_{\kappa (l),\uparrow }a_{\kappa (l),\downarrow })%
\end{array}
\label{field of Cooper pairs}
\end{equation}%
are
\begin{equation}
\omega _{\mathrm{\zeta }_{\mathrm{c}_{\beta }}}(\Phi _{\kappa (l)})=2%
\mathop{\rm Re}\{\mathrm{c}_{\beta }\}\mathrm{\ and\ }\omega _{\mathrm{\zeta
}_{\mathrm{c}_{\beta }}}(\Psi _{\kappa (l)})=2\mathop{\rm Im}\{\mathrm{c}%
_{\beta }\}  \label{field of Cooper pairsbis}
\end{equation}%
for any pure state $\omega _{\mathrm{\zeta }_{\mathrm{c}_{\beta }}}$ of $%
\mathit{\Omega }_{\beta }$ and $l\in \mathbb{N}$, where we recall that $%
\mathrm{c}_{\beta }:=\mathrm{r}_{\beta }^{1/2}e^{i\phi }$ is some maximizer
of the first variational problem given in Theorem \ref{BCS theorem 1}. In
particular, $\omega (\Phi _{\kappa (l)})\neq 0$ or $\omega (\Psi _{\kappa
(l)})\neq 0$\ for any pure state $\omega \in \mathit{\Omega }_{\beta }$ is a
manifestation of the breakdown of the $U(1)$--gauge symmetry.

Unfortunately, the operators $\Phi _{\kappa (l)}$ and $\Psi _{\kappa (l)}$
do not correspond to any experiment, as they are not gauge invariant. More
generally, experiments only \textquotedblleft see\textquotedblright\ the
restriction of states $\omega _{\mathrm{\zeta }_{\mathrm{c}_{\beta }}}$ to
the subalgebra of gauge invariant elements. Consequently, the next step is
to prove the so--called \textit{off diagonal long range order} (ODLRO)
property proposed by Yang \cite{ODLRO} to define the superconducting phase.
Indeed, one detects the presence of $U(1)$--gauge symmetry breaking by
considering the asymptotics, as $|l-m|\rightarrow \infty $, of the ($U(1)$%
--gauge symmetric) Cooper pair correlation function
\begin{equation}
G_{\omega }(l,m):=\omega (a_{\kappa (l),\uparrow }^{\ast }a_{\kappa
(l),\downarrow }^{\ast }a_{\kappa (m),\downarrow }a_{\kappa (m),\uparrow })
\label{Cooper-pair-correlation-function}
\end{equation}%
associated with some state $\omega $. In particular, if $G_{\omega }(l,m)$
converges to some fixed non--zero value whenever $|l-m|\rightarrow \infty $,
the state $\omega $ shows \textit{off diagonal long range order} (ODLRO).
This property can directly be analyzed for equilibrium states from our next
statement.\

\begin{theorem}[Cooper pair correlation function]
\label{Theorem equilibrium state 3}\mbox{ }\newline
For any $\beta ,\gamma >0$ and $\mu ,\lambda ,h\in \mathbb{R}$ away from any
critical point, the Cooper pair correlation function $G_{\omega _{N}}(l,m)$
associated with the local Gibbs state $\omega _{N}$ converges for fixed $%
l\neq m$ towards
\begin{equation*}
\lim\limits_{N\rightarrow \infty }G_{\omega _{N}}\left( l,m\right)
=G_{\omega }\left( l,m\right) =\mathrm{r}_{\beta },
\end{equation*}%
for any equilibrium state $\omega \in \mathit{\Omega }_{\beta }$, and with $%
\mathrm{r}_{\beta }$ being the solution of (\ref{BCS pressure 2}).
\end{theorem}

\noindent \textit{Proof:} By similar arguments as in the proof of Theorem %
\ref{Theorem equilibrium state 4 U(1) broken}, if $G_{\omega }\left(
l,m\right) =\mathrm{r}_{\beta }$ for all equilibrium states $\omega $, then
\begin{equation*}
\lim\limits_{N\rightarrow \infty }G_{\omega _{N}}\left( l,m\right) =\mathrm{r%
}_{\beta }.
\end{equation*}%
By permutation invariance of $\omega \in \mathit{\Omega }_{\beta }$, note
that
\begin{equation}
G_{\omega }(l,m)=G_{\omega }(1,2)  \label{eq order parameter 9}
\end{equation}%
for any $l\not=m.$ If $\omega =\omega _{\mathrm{\zeta }_{\mathrm{c}_{\beta
}}}$ is an extremal equilibrium state, then one clearly has
\begin{equation*}
G_{\omega _{\mathrm{\zeta }_{\mathrm{c}_{\beta }}}}(1,2)=\mathrm{\zeta }_{%
\mathrm{c}_{\beta }}(a_{\uparrow }^{\ast }a_{\downarrow }^{\ast })\mathrm{%
\zeta }_{\mathrm{c}_{\beta }}(a_{\downarrow }a_{\uparrow })=|\mathrm{c}%
_{\beta }|^{2}=\mathrm{r}_{\beta }.
\end{equation*}%
On the other hand, the set $\mathit{\Omega }_{\beta }$ of equilibrium states
for fixed parameters $\beta ,\gamma >0$, and $\mu ,\lambda ,h\in \mathbb{R}$
is weak$^\ast $--compact. In particular, if $\omega \in \mathit{\Omega }%
_{\beta }$ is not extremal, the function $G_{\omega }(1,2)$ is given, up to
arbitrarily small errors, by convex sums of the form
\begin{equation}
\sum\limits_{j=1}^{k}\lambda _{j}G_{\omega ^{(j)}}(1,2),\quad \lambda
_{1},\ldots ,\lambda _{k}\geq 0,\quad \lambda _{1}+\ldots +\lambda _{k}=1,
\label{eq order parameter 10}
\end{equation}%
where $\{\omega ^{(j)}\}_{j=1,...,k}$ are extremal equilibrium states. Since
any weak$^\ast $--limit $\omega _{\infty }$ of local Gibbs states $\omega
_{N}$ (\ref{BCS gibbs state Hn}) is an equilibrium state (see proof of Lemma %
\ref{BCS Lemma 3}), the theorem is then a consequence of (\ref{eq order
parameter 9})--(\ref{eq order parameter 10}).\hfill $\Box $

Since
\begin{eqnarray*}
&&\frac{1}{N^{2}}\sum\limits_{l,m=1}^{N}\omega _{N}\left( a_{\kappa
(l),\uparrow }^{\ast }a_{\kappa (l),\downarrow }^{\ast }a_{\kappa
(m),\downarrow }a_{\kappa (m),\uparrow }\right) \\
&=&\frac{N(N-1)}{N^{2}}\omega _{N}\left( a_{\kappa (1),\uparrow }^{\ast
}a_{\kappa (1),\downarrow }^{\ast }a_{\kappa (2),\downarrow }a_{\kappa
(2),\uparrow }\right) +\mathcal{O}(N^{-1}),
\end{eqnarray*}%
note that this theorem implies Theorem \ref{BCS theorem 2-0}.

Therefore, away from any critical point, if an equilibrium state shows ODLRO
then all pure equilibrium states break the $U(1)$--gauge symmetry.
Conversely, if all pure equilibrium states break the $U(1)$--gauge symmetry,
then all equilibrium state show ODLRO. This is due to the fact that the
order parameter $\mathrm{r}_{\beta }$ is unique away from any critical point.
In particular, from Section \ref{section variational problem}, at
sufficiently small inverse temperature $\beta $ there is no ODLRO and $%
\mathit{\Omega }_{\beta }=\{\omega _{\mathrm{\zeta }_{0}}\}$, whereas for
sufficiently large $\beta $ and $\gamma $ all equilibrium states show ODLRO.

For any $\beta ,\gamma >0$ and real numbers $\mu ,\lambda ,h$ at some
critical point, this property is not satisfied in general. There are indeed
cases where the phase transition is of first order, cf. figure \ref%
{order-parameter-temp-lambda.eps}. In this situation, $0$ and some $\mathrm{r%
}_{\beta }>0$ are maximizers at the same time, and hence, there are some
equilibrium states breaking the $U(1)$--gauge symmetry and other equilibrium
states which do not show ODLRO in this specific situation.

Observe now that the superconducting phase is not only characterized by
ODLRO and the breakdown of the $U(1)$--gauge symmetry. Indeed, the
two--point correlation function determines its type: s--wave, d--wave,
p--wave, etc. In fact, for any extremal equilibrium state $\omega =\omega _{%
\mathrm{\zeta }_{\mathrm{c}_{\beta }}}$, $x,y\in \mathbb{Z}^{d}$ and $%
s_{1},s_{2}\in \{\uparrow ,\downarrow \}$, one clearly has
\begin{equation*}
\omega _{\mathrm{\zeta }_{\mathrm{c}_{\beta
}}}(a_{x,s_{1}}a_{y,s_{2}})=\left\{
\begin{array}{l}
\mathrm{\zeta }_{\mathrm{c}_{\beta }}(a_{x,s_{1}})\mathrm{\zeta }_{\mathrm{c}%
_{\beta }}(a_{y,s_{2}})\mathrm{\quad if\ }x\neq y \\
\mathrm{\zeta }_{\mathrm{c}_{\beta }}(a_{x,s_{1}}a_{x,s_{2}})\mathrm{\quad
\quad \quad if\ }x=y%
\end{array}%
\right. =\left\{
\begin{array}{l}
0\mathrm{\quad \quad \ if\ }x\neq y. \\
0\mathrm{\quad \quad \ if\ }x=y,\mathrm{\ }s_{1}=s_{2}. \\
\mathrm{c}_{\beta }\mathrm{\quad \quad if\ }x=y,\mathrm{\ }s_{1}\neq s_{2}.%
\end{array}%
\right.
\end{equation*}%
As a consequence, for any equilibrium state $\omega \in \mathit{\Omega }%
_{\beta }$, we have $\omega(a_{x,s_{1}}a_{y,s_{2}})=%
\omega(a_{0,s_{1}}a_{0,s_{2}})\delta _{x,y}$ and we obtain a \textit{s--wave}
superconducting phase. In particular, Theorem \ref{BCS s-wave-thm} is a
simple consequence of this last equalities combined with (\ref{gauge
invariance equation 0}), (\ref{gauge invariance equation 1}) and the fact
that any weak$^\ast $--limits $\omega _{\infty ,\alpha ,\phi }\in \mathit{%
\Omega }_{\beta ,\alpha ,\phi }$ of local Gibbs states $\omega _{N,\alpha
,\phi }$ (\ref{local Gibbs states avec alpha}) are equilibrium states (see
again the proof of Lemma \ref{BCS Lemma 3}).

Now we would like to pursue this analysis of equilibrium states by showing
that their definition is in accordance with results of Theorems \ref{BCS
theorem 2-1}, \ref{BCS theorem 2-2} and \ref{BCS theorem 2-3}. This
statement is given in the next theorem.

\begin{theorem}[Uniqueness of densities for equilibrium states]
\label{Theorem equilibrium state 4}\mbox{ }\newline
Take $\beta ,\gamma >0$ and real numbers $\mu ,\lambda ,h$ away from any
critical point. Then, for any equilibrium state $\omega \in \mathit{\Omega }%
_{\beta }$ and $l\in \mathbb{N},$ all densities are uniquely defined:\newline
(i) The electron density is equal to
\begin{equation*}
\underset{N\rightarrow \infty }{\lim }\left\{ \dfrac{1}{N}%
\sum\limits_{l^{\prime }=1}^{N}\omega _{N}\left( n_{\kappa (l^{\prime
}),\uparrow }+n_{\kappa (l^{\prime }),\downarrow }\right) \right\} =\omega
(n_{\kappa (l),\uparrow }+n_{\kappa (l),\downarrow })=\mathrm{d}_{\beta },
\end{equation*}%
cf. Theorem \ref{BCS theorem 2-1}.\newline
(ii) The magnetization density is equal to
\begin{equation*}
\underset{N\rightarrow \infty }{\lim }\left\{ \dfrac{1}{N}%
\sum\limits_{l^{\prime }=1}^{N}\omega _{N}\left( n_{\kappa (l^{\prime
}),\uparrow }-n_{\kappa (l^{\prime }),\downarrow }\right) \right\} =\omega
(n_{\kappa (l),\uparrow }-n_{\kappa (l),\downarrow })=\mathrm{m}_{\beta },
\end{equation*}%
cf. Theorem \ref{BCS theorem 2-2}.\newline
(iii) The Coulomb correlation density is equal to
\begin{equation*}
\underset{N\rightarrow \infty }{\lim }\left\{ \dfrac{1}{N}%
\sum\limits_{l^{\prime }=1}^{N}\omega _{N}\left( n_{\kappa (l^{\prime
}),\uparrow }n_{\kappa (l^{\prime }),\downarrow }\right) \right\} =\omega
(n_{\kappa (l),\uparrow }n_{\kappa (l),\downarrow })=\mathrm{w}_{\beta },
\end{equation*}%
cf. Theorem \ref{BCS theorem 2-3}.
\end{theorem}

\noindent \textit{Proof:} Suppose first that $\omega \in \mathit{\Omega }%
_{\beta }$ is pure. Then, from Corollary \ref{Theorem equilibrium state 2}
it follows that
\begin{equation*}
\omega \left( n_{\kappa (l),\uparrow }+n_{\kappa (l),\downarrow }\right)
=\omega _{\mathrm{\zeta }_{\mathrm{c}_{\beta }}}\left( n_{\kappa
(l),\uparrow }+n_{\kappa (l),\downarrow }\right) ,
\end{equation*}%
with $\mathrm{c}_{\beta }=\mathrm{r}_{\beta }^{1/2}e^{i\phi }$ for some $%
\phi \in \lbrack 0,2\pi )$. Thus, by using the gauge invariance of the map $%
c\mapsto p(c)$ we directly get
\begin{equation}
\omega \left( n_{\kappa (l),\uparrow }+n_{\kappa (l),\downarrow }\right)
=\partial _{\mu }p(\beta ,\mu ,\lambda ,\gamma ,h;\mathrm{c}_{\beta
})=\partial _{\mu }p(\beta ,\mu ,\lambda ,\gamma ,h;\mathrm{r}_{\beta
}^{1/2})=\mathrm{d}_{\beta }.  \label{sans griffiths 1}
\end{equation}%
At fixed parameters $\beta ,\gamma >0$, $\mu ,\lambda ,h\in \mathbb{R}$,
recall that the set $\mathit{\Omega }_{\beta }$ of equilibrium states is
weak$^\ast $--compact. In particular, if $\omega \in \mathit{\Omega }_{\beta
}$ is not pure, it is the weak$^\ast $--limit of convex combinations of pure
states. Therefore, we obtain (\ref{sans griffiths 1}) for any $\omega \in
\mathit{\Omega }_{\beta }.$ Similarly one gets%
\begin{equation}
\omega (n_{\kappa (l),\uparrow }-n_{\kappa (l),\downarrow })=\mathrm{m}%
_{\beta }\quad \mathrm{\ and\ }\quad \omega (n_{\kappa (l),\uparrow
}n_{\kappa (l),\downarrow })=\mathrm{w}_{\beta },  \label{sans griffiths 2}
\end{equation}%
for any equilibrium state $\omega \in \mathit{\Omega }_{\beta }$ and $l\in
\mathbb{N}$. Moreover, since any weak$^\ast $--limit $\omega _{\infty }$ of
local Gibbs states $\omega _{N}$ (\ref{BCS gibbs state Hn}) is an
equilibrium state, i.e., $\omega _{\infty }\in \mathit{\Omega }_{\beta }$,
we therefore deduce from (\ref{sans griffiths 1})-(\ref{sans griffiths 2}),
exactly as in the proof of Theorem \ref{Theorem equilibrium state 4 U(1)
broken}, the existence of the limits in the statements (i)-(iii).\hfill $%
\Box $

Observe that the weak$^\ast $--limit $\omega _{\infty }\in \mathit{\Omega }%
_{\beta }$ of local Gibbs states $\omega _{N}$ (\ref{BCS gibbs state Hn})
can easily be performed, \textit{even at critical points}, by using the
decomposition theory for states \cite{BrattelliRobinsonI}:

\begin{theorem}[Asymptotics of the local Gibbs state $\protect\omega _{N}$
as $N\rightarrow \infty $]
\label{Theorem equilibrium state 4bis}\mbox{ }\newline
Recall that for any $\phi \in \lbrack 0,2\pi )$, $\mathrm{c}_{\beta }:=%
\mathrm{r}_{\beta }^{1/2}e^{i\phi }$ is a maximizer of the first variational
problem given in Theorem \ref{BCS theorem 1}, whereas the states $\mathrm{%
\zeta }_{c}$ and $\omega _{\zeta }$ are respectively defined by (\ref%
{Gibbs.state nu}) and (\ref{shifft-invariant gibbs states}). Take any $\beta
,\gamma >0$, $\mu ,\lambda ,h\in \mathbb{R}$, and let $N\rightarrow \infty $%
. \newline
(i) Away From any critical point, the local Gibbs state $\omega _{N}$
converges in the weak$^\ast $--topology towards the equilibrium state%
\begin{equation}
\omega _{\infty }\left( \cdot \right) =\frac{1}{2\pi }\underset{0}{\overset{%
2\pi }{\int }}\omega _{\mathrm{\zeta }_{\mathrm{c}_{\beta }}}\left( \cdot
\right) \mathrm{d}\phi .  \label{Gibbs.lim.1}
\end{equation}%
(ii) For each weak$^\ast $ limit point $\omega _{\infty }$ of local Gibbs
states $\omega _{N}$ with parameters $(\beta _{N},\gamma _{N},\mu
_{N},\lambda _{N},h_{N})$ converging to any critical point $(\beta ,\gamma
,\mu ,\lambda ,h)\in \partial \mathcal{S}$ (\ref{critical point open set}),
there is $\tau \in \lbrack 0,1]$ such that%
\begin{equation*}
\omega _{\infty }\left( \cdot \right) =\left( 1-\tau \right) \omega _{%
\mathrm{\zeta }_{0}}\left( \cdot \right) +\frac{\tau }{2\pi }\underset{0}{%
\overset{2\pi }{\int }}\omega _{\mathrm{\zeta }_{\mathrm{c}_{\beta }}}\left(
\cdot \right) \mathrm{d}\phi .
\end{equation*}
\end{theorem}

\noindent \textit{Proof:} By $U(1)$--gauge symmetry of the Hamiltonians $%
\mathrm{H}_{N}$ (\ref{Hamiltonian BCS-Hubbard}) recall that any
weak$^\ast $--limit $\omega _{\infty }$ of local Gibbs states $\omega _{N}$ (\ref{BCS
gibbs state Hn}) is a $U(1)$--invariant equilibrium state. So, in order to
prove the first part of the Theorem it suffices to show that the equilibrium
state given in (i) is the unique $U(1)$--invariant state in $\mathit{\Omega }%
_{\beta }$. If the solution $\mathrm{r}_{\beta }$ of (\ref{BCS pressure 2})
is zero, then this follows immediately from Corollary \ref{Theorem
equilibrium state 2}.

Let $\mathrm{r}_{\beta }>0$ be the unique maximizer of (\ref{BCS pressure 2}%
), i.e., $\mathrm{c}_{\beta }:=\mathrm{r}_{\beta }^{1/2}e^{i\phi }\neq 0$
for any $\phi \in \lbrack 0,2\pi )$. Let
\begin{equation*}
\partial \mathit{\Omega }_{\beta }=\left\{ \omega _{\mathrm{\zeta }}:\mathrm{%
\zeta }\in \mathcal{G}_{\beta }\right\}
\end{equation*}%
be the set of all extremal states of $\mathit{\Omega }_{\beta }$, see (\ref%
{definition of one-site equilibrium state}) for the definition of the set $%
\mathcal{G}_{\beta }$ of one--site equilibrium states. Observe that the
closed convex hull of $\partial \mathit{\Omega }_{\beta }$ is precisely $%
\mathit{\Omega }_{\beta }$ and that $\partial \mathit{\Omega }_{\beta }$ is
the image of the torus $[0,2\pi )$ under the continuous map $\phi \mapsto
\omega _{\mathrm{\zeta }_{\mathrm{c}_{\beta }}}$, with $\mathrm{c}_{\beta }:=%
\mathrm{r}_{\beta }^{1/2}e^{i\phi }$. This last map defines a homeomorphism
between the torus and $\partial \mathit{\Omega }_{\beta }$. In particular,
the set $\partial \mathit{\Omega }_{\beta }$ is compact and for each
equilibrium state $\omega \in \mathit{\Omega }_{\beta }$ there is a uniquely
defined probability measure $\mathrm{d}\mathfrak{\hat{m}}_{\omega }$ on the
torus such that%
\begin{equation}
\omega \left( A\right) =\underset{0}{\overset{2\pi }{\int }}\omega _{\mathrm{%
\zeta }_{\mathrm{c}_{\beta }}}\left( A\right) \mathrm{d}\mathfrak{\hat{m}}%
_{\omega }\left( \phi \right) ,\mathrm{\ for\ all\ }A\in \mathcal{U}.
\label{decomposition theory 2}
\end{equation}%
See, e.g., Proposition 1.2 of \cite{Phe}. By $U(1)$--invariance of $\omega
_{\infty }$, for any $n\in \mathbb{N}$ one has from (\ref{decomposition
theory 2}) that%
\begin{equation*}
\omega _{\infty }\left( \underset{l=1}{\overset{n}{\prod }}a_{\kappa
(l),\uparrow }a_{\kappa (l),\downarrow }\right) =\mathrm{r}_{\beta }^{n/2}%
\underset{0}{\overset{2\pi }{\int }}e^{in\phi }\mathrm{d}\mathfrak{\hat{m}}%
_{\omega _{\infty }}\left( \phi \right) =0.
\end{equation*}%
Therefore, if $\mathrm{r}_{\beta }>0,$ there is a unique probability measure
allowing the $U(1)$--gauge symmetry of $\omega _{\infty }$: $\mathrm{d}%
\mathfrak{\hat{m}}_{\omega _{\infty }}\left( \phi \right) $ must be the
uniform probability measure on $[0,2\pi )$.

From Lemma \ref{lemma cardinality} the cardinality of set of maximizers of (%
\ref{BCS pressure 2}) is at most $2$. Indeed, away from any critical point, it
is $1$ whereas at a critical point it can be either $1$ (second order phase
transition) or $2$ (first order phase transition). For more details, see
Section \ref{section variational problem}. In both cases, we can use the
same arguments as above. By similar estimates as in the proof of Lemma \ref%
{BCS Lemma 3} it immediately follows that all limit points of the Gibbs
states $\omega _{N}$ with parameters $(\beta _{N},\gamma _{N},\mu
_{N},\lambda _{N},h_{N})$ converging to $(\beta ,\gamma ,\mu ,\lambda ,h)\in
\partial \mathcal{S}$ as $N\rightarrow \infty $, belongs to $\mathit{\Omega }%
_{\beta }=\mathit{\Omega }_{\beta }(\mu ,\lambda ,\gamma ,h)$. Since the set
of all $U(1)$--invariant equilibrium states from $\mathit{\Omega }_{\beta }$
is $\{\omega ^{(\tau )}$\ for\ any $\tau \in \lbrack 0,1]\}$ with%
\begin{equation}
\omega ^{(\tau )}\left( \cdot \right) :=\left( 1-\tau \right) \omega _{%
\mathrm{\zeta }_{0}}\left( \cdot \right) +\frac{\tau }{2\pi }\underset{0}{%
\overset{2\pi }{\int }}\omega _{\mathrm{\zeta }_{\mathrm{c}_{\beta }}}\left(
\cdot \right) \mathrm{d}\phi ,  \label{decomposition theory 3}
\end{equation}%
we obtain the second statement (ii).\hfill $\Box $

This theorem is a generalization of results obtained for the strong coupling%
\footnote{%
See (\ref{Hamiltonian BCS-Hubbard}) with $\lambda =0$ and $h=0$.} BCS model
\cite{Thir68}. Note however, that Thirring's analysis \cite{Thir68} of the
asymptotics of local Gibbs states comes from explicit computations, whereas
we use the structure of sets of states, as explained for instance in \cite{S}%
.

Observe that Theorem \ref{phase.mix.Th} is a simple consequence of Theorem %
\ref{Theorem equilibrium state 4bis}. Indeed, assume for instance that the
order parameter $\mathrm{r}_{\beta }=\mathrm{r}_{\beta }(\mu ,\lambda
,\gamma ,h)$ and the electron density per site $\mathrm{d}_{\beta }=\mathrm{d%
}_{\beta }(\mu ,\lambda ,\gamma ,h)$ jumps respectively from $\mathrm{r}%
_{\beta }^{-}=0$ to $\mathrm{r}_{\beta }^{+}$ and from $\mathrm{d}_{\beta
}^{-}$ to $\mathrm{d}_{\beta }^{+}$ by crossing a critical chemical
potential $\mathrm{\mu }_{\beta }^{(c)}$ at fixed parameters $(\beta
,\lambda ,\gamma ,h)$. An example of such behavior is given in figure \ref%
{Mott-Insulator-1.eps} for an electron density smaller than one. If $\rho
\in \lbrack \mathrm{d}_{\beta }^{-},\mathrm{d}_{\beta }^{+}]$, then the
unique solution $\mu _{N,\beta }=\mu _{N,\beta }(\rho ,\lambda ,\gamma ,h)$
of (\ref{mu fixed particle density}) must converge towards $\mathrm{\mu }%
_{\beta }^{(c)}$ as $N\rightarrow \infty $. Meanwhile, at fixed $(\beta ,%
\mathrm{\mu }_{\beta }^{(c)},\lambda ,\gamma ,h)$
\begin{equation*}
\omega _{\mathrm{\zeta }_{0}}\left( n_{\uparrow }+n_{\downarrow }\right) =%
\mathrm{d}_{\beta }^{-}\mathrm{\ and\ }\omega _{\mathrm{\zeta }_{\mathrm{c}%
_{\beta }^{+}}}\left( n_{\uparrow }+n_{\downarrow }\right) =\mathrm{d}%
_{\beta }^{+},
\end{equation*}%
with $\mathrm{c}_{\beta }^{+}:=\sqrt{\mathrm{r}_{\beta }^{+}}e^{i\phi }$ and
$\phi \in \lbrack 0,2\pi ).$ Any weak$^\ast $--limit $\omega _{\infty }$ of
local Gibbs states $\omega _{N}$ satisfies per construction
\begin{equation*}
\omega _{\infty }\left( n_{\uparrow }+n_{\downarrow }\right) =\rho
\end{equation*}%
and has the form $\omega ^{(\tau )}\left( \cdot \right) $ (\ref%
{decomposition theory 3}), by Theorem \ref{Theorem equilibrium state 4bis}.
Hence, the Gibbs state $\omega _{N}$ converges in the weak$^\ast $--topology
towards $\omega ^{(\tau _{\rho })}\left( \cdot \right) $ with $\tau _{\rho }$
defined in Theorem \ref{phase.mix.Th}. Indeed, the existence of the limits
(i)--(iii) in Theorem \ref{phase.mix.Th} follows from the uniqueness of the
limiting equilibrium state with fixed electron density $\rho \in \lbrack
d_{\beta }^{-},d_{\beta }^{+}]$.

We give now various important properties of densities in ground states,
i.e., for $\beta =\infty $, which immediately follow from Theorem \ref%
{Theorem equilibrium state 4}. Recall that the set $\mathit{\Omega }_{\infty
}$ of ground states is the set of all weak$^\ast $ limit points as $%
n\rightarrow \infty $ of all equilibrium state sequences $\{\omega
^{(n)}\}_{n\in \mathbb{N}}$ with diverging inverse temperature $\beta
_{n}\rightarrow \infty $.

Take $\gamma >0$ and parameters $\mu ,\lambda ,h$ such that $|\mu -\lambda
|\not=\lambda +|h|$. Then the electron and Coulomb correlation densities
equal respectively%
\begin{equation}
\mathrm{d}:=\omega (n_{\kappa (l),\uparrow }+n_{\kappa (l),\downarrow })=%
\mathrm{d}_{\infty }\quad \mathrm{\ and\ \quad} \mathrm{w}:=\omega
(n_{\kappa (l),\uparrow }n_{\kappa (l),\downarrow })=\mathrm{w}_{\infty },
\label{density vector 1}
\end{equation}%
for any ground state $\omega \in \mathit{\Omega }_{\infty }$ and $l\in
\mathbb{N}$, cf. Corollaries \ref{BCS theorem 2-1bis} and \ref{BCS theorem
2-3bis}.

If additionally $\gamma >\Gamma _{|\mu -\lambda |,\lambda +|h|}$, we are in
the superconducting phase for ground states, cf. Corollary \ref{BCS theorem
2-0bis}.\ Indeed, for any $\varphi \in \lbrack 0,2\pi )$, there is a ground
state $\omega \in \mathit{\Omega }_{\infty }$ such that for any $l\in
\mathbb{N}$,
\begin{equation*}
\omega (a_{\kappa (l),\downarrow }a_{\kappa (l),\uparrow })=\mathrm{r}_{\max
}^{1/2}e^{i\varphi }.
\end{equation*}%
In the superconducting phase, from Corollary \ref{BCS theorem 2-3bis} we
observe that $\mathrm{d}_{\infty }=2\mathrm{w}_{\infty }$, whereas the
magnetization density equals
\begin{equation}
\mathrm{m}:=\omega (n_{\kappa (l),\uparrow }-n_{\kappa (l),\downarrow })=%
\mathrm{m}_{\infty }=0,  \label{density vector 2}
\end{equation}%
for any superconducting state $\omega \in \mathit{\Omega }_{\infty }$ and $%
l\in \mathbb{N}$. This is the Mei{\ss }ner effect, see Corollary \ref{BCS
theorem 2-2bis}. On the other hand, the Cauchy--Schwarz inequality for the
states implies the inequalities%
\begin{equation}
0\leq \omega \left( n_{\kappa (l),\uparrow }n_{\kappa (l),\downarrow
}\right) \leq \sqrt{\omega \left( n_{\kappa (l),\uparrow }\right) }\sqrt{%
\omega \left( n_{\kappa (l),\downarrow }\right) }  \label{ineq paring}
\end{equation}%
for any $l\in \mathbb{N}$ and $\omega \in E_{\mathcal{U}}^{+}$. In fact, in
the superconducting phase the second inequality of (\ref{ineq paring}) is an
equality for any $\omega \in \mathit{\Omega }_{\infty }$. Indeed, (\ref%
{density vector 2}) and Corollary \ref{BCS theorem 2-3bis} yield
\begin{equation}
\omega (n_{\kappa (l),\uparrow }n_{\kappa (l),\downarrow })=\omega
(n_{\kappa (l),\uparrow })=\omega (n_{\kappa (l),\downarrow }),
\label{ineq paring0}
\end{equation}%
for any $\omega \in \mathit{\Omega }_{\infty }$ and $l\in \mathbb{N}$. It
shows that 100\% of electrons form Cooper pairs in superconducting ground
states.

In the case where $h\neq 0$ with $\gamma >\Gamma _{|\mu -\lambda |,\lambda
+|h|}$ and $|\mu -\lambda |\not=\lambda +|h|$, the\ density vector $(\mathrm{%
d},\mathrm{m},\mathrm{w})$ defined by (\ref{density vector 1}) and (\ref%
{density vector 2}) is also unique as in the superconducting phase. It
equals $(\mathrm{d}_{\infty },\mathrm{m}_{\infty },\mathrm{w}_{\infty })$,
see Corollaries \ref{BCS theorem 2-1bis}, \ref{BCS theorem 2-2bis} and \ref%
{BCS theorem 2-3bis}. However, if $h=0$ with $\gamma <\Gamma _{|\mu -\lambda
|,\lambda }$, or $\gamma =\Gamma _{|\mu -\lambda |,\lambda +|h|}$, or $|\mu
-\lambda |=\lambda +|h|$, then the density vector $(\mathrm{d},\mathrm{m},%
\mathrm{w})$ belongs, in general, to a non trivial convex set. In other
words, there are phase transitions involving to these densities. In
particular, even in the case $h=0$ where the Hamiltonian $\mathrm{H}_{N}$ (%
\ref{Hamiltonian BCS-Hubbard}) is spin invariant, there are ground states
breaking the spin $SU(2)$--symmetry.

For instance, take $\beta ,\gamma >0$ and parameters $\mu ,\lambda $ such
that $|\mu -\lambda |<\lambda $ and $\gamma <\Gamma _{|\mu -\lambda
|,\lambda }$. Then for any $\omega \in \mathit{\Omega }_{\infty }$ and $l\in
\mathbb{N}$, the electron density equals $\mathrm{d}=\mathrm{d}_{\infty }=1$%
, whereas the Coulomb correlation density is $\mathrm{w}=\mathrm{w}_{\infty
}=0$. In particular, the first inequality of (\ref{ineq paring}) is an
equality showing that 0\% of electrons forms Cooper pairs. But, even if
the magnetic field vanishes, i.e., $h=0,$ for any $x\in (-1,1)$ there exists
a ground state $\omega ^{\left( x\right) }\in \mathit{\Omega }_{\infty }$
with magnetization density $\mathrm{m}=x$ (see (\ref{density vector 2}) for
the definition of $\mathrm{m}$).

Therefore, all the thermodynamics of the strong coupling BCS--Hubbard model
discussed in Sections \ref{Section BCS phase transition}--\ref{Section Mott
insulator} is encoded in the notion of equilibrium and ground states $\omega
\in \mathit{\Omega }_{\beta }$ with $\beta \in (0,\infty ]$. However, there
is still an important open question related to the thermodynamics of this
model. It concerns the problem of fluctuations of the Cooper pair condensate
density (Theorem \ref{BCS theorem 2-0}) or Cooper fields $\Phi _{\kappa (l)}$
and $\Psi _{\kappa (l)}$ (\ref{field of Cooper pairs}) as a function of the
temperature. Unfortunately, no result in that direction are known as soon as
the thermodynamic limit is concerned. We prove however a simple statement
about fluctuations of Cooper fields for pure states from $\mathit{\Omega }%
_{\beta }$ in the limit $\gamma \beta \rightarrow \infty $.

\begin{theorem}[Fluctuations of Cooper fields]
\label{Theorem equilibrium state 5}\mbox{ }\newline
Take $\beta ,\gamma >0$ and real numbers $\mu ,\lambda ,h$ away from any
critical point. Then, for any pure state $\omega _{\mathrm{\zeta }_{\mathrm{c%
}_{\beta }}}\in \mathit{\Omega }_{\beta }$ and $l\in \mathbb{N}$, the
fluctuations of Cooper fields $\Phi _{\kappa (l)}$ and $\Psi _{\kappa (l)}$ (%
\ref{field of Cooper pairs}) are bounded by
\begin{equation*}
\begin{array}{c}
0\leq \omega _{\mathrm{\zeta }_{\mathrm{c}_{\beta }}}\left( \{\Phi _{\kappa
(l)}-\omega _{\mathrm{\zeta }_{\mathrm{c}_{\beta }}}(\Phi _{\kappa
(l)})\}^{2}\right) \leq 2\gamma ^{-1}\beta ^{-1}, \\
0\leq \omega _{\mathrm{\zeta }_{\mathrm{c}_{\beta }}}\left( \{\Psi _{\kappa
(l)}-\omega _{\mathrm{\zeta }_{\mathrm{c}_{\beta }}}(\Psi _{\kappa
(l)})\}^{2}\right) \leq 2\gamma ^{-1}\beta ^{-1},%
\end{array}%
\end{equation*}%
i.e., they vanish in the limit $\gamma \beta \rightarrow \infty $.
\end{theorem}

\noindent \textit{Proof:} Recall that properties of pure states are
characterized in Corollary \ref{Theorem equilibrium state 2}, i.e., they are
product states $\omega _{\mathrm{\zeta }_{\mathrm{c}_{\beta }}}$ with the
one--site state $\mathrm{\zeta }_{\mathrm{c}_{\beta }}$ being defined in (%
\ref{Gibbs.state nu}). In particular, they satisfy (\ref{field of Cooper
pairsbis}). Now, to avoid triviality, assume that $\mathrm{c}_{\beta }:=%
\mathrm{r}_{\beta }^{1/2}e^{i\phi }\neq 0$ and let $\mathfrak{f}(\tau )$ be
the function defined for any $\tau \in \mathbb{R}$ by%
\begin{equation*}
\mathfrak{f}\left( \tau \right) :=-\gamma |\mathrm{c}_{\beta }+\tau |^{2}+p(%
\mathrm{c}_{\beta }+\tau ).
\end{equation*}%
Since $\mathrm{c}_{\beta }\neq 0$ is a maximizer of the function $-\gamma
|c|+p(c)$ of $c\in \mathbb{C}$, one has $\partial _{\tau }^{2}\mathfrak{f}%
\left( 0\right) \leq 0$, i.e., $\partial _{\tau }^{2}p(\mathrm{c}_{\beta
}+\tau )|_{\tau =0}\leq 2\gamma $. From straightforward computations,
observe that $p(\mathrm{c}_{\beta }+\tau )$ is a convex function of $\tau
\in \mathbb{R}$ with
\begin{equation*}
\beta ^{-1}\gamma ^{-2}\{\partial _{\tau }^{2}p(\mathrm{c}_{\beta }+\tau
)\}|_{\tau =0}=\omega _{\mathrm{\zeta }_{\mathrm{c}_{\beta }}}\left( \{\Phi
_{\kappa (l)}-\omega _{\mathrm{\zeta }_{\mathrm{c}_{\beta }}}(\Phi _{\kappa
(l)})\}^{2}\right) \geq 0.
\end{equation*}%
From this last equality combined with $\{\partial _{\tau }^{2}p(\mathrm{c}%
_{\beta }+\tau )\}|_{\tau =0}\leq 2\gamma $, we deduce the theorem for $\Phi
_{\kappa (l)}$. Moreover, from similar arguments using the function $%
\mathfrak{\hat{f}}\left( \tau \right) :=\mathfrak{f}\left( i\tau \right) $
instead of $\mathfrak{f}$, the fluctuations of the Cooper field $\Psi
_{\kappa (l)}$ are also bounded by $2\gamma ^{-1}\beta ^{-1}$. \hfill $\Box $

From Theorem \ref{Theorem equilibrium state 5}, note that Cooper fields are $%
c$--numbers in the corresponding GNS--representation \cite%
{BrattelliRobinsonI} of pure ground states defined as weak$^\ast $--limits
of pure equilibrium states:

\begin{corollary}[Cooper fields for pure ground states]
\label{coro c-number}\mbox{  }\newline
Let $\omega \in \mathit{\Omega }_{\infty }$ be any weak$^\ast $--limit of
pure equilibrium states and let $(\psi ,\mathrm{\pi },\mathcal{H})$ be the
corresponding GNS--representation of $\omega $ on bounded operators on the
Hilbert space $\mathcal{H}$ with cyclic vacuum $\psi $. Then $\omega $ is
pure and for any $l\in \mathbb{N}$, $\mathrm{\pi }(\Phi _{\kappa
(l)})=\omega (\Phi _{\kappa (l)}) \mathbb{I}_{\mathcal{H}}$ and $\mathrm{\pi
}(\Psi _{\kappa (l)})=\omega (\Psi _{\kappa (l)})\mathbb{I}_{\mathcal{H}}$.
\end{corollary}

\noindent \textit{Proof:} A pure equilibrium state is a product state (\ref%
{shifft-invariant gibbs states}) and any weak$^\ast $--limit of product
states in $E_{\mathcal{U}}^{S,+}$ is also a product state. Thus, by Lemma %
\ref{BCS Lemma 4}, any ground state $\omega \in \mathit{\Omega }_{\infty }$
defined as the weak$^\ast $--limit of pure equilibrium states is extremal in
$E_{\mathcal{U}}^{S,+}$ and hence extremal in $\mathit{\Omega }_{\infty }$.
Clearly, for such ground state, $\mathrm{\pi }(\omega (\Phi _{\kappa
(l)}))=\omega (\Phi _{\kappa (l)})\mathbb{I}_{\mathcal{H}}$ for any $l\in
\mathbb{N}$. Let $\tilde{\Phi}:=\Phi _{\kappa (l)}-\omega \left( \Phi
_{\kappa (l)}\right) $. From Theorem \ref{Theorem equilibrium state 5}
combined with the Cauchy--Schwarz inequality we obtain for any $A\in
\mathcal{U}$ that%
\begin{eqnarray*}
\left\Vert \mathrm{\pi }(\tilde{\Phi})\mathrm{\pi }(A)\psi \right\Vert _{%
\mathcal{H}}^{2} &=&\omega (A^{\ast }\tilde{\Phi}\tilde{\Phi}A)\leq \Vert
A\Vert \sqrt{\omega \left( \tilde{\Phi}(\tilde{\Phi}AA^{\ast }\tilde{\Phi}%
^{2})\right) } \\
&\leq &\Vert A\Vert ^{2}\Vert \tilde{\Phi}\Vert ^{3/2}[\omega (\tilde{\Phi}%
^{2})]^{1/4}=0.
\end{eqnarray*}%
From the cyclicity of $\psi $, it follows that $\mathrm{\pi }(\Phi _{\kappa
(l)})=\omega (\Phi _{\kappa (l)})\mathbb{I}_{\mathcal{H}}.$ The proof of $%
\mathrm{\pi }(\Psi _{\kappa (l)})=\omega (\Psi _{\kappa (l)})\mathbb{I}_{%
\mathcal{H}}$ is also performed in the same way. We omit the details. \hfill
$\Box $

In particular, for such pure ground states $\omega $ in $\mathit{\Omega }%
_{\infty }$, correlation functions can explicitly be computed at any order
in Cooper fields. For instance, for all $N\in \mathbb{N}$, all $%
k_{j},l_{j}\in \mathbb{N}$, $m_{j},n_{j}\in \mathbb{N}_0$, $j=1,\ldots ,N$,
and any $A_{n}\in \mathcal{U}$, $n=1,\ldots ,N+1$, one has%
\begin{eqnarray*}
&&\omega \left( A_{1}\Phi _{\kappa (k_{1})}^{m_{1}}\Psi _{\kappa
(l_{1})}^{n_{1}}A_{2}\ldots A_{N}\Phi _{\kappa (k_{N})}^{m_{N}}\Psi _{\kappa
(l_{N})}^{n_{N}}A_{N+1}\right) \\
&=&\omega (\Phi _{\kappa (k_{1})}^{m_{1}})\omega (\Psi _{\kappa
(l_{1})}^{n_{1}})\ldots \omega (\Phi _{\kappa (k_{N})}^{m_{N}})\omega (\Psi
_{\kappa (l_{N})}^{m_{N}})\;\omega \left( A_{1}\ldots A_{N+1}\right) .
\end{eqnarray*}

\section{Analysis of the variational problem\label{section variational
problem}}

The variational problem (\ref{BCS pressure 2}) is quite explicit but for the
reader convenience, we collect here some properties of its solution $\mathrm{%
r}_{\beta }$ w.r.t. $\beta ,\gamma >0$ and $\mu ,\lambda ,h\in \mathbb{R}.$
We show in particular that $\mathrm{r}_{\beta }>0$ exists in a non--empty
domain of $(\beta ,\gamma ,\mu ,\lambda ,h)$ with some monotonicity
properties as well as the existence of both first and second order phase
transitions. We conclude this section by giving the asymptotics of $\mathrm{r%
}_{\beta }$ as $\beta \rightarrow \infty $, i.e., by proving Corollary \ref%
{BCS theorem 2-0bis}. \newline

\noindent \textbf{1.} We start by showing that $\mathrm{r}_{\beta }=0$ for
sufficiently small inverse temperatures $\beta $ at fixed $\gamma $, $\mu $,
$\lambda $ and $h$. Indeed, for any $r\geq 0$ one computes that%
\begin{equation}
\partial _{r}f\left( r\right) =\gamma \left( \frac{\gamma \sinh \left( \beta
g_{r}\right) }{2g_{r}\left( e^{\lambda \beta }\cosh \left( \beta h\right)
+\cosh \left( \beta g_{r}\right) \right) }-1\right) ,
\label{pressure derivee 1}
\end{equation}%
cf. Theorem \ref{BCS theorem 1}. Direct estimations show that if $0<\beta
<2\gamma ^{-1}$, then $\partial _{r}f(r)<0$ for any $r\geq 0,$ i.e., $%
\mathrm{r}_{\beta }=0.$\newline

\noindent \textbf{2.} Fix now $\beta >0$ and $\mu ,\lambda ,h\in \mathbb{R}$%
, then $\mathrm{r}_{\beta }>0$ for sufficiently large coupling constants $%
\gamma .$ Indeed, for large enough $\gamma >0$ there is, at least, one
strictly positive solution $\mathrm{\tilde{r}}_{\beta }>0$ of (\ref{BCS gap
equation}). Since direct computations using again (\ref{BCS gap equation})
imply that
\begin{equation*}
\frac{d}{d\gamma}\left\{ f\left( \beta ,\mu ,\lambda ,\gamma ,h;\mathrm{%
\tilde{r}}_{\beta }(\gamma)\right) -f\left( \beta ,\mu ,\lambda ,\gamma
,h;0\right) \right\} =\mathrm{\tilde{r}}_{\beta }(\gamma)>0,
\end{equation*}%
and
\begin{equation*}
f\left( \beta ,\mu ,\lambda ,\gamma ,h;\mathrm{\tilde{r}}_{\beta }\right)
-f\left( \beta ,\mu ,\lambda ,\gamma ,h;0\right) =\mathcal{O}\left( \gamma
\right) \mathrm{\ as\;}\gamma \rightarrow \infty ,
\end{equation*}%
for any fixed $\beta >0$ and $\mu ,\lambda ,h\in \mathbb{R}$, there is a
unique $\gamma _{c}>2|\lambda -\mu |$ such that $f(\mathrm{\tilde{r}}_{\beta
})>f(0)$, i.e., $\mathrm{r}_{\beta }>0$ for $\gamma >\gamma _{c}$. The
domain of parameters $(\beta ,\mu ,\lambda ,\gamma ,h)$ where $\mathrm{r}%
_{\beta }$ is strictly positive is therefore non--empty, cf. figures \ref%
{order-parameter-temp-lambda.eps}--\ref{order-parameter-zerp-temp.eps}.%
\newline

\noindent \textbf{3.} To get an intuitive idea of the behavior of the
function $f\left( r\right) $ (cf. Theorem \ref{BCS theorem 1}), we analyze
the cardinality of the set $\mathfrak{S}$ of strictly positive solutions of
the gap equation (\ref{BCS gap equation}):

\begin{lemma}[Cardinality of the set $\mathfrak{S}$]
\label{lemma cardinality}\mbox{  }\newline
If $\beta \gamma \leq 6$, the gap equation (\ref{BCS gap equation}) has at
most one strictly positive solution, whereas it has, at most, two strictly
positive solutions when $\beta \gamma >6$.
\end{lemma}

\noindent \textit{Proof:} From (\ref{pressure derivee 1}), any strictly
positive maximizer $\mathrm{r}_{\beta }>0$ of (\ref{BCS pressure 2}) is
solution of the equation%
\begin{equation}
\mathfrak{h}_{1}\left( g_{r}\right) =0 , \quad \mathrm{\ with}\quad
\mathfrak{h}_{1}\left( x\right) :=\frac{\gamma }{2x}\sinh \left( \beta
x\right) -e^{\lambda \beta }\cosh \left( \beta h\right) -\cosh \left( \beta
x\right) .  \label{BCS gap equationbis}
\end{equation}%
This last equation is equivalent to the gap equation (\ref{BCS gap equation}%
). For any $x>0$, observe that
\begin{equation}
\partial _{x}\mathfrak{h}_{1}\left( x\right) =\frac{\beta \gamma }{2x}\cosh
\left( x\beta \right) -\left( \frac{\gamma }{2x^{2}}+\beta \right) \sinh
\left( x\beta \right) =0  \label{newequation 1}
\end{equation}%
if and only if%
\begin{equation}
(2\beta ^{-1}\gamma ^{-1})^{1/2}y=\sqrt{\frac{y}{\tanh (y)}-1}%
=:C(y),\;\;y=\beta x>0.  \label{newequation 2}
\end{equation}%
The map $y\mapsto C(y)$ is strictly concave for $y>0$, $C(0)=0$, and $%
\partial _{y}C(0)=(2/6)^{1/2}$. Therefore, if $\beta \gamma >6$ there is a
unique strictly positive solution $\widetilde{y}=\beta \widetilde{x}>0$ of (%
\ref{newequation 2}), and there is no strictly positive solution of (\ref%
{newequation 2}) when $\beta \gamma <6$. Since $\mathfrak{h}_{1}(0)$ could
be negative in some cases and $\mathfrak{h}_{1}\left( x\right) $ diverges
exponentially to $-\infty $ as $x\rightarrow \infty $, the cardinality of
set of strictly positive solutions of the gap equation (\ref{BCS gap
equation}) is at most two if $\beta \gamma >6$, or at most one if $\beta
\gamma \leq 6$.\hfill $\Box $

\begin{figure*}[hbtp]
\begin{center}
\mbox{
\leavevmode
\subfigure
{ \includegraphics[angle=0,scale=1,clip=true,width=3.8cm]{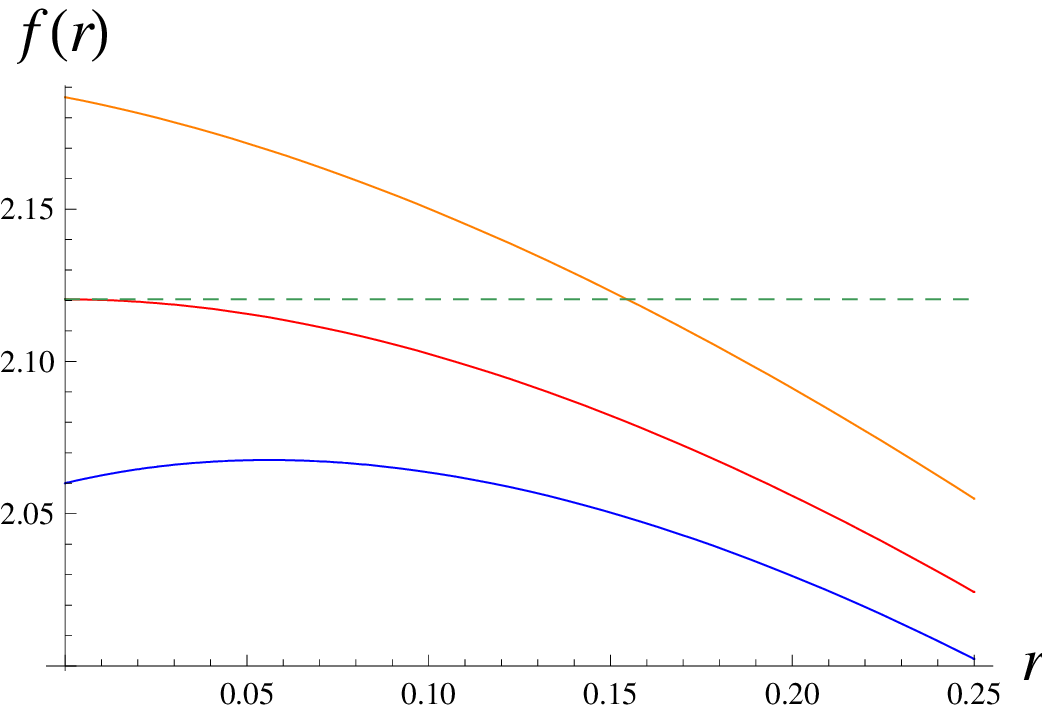} }

\leavevmode
\subfigure
{ \includegraphics[angle=0,scale=1,clip=true,width=3.8cm]{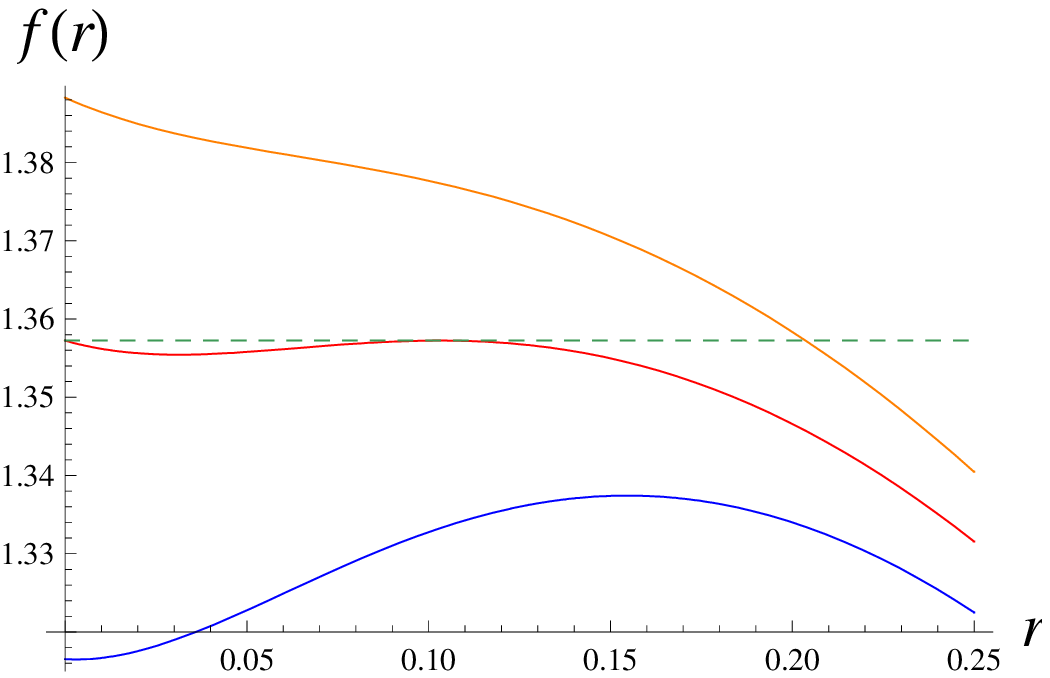} }

\leavevmode
\subfigure
{ \includegraphics[angle=0,scale=1,clip=true,width=3.8cm]{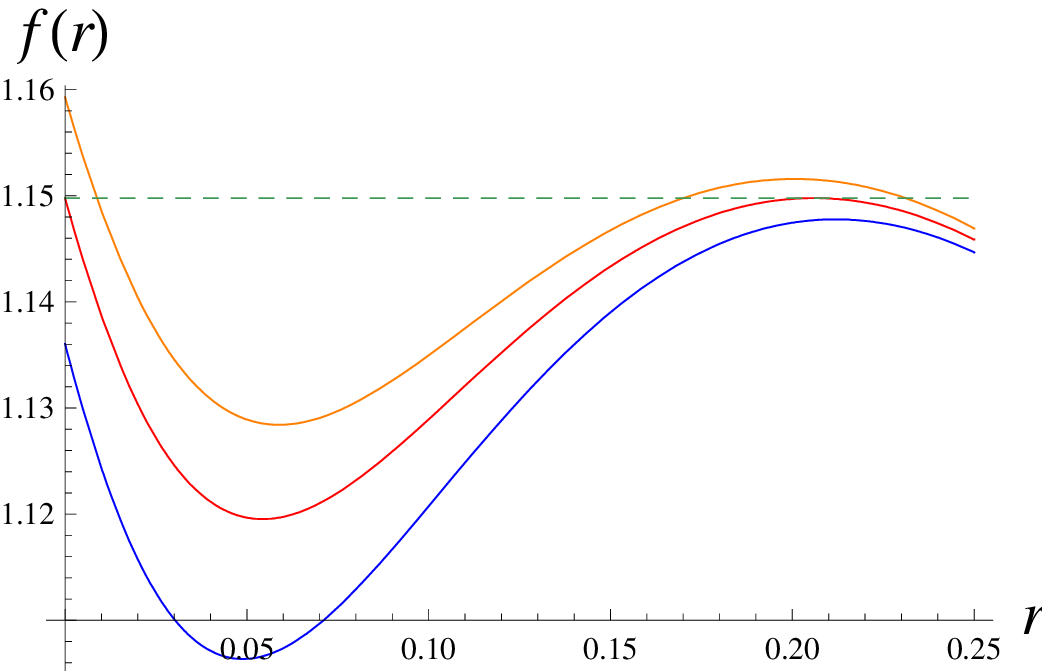} }
}
\end{center}
\caption{\emph{Illustrations of the function $f\left( r\right) $ for $r\in \lbrack 0,1/4]$
at $(\mu ,\gamma ,h)=(1,2.6,0)$ with inverse temperatures $\beta =\beta
_{c}-0.3$ (orange line), $\beta =\beta _{c}$ (red line), $\beta =\beta
_{c}+0.5$ (blue line), and with coupling constants $\lambda =0$ (left
figure), $\lambda =0.45$ (figure on the center) and $\lambda =0.575\,$
(right figure). Here $\beta _{c}=\theta _{c}^{-1}$ is the critical inverse
temperature which, from left to right, equals $2.04$, $3.46$ and $6.35$
respectively.}}
\label{fonction.eps}
\end{figure*}

Consequently, if the gap equation (\ref{BCS gap equation}) has no solution,
then $f(r)$ is strictly decreasing for any $r\geq 0$. If the gap equation (%
\ref{BCS gap equation}) has one unique solution $\mathrm{r}_{\beta }>0,$ the
function $f(r)$ is increasing until its (strictly positive) maximizer $%
\mathrm{r}_{\beta }>0$ and decreasing next for $r\geq \mathrm{r}_{\beta }.$
Finally, when there are two strictly positive solutions of (\ref{BCS gap
equation}), the lower one must be one local minimum whereas the larger
solution must be a local maximum. In this case the function $f(r)$ decreases
for $r\geq 0$ until its local minimum, then increases until its local
maximum, and finally decreases again to diverge towards $-\infty .$ Note
that none of these cases can be excluded, i.e., they all appear depending on
$\beta ,\gamma >0$ and $\mu ,\lambda ,h\in \mathbb{R}$. See figures \ref%
{order-parameter-temp-lambda.eps} and \ref{fonction.eps}.%
%
%
\newline

\noindent \textbf{4.} We study now the dependence of $\mathrm{r}_{\beta }>0\
$w.r.t. variations of each parameter. So, let us fix the parameters $\{\beta
,\mu ,\lambda ,\gamma ,h\}\backslash \{\nu \}$ with $\nu =\beta ,$ $\mu ,$ $%
\lambda ,$ $\gamma ,$ or $h$ and consider the function $\xi \left( r,\nu
\right) :=\partial _{r}f\left( r,\nu \right) $ for $r\geq 0$ and $\nu $ in
the open set of definition of $f(r,\nu )=f(\beta ,\mu ,\lambda ,\gamma ,h;r)$%
, see (\ref{pressure derivee 1}). Recall that $\mathrm{r}_{\beta }>0$ is a
solution at $\nu =\nu _{0}$ of the gap equation (\ref{BCS gap equation}),
i.e., $\xi (\mathrm{r}_{\beta },\nu _{0})=0.$

Straightforward computations imply that
\begin{equation}
\partial _{r}^{2}f\left( r\right) =\frac{\gamma ^{4}\beta }{4g_{r}^{2}\left(
e^{\lambda \beta }\cosh \left( \beta h\right) +\cosh \left( \beta
g_{r}\right) \right) }\mathfrak{h}_{2}\left( g_{r}\right) ,
\label{pressure derivee 2}
\end{equation}%
for any $r>0$ with
\begin{equation}
\mathfrak{h}_{2}\left( x\right) :=\frac{e^{\lambda \beta }\cosh \left( \beta
h\right) \cosh \left( \beta x\right) +1}{e^{\lambda \beta }\cosh \left(
\beta h\right) +\cosh \left( \beta x\right) }-\frac{\sinh \left( \beta
x\right) }{\beta x}.  \label{pressure derivee 2bis}
\end{equation}%
It yields that there is at most one strictly positive solution, $\mathrm{%
\tilde{r}}\geq 0$ of $\partial _{r}\xi (r,\nu _{0})=0$ for each fixed set of
parameters. For instance, if $e^{\lambda \beta }\cosh (\beta h)\leq 1$, then
it is straightforward to check that $\partial _{r}\xi \left( r,\nu
_{0}\right) <0$ for any $r>0$. In the situation where the gap equation (\ref%
{BCS gap equation}) has two strictly positive solutions, $\mathrm{r}_{\beta
}>0$ cannot solve $\partial _{r}\xi (r,\nu _{0})=0$, since in this case the
equation $\mathfrak{h}_{2}(x)=0$ would have at least two strictly positive
solutions, as $\mathrm{r}_{\beta }$ is a maximizer.

Consequently, to simplify our study we restrict on the very large set of
parameters where $\partial _{r}\xi (\mathrm{r}_{\beta },\nu _{0})\neq 0.$ In
this case, the differential $d\xi $ has maximal rank at $(\mathrm{r}_{\beta
},\nu _{0})$ and from the implicit function theorem, there are $\varepsilon
>0$ and a smooth and strictly positive function\footnote{%
If $\nu =\beta ,$ then of course $\mathrm{r}_{\beta }(\nu ):=\mathrm{r}_{\nu
}.$} $\mathrm{r}_{\beta }(\nu )>0$ defined on the ball $B_{\varepsilon }(\nu
_{0})$ centered on the point $\nu _{0}$ and with radius $\varepsilon $ such
that $\xi (\nu ,\mathrm{r}_{\beta }(\nu ))=0$ for any $\nu \in B_{\epsilon
}(\nu _{0})$. By continuity of the function $\partial _{r}\xi $ we can
choose $\varepsilon >0$ such that $\partial _{r}\xi (\nu ,\mathrm{r}_{\beta
}(\nu ))$ does not change its sign for $\nu \in B_{\epsilon }(\nu _{0})$.
Thus $\mathrm{r}_{\beta }(\nu )$ describes the evolution of the solution of (%
\ref{BCS pressure 2}) for $\nu \in B_{\epsilon }(\nu _{0})$. If $\mathrm{r}%
_{\beta }=\mathrm{r}_{\beta }(\nu _{0})>0$ is the unique maximizer of (\ref%
{BCS pressure 2}) with $\partial _{r}\xi (\mathrm{r}_{\beta },\nu _{0})\neq
0 $, then the function $\mathrm{r}_{\beta }(\nu )$ describes the smooth
evolution of the Cooper pair condensate density w.r.t. small perturbations
of $\nu _{0}$. Observe that%
\begin{equation*}
\partial _{\nu }\xi \left( \mathrm{r}_{\beta }\left( \nu \right) ,\nu
\right) =\left\{ \partial _{\nu }\mathrm{r}_{\beta }\left( \nu \right)
\right\} \left\{ \partial _{r}\xi \left( r,\nu \right) \right\} |_{r=\mathrm{%
r}_{\beta }\left( \nu \right) }+\left\{ \partial _{\nu }\xi \left( r,\nu
\right) \right\} |_{r=\mathrm{r}_{\beta }\left( \nu \right) }=0
\end{equation*}%
and $\left\{ \partial _{r}\xi \left( r,\nu _{0}\right) \right\} |_{r=\mathrm{%
r}_{\beta }\left( \nu _{0}\right) }<0$ because $\mathrm{r}_{\beta }$ is a
maximizer. Consequently, one obtains
\begin{equation*}
\mathrm{sgn}\left\{ \partial _{\nu }\mathrm{r}_{\beta }\left( \nu
_{0}\right) \right\} =\mathrm{sgn}\left\{ \left\{ \partial _{\nu }\partial
_{r}f\left( r,\nu _{0}\right) \right\} |_{r=\mathrm{r}_{\beta }\left( \nu
_{0}\right) }\right\} .
\end{equation*}%
In other words, the function $\mathrm{r}_{\beta }(\nu )$ of $\nu \in
B_{\epsilon }(\nu _{0})$ is either increasing if
\begin{equation*}
\left\{ \partial _{\nu }\partial _{r}f\left( r,\nu _{0}\right) \right\} |_{r=%
\mathrm{r}_{\beta }\left( \nu _{0}\right) }>0,
\end{equation*}%
or decreasing if
\begin{equation*}
\left\{ \partial _{\nu }\partial _{r}f\left( r,\nu _{0}\right) \right\} |_{r=%
\mathrm{r}_{\beta }\left( \nu _{0}\right) }<0,
\end{equation*}%
as soon as $\mathrm{r}_{\beta }>0$ is the unique maximizer of (\ref{BCS
pressure 2}) with $\partial _{r}\xi (\mathrm{r}_{\beta },\nu _{0})\neq 0$.
\newline

\noindent \textbf{5.} By applying this last result respectively to $\nu
_{0}=\gamma >\Gamma _{|\mu -\lambda |,\lambda +|h|}$ (Corollary \ref{BCS
theorem 2-0bis}) and $\nu _{0}=h\in \mathbb{R}$, we obtain that $\mathrm{r}%
_{\beta }>0$ is an increasing function of $\gamma >0$ and a decreasing
function of $|h|$ because via (\ref{BCS gap equation}) one has
\begin{equation*}
\left\{ \partial _{\gamma }\partial _{r}f\left( r,\gamma \right) \right\}
|_{r=\mathrm{r}_{\beta }}>4\gamma ^{-2}\left( \mu -\lambda \right) ^{2}\geq 0
\end{equation*}%
at fixed parameters $(\beta ,\mu ,\lambda ,h)$ and
\begin{equation*}
\left\{ \partial _{h}\partial _{r}f\left( r,h\right) \right\} |_{r=\mathrm{r}%
_{\beta }}=-\frac{2g_{\mathrm{r}_{\beta }}\beta e^{\lambda \beta }\sinh
\left( \beta h\right) }{\sinh \left( \beta g_{\mathrm{r}_{\beta }}\right) }
\end{equation*}%
at fixed $(\beta ,\mu ,\lambda ,\gamma ).$\newline

\noindent \textbf{6.} If $\gamma >\Gamma _{|\mu -\lambda |,\lambda +|h|},$
for any fixed $(\beta ,\gamma ,\lambda ,h)$ the order parameter $\mathrm{r}%
_{\beta }>0$ is a decreasing function of $|\mu -\lambda |$ under the
condition that $e^{\lambda \beta }\cosh \left( \beta h\right) \leq 1$, as
\begin{equation*}
\left\{ \partial _{\mu }\partial _{r}f\left( r,\mu \right) \right\} |_{r=%
\mathrm{r}_{\beta }}=\frac{\gamma ^{2}\beta \left( \mu -\lambda \right) }{%
2g_{r}^{2}\left( e^{\lambda \beta }\cosh \left( \beta h\right) +\cosh \left(
\beta g_{\mathrm{r}_{\beta }}\right) \right) }\mathfrak{h}_{2}\left( g_{%
\mathrm{r}_{\beta }}\right) ,
\end{equation*}%
cf. (\ref{pressure derivee 2bis}). If $e^{\lambda \beta }\cosh \left( \beta
h\right) >1$, the behavior of $\mathrm{r}_{\beta }>0$ is not anymore
monotone as a function of $|\mu -\lambda |$ ($\lambda $ being fixed), cf.
figure \ref{Mott-Insulator-1.eps}.

The behavior of $\mathrm{r}_{\beta }$ as a function of $\lambda $ or $\beta $
is also not clear in general. But, at least as a function of the inverse
temperature $\beta >0$, we can give simple sufficient conditions to get its
monotonicity. Indeed, direct computations show that
\begin{eqnarray*}
\left\{ \partial _{\beta }\partial _{r}f\left( r,\beta \right) \right\} |_{r=%
\mathrm{r}_{\beta }} &=&\left( \gamma +2\lambda \right) g_{\mathrm{r}_{\beta
}}\frac{\cosh \left( \beta g_{\mathrm{r}_{\beta }}\right) }{\sinh \left(
\beta g_{\mathrm{r}_{\beta }}\right) }-\left( \lambda \gamma +2g_{\mathrm{r}%
_{\beta }}^{2}\right) \\
&&-2hg_{\mathrm{r}_{\beta }}\frac{e^{\lambda \beta }\sinh \left( \beta
h\right) }{\sinh \left( \beta g_{\mathrm{r}_{\beta }}\right) }.
\end{eqnarray*}%
By combining this last equality with (\ref{BCS gap equation}), we then get
that
\begin{equation}
\{\partial _{\beta }\partial _{r}f(r,\beta )\}|_{r=\mathrm{r}_{\beta }}\geq 0
\label{encore new4bisbis}
\end{equation}%
with $\mathrm{r}_{\beta }>0$ if and only if%
\begin{equation}
g_{\mathrm{r}_{\beta }}^{2}\leq \frac{\gamma \left( \gamma \cosh \left(
\beta g_{\mathrm{r}_{\beta }}\right) -2e^{\lambda \beta }\cosh \left( \beta
h\right) \left( \lambda +h\tanh \left( \beta h\right) \right) \right) }{%
4\left( \cosh \left( \beta g_{\mathrm{r}_{\beta }}\right) +e^{\lambda \beta
}\cosh \left( \beta h\right) \right) }.  \label{encore new4}
\end{equation}%
From (\ref{BCS gap equation}) combined with $\tanh (x)<1$, we also have%
\begin{equation}
g_{\mathrm{r}_{\beta }}^{2}<\frac{\gamma ^{2}\cosh ^{2}\left( \beta g_{%
\mathrm{r}_{\beta }}\right) }{4\left( \cosh \left( \beta g_{\mathrm{r}%
_{\beta }}\right) +e^{\lambda \beta }\cosh \left( \beta h\right) \right) ^{2}%
}.  \label{encore new4bis}
\end{equation}%
Therefore, a sufficient condition to satisfy the inequality (\ref{encore
new4}) is obtained by bounding the r.h.s. of (\ref{encore new4bis}) with the
r.h.s. of (\ref{encore new4}). From (\ref{BCS gap equation}) this implies
the condition
\begin{equation*}
g_{\mathrm{r}_{\beta }}\geq \left( \lambda +h\tanh \left( \beta h\right)
\right) \tanh \left( \beta g_{\mathrm{r}_{\beta }}\right) ,
\end{equation*}%
under which $\mathrm{r}_{\beta }$ is an increasing function of $\beta >0$.
This inequality is also equivalent to
\begin{equation*}
g_{\mathrm{r}_{\beta }}\leq \tanh \left( \beta g_{\mathrm{r}_{\beta
}}\right) \left( \frac{\gamma }{2}-\frac{e^{\lambda \beta }\cosh \left(
\beta h\right) }{\cosh \left( \beta g_{\mathrm{r}_{\beta }}\right) }\left(
\lambda +h\tanh \left( \beta h\right) \right) \right) .
\end{equation*}%
In particular, by using again the gap equation (\ref{BCS gap equation}), if
\begin{equation*}
\gamma >2\left( \lambda +h\tanh \left( \beta h\right) \right) \left( 1+\frac{%
e^{\lambda \beta }\cosh \left( \beta h\right) }{\cosh \left( \beta g_{%
\mathrm{r}_{\beta }}\right) }\right) ,
\end{equation*}%
then $\mathrm{r}_{\beta }>0$ is an increasing function of $\beta >0$. Since $%
\tanh x\leq 1,$ another sufficient condition to get (\ref{encore new4bisbis}%
) is $\lambda+|h| \leq g_{\mathrm{r}_{\beta }}$. In particular, if $\lambda
<|\mu -\lambda |$ and $\gamma >\Gamma _{|\mu -\lambda |,\lambda +|h|} $ with
$h$ sufficiently small, then $\mathrm{r}_{\beta }>0$ is again an increasing
function of $\beta >0.$

Therefore, the domain of $(\mu ,\lambda ,\gamma ,h)$ where $\mathrm{r}%
_{\beta }>0$ is proven to be an increasing function of $\beta >0$ is rather
large. Actually, from a huge number of numerical computations, we conjecture
that $\mathrm{r}_{\beta }>0$ is always an increasing function of $\beta >0.$
In other words, this conjecture implies that the condition expressed in
Corollary \ref{BCS theorem 2-0bis} on $(\mu ,\lambda ,\gamma ,h)$ should be
necessary to obtain a superconductor at a fixed temperature.\newline

\noindent \textbf{7.} Observe that the order of the phase transition depends
on the parameters. For instance, assume $\lambda \leq 0$, $h=0$ and $\gamma
>\Gamma _{|\mu -\lambda |,\lambda }$. Then, at any inverse temperature $%
\beta >0$ it follows from (\ref{pressure derivee 2}) that $f(r)$ is a
strictly concave function of $r>0$. This property justifies the existence
and uniqueness of the inverse temperature $\beta _{c}$ solution of the
equation
\begin{equation*}
\frac{\tanh \left( \beta |\mu -\lambda |\right) }{|\mu -\lambda |}=\frac{2}{%
\gamma }\left( 1+\frac{e^{\lambda \beta }}{\cosh \left( \beta |\mu -\lambda
|\right) }\right) ,
\end{equation*}%
i.e., (\ref{BCS gap equation}) for $\lambda \leq 0$, $h=0$ and $r=0$. In
particular, $\beta _{c}$ is such that the Cooper pair condensate density
continuously goes from $\mathrm{r}_{\beta }=0$ for $\beta \leq \beta _{c}$
to $\mathrm{r}_{\beta }>0$ for $\beta >\beta _{c}$. In this case the
superconducting phase transition is of second order, cf. figure \ref%
{order-parameter-temp-lambda.eps}.

The appearance of a first order phase transition at some fixed $(\mu
,\lambda ,\gamma ,h)$ is also not surprising. Indeed, recall that the
function $f(r)$ may have a local minimum and a local maximum, see
discussions below Lemma \ref{lemma cardinality}. For instance, assume now $%
\lambda =\mu >0$, $h=0$ and $4\lambda =\Gamma _{0,\lambda }<\gamma \leq
6\lambda $. Then, from (\ref{pressure derivee 1}) for $r=0$,
\begin{equation*}
\partial _{r}f\left( 0\right) =\frac{\gamma }{e^{\lambda \beta }+1}\left(
\frac{\gamma \beta }{2}-\left( e^{\lambda \beta }+1\right) \right) .
\end{equation*}%
Since by explicit computations
\begin{equation*}
\underset{x>0}{\min }\left\{ \frac{e^{x}+1}{x}\right\} >3,
\end{equation*}%
it follows that $\partial _{r}f(0)<0$ for any $\beta >0$ whenever $\lambda
=\mu >0$, $h=0$ and $0<\gamma \leq 6\lambda $. Therefore, as soon as there
is a superconducting phase transition, for instance if $4\lambda <\gamma
\leq 6\lambda $ (cf. Corollary \ref{BCS theorem 2-0bis}), the function $%
\mathrm{r}_{\beta }$ of $\beta >0$ must be discontinuous at the critical
point. This case is an example of a first order superconducting phase
transition. Numerical illustrations of a similar first order phase
transition are also given in figure \ref{order-parameter-temp-lambda.eps}.%
\newline

\noindent \textbf{8.} We conclude this section by a computation of the
asymptotics of the order parameter $\mathrm{r}_{\beta }$ as $\beta
\rightarrow \infty $. We prove in particular Corollary \ref{BCS theorem
2-0bis}.

From (\ref{definition r max}), we already know that $\mathrm{r}_{\beta }=0$
for any $\gamma \leq 2|\tilde{\mu}_{\lambda }|$ with $\tilde{\mu}_{\lambda
}:=\mu -\lambda .$ Therefore, we consider here that $\gamma >2|\tilde{\mu}%
_{\lambda }|$ and we look for the domain where the parameter $\mathrm{r}%
_{\beta }$ is strictly positive in the limit $\beta \rightarrow \infty $.
Recall that $\mathrm{r}_{\beta }$ is solution of the variational problem (%
\ref{BCS pressure 2}), i.e.,
\begin{equation}
\frac{1}{\beta }\ln 2+\underset{r\geq 0}{\sup }f\left( r\right) =-\gamma
\mathrm{r}_{\beta }+\frac{1}{\beta }\ln \left\{ e^{\beta h}+e^{-\beta
h}+e^{\beta \left( g_{\mathrm{r}_{\beta }}-\lambda \right) }+e^{-\beta
\left( g_{\mathrm{r}_{\beta }}+\lambda \right) }\right\} .
\label{BCS particle density and pressure 2bisbis}
\end{equation}%
When $\beta \rightarrow \infty $ the last exponential term can always be
neglected for our analysis since $g_{\mathrm{r}_{\beta }}\geq 0$.

Now, assume first that $g_{0}=|\tilde{\mu}_{\lambda }|>\lambda +|h|.$ Then $%
g_{r}>\lambda +|h|$ for any $r\geq 0$ and when $\beta \rightarrow \infty $
the function $f\left( r\right) $ converges to
\begin{equation*}
w\left( r\right) :=-\gamma r+g_{r}-\lambda .
\end{equation*}%
In particular, the order parameter $\mathrm{r}_{\beta }$ converges towards
the unique maximizer $\mathrm{r}_{\max }$ (\ref{definition r max}) of the
function $w\left( r\right) $ for $r\geq 0,$ i.e.,
\begin{equation}
\mathrm{r}_{\infty }:=\underset{\beta \rightarrow \infty }{\lim }\mathrm{r}%
_{\beta }=\mathrm{r}_{\max },  \label{eq order parameter 4}
\end{equation}%
for any $\gamma >2|\tilde{\mu}_{\lambda }|$ and real numbers $\mu ,\lambda
,h $ satisfying $|\tilde{\mu}_{\lambda }|>\lambda +|h|$.

Assume now that $|\tilde{\mu}_{\lambda }|\leq \lambda +|h|$ and let $\mathrm{%
r}_{\min }$ be the solution of $g_{r}=\lambda +|h|$, i.e.,
\begin{equation}
\mathrm{r}_{\min }:=\gamma ^{-2}\left( \left( \lambda +|h|\right) ^{2}-%
\tilde{\mu}_{\lambda }^{2}\right) \geq 0.  \label{definition r min}
\end{equation}%
Then, for any $r\in \lbrack 0,\mathrm{r}_{\min }]$
\begin{equation*}
f\left( r\right) =-\gamma r+\left\vert h\right\vert +o\left( 1\right)
\mathrm{\ as\ }\beta \rightarrow \infty .
\end{equation*}%
In particular, since $\gamma >0$,
\begin{equation}
\underset{0\leq r\leq \mathrm{r}_{\min }}{\sup }f\left( r\right) =f\left(
\delta \right) =\left\vert h\right\vert +o\left( 1\right) ,\quad \mathrm{\
with\ }\delta =o\left( 1\right) \mathrm{\ as\ }\beta \rightarrow \infty .
\label{eq order parameter 5}
\end{equation}%
The solution $\mathrm{r}_{\beta }$ of the variational problem (\ref{BCS
particle density and pressure 2bisbis}) converges either to $0$, or to some
strictly positive value $\mathrm{r}_{\infty }>\mathrm{r}_{\min }.$ In the
case where $\mathrm{r}_{\infty }>\mathrm{r}_{\min }$, we would have
\begin{equation}
f\left( \mathrm{r}_{\infty }\right) =w\left( \mathrm{r}_{\infty }\right)
+o\left( 1\right) \mathrm{\ as\ }\beta \rightarrow \infty .
\label{eq order parameter 6}
\end{equation}%
Now, if $|\tilde{\mu}_{\lambda }|\leq \lambda +|h|$ and $\gamma \leq
2(\lambda +|h|),$ then $\mathrm{r}_{\min }\geq \mathrm{r}_{\max }$, cf. (\ref%
{definition r max}) and (\ref{definition r min}). In this regime,
straightforward computations show that
\begin{equation}
\left\vert h\right\vert -\underset{r\geq \mathrm{r}_{\min }}{\sup }w\left(
r\right) =\left\vert h\right\vert -w\left( \mathrm{r}_{\min }\right) =\gamma
^{-1}\left( \left( \left\vert h\right\vert +\lambda \right) ^{2}-\tilde{\mu}%
_{\lambda }^{2}\right) \geq 0.  \label{eq order parameter 4-1bis}
\end{equation}%
In other words, the order parameter $\mathrm{r}_{\beta }$ converges towards
\begin{equation}
\mathrm{r}_{\infty }:=\underset{\beta \rightarrow \infty }{\lim }\mathrm{r}%
_{\beta }=0,  \label{eq order parameter 4-1}
\end{equation}%
for any $\gamma \leq 2(\lambda +|h|)$ and real numbers $\mu ,\lambda ,h$
satisfying $|\tilde{\mu}_{\lambda }|\leq \lambda +|h|$.

However, if $|\tilde{\mu}_{\lambda }|\leq \lambda +|h|$ and $\gamma
>2(\lambda +|h|),$ then $\mathrm{r}_{\min }<\mathrm{r}_{\max }$. In
particular one gets
\begin{equation}
\left\vert h\right\vert -\underset{r\geq \mathrm{r}_{\min }}{\sup }w\left(
r\right) =\left\vert h\right\vert -w\left( \mathrm{r}_{\max }\right) =-\frac{%
1}{4\gamma }\left( \gamma -\tilde{\Gamma}_{|\tilde{\mu}_{\lambda }|,\lambda
+|h|}\right) \left( \gamma -\Gamma _{|\tilde{\mu}_{\lambda }|,\lambda
+|h|}\right) ,  \label{eq98}
\end{equation}
with $\Gamma _{x,y}\geq 2y$ defined for any $x\in \mathbb{R}_{+}$ and $y\in
\mathbb{R}$ in Corollary \ref{BCS theorem 2-0bis} and
\begin{equation*}
\tilde{\Gamma}_{|\tilde{\mu}_{\lambda }|,\lambda +|h|}:=2\left( \lambda
+\left\vert h\right\vert -\sqrt{\left( \lambda +\left\vert h\right\vert
\right) ^{2}-\tilde{\mu}_{\lambda }^{2}}\right) \leq 2\left\vert \tilde{\mu}%
_{\lambda }\right\vert .
\end{equation*}%
In particular,
\begin{equation}
\underset{r\geq \mathrm{r}_{\min }}{\sup }w\left( r\right) =w\left( \mathrm{r%
}_{\max }\right) >\left\vert h\right\vert ,  \label{eq order parameter 8}
\end{equation}%
for any $\gamma >\Gamma _{|\tilde{\mu}_{\lambda }|,\lambda +|h|}\geq 2|%
\tilde{\mu}_{\lambda }|.$ Therefore, by combining (\ref{eq order parameter 5}%
) with (\ref{eq order parameter 6}) and (\ref{eq order parameter 8}), we
obtain
\begin{equation}
\mathrm{r}_{\infty }:=\underset{\beta \rightarrow \infty }{\lim }\mathrm{r}%
_{\beta }=\mathrm{r}_{\max },  \label{eq order parameter 4-2}
\end{equation}%
for any $\gamma >\Gamma _{|\tilde{\mu}_{\lambda }|,\lambda +|h|}$ and real
numbers $\mu ,\lambda ,h$ satisfying $|\tilde{\mu}_{\lambda }|\leq \lambda
+|h|$.

Finally, if $\gamma =\Gamma _{|\tilde{\mu}_{\lambda }|,\lambda +|h|}$ and $|%
\tilde{\mu}_{\lambda }|<\lambda +|h|,$ observe that (\ref{eq98}) is zero.
So, we analyze the next order term to know which number, $0$ or $\mathrm{r}%
_{\max },$ maximizes the function $f\left( r\right) $ when $\beta
\rightarrow \infty $. On the one hand, straightforward estimations imply that%
\begin{equation}
f\left( 0\right) -\left\vert h\right\vert =\beta ^{-1}\left( e^{-\beta
\left( \lambda +\left\vert h\right\vert -|\tilde{\mu}_{\lambda }|\right)
}+e^{-2\beta \left\vert h\right\vert }\right) \left( 1+o\left( 1\right)
\right) \mathrm{\ as\ }\beta \rightarrow \infty .
\label{eq order parameter 4-2+1}
\end{equation}%
On the other hand, if $\gamma =\Gamma _{|\tilde{\mu}_{\lambda }|,\lambda
+|h|}$ with $|\tilde{\mu}_{\lambda }|<\lambda +|h|$, then by using (\ref%
{definition r max}) one obtains
\begin{equation}
f\left( \mathrm{r}_{\max }\right) -\left\vert h\right\vert =\beta
^{-1}e^{-\beta \sqrt{\left( \lambda +\left\vert h\right\vert \right) ^{2}-%
\tilde{\mu}_{\lambda }^{2}}}\left( 1+o\left( 1\right) \right) \mathrm{\ as\ }%
\beta \rightarrow \infty .  \label{eq order parameter 4-2+2}
\end{equation}%
Therefore, if $\gamma =\Gamma _{|\tilde{\mu}_{\lambda }|,\lambda +|h|}$ and $%
|\tilde{\mu}_{\lambda }|<\lambda +|h|$, it is trivial to check from (\ref{eq
order parameter 4-2+1})-(\ref{eq order parameter 4-2+2}) that $f(0)>f(%
\mathrm{r}_{\max })$ when $\beta \rightarrow \infty $.

Consequently, the limits (\ref{eq order parameter 4}), (\ref{eq order
parameter 4-1}) and (\ref{eq order parameter 4-2}) together with (\ref%
{definition r max}) imply Corollary \ref{BCS theorem 2-0bis} for any $\gamma
\neq \Gamma _{|\mu -\lambda |,\lambda +|h|}$, whereas if $\gamma =\Gamma
_{|\mu -\lambda |,\lambda +|h|}$, the order parameter $\mathrm{r}_{\beta }$
converges to $\mathrm{r}_{\infty }=0$.

\section{Appendix: Griffiths arguments\label{section proof griffiths}}

As we have an explicit representation of the pressure, it can be verified in
some cases that $\mathrm{r}_{\beta }$ is a $C^{1}$--function\footnote{%
For instance, for special choices of parameters one could check that $%
\partial _{r}\xi (\mathrm{r}_{\beta },\nu _{0})\neq 0$, see Section \ref%
{section variational problem}.} of parameters implying that $\mathrm{p}%
\left( \beta ,\mu ,\lambda ,\gamma ,h\right) $ is differentiable w.r.t.
parameters. In this particular situation, the proofs of Theorems \ref{BCS
theorem 2-0}, \ref{BCS theorem 3}, \ref{BCS theorem 2-1}, \ref{BCS theorem
2-2}, \ref{BCS theorem 2-3} and \ref{BCS theorem 2-4} done in Section \ref%
{equilibirum.paragraph} could also be performed without our notion of
equilibrium states by using Griffiths arguments \cite%
{BruZagrebnov8,Griffiths1,HeppLieb}, which are based on convexity properties
of the pressure. We explain it shortly and we conclude by a discussion of an
alternative proof of Theorem \ref{BCS theorem 3}.

\begin{remark}
\label{remark griffiths copy(1)} Our method gives access to all correlation functions at once (cf. Theorem \ref{Theorem
equilibrium state 4bis}). It is generalized in \cite%
{BruPedra2} to all translation invariant Fermi systems. However, computing all correlation functions with
Griffiths arguments \cite%
{BruZagrebnov8,Griffiths1,HeppLieb} requires the differentiability of the pressure w.r.t. any perturbation as well as the computation of its corresponding derivative.
This is generally a very hard task, for instance for correlation functions involving many lattice points.
\end{remark}

\noindent \textbf{1.} Take self--adjoint operators $\mathfrak{P}_{N}\ $%
acting on the fermionic Fock space and assume the existence of the (infinite
volume) grand--canonical pressure
\begin{equation*}
\mathrm{p}_{\varepsilon }\left( \beta ,\mu ,\lambda ,\gamma ,h\right) :=%
\underset{N\rightarrow \infty }{\lim }\mathrm{p}_{N,\varepsilon }\left(
\beta ,\mu ,\lambda ,\gamma ,h\right)
\end{equation*}%
for any fixed $\varepsilon $ in a neighborhood $\mathcal{V}$ of $0.$ In this
case, observe that the finite volume pressure
\begin{equation*}
\mathrm{p}_{N,\varepsilon }\left( \beta ,\mu ,\lambda ,\gamma ,h\right) :=%
\frac{1}{\beta N}\ln \mathrm{Trace}\left( e^{-\beta \left( \mathrm{H}%
_{N}-\varepsilon \mathfrak{P}_{N}\right) }\right)
\end{equation*}%
is convex as a function of $\varepsilon \in \mathcal{V}$ and
\begin{equation*}
\partial _{\varepsilon }\mathrm{p}_{N,0}=N^{-1}\omega _{N}\left( \mathfrak{P}%
_{N}\right) .
\end{equation*}%
Consequently, the point-wise convergence of the function $\mathrm{p}%
_{N,\varepsilon }$ towards $\mathrm{p}_{\varepsilon }$ implies that
\begin{equation}
\underset{N\rightarrow \infty }{\lim {\inf }}\left\{ \underset{\varepsilon
\rightarrow 0^{-}}{\lim }\partial _{\varepsilon }\mathrm{p}_{N,\varepsilon
}\right\} \geq \underset{\varepsilon \rightarrow 0^{-}}{\lim }\partial
_{\varepsilon }\mathrm{p}_{\varepsilon }\mathrm{\ and\ }\underset{%
N\rightarrow \infty }{\lim {\sup }}\left\{ \underset{\varepsilon \rightarrow
0^{+}}{\lim }\partial _{\varepsilon }\mathrm{p}_{N,\varepsilon }\right\}
\leq \underset{\varepsilon \rightarrow 0^{+}}{\lim }\partial _{\varepsilon }%
\mathrm{p}_{\varepsilon },  \label{critical point}
\end{equation}%
see Griffiths lemma \cite{Griffiths1,HeppLieb} or \cite[Appendix C]%
{BruZagrebnov8}. In particular, one gets
\begin{equation}
\underset{N\rightarrow \infty }{\lim }\left\{ \partial _{\varepsilon }%
\mathrm{p}_{N,0}\right\} =\underset{N\rightarrow \infty }{\lim }\left\{
N^{-1}\omega _{N}\left( \mathfrak{P}_{N}\right) \right\} =\partial
_{\varepsilon }\mathrm{p}_{\varepsilon =0},  \label{griffiths equation 1}
\end{equation}%
under the assumption that $\mathrm{p}_{\varepsilon }$ is differentiable at $%
\varepsilon =0.$\newline

\noindent \textbf{2.} Therefore, by taking
\begin{equation*}
\mathfrak{P}_{N}=\sum_{x,y\in \Lambda _{N}}a_{x,\uparrow }^{\ast
}a_{x,\downarrow }^{\ast }a_{y,\downarrow }a_{y,\uparrow },
\end{equation*}%
we obtain from (\ref{griffiths equation 1}) that
\begin{equation*}
\underset{N\rightarrow \infty }{\lim }\left\{ \frac{1}{N^{2}}\sum_{x,y\in
\Lambda _{N}}a_{x,\uparrow }^{\ast }a_{x,\downarrow }^{\ast }a_{y,\downarrow
}a_{y,\uparrow }\right\} =\partial _{\gamma }\mathrm{p}\left( \beta ,\mu
,\lambda ,\gamma ,h\right) ,
\end{equation*}%
as soon as the (infinite volume) pressure $\mathrm{p}\left( \beta ,\mu
,\lambda ,\gamma ,h\right) $ has continuous derivative w.r.t. $\gamma >0$.
Combined with Theorem \ref{BCS theorem 1} and (\ref{BCS gap equation}) we
would obtain Theorem \ref{BCS theorem 2-0}. Meanwhile, Theorem \ref{BCS
theorem 2-1}, \ref{BCS theorem 2-2}, \ref{BCS theorem 2-3} and \ref{BCS
theorem 2-4} could have been deduced in the same way from (\ref{griffiths
equation 1}) combined with explicit computations using (\ref{BCS gap
equation}). \newline

\noindent \textbf{3.} A direct proof of Theorem \ref{BCS theorem 3} using
Griffiths arguments is more delicate. One uses similar arguments as in \cite%
{BruZagrebnov8,Ginibre}. We give them for the interested reader.

For any $\phi \in \left[ 0,2\pi \right) $, first recall that the pressure $%
\mathrm{p}_{\alpha ,\phi }$ associated with $\mathrm{H}_{N,\alpha ,\phi }$ (%
\ref{perturbed hamiltonian}) in the thermodynamic limit is given by (\ref%
{pressure avec alpha}), which equals (\ref{pressure avec alphabis}).
Additionally, if the parameters $\beta ,$ $\mu ,$ $\lambda ,$ $\gamma ,$ and
$h$ are such that (\ref{BCS pressure 2}) has a unique maximizer $\mathrm{r}%
_{\beta }$, then the variational problem (\ref{pressure avec alphabis}) has
a unique maximizer $\mathrm{c}_{\beta ,\alpha ,\phi }\in e^{i\phi }\mathbb{R}
$ for $\alpha >0$ sufficiently small, and $\mathrm{c}_{\beta ,\alpha ,\phi }$
converges to $\mathrm{r}_{\beta }^{1/2} e^{i \phi}$ as $\alpha \rightarrow 0$%
, see proof of Theorem \ref{Theorem equilibrium state 4 U(1) broken}.

Now, let us denote by%
\begin{equation*}
\mathfrak{N}_{N}:=\sum\limits_{x\in \Lambda _{N}}\left( n_{x,\uparrow
}+n_{x,\downarrow }\right)
\end{equation*}%
the full particle number operator. By straightforward computations observe
that
\begin{equation}
\left[ a_{x,\uparrow },\mathfrak{N}_{N}\right] =a_{x,\uparrow }\mathrm{\
and\ }\left[ a_{x,\downarrow },\mathfrak{N}_{N}\right] =a_{x,\downarrow },
\label{order parameter eq 00}
\end{equation}%
for any lattice site labelled by $x\in \Lambda _{N}$, where $[A,B]:=AB-BA$.
Therefore the unitary operator $U_{\phi }:=e^{-\frac{i\phi }{2}\mathfrak{N}%
_{N}}$ realizes a global gauge transformation because one deduces from (\ref%
{order parameter eq 00}) that%
\begin{equation}
U_{\phi }a_{x,\uparrow }U_{\phi }^{\ast }=e^{\frac{i\phi }{2}}a_{x,\uparrow
}\;\; \mathrm{\ and\ } \;\; U_{\phi }a_{x,\downarrow }U_{\phi }^{\ast }=e^{%
\frac{i\phi }{2}}a_{x,\downarrow }.  \label{order parameter eq 01}
\end{equation}%
In particular the unitary transformation of the Hamiltonian $\mathrm{H}%
_{N,\alpha ,\phi }$ (\ref{perturbed hamiltonian}) equals
\begin{equation*}
U_{\phi }\mathrm{H}_{N,\alpha ,\phi }U_{\phi }^{\ast }=\mathrm{H}_{N,\alpha
,0}.
\end{equation*}%
It implies on the corresponding Gibbs states (\ref{local Gibbs states avec
alpha}) that
\begin{equation}
\omega _{N,\alpha ,\phi }\left( \mathfrak{B}_{N}\right) =e^{i\phi }\omega
_{N,\alpha ,0}\left( \mathfrak{B}_{N}\right) ,
\label{gauge transformation gibbs state}
\end{equation}
with the operator $\mathfrak{B}_{N}$ be defined by
\begin{equation*}
\mathfrak{B}_{N}:=\sum\limits_{x\in \Lambda _{N}}a_{x,\downarrow
}a_{x,\uparrow }.
\end{equation*}%
In other words, it suffices to prove Theorem \ref{BCS theorem 3} for $\phi
=0 $.

Take $\phi =0$. Observe that
\begin{equation}
0=\omega _{N,\alpha ,0}\left( \left[ \mathrm{H}_{N,\alpha ,0},\mathfrak{N}%
_{N}\right] \right) =\alpha \omega _{N,\alpha ,0}\left( \mathfrak{B}_{N}-%
\mathfrak{B}_{N}^{\ast }\right) .  \label{order parameter eq 1}
\end{equation}%
Additionally, by using the positive semidefinite Bogoliubov--Duhamel scalar
product
\begin{equation*}
\left( X,Y\right) _{\mathrm{H}_{N,\alpha ,0}}:=\beta ^{-1}e^{-\beta N\mathrm{%
p}_{N,\alpha ,0}\left( \beta ,\mu ,\lambda ,\gamma ,h\right)
}\int_{0}^{\beta }\mathrm{Trace}\left( e^{-\left( \beta -\tau \right)
\mathrm{H}_{N,\alpha ,0}}X^{\ast }e^{-\tau \mathrm{H}_{N,\alpha ,0}}Y\right)
\mathrm{d}\tau
\end{equation*}%
w.r.t. the Hamiltonian $\mathrm{H}_{N,\alpha ,0}$ (see, e.g., \cite%
{BrattelliRobinson,BruZagrebnov8,Ginibre}), one gets that%
\begin{eqnarray}
0 &\leq &\beta \left( \left[ \mathfrak{N}_{N},\mathrm{H}_{N,\alpha ,0}\right]
,\left[ \mathfrak{N}_{N},\mathrm{H}_{N,\alpha ,0}\right] \right) _{\mathrm{H}%
_{N,\alpha ,0}}  \notag \\
&=&\omega _{N,\alpha ,0}\left( \left[ \mathfrak{N}_{N},\left[ \mathrm{H}%
_{N,\alpha ,0},\mathfrak{N}_{N}\right] \right] \right) =\alpha \omega
_{N,\alpha ,0}\left( \mathfrak{B}_{N}+\mathfrak{B}_{N}^{\ast }\right) .
\label{order parameter eq 2}
\end{eqnarray}%
So, by combining (\ref{order parameter eq 1}) with (\ref{order parameter eq
2}) it follows that
\begin{equation*}
\omega _{N,\alpha ,0}\left( \mathfrak{B}_{N}\right) =\omega _{N,\alpha
,0}\left( \mathfrak{B}_{N}^{\ast }\right) \geq 0
\end{equation*}%
for any $\alpha \geq 0$. In particular $\omega _{N,\alpha ,0}\left(
\mathfrak{B}_{N}\right) =\omega _{N,\alpha ,0}\left( \mathfrak{B}_{N}^{\ast
}\right) $ is a real number.

The function $\mathrm{p}_{N,\alpha ,0}$ is a convex function of $\alpha \geq
0$ because
\begin{eqnarray*}
&&\beta \Big(\left\{ \left( \mathfrak{B}_{N}+\mathfrak{B}_{N}^{\ast }\right)
-\omega _{N,\alpha ,0}\left( \mathfrak{B}_{N}+\mathfrak{B}_{N}^{\ast
}\right) \right\} ,\left\{ \left( \mathfrak{B}_{N}+\mathfrak{B}_{N}^{\ast
}\right) -\omega _{N,\alpha ,0}\left( \mathfrak{B}_{N}+\mathfrak{B}%
_{N}^{\ast }\right) \right\} \Big)_{\mathrm{H}_{N,\alpha ,0}} \\
&=&\partial _{\alpha }^{2}\mathrm{p}_{N,\alpha ,0}\left( \beta ,\mu ,\lambda
,\gamma ,h\right) .
\end{eqnarray*}%
Then, under the assumption that $\mathrm{p}_{\alpha ,0}$ is differentiable
at $\alpha =0$ away from any critical point, the equations (\ref{griffiths
equation 1}), with
\begin{equation*}
\mathfrak{P}_{N}=\mathfrak{B}_{N}+\mathfrak{B}_{N}^{\ast }
\end{equation*}%
and (\ref{pressure avec alphabis}), imply that
\begin{eqnarray*}
\underset{N\rightarrow \infty }{\lim }\left( \frac{1}{N}\omega _{N,\alpha
,0}\left( \mathfrak{B}_{N}+\mathfrak{B}_{N}^{\ast }\right) \right) &=&%
\underset{N\rightarrow \infty }{\lim }\partial _{\alpha }\left( \frac{1}{%
\beta N}\ln \mathrm{Trace}\left( e^{-\beta \mathrm{H}_{N,\alpha ,0}}\right)
\right) \\
&=&\partial _{\alpha }\mathrm{p}_{\alpha ,0}\left( \beta ,\mu ,\lambda
,\gamma ,h\right) \\
&=&\mathrm{\zeta }_{\mathrm{c}_{\beta ,\alpha ,0}}\left( a_{\downarrow
}^{\ast }a_{\uparrow }^{\ast }+a_{\uparrow }a_{\downarrow }\right) ,
\end{eqnarray*}%
for any $\alpha >0$ sufficiently small and with $\mathrm{\zeta }_{c}(\cdot )$
defined for any $c\in \mathbb{C}$ by (\ref{Gibbs.state nu}).

Returning back to the original Hamiltonian $\mathrm{H}_{N,\alpha ,\phi }$ (%
\ref{perturbed hamiltonian}) for any $\phi \in \left[ 0,2\pi \right) $, we
conclude from (\ref{gauge transformation gibbs state}) combined with the
last equalities that%
\begin{equation*}
\underset{N\rightarrow \infty }{\lim }\left\{ \dfrac{1}{N}\sum\limits_{x\in
\Lambda _{N}}\omega _{N,\alpha ,\phi }\left( a_{x,\uparrow }a_{x,\downarrow
}\right) \right\} =\frac{e^{i\phi }}{2}\mathrm{\zeta }_{\mathrm{c}_{\beta
,\alpha ,0}}\left( a_{\downarrow }^{\ast }a_{\uparrow }^{\ast }+a_{\uparrow
}a_{\downarrow }\right) .
\end{equation*}%
Therefore, by taking the limit $\alpha \rightarrow 0$, Theorem \ref{BCS
theorem 3} would follow if one additionally checks that $\mathrm{p}_{\alpha
,0}$ is differentiable at $\alpha =0$ away from any critical point.

\section*{Acknowledgments}

\noindent We are very grateful to Volker Bach and Jakob Yngvason for their
hospitality at the Erwin Schr\"{o}dinger International Institute for
Mathematical Physics, at the Physics University of Vienna, and at the
Institute of Mathematics of the Johannes Gutenberg--University that allowed
us to work on different aspects of the present paper. We also thank N. S.
Tonchev and V.A. Zagrebnov for giving us relevant references, as well as the
referee for having helped us to improve the paper. Additionally, J.-B.B.
especially thanks the mathematical physics group of the
Department of Physics  of the  University
of Vienna for the very nice working environment.

\end{document}